\documentclass[11pt]{article}
\usepackage{graphicx}
\usepackage{amssymb}
\usepackage{amsmath}
\usepackage{graphicx}
\usepackage{amstext}
\usepackage{esint}
\usepackage[utf8]{inputenc}
\usepackage{color}
\usepackage{float}
\usepackage{cite}
 \numberwithin{equation}{section}

\renewcommand{\vec}[1]{{\bf #1}}       %%%  vectors in bold
\def\beq{\begin{eqnarray}}    %%%  begequation/eqnarray
\def\eeq{\end{eqnarray}}      %%%  endequation/eqnarray
\newcommand{\rL}{\rho_\Lambda}

\newcommand{\CC}{\Lambda}

\newcommand{\Omo}{\Omega_{m}}

\newcommand{\OLo}{\Omega_{\Lambda}}

\newcommand{\rco}{\rho^0_{c}}

\newcommand{\rLo}{\rho_{\CC}}

\newcommand{\rDE}{\rho_{\rm DE}}

\newcommand{\nueff}{\nu_{\rm eff}}

\newcommand{\oD}{\omega_{\rm BD}}
\newcommand{\newtext}[1]{{\textcolor{black}{#1}}}
\newcommand{\joantext}[1]{{\textcolor{black}{#1}}}
\newcommand{\dpsi}{\dot{\psi}}
\newcommand{\ddpsi}{\ddot{\psi}}
\newcommand{\wBD}{\omega_{\rm BD}}

\newcommand{\fracdpsipsi}{\frac{\dpsi}{\psi}}

\newcommand{\xr}{x_r}
\newcommand{\xm}{x_m}
\newcommand{\xl}{x_\Lambda}
\newcommand{\xp}{x_\psi}
\newcommand{\w}{\omega_{\rm BD}}

\newcommand{\p}{\prime}
\newcommand{\pp}{{\prime\prime}}

\newcommand{\Om}{\tilde{\Omega}_m}

\newcommand{\eBD}{\epsilon_{BD}}
\newcommand{\mPl}{m_{\rm Pl}}

\newcommand{\Geff}{G_{\rm eff}}
\newcommand{\dvphi}{\Delta\varphi}
\newcommand{\weff}{w_{\rm eff}}
\newcommand{\rvphi}{\rho_{\varphi}}
\newcommand{\pvphi}{p_{\varphi}}

\begin{document}
%\pubblock

%\today

%\vspace{1cm}

 \hyphenation{nu-cleo-syn-the-sis u-sing si-mu-la-te ma-king
cos-mo-lo-gy know-led-ge e-vi-den-ce stu-dies be-ha-vi-or
res-pec-ti-ve-ly appro-xi-ma-te-ly gra-vi-ty sca-ling
ge-ne-ra-li-zed re-mai-ning va-cu-um}

%%%%%%%%%%%%%%%%%%%%%%%%%%%%%%%%%%%%%%%%%%%%%%%%%%%%%%%%%

%\newpage

%%%%%%%%%%%%%%%%%%%%%%%%%%%%%%%%%%%%%%%%%%%%%%%%%%%%%%%%%
%\flushright{UB-ECM-PF-06/24 }\\

\begin{center}
{\bf \LARGE  Brans-Dicke cosmology with a $\CC$- term: a possible solution to $\CC$CDM tensions\footnote{Dedicated to Roberto D. Peccei}} \vskip 2mm

 \vskip 8mm

\textbf{\large Joan Sol\`a Peracaula$^\dagger$, Adri\`a G\'omez-Valent$^\ddagger$, Javier de Cruz P\'erez $^\dagger$\\ and Cristian Moreno-Pulido$^\dagger$}

\vskip 0.5cm
$^\dagger$ Departament de F\'isica Qu\`antica i Astrof\'isica, and Institute of Cosmos Sciences,\\ Universitat de Barcelona,
Av. Diagonal 647, E-08028 Barcelona, Catalonia, Spain\\
$^\ddagger$ Institut f\"{u}r Theoretische Physik, Ruprecht-Karls-Universit\"{a}t Heidelberg,
Philosophenweg 16, D-69120 Heidelberg, Germany

\vskip0.3cm

{ E-mails:   sola@fqa.ub.edu, gomez-valent@thphys.uni-heidelberg.de, decruz@fqa.ub.edu, cristian.moreno@fqa.ub.edu}

 \vskip2mm

\end{center}
\vskip 7mm

\begin{quotation}
\noindent {\large\it \underline{Abstract}}.
We present a full-fledged analysis of Brans-Dicke cosmology with a cosmological
constant and cold dark matter (BD-$\CC$CDM  for short). We extend the scenarios where the current cosmological value of the BD-field is restricted by the local astrophysical domain to scenarios
where that value is fixed only by the cosmological observations, {which should be more natural in view of the possible existence of local screening mechanims}. Our analysis includes
both the background and perturbations equations in different gauges. We find that the
BD-$\CC$CDM is favored by the overall cosmological data as compared to
the concordance GR-$\CC$CDM model, namely data on distant
supernovae, cosmic chronometers, local measurements of the Hubble parameter,  baryonic acoustic oscillations, Large-Scale Structure
formation and the cosmic microwave background under  full Planck 2018 CMB likelihood. {We also test the impact of Strong and Weak-Lensing data on our results, which can be significant}. We
find that the BD-$\CC$CDM can mimic effective quintessence with a significance  {of about  $3.0-3.5\sigma$  c.l.  (depending on the lensing datasets).}
The fact that  the BD-$\CC$CDM behaves effectively as a Running Vacuum Model (RVM) when viewed from the GR perspective
helps to alleviate some of the existing tensions with the data, such as the $\sigma_8$ excess predicted  by
GR-$\CC$CDM.  {On the other hand,  the BD-$\CC$CDM model has a crucial bearing on the acute $H_0$-tension with the local measurements, which is rendered virtually harmless owing to the small increase of the effective value of the gravitational constant  with the expansion.  The simultaneous alleviation of the two tensions is a most remarkable feature of  BD-gravity with a cosmological constant in the light of the current observations, and hence goes in support of  BD-$\CC$CDM against GR-$\CC$CDM.}

\end{quotation}
\vskip 3mm

\begin{quotation}
\end{quotation}

\newpage

%\tableofcontents

\newpage

\tableofcontents

%%%%%%%%%%%%%%%%%%%%%%%%%%%%%%%%%%%%%%%%%%%%%%%%%%%%%%%%%%%%%%%%%
%%%%%%%%%%%%%%%%%%%%%%%%%%%%%%%%%%%%%%%%%%%%%%%%%%%%%%%%%%%%%%%%%
%%%%%%%%%%%%%%%%%%%%%%%%%%%%%%%%%%%%%%%%%%%%%%%%%%%%%%%%%%%%%%%%%

\section{Introduction}\label{intro}

For slightly more than 20 years we know with a substantial degree of certainty that the universe is in accelerated expansion\,\cite{Riess:1998cb, Perlmutter:1998np}. That knowledge is, however, a pure kinematical result based on assuming General Relativity (GR), among other things. It does not mean that we really understand the primary (dynamical) cause for such an acceleration or that we are in position to propose a cosmological model explaining  the speeding up of the cosmos at the level of fundamental physics, say quantum field theory (QFT) in curved spacetime, quantum gravity or string theory. The overarching canonical picture of our universe, formulated in the context of the GR paradigm, is to assume that the cosmic acceleration is caused by a rigid cosmological constant (CC) term, $\CC$, in Einstein's equations, whose value has been pinned down  from a large set of cosmological observations, which by themselves also point to the existence of large amounts of dark matter (DM) --  apart from the conventional baryons \cite{Aghanim:2018eyx}. We call such an overall picture of the universe the  ``concordance (or standard) cosmological model'' \cite{peebles:1993}, or simply GR-$\CC$CDM, where we append GR explicitly in the name because we wish to study the $\CC$CDM model also from a different perspective to that of the GR paradigm.

Even though the  $\CC$-term is a key ingredient of our current cosmological picture, unfortunately we do not know what is its origin at a deep theoretical level. This fundamental failure may well lie  at the root of the so-called `old' Cosmological Constant Problem (CCP)\,\cite{Weinberg:1988cp}, which, in essence,  can be formulated by recognizing our complete  inability to perform at present not even a  ballpark estimate of the value of  $\CC$, as  all our theoretical predictions fail by many orders of magnitude. What we do know for certain, however, and with remarkable observational precision, is the measured value of the vacuum energy density associated to $\CC$, which is positive (hence generating a repulsive force) and of order $\rLo=\CC/(8\pi G_N)\sim 10^{-47}$ GeV$^4$ ($G_N$ being Newton's constant). Expressed in units of the critical density, we have $\OLo \equiv\rLo/\rco\simeq 0.70$ (the superindex '$0$' denotes current value). We attribute the accelerated expansion of the universe to the existence of such a repulsive vacuum-like force, which works against the attractive gravitational interaction and tends to push the clusters of galaxies apart at a speed continuously increasing with the cosmic expansion. Simple as it can be, such a straightforward explanation for the cosmic acceleration is nevertheless  troublesome since it clashes violently with the well known breaking of the symmetries at the classical and quantum level, what aggravates further the old CCP --  see e.g. \,\cite{Sola:2013gha} for a review.

Thus, in spite of the many virtues of the $\CC$CDM,  the longstanding (and unaccounted) constancy and small value of $\rLo$ (as compared to standard  QFT predictions) leads to an unsatisfactory theoretical picture which has motivated a variety of alternative explanations for the cosmic acceleration beyond the $\CC$-term.  Such theories include quintessence and phantom fields and they go under the generic name of dark energy (DE), see e.g.\,\cite{Sahni:1999gb,Peebles:2002gy,Padmanabhan:2002ji,Copeland:2006wr,Amendola:2015ksp} and references therein. There were also proposals in the past (previous to quintessence ideas and the like)  where one attempted to use scalar fields with the purpose of explaining not only a possible time evolution, but even the value itself of $\CC$  (thought to be zero at that time) on purely dynamical grounds, see e.g. \cite{gibbons1985very,Abbott:1984qf,Peccei:1987mm, Barr:1986ya ,Ford:1987de}.  However, the idea that the DE could be not just the CC of Einstein's equations but a dynamical variable, or some appropriate function of the cosmic time, has been explored since long ago and sometimes on purely phenomenological grounds\,\cite{Overduin:1998zv}, see particularly\,\cite{Ozer:1985ws, Ozer:1985wr, Bertolami:1986bg, Freese:1986dd, Carvalho:1991ut}. The main aim in most papers trading the  rigid $\CC$-term for a dynamical quantity was not so much to cope with the old CCP, which personifies the hard core of the problem,  but to address a different aspect of the CCP called the coincidence problem\,\cite{1997cpp..conf..123S ,Steinhardt:2003st}, namely the perplexity which may produce to our mind  the fact that the value of the (rapidly evolving) matter density  ($\Omo(a)\sim a^{-3}$) right now turns out to be just of the same order of magnitude as the current vacuum energy density (allegedly constant throughout the entire cosmic history): $\Omega_m\simeq\OLo={\cal O}(1)$ in our days.

Notwithstanding the severity of the CCP associated with the $\CC$-term in Einstein's equations, the problem actually affects all forms of dark energy. The large class of DE models pervading  the cosmology market\,\cite{Amendola:2015ksp} do not actually offer a real solution to it either. Thus, we stick here to the $\CC$-term and keep on  in  standby  -- hopefully provisional -- the eventual solution of the hard theoretical conundrums mentioned above.  On pure empirical  grounds, the concordance  $\CC$CDM model can be considered a phenomenologically acceptable model which is built upon the measurement  of the $\CC$-term (with the associated vacuum energy density $\rLo$), as well as upon the existence of the as yet undetected DM and, of course, of the baryonic matter (the known structural component of the galaxies and of ourselves). It assumes that the spacetime metric satisfies the Cosmological Principle and hence adopts  the  Friedmann-Lema\^{\i}tre-Robertson-Walker (FLRW) form.  We will also assume a spatially flat three-dimensional background.

The standard  $\CC$CDM has remained robust and unbeaten for a long time. It is roughly consistent with a large body of observations, including the high precision data from the cosmic microwave background (CMB) anisotropies\,\cite{Aghanim:2018eyx}.  Nonetheless, the observational situation in recent years does not seem to paint a fully rosy picture of the phenomenological status of the $\CC$CDM anymore. This is particularly true in what concerns two main observables\,\cite{Verde:2019ivm,Macaulay:2013swa,Nesseris:2017vor,DiValentino:2020zio,DiValentino:2020vvd}.  On the first place, there is a most worrisome tension between the disparate current values of the Hubble parameter $H_0$ obtained independently from measurements of the local and the early universe (CMB).  {Here the discordance is between the predicted value from the CMB ($H_0=67.4\pm 0.5$ km/s/Mpc\,\cite{Aghanim:2018eyx}) and the local (distance ladder) measurements\cite{Riess:2016jrr,Riess:2018uxu,Riess:2019cxk,Reid:2019tiq}. The mismatch appears  in fact as a rather  acute  and preoccupying one,  especially after its persistence in time and its tendency to worsen from the original\,\cite{Riess:2018uxu, Riess:2016jrr} to later  measurements \cite{Riess:2019cxk}, which brought it to $4.4\sigma$. Despite it recently decreased a bit to $4.2\sigma$\,\cite{Reid:2019tiq}) it is still very significant.  Actually, when the most recent SH0ES determination of $ H_0= (73.5\pm 1.4)$ km/s/Mpc\,\cite{Reid:2019tiq}  is combined with  a very different kind of independent observations, such as the Strong-Lensing results from the H$0$LICOW collab.\cite{Wong:2019kwg},  which lead to  $H_0=(73.3^{+1.7}_{-1.8})$ km/s/Mpc, the combination produces $ H_0= (73.42\pm 1.09)$ km/s/Mpc  and this result amounts to an astounding  $\sim 5\sigma$ tension \textit{w.r.t.} the aforementioned Planck 2018 result}.

In another vein, we have some trouble in the structure formation data. The concordance model predicts a value of $\sigma_8$ (the current matter density rms fluctuations within spheres of radius $8h^{-1}$~Mpc, with $h\simeq 0.7$ the reduced Hubble constant)  in excess by  $2-3\sigma$ over the direct data values at low redshifts\,\cite{Macaulay:2013swa,Nesseris:2017vor,Kazantzidis:2018rnb,DiValentino:2018gcu,SolaPeracaula:2018xsi,Skara:2019usd,DiValentino:2020vvd}; more specifically, the tension is in between measurements of the amplitude of the power spectrum of density perturbations inferred using CMB data against those directly measured by Large-Scale Structure (LSS) formation on smaller scales, from redshift space distortions (RSD) {(see e.g. \cite{Gil-Marin:2016wya}) and Weak-Lensing (WL) data \cite{Hildebrandt:2016iqg,Joudaki:2017zdt,Kohlinger:2018sxx,Wright:2020ppw}.} Whether all these tensions are the result of yet unknown systematic errors or really hint to some underlying new physics is still unclear, e.g. the $\sigma_8$-tension might be related to issues on the Planck internal consistency of the gravitational lensing amplitude in CMB data\,\cite{DiValentino:2018gcu}. There remains, however,  strongly the possibility that these discrepancies may just be the signal of a deviation from the $\CC$CDM model. If so, any new physics introduced to explain the severe $H_0$-tension  should not aggravate the $\sigma_8$ one in a noticeable way.  {This ``golden rule'' to ameliorate the acute $H_0$-tension will be our guiding principle, and as we shall see it is not preserved by many existing attempts in the literature.}

 It is actually on these problems of very practical nature where we would like to focus our attention for the rest of the paper.
Of  paramount importance in this respect is the  observation that the basic gravitational paradigm nourishing the  $\CC$CDM model and most of its DE extensions in the literature is still Einstein's GR. This may represent a fundamental limitation to the possible solution of the mentioned tensions, even more so if we take into account that these problems can be affected by assumptions on the basic parameters of gravitation.
As it is well-known, in the sixties a very significant revolution occurred in the gravity context, which implied a fundamental change of the conceptual construct on gravitation. In it,  $G$ was boldly assumed to be a dynamical variable rather than a constant of Nature. This proposal clearly departs from the strict GR context. It actually traces back to early ideas in the thirties on the possibility of a time-evolving gravitational constant $G$ by Milne\,\cite{mccrea_1935} and the suggestion by Dirac of the large number hypothesis \cite{dirac1937cosmological , dirac1938cosmological}, which led him also to propose the time evolution of $G$. Along similar lines, Jordan and Fierz speculated that the fine structure constant $\alpha_{\rm em}$ together with $G$ could be both space and time dependent\,\cite{jordan1937naturwiss, jordan1939ZPhys, jordan1955schwerkraft,Fierz:1956zz}. Finally, $G$ was formally associated to the existence of a dynamical scalar field $\psi\sim 1/G$ coupled to the curvature.  Such was the famous gravity formulation originally proposed by Brans and Dicke (``BD'' for short) \cite{BransDicke1961, brans1962mach, dicke1962physical}, which was the first historical attempt to extend GR in order to accommodate variations in the Newtonian coupling $G$. Subsequently these ideas were generalized in the form of scalar-tensor theories\,\cite{Bergmann:1968ve, Nordtvedt:1970uv , Wagoner:1970vr}, and thereafter further extended in multifarious ways  still compatible with the weak form of the Equivalence Principle\,\cite{Horndeski:1974wa,Fujii:2003pa}, see e.g. \cite{Sotiriou:2008rp,Capozziello:2011et,Clifton:2011jh,Will:2014kxa} for a review.

As already mentioned, in this work we wish to stick firmly to the $\CC$-term as the simplest provisional explanation for  the cosmic acceleration. But we want to do it on the grounds of  what we may call the BD-$\CC$CDM model, i.e. a cosmological framework still encompassing all the phenomenological ingredients of the concordance $\CC$CDM  model,  in particular a strict cosmological constant term $\CC$ and dark matter along with baryons, but now all of them ruled over by a different gravitational paradigm: BD-gravity\,\cite{BransDicke1961, brans1962mach, dicke1962physical}, instead of GR.  Our intention is to combine the dynamical character of $G$ in the context of BD-gravity with a cosmological term $\CC$, aiming at finding a form of (effective) dynamical dark energy (DDE) capable to overcome the mentioned tensions.  Not all forms of DDE can make it.  Recent analyses comparing different models of dark energy using  similar data can be found e.g.  in\,\cite{Sola:2018sjf,Gomez-Valent:2020mqn}, and references therein. However, as we will discuss below and throughout the paper the simultaneous solution/alleviation of the two tensions requires a very particular form of DDE which is  ``aligned'' with the kind of observables to be adjusted.

Indeed, the key observation for suspecting that such program is really feasible in the specific context of the BD-$\CC$CDM, and specially in what concerns the critical trouble with the local Hubble parameter $H_0$,  lies on the fact that the BD-$\CC$CDM can naturally implement the aforesaid golden rule for safely quenching the two tensions.  {In fact, for a given critical density, the effective increase of  the $G$  value within the BD model over the local gravity value, $G_N$, in combination with a negative value of the BD-parameter $\oD\equiv 1/\eBD$  permits to reduce the $\sigma_8$-tension.  Both benefits can be gained at no expense to the accurate values of the  CMB observables.}  Such unusual ability seems to be a unique gift of the BD-$\CC$CDM model and can be accomplished with  minimum effort, i.e. with a minimal number of extra parameters. Details will be generously provided throughout the analysis presented in this paper.   We are not aware, to the best of our knowledge,  of any other fundamental model in the literature capable of such achievement.

That the above picture can be possible at all, stems in part  from the following observation.  If one tries to encapsulate the slow evolution of the BD-field in terms of the current GR framework (in which $G$  and $\CC$ remain both constant), the effective theory that emerges is a variant of the $\CC$CDM framework in which $\rho_\Lambda$ acquires a mild time-evolving component of a very specific nature. The resulting $\CC$CDM-like model appears  as if $\rL=\CC/(8\pi G)$ were a dynamical vacuum energy density composed of a constant term plus a small dynamical term  $\sim\nu H^2$ ($|\nu|\ll1$). This form is well-known in the literature  and goes under the name of Running Vacuum Model (RVM) see \cite{Sola:2013gha,sola2015cosmology,Gomez-Valent:2017tkh} and references therein.  It can describe in an effective way the current as well as the early universe through additional (even) powers of the Hubble rate\,\cite{lima2013expansion , perico2013complete}; see e.g. \cite{peracaula2020particle}  for a thorough thermodynamical study of the extended RVM model.  Recently the RVM structure for the vacuum energy density has been shown to appear from generic features of the bosonic string effective action\cite{Basilakos:2019acj,Basilakos:2020qmu} and also from explicit calculation of the adiabatically renormalized energy-momentum tensor  in QFT in curved spacetime\,\cite{Moreno-Pulido:2020anb}.
The fact that the BD-$\CC$CDM model  mimics the RVM is very useful  since  the latter turns out to be  a rather successful  framework for improving the overall fit to the cosmological data as compared to the conventional $\CC$CDM, see e.g. the most recent detailed  analyses\,\cite{Sola:2016ecz , Sola:2017jbl , Gomez-Valent:2018nib ,Gomez-Valent:2017idt,Sola:2017znb , sola2017first , Sola:2016hnq , Sola:2015wwa , Geng:2017apd,Rezaei:2019xwo,Geng:2020mga} and some older ones \cite{Gomez-Valent:2014rxa , Gomez-Valent:2014fda,Grande:2011xf , Basilakos:2009wi} and references therein.  In particular, in \cite{Sola:2017znb,Sola:2017jbl,Gomez-Valent:2018nib,Sola:2016ecz,Gomez-Valent:2017idt} it was shown that the RVM can help to alleviate the $\sigma_8$-tension, but the $H_0$ one remained unaccounted  {since in that form of DDE the involved $G$ remains fixed at the value $G_N$ (i.e. the RVM was still GR-like). When $G$ becomes slightly dynamical and  we allow for  the cosmological values of $G$ to be larger than the one measured at the Earth,  one can adjust better the $H_0$ value. We may  even combine this feature with the DDE behavior of  BD-$\CC$CDM (when viewed from the point of view of a departure from GR) for solving the $\sigma_8$-tension as well.}  A first hint that Brans-Dicke gravity could lead to such a promising (BD-like) version of the RVM,  was put forward in \cite{Peracaula:2018dkg}. Further elaboration on this idea and a first comparison with data was subsequently given in  \cite{Perez:2018qgw}, and finally a more sophisticated study using a full Boltzmann code for the CMB part was presented  in\,\cite{Sola:2019jek}.

Many alternative possibilities have been addressed in the literature to improve the overall fit to the cosmological data and possibly to relax the main tensions through DDE proposals and of other nature. Any attempt at condensing them here would be surely incomplete and unavoidably biased, but a few of them can be listed, see e.g. \,\cite{Zhao:2017cud , DiValentino:2019dzu , DiValentino:2017rcr , DiValentino:2017iww , DiValentino:2017zyq , DiValentino:2016hlg , Martinelli:2019dau , Salvatelli:2014zta,Costa:2016tpb, Anand:2017wsj , An:2017crg , Li:2015vla , Li:2014cee , Li:2014eha , Hazra:2018opk , Yan:2019gbw,Liao:2020zko,WangChen2020,Jedamzik:2020krr,Vagnozzi:2019ezj,Calderon:2020hoc}. In general these attempts are purely phenomenological,  e.g. those using phantom equation of state for the DE since they have no consistent theoretical support and usually they further spoil the $\sigma_8$-tension \,\cite{DiValentino:2019jae,Alestas:2020mvb}. This drawback was repeatedly emphasized in different existing analyses in the literature\,\cite{Sola:2017jbl , Gomez-Valent:2018nib , Sola:2016ecz , Sola:2017znb , sola2017first , Sola:2016hnq,Sola:2018sjf}. There are also attempts to introduce early dark energy (EDE), but different works seem to reach different conclusions on the effectiveness of these models to improve the tensions, see e.g. \cite{Poulin:2018cxd,Hill:2020osr,Chudaykin:2020acu,Braglia:2020bym}.

We would also like to remark recent studies on testing  BD theories and on attempts at mitigating the tensions using particular potentials in the BD framework, see e.g. \cite{Umilta:2015cta,Ballardini:2016cvy,Rossi:2019lgt}. At the same time, there are different works trying to loosen the  GR-$\CC$CDM tensions by considering the possibility of a variable Newton's constant $G$. Usually these models are essentially GR-like, in the sense that {the basic term of the gravitational action still has the form of the Hilbert-Einstein term, plus a non-minimal coupling of curvature with a scalar field.}  This type of models and variations thereabout are  valuable and have been recently used to try to mitigate the $H_0$ or the $\sigma_8$ tensions\,\cite{Ballesteros:2020sik,Braglia:2020iik,Ballardini:2020iws,Bertini:2019xws,Rodrigues:2015hba}, but more difficult is to try to fulfill the mentioned golden rule -- which requires not to aggravate one of the two tensions when improving the other. In point of fact, in some cases the $\sigma_8$ one actually gets significantly worse. We further comment on these models in our section of discussion.

Building up upon our first complete analysis performed in the context of Planck 2015 data\,\cite{Sola:2019jek}, here we present a comprehensive study  of BD-$\CC$CDM cosmology vis-\`a-vis  observations  using a significant amount of  new and updated data, in particular  we make use of the full Planck 2018 likelihood as well as of additional datasets  (e.g. on Strong-Lensing and updated structure formation data) which prove quite revealing.  Furthermore, we discuss here at length many analytical and numerical details of the calculation and the possible implications that the BD-$\CC$CDM model can have on the current cosmological observations, most particularly on the  $\sigma_8$ and $H_0$ tensions. {We show that the $H_0$-tension  can be significantly reduced in this context; and, interestingly enough, this can be achieved without detriment of the  $\sigma_8$-tension, hence fulfilling the golden rule mentioned above.} We also find that an effective signature for dynamical DE may ultimately emerge from the BD-$\CC$CDM dynamics, and such signature is able to mimic quintessence at a confidence level in between $3-4\sigma$ level, depending on the datasets used. If a successful phenomenological description of the data and the loosening of the tensions could be reconfirmed with the advent of new and more precise data and new analyses in the future, it would be tantalizing to suggest the possibility that the underlying fundamental theory of gravity might actually be BD rather than GR. But there is still a long way to follow, of course.

The layout of this paper reads as follows. In Section \ref{sec:BDgravity} we introduce the basics of the Brans-Dicke (BD) model. \joantext{In particular, we discuss the introduction of the cosmological term in this theory and the notion of effective gravitational coupling.} Section \ref{sec:preview} is a preview of the whole paper, and in this sense it is a very important section where the reader can get a road map of the basic results of the paper and above all an explanation of why BD-cosmology with a CC can be a natural and efficient solution to the $H_0$ and $\sigma_8$ tensions.  Section \ref{sec:EffectiveEoS} shows how to parametrize  BD-cosmology as a departure from GR-cosmology. We show that BD with a CC appears as GR with an effective equation of state (EoS) for the dark energy which behaves quintessence-like at a high confidence level.  The remaining of the paper presents the technical details and the numerical analysis, as well as complementary discussions. Thus, Section \ref{sec:StructureFormation} discusses the perturbations equations for BD-gravity in the linear regime (leaving for Appendixes \ref{AppendixC} and \ref{AppendixD} an extended discussion with more technical details  in different gauges); Section \ref{sec:Mach} defines four scenarios for the BD-cosmology in the light of  Mach's Principle; Section \ref{sec:MethodData} carefully describes the data used from the various cosmological sources; Section \ref{sec:NumericalAnalysis} presents the numerical analysis and results. \newtext{In Section \ref{sec:Discussion} we perform a detailed discussion of the obtained results and we include a variety of extended considerations, in particular we assess the impact of massive neutrinos in the BD-$\CC$CDM framework. Finally, in Section \,\ref{sec:Conclusions} we deliver our conclusions}. Four appendices at the end provide additional complementary material. In Appendix \ref{AppendixA} we compute semi-analytical solutions to the BD equations in different epochs, which are helpful to further understand the numerical results. We also recall the reader at this point why BD-cosmology mimics the Running Vacuum Model (RVM); in Appendix \ref{AppendixB} we compute the fixed points of the BD-cosmology with a cosmological constant.  The aforementioned Appendices  \ref{AppendixC} and  \ref{AppendixD} provide the perturbations equations in the synchronous and conformal Newton gauges, respectively, and illustrate the correspondence between the two.

%\newpage

\section{BD-$\CC$CDM: Brans-Dicke gravity with a cosmological constant}\label{sec:BDgravity}
Since the appearance of GR, more than one hundred years ago, many alternative theories of gravity have arisen, see e.g. \cite{Sotiriou:2008rp,Capozziello:2011et,Clifton:2011jh,Will:2014kxa} and references therein. The most important one, however, was proposed by Brans and Dicke almost sixty years ago\,\cite{BransDicke1961}. This theoretical framework  contains an additional gravitational  \textit{d.o.f.}  as compared to GR, and as a consequence it is different from GR in a fundamental way, see the previously cited reviews. The new \textit{d.o.f.}  is a scalar field, $\psi$, coupled to the Ricci scalar of curvature, $R$.  BD-gravity is indeed the first historical attempt to extended GR to accommodate variations in the Newtonian coupling $G$. A generalization of it has led to a wide panoply of scalar-tensor theories since long ago\,\cite{Bergmann:1968ve, Nordtvedt:1970uv, Wagoner:1970vr, Horndeski:1974wa}.  The theory is also characterized by  a   (dimensionless) constant parameter, $\oD$, in front of the kinetic term of $\psi$.

\subsection{Action and field equations}

In our study we will consider the original BD-action extended with a cosmological constant density term, $\rL$, as it is essential to mimic the conventional $\CC$CDM model based on GR and reproduce its main successes. In this way we obtain what we have called the `BD-$\CC$CDM model' in the introduction, i.e. the version of the  $\CC$CDM within the BD paradigm.
The BD action reads,  in the Jordan frame, as follows\,\footnote{We use natural units, therefore $\hbar=c=1$ and $G_N=1/\mPl^2$, where $\mPl\simeq 1.22\times 10^{19}$ GeV is the Planck mass. As for the geometrical quantities, we use space dominant signature of the metric $(-, +,+,+ )$, Riemann curvature tensor
$R^\lambda_{\,\,\,\,\mu \nu \sigma} = \partial_\nu \, \Gamma^\lambda_{\,\,\mu\sigma} + \Gamma^\rho_{\,\, \mu\sigma} \, \Gamma^\lambda_{\,\, \rho\nu} - (\nu \leftrightarrow \sigma)$, Ricci tensor $R_{\mu\nu} = R^\lambda_{\,\,\,\,\mu \lambda \nu}$, and Ricci scalar $R = g^{\mu\nu} R_{\mu\nu}$.  Overall, these correspond to the $(+, +, +)$ conventions in the popular classification by Misner, Thorn and Wheeler\,\cite{MTW:1974}.}:
\begin{eqnarray}
S_{\rm BD}=\int d^{4}x\sqrt{-g}\left[\frac{1}{16\pi}\left(R\psi-\frac{\oD}{\psi}g^{\mu\nu}\partial_{\nu}\psi\partial_{\mu}\psi\right)-\rL\right]+\int d^{4}x\sqrt{-g}\,{\cal L}_m(\chi_i,g_{\mu\nu})\,. \label{eq:BDaction}
\end{eqnarray}
The (dimensionless) factor  in front of the kinetic term of $\psi$, i.e. $\oD$, will be referred to as the BD-parameter. While it is true that this parameter is not restricted to be a constant, throughout this paper we will consider just the canonical option  $\oD=$const.
The last term of (\ref{eq:BDaction}) stands for the matter action $S_{m}$, which is constructed from the Lagrangian density of the matter fields, collectively denoted as $\chi_i$. There is no potential for the BD-field $\psi$ in the original BD-theory, but we admit the presence of a CC term associated to $\rL$. \newtext{By not introducing any specific potential we keep the number of additional parameters to the minimum.}

The Brans-Dicke field, $\psi$, has dimension $2$ in natural units (i.e. mass dimension squared), in contrast to the dimension $1$ of ordinary scalar fields. This is because we wish the effective value of $G$ at any time to be given directly by $1/\psi$. It goes without saying that  $\psi$ must be a field variable evolving very slowly with time.

The field equations of motion ensue after performing variation of the action \eqref{eq:BDaction}  with respect to both the metric and the scalar field $\psi$.  While the  first variation yields
\begin{equation}\label{eq:BDFieldEquation1}
\psi\,G_{\mu\nu}+\left(\Box\psi +\frac{\oD}{2\psi}\left(\nabla\psi\right)^2\right)\,g_{\mu\nu}-\nabla_{\mu}\nabla_{\nu}\psi-\frac{\oD}{\psi}\nabla_{\mu}\psi\nabla_{\nu}\psi=8\pi\left(\,T_{\mu\nu}-g_{\mu\nu}\rL\right)\,,
\end{equation}
the second variation gives the wave equation for $\psi$, which  depends on the curvature scalar $R$:
\begin{equation}\label{eq:BDFieldEquation2a}
\Box\psi-\frac{1}{2\psi}\left(\nabla\psi\right)^2+\frac{\psi}{2\oD}\,R=0\,.
\end{equation}
To simplify the notation, we have written  $(\nabla\psi)^2\equiv g^{\mu\nu}\nabla_{\mu}\psi\nabla_{\nu}\psi$.
In the first field equation, $G_{\mu\nu}=R_{\mu\nu}-(1/2)Rg_{\mu\nu}$ is the Einstein tensor, and  on its {\it r.h.s.} $T_{\mu \nu}=-(2/\sqrt{-g})\delta S_{m}/\delta g^{\mu\nu}$ is the energy-momentum tensor of matter.
We can take the trace of Eq.\,(\ref{eq:BDFieldEquation1}) to eliminate $R$ from (\ref{eq:BDFieldEquation2a}), what leads to a most compact result for the wave equation of $\psi$:
\begin{equation}\label{eq:BDFieldEquation2}
\Box\psi=\frac{8\pi}{2\oD+3}\,\left(T-4\rL\right)\,,
\end{equation}
where  $T\equiv T^{\mu}_{\mu}$ is the trace of the energy-momentum tensor for the matter part (relativistic and nonrelativistic). The total energy-momentum tensor as written on the \textit{r.h.s.} of (\ref{eq:BDFieldEquation1})  is the sum of the matter and vacuum parts and adopts the perfect fluid form:
\begin{equation} \label{eq:EMT}
\tilde{T}_{\mu\nu} =T_{\mu\nu}  -\rho_\CC g_{\mu\nu}=p\,g_{\mu\nu}+(\rho + p)u_\mu{u_\nu}\,,
\end{equation}
{with $\rho \equiv \rho_m + \rho_\gamma+\rho_\nu + \rho_\Lambda$ and $p \equiv p_m + p_\gamma + p_\nu + p_\Lambda$. The matter part $\rho_m \equiv \rho_b + \rho_{cdm}$, contains the pressureless contribution from baryons and cold dark matter. Photons are of course relativistic, so $p_\gamma=\rho_\gamma/3$. The functions $\rho_\nu$ and $p_\nu$ include the effect of massive and massless neutrinos, and therefore must be computed numerically.}

As in GR, we have included a constant vacuum energy density, $\rL$, in the BD-action (\ref{eq:BDaction}), with the usual equation of state $p_\Lambda=-\rL$. The quantum matter fields usually induce an additional, and very large, contribution to $\rL$. This is of course the origin of the Cosmological Constant Problem mentioned in the introduction\, \cite{Weinberg:1988cp,Sola:2013gha,Sahni:1999gb,Peebles:2002gy,Padmanabhan:2002ji,Copeland:2006wr,Amendola:2015ksp}\footnote{A recent proposal to alleviate the CCP within BD-gravity was made in \cite{Peracaula:2018dkg}.  Interestingly, the Higgs potential itself can be motivated in BD-gravity theories\,\cite{Sola:2016our,Peracaula:2018dkg}. }.

Let us write down the field equations in the flat FLRW metric,  $ds^2=-dt^2 + a^2\delta_{ij}dx^idx^j$. Using the total density $\rho$ and pressure $p$ as indicated above, Eq.\,(\ref{eq:BDFieldEquation1}) renders the two independent equations
\begin{equation}
3H^2 + 3H\fracdpsipsi -\frac{\wBD}{2}\left(\fracdpsipsi\right)^2 = \frac{8\pi}{\psi}\rho\label{eq:Friedmannequation}
\end{equation}
and
\begin{equation}
2\dot{H} + 3H^2 + \frac{\ddpsi}{\psi} + 2H\frac{\dpsi}{\psi} + \frac{\wBD}{2}\left(\frac{\dpsi}{\psi}\right)^2 = -\frac{8\pi}{\psi}p\,,\label{eq:pressureequation}
\end{equation}
whereas (\ref{eq:BDFieldEquation2}) yields
\begin{equation}\label{eq:FieldeqPsi}
\ddpsi +3H\dpsi = \frac{8\pi}{2\wBD +3}(\rho - 3p)\,.
\end{equation}
Here dots stand for derivatives with respect to the cosmic time and $H=\dot{a}/a$ is the Hubble rate. For constant $\psi=1/G_N$, the first two equations reduce to the Friedmann and pressure equations of GR, and the third requires $\oD\to\infty$ for consistency (except in the pure radiation-dominated epoch, where $\rho - 3p=0$) .  The connection between GR and  the $\oD\to\infty$ limit is sometimes not as straightforward as one might  naively think\,\cite{Faraoni:1998yq,Faraoni:1999yp}. We will have due occasion in this work to appraise the significance of this important observation (cf. de BD scenarios described in Sec.\,\ref{sec:Mach}).

By combining the above equations we expect to find a local covariant conservation law, similar to GR.  This is because there is no interaction between matter and the BD-field.  Although the details are more involved than in GR,  the result can  be obtained  upon  straightforward calculation of the covariant derivative on both sides of Eq.\,(\ref{eq:BDFieldEquation1}) and using the Bianchi identity satisfied by $G_{\mu\nu}$ and the field equation of motion for $\psi$. The final result turns out to be the same:
\begin{equation}\label{eq:FullConservationLaw}
\dot{\rho} + 3H(\rho + p)=\sum_{N} \left[\dot{\rho}_N + 3H(\rho_N + p_N)\right] = 0\,,
\end{equation}
where the sum is over all components, i.e. baryons, dark matter, neutrinos, photons and vacuum.

Here we take the point of view that all of the matter components are separately conserved in the main periods of the cosmic evolution.  In particular, the vacuum component $\rL$ obviously does not contribute in the sum since it is assumed to be constant and $\rho_\CC + p_\CC=0$ .

 Hereafter,  for convenience,  we will  use a dimensionless BD-field, $\varphi$, and the inverse of the BD-parameter, according to the following definitions:
\begin{equation}\label{eq:definitions}
\varphi(t) \equiv G_N\psi(t)\,,\qquad  \qquad\epsilon_{BD} \equiv \frac{1}{\omega_{BD}}\,.
\end{equation}
In this expression, $G_N = 1/\mPl^2$, with $\mPl$ the Planck mass as defined previously; $G_N$ gives the local value of the gravitational coupling, e.g. obtained from Cavendish-like  (torsion balance) experiments. Note that a nonvanishing value of $\eBD$ entails a deviation from GR. Being $\varphi(t)$  a dimensionless quantity,  we can recover GR by enforcing the simultaneous limits   $\epsilon_{BD} \rightarrow 0$ \textit{and} $\varphi\to 1 $.  We emphasize that it is not enough to set $\epsilon_{BD} \rightarrow 0$.  In this partial limit, we can only insure that $\varphi$ (and $\psi$, of course) does not evolve, but it does not fix its constant value. As we will see later on,  this feature can be important in our analysis. Using \eqref{eq:definitions}, we can see that \eqref{eq:BDaction} can be rewritten
\begin{eqnarray}
S_{\rm BD}=\int d^{4}x\sqrt{-g}\left[\frac{1}{16\pi G_N}\left(R\varphi-\frac{\oD}{\varphi}g^{\mu\nu}\partial_{\nu}\varphi\partial_{\mu}\varphi - 2\Lambda\right)\right]+\int d^{4}x\sqrt{-g}\,{\cal L}_m(\chi_i,g_{\mu\nu})\,, \label{eq:BDactionvarphi}
\end{eqnarray}
where $\CC$ is the cosmological constant, which is related with the associated vacuum energy density as $\rL=\CC/(8\pi G_N)$.

 \subsection{Cosmological constant and vacuum energy in BD theory}

 \newtext{There are several ways to introduce the cosmological constant in the BD framework, so  a few comments are in order at this point, see e.g. the comprehensive exposition\,\cite{Fujii:2003pa} and the works \cite{mathiazhagan1984inflationary, La1989, Weinberg1989, barrow1990extended} in the context of inflation. We can sum up the situation by mentioning  essentially three ways.
In one of them, the BD-action \eqref{eq:BDaction} is obtained upon promoting $G_N$ in the Einstein-Hilbert (EH) action with the $\rL$ term into a dynamical scalar field $1/\psi$,  adding the corresponding kinetic term and  keeping $\rho_\Lambda=const.$ The CC term $\Lambda$ is then related with the vacuum energy density through $\rho_\Lambda=\Lambda/(8\pi G_N )$. The EoS of the vacuum fluid is defined as $p_\Lambda=-\rho_\Lambda$. With these definitions, which we adopt throughout this work, the CC is not directly coupled to the BD-field. The latter, therefore, has a trivial (constant) potential, and the late-time acceleration source behaves as in the GR-$\Lambda$CDM model.}

\newtext{Alternatively, one could also adopt the EH action with CC and promote $G_N$ to a dynamical scalar, adding the corresponding kinetic term as before, but now keeping $\Lambda=const.$ instead of $\rho_\Lambda=const$. In this case, $\Lambda$ is linearly coupled to $\psi$. The potential energy density for the scalar field takes the form $\rho_\Lambda(\psi)= \psi \Lambda/(8\pi)$, so it is time-evolving. The coupling between the cosmological constant and the BD-field modifies the equations of motion. Thus, it may also alter the physics with respect to the option \eqref{eq:BDaction}, at least when the dynamics of the scalar field is not negligible \cite{Nariai:1969vh,endo1977cosmological, Uehara1982, lorenz1984exact, barrow1990extended, romero1992brans, Tretyakova2012}.}

\newtext{A third possibility, of course, is to consider more general potentials, but they do not have a direct interpretation of a  CC term as in the  original GR action with a cosmological term, see e.g. \cite{esposito2001scalar,Alsing:2011er, Ozer:2017oik,Faraoni:2004pi}.  Let us briefly explain why. If one starts from a general potential for the BD-field in the action, say some  arbitrary function of the BD-field $U(\psi)$ (not carrying along any additive constant)  in place of the constant term $\rL$ in \eqref{eq:BDaction}, then it is not so straightforward to  generate a CC term and still remain in a pure BD framework. For if one assumes that $\psi$ develops a vacuum expectation value\,\footnote{\newtext{We may assume, for the sake of the argument, that it is a classical ground state, since the BD-field is supposed to be part of the external gravitational field. Recall that we do not assume gravity  being a quantized theory here but just a set of background fields, in this case composed of the metric components $g_{\mu\nu}$ and $\psi$ (or equivalently $\varphi$)}.}, then the theory (when written in terms of the  fluctuating field around the ground state) would split into a non-minimal term coupled to curvature and a conventional EH term, so it would not be a pure BD theory.}

\newtext{This point of view is perfectly possible and has been considered by other authors, on which we shall comment in some detail in Sec.\,\ref{sec:Discussion}. However, proceeding in this way would lead us astray from our scientific leitmotif in this work (which is, of course, to remain fully within the BD paradigm).  Yet there is another option which preserves our BD philosophy, which is to perform a conformal transformation to the Einstein frame, where one can define a strict CC term.  Then we could impose that the effective potential in that frame, $V$,  is a constant. However, once this is done, the original potential in the (Jordan) BD frame, $U$, would no longer be constant since it would be proportional to $V$ times $\varphi^2$, with $\varphi$ defined as in \eqref{eq:definitions}.  Conversely, one may assume $U=\rL=const$. in the Jordan frame (as we actually did) and then the vacuum energy density will appear mildly time-evolving in the Einstein frame (for sufficiently large $\wBD$, of course).  This last option is actually  another way to understand why the vacuum energy density can be perceived as a slightly time-evolving quantity when the BD theory is viewed from the GR standpoint. It is also the reason behind the fact that the BD-$\CC$CDM framework mimics the so-called running vacuum model (RVM), see  Appendix \ref{sec:RVMconnection} and references therein  for details.}

\newtext{As indicated, in this work we opt for considering the definition provided in \eqref{eq:BDaction}, as we wish  to preserve the exact canonical form of the late-time acceleration source that is employed in the GR-$\Lambda$CDM model. At the same time we exploit the connection of the BD framework with the RVM and its well-known successful phenomenological properties, see e.g.\,\cite{Sola:2016ecz , Sola:2017jbl , Gomez-Valent:2018nib ,Gomez-Valent:2017idt,Sola:2017znb , sola2017first , Sola:2016hnq , Sola:2015wwa }.  In addition, in Sec.\,\ref{sec:EffectiveEoS} we  consider a direct parametrization of the departures of BD-$\Lambda$CDM  from  GR-$\Lambda$CDM. }

 %%%%%%%%%%%%%%%%%%%%%%%%%%%%%%%%%%%%%%%%%%%%%%%%%%%%%%%%%%%%%%%%%%%%%%%%%%%%%

 \subsection{Effective gravitational strength}
From \eqref{eq:BDactionvarphi} it follows that the quantity
\begin{equation}\label{eq:Gvarphi}
G(\varphi)=\frac{G_N}{\varphi}
\end{equation}
constitutes the effective gravitational coupling at the level of the BD-action. We will argue that $G(\varphi) $ is larger than $G_N$ because $\varphi<1$ (as it will follow from our analysis).  The gravitational field between two tests masses, however, is \textit{not} yet $G(\varphi) $ but the quantity  $G_{\rm eff}(\varphi)$  computed below.
 %effective $G$ felt by the structure formation data  is not the same as the effective $G(\varphi)$ at the action level,  being the former smaller than the latter.}

{Let us remark that if one would like to rewrite the BD action in terms of a canonically normalized scalar field $\phi$ (of dimension $1$) having  a non-minimal coupling to curvature of the form  $\frac12\xi\phi^2 R$, it would suffice to redefine the BD-field as  $\psi=8\pi \xi\,\phi^2$   with $\xi=1/(4\oD)=\eBD/4$, and then the scalar part of the action takes on the usual form}
\begin{eqnarray}
S_{\rm BD}=\int d^{4}x\sqrt{-g}\left(\frac12\,\xi\phi^2 R-\frac12 g^{\mu\nu}\partial_{\nu}\phi\partial_{\mu}\phi-\rL\right)+\int d^{4}x\sqrt{-g}\,{\cal L}_m(\chi_i,g_{\mu\nu})\,. \label{eq:BDaction2}
\end{eqnarray}
This alternative expression allows us to immediately connect with the usual parametrized post-Newtonian (PN) parameters, which restrict the deviation of the scalar-tensor theories of gravity with respect to GR\,\cite{Boisseau:2000pr, Clifton:2011jh,Will:2014kxa}.  Indeed, if we start from the generic scalar-tensor  action
\begin{eqnarray}
S_{\rm }=\int d^{4}x\sqrt{-g}\left(\frac12\, F(\phi) R-\frac12 g^{\mu\nu}\partial_{\nu}\phi\partial_{\mu}\phi-V(\phi)\right)+\int d^{4}x\sqrt{-g}\,{\cal L}_m(\chi_i,g_{\mu\nu})\,, \label{eq:BDactionST}
\end{eqnarray}
we can easily recognize from \eqref{eq:BDaction2} that $F(\phi)=\xi\phi^2$, and that the potential $V(\phi)$ is just replaced by the  CC density $\rL$.  {In this way we can easily apply the well-known formulae of the PN formalism. We find the following values for the main PN parameters $\gamma^{\rm PN}$ and $\beta^{\rm PN}$ in our case (both being equal to $1$ in strict GR)}:
\begin{equation}\label{eq:gammaPPN}
1-\gamma^{\rm PN} =\frac{F^{\prime}(\phi)^2}{F(\phi)+2F^{\prime}(\phi)^2}= \frac{4\xi}{1+ 8\xi}=\frac{\eBD}{1+ 2\eBD}\simeq \eBD +   {\cal O}(\eBD^2)
\end{equation}
and
\begin{equation}\label{eq:betaPPN}
 1-\beta^{\rm PN}=-\frac14\, \frac{F(\phi) F^{\prime}(\phi)}{2F(\phi)+3F^{\prime}(\phi)^2}\,\frac{d \gamma^{\rm PN}}{d\phi}  =0\,,
\end{equation}
where {the primes refer here to derivatives with respect to $\phi$}. We are neglecting terms of $ {\cal O}(\eBD^2)$  and, in the second expression,  we use the fact that  $d\gamma_{\rm PN}/d\phi=0$ since $\oD=$ const. (hence $\eBD=$ const. too) in our case.  {Therefore, in BD-gravity, $\gamma^{\rm PN}$ deviates from $1$  an amount given precisely by $\eBD$ (in linear order), whereas $\beta^{\rm PN}$ undergoes no deviation at all. Furthermore, the effective gravitational strength between two test masses in the context of the scalar-tensor framework \eqref{eq:BDactionST} is well-known\,\cite{Boisseau:2000pr,Will:2014kxa,Clifton:2011jh}. In our case it  leads to the following result:}
\begin{equation}\label{eq:LocalGN}
G_{\rm eff}(\varphi) =\frac{1}{8\pi F(\phi)}\frac{2F(\phi) + 4F^{\prime}(\phi)^2}{2F(\phi)+3F^{\prime}(\phi)^2}=\frac{1}{8\pi\xi\phi^2} \frac{1+8\xi}{1+6\xi}= G(\varphi) \frac{2+4\eBD}{2+3\eBD}\,,
\end{equation}
where $G(\varphi)$ is the effective gravitational coupling in the BD action, as indicated in Eq. \eqref{eq:Gvarphi}.
Expanding linearly in $\eBD$, we find
\begin{equation}\label{eq:LocalGN2}
G_{\rm eff}(\varphi) =G(\varphi)\left[1+\frac12\,\eBD + {\cal O}(\eBD^2)\right] \,.
\end{equation}
{We confirm from the above two equations that the physical gravitational field undergone by two tests masses is not just determined by the effective  $G(\varphi)$ of the action but by  $G(\varphi)$ times a correction factor which depends on $\eBD$ and is larger (smaller) than $G(\varphi)$  for  $\eBD>0\,   (\eBD<0 )$.}

{From the exact formula \eqref{eq:LocalGN} we realize that if the local gravitational constraint ought to hold strictly, i.e. $G_{\rm eff}\to G_N$,  such formula would obviously enforce}
\begin{equation}\label{eq:LC}
\varphi=\frac{2+4\eBD}{2+3\eBD}\,.
\end{equation}
Due to Eq.\,\eqref{eq:gammaPPN}, the bound obtained from the Cassini mission\,\cite{Bertotti:2003rm}, $\gamma^{\rm PN}-1=(2.1\pm 2.3)\times 10^{-5}$, translates directly into a constraint on  $\eBD\simeq (-2.1\pm 2.3)\times 10^{-5}$ (in linear order), what implies $\oD\gtrsim 10^4$ for the BD-parameter. Thus, if considered together with the assumption $\varphi\simeq 1$ we would be left with very little margin for departures of $\Geff$ from $G_N$. However,  as previously  indicated,  we will not apply these local astrophysical bounds in most of our analysis since  we will assume that $\eBD$ may not be constrained in the cosmological domain and that the cosmological value of the gravitational coupling $G(\varphi)$  is different from $G_N$  owing to some possible variation of the BD-field $\varphi$ at the cosmological level. This can still be compatible with the local astrophysical constraints provided that we assume the existence of a screening mechanism in the local range which `renormalizes' the value of $\wBD$ and makes it appear much higher than its `bare' value (the latter being accessible only at the cosmological scales, where matter is much more diluted and uninfluential) --- cf. Sec.\,\ref{sec:Mach} for details on the various possible BD scenarios.  We know that this possibility remains open and hence it must be explored\,\cite{Avilez:2013dxa}, see also \cite{Amendola:2015ksp,Clifton:2011jh}  and references therein.

Henceforth we shall stick to the original BD-form (\ref{eq:BDaction}) of the action  since the field $\psi$ (or, alternatively, its dimensionless companion $\varphi$)  is directly related to the effective gravitational coupling and the  $\oD$ parameter can be ascribed as being part of the kinetic term of $\psi$.  In contrast, the form (\ref{eq:BDaction2}) involves both $\phi$ and $\xi=1/(4\oD)$ in the definition of the gravitational strength.

\section{Why  BD-$\CC$CDM alleviates at a time the $H_0$ and $\sigma_8$ tensions?  Detailed preview and main results  of our analysis.}\label{sec:preview}

The field $\varphi$ and the parameter $\epsilon_{BD}$ defined in the previous section, Eq.\,\eqref{eq:definitions}, are the fundamental new ingredients of BD-gravity as compared to GR  in the context of our analysis.  Any departure of $\varphi$ from one and/or of  $\epsilon_{BD}$ from zero should reveal an extra effect of the BD-$\CC$CDM model as compared to the  conventional GR-$\CC$CDM one.   We devote this section to study the influence of  $\varphi$ and $\epsilon_{BD}$ on the various observables we use in this work to constrain the BD model.  This preliminary presentation will serve as a preview of the results presented in the rest of the paper and will allow us to anticipate why BD-$\Lambda$CDM is able to alleviate so efficiently both of the $H_0$ and $\sigma_8$ tensions that are found in the context of the traditional  GR-$\Lambda$CDM framework.

 Interestingly, many Horndeski theories \cite{Horndeski:1974wa} reduce in practice to BD at cosmological scales \cite{Avilez:2013dxa}, so the ability of  BD-$\Lambda$CDM to describe the wealth of current observations can also be extended to other, more general, scalar-tensor theories of gravity. Hence, it is crucial to clearly identify the reasons why BD-$\Lambda$CDM leads to such an improvement in the description of the data. Only later on (cf. Sec. \ref{sec:NumericalAnalysis}) we will fit in detail the overall  data to the BD-$\Lambda$CDM model and will display the final numerical results. Here, in contrast, we will endeavour  to show why BD-gravity has specific clues to the solution which are not available to GR.

We can show this in two steps. First, we analyze what happens when we set $\epsilon_{BD}=0$  at fixed values of $\varphi$ different from $1$.  From Eq.\,\eqref{eq:LocalGN} we can see that this scenario means to stick to the standard GR picture, but assuming that the effective Newtonian coupling can act at cosmological scales with a (constant) value $G_{\rm eff}=G(\varphi)$  different from the local one $G_N$.   In a second stage,  we  study the effect of the time dependence of $\varphi$  (triggered by a nonvanishing value of $\epsilon_{BD}$), i.e. we will exploit the departure of $G_{\rm eff}$  in  Eq.\,\eqref{eq:LocalGN} from $G_N$ caused by $\eBD\neq0$ \textit{and} a variable $G(\varphi)$.  It will become clear  from this two-step procedure why BD-gravity has the double ability to reduce the two $\CC$CDM tensions in an harmonic way. On the one hand, a value of $\varphi$ below $1$ in the late universe increases the value of $G_{\rm eff}$  and hence of $H_0$, so it should be  able to significantly  reduce the $H_0$-tension;  and on the other hand {the dynamics of $\varphi$ triggered by a finite (but negative) value of $\epsilon_{BD}$ helps to suppress the structure formation processes in the universe, since it enhances the Hubble friction and also leads to a decrease of the Poisson term in the perturbations equation.} {The upshot is that the $\sigma_8$-tension becomes reduced as well.} \newtext{Let us note that the lack of use of LSS data may lead to a different conclusion, in particular to $\eBD>0$, see e.g. \cite{Yadav:2019fjx}. This reference, in addition, uses only an approximate treatment of the CMB data through distance priors.}

\subsection{Role of $\varphi$, and the $H_0$-tension}\label{sec:rolesvarphiH0}

Let us start, then, by studying how the observables change with $\varphi$ when $\epsilon_{BD}=0$,  for fixed values of the current energy densities. In the context of BD-gravity, as well as in GR,  if the energy densities are fixed at present we can fully determine their cosmological evolution, since all the species are self-conserved, as discussed in Sec. \ref{sec:BDgravity}. In BD-gravity, with $\epsilon_{BD}=0$, the Hubble function takes the  form
\begin{equation}\label{eq:H1}
H^2(a) = \frac{8\pi G_N}{3\varphi}\rho(a)\,,
\end{equation}
where  $\varphi=const$. We have just removed the time derivatives of the scalar field in the Friedmann equation \eqref{eq:Friedmannequation}.  We define $H_0$ from the value of the previous expression at $a=1$ (current value). Recall from the previous section that $\rho$ is the sum of all the energy density contributions, namely $\rho \equiv \rho_m + \rho_\gamma + \rho_\nu + \rho_\Lambda$. Therefore,
\begin{equation}\label{eq:H1b}
H_0^2 = \frac{8\pi G_N}{3\varphi}\rho^0=\frac{8\pi G_N}{3\varphi}\left( \rho_m^0+\rho_\gamma^0+ \rho_\nu^{0}+\rL\right)\,.
\end{equation}
$\rho^0=\rho(a=1)$ is the total energy density at present, $\rho_\gamma^0$ is the corresponding density of photons and $\rho_\nu^{0}$ that of neutrinos, and finally $\rL$ is the original cosmological constant density in the BD-action \eqref{eq:BDaction}.
Using \eqref{eq:H1b} it proves now useful  to rewrite \eqref{eq:H1} in the alternative way:
\begin{equation}\label{eq:H2}
H^2(a)= H_0^2\left[1+\tilde{\Omega}_{m}(a^{-3}-1)+\tilde{\Omega}_\gamma(a^{-4}-1)+\tilde{\Omega}_\nu(a)-\tilde{\Omega}_\nu\right]\,,
\end{equation}
where we use the modified cosmological parameters (more appropriate for the BD theory):
\begin{equation}\label{eq:tildeOmegues}
\tilde{\Omega}_i\equiv\frac{\rho^{0}_i}{\rho^{0}}=\frac{\Omega_i}{\varphi}\,,\ \ \ \ \ \ \ \ \  \rho^{0}=\frac{3H_0^2}{8\pi G_N}\,\varphi=\rco\varphi\,.
\end{equation}
The tilde is to distinguish the modified $\tilde{\Omega}_i$ from the usual cosmological parameters $\Omega_i=\rho_i^{0}/\rco$ employed in the GR-$\Lambda$CDM model.  In addition,  $\tilde{\Omega}_{m}=\tilde{\Omega}_{cdm}+\tilde{\Omega}_b$ is the sum of the contributions from CDM and baryons; and $\tilde{\Omega}_\gamma$ and $\tilde{\Omega}_\nu$ are the current values for the photons and neutrinos. For convenience, we also define $\tilde{\Omega}_r=\tilde{\Omega}_\gamma+\tilde{\Omega}_\nu$.  We remark that he current total energy density $ \rho^{0}$ is related to the usual critical density $\rco=3H_0^2/(8\pi G_N)$  through a factor of $\varphi$, as indicated above.
The modified parameters obviously satisfy the canonical sum rule
\begin{equation}\label{eq:SumRuleBD}
\tilde{\Omega}_m+\tilde{\Omega}_r+\tilde{\Omega}_\CC=1\,.
\end{equation}

The form of \eqref{eq:H2} is completely analogous to the one found in GR-$\Lambda$CDM since in BD-$\CC$CDM the $\tilde{\Omega}_i$'s represent the fraction of energy carried by the various species in the current universe, as the $\Omega_i$'s do in GR, so from the physical point of view the $\tilde{\Omega}_i$'s in BD and the $\Omega_i$'s in GR contain the same information\footnote{For $|\eBD|\neq 0$ and small, the $\tilde{\Omega}_i$ parameters defined in Eq.\,\eqref{eq:tildeOmegues} receive a correction of ${\cal O}(\eBD)$, see Sec.\,\ref{sec:eBDands8}.}. $H_0$ represents in both cases the current value of the Hubble function. Nevertheless, there is a very important (although maybe subtle) difference, namely: in BD-$\Lambda$CDM there does not exist a  one-to-one correspondence between $H_0$ and $\rho^{0}$. In contradistinction to GR-$\Lambda$CDM, in the BD version of the concordance model the value of $\varphi$ can modulate $H_0$ for a given amount of the total (critical) energy density. In other words, given a concrete value of $H_0$ there is a $100\%$ degeneracy between $\varphi$ and $\rho^{0}$. This degeneracy is broken by the data, of course. The question we want to address is precisely whether there is still room for an increase of $H_0$ with respect to the value found in the GR-$\Lambda$CDM model once the aforementioned degeneracy is broken by observations. We will see that this is actually the case by analyzing what is the impact that $\varphi$ has on each observable considered in our analyses.
\newline
\newline
{\bf Supernovae data}
\newline
\newline
\noindent
In the case  of Type Ia  Supernovae data (SNIa)  we fit observational points on their apparent magnitudes $m(z)= M +5\log_{10}[D_L(z)/10{\rm pc}]$, where $M$ is the absolute magnitude of the SNIa and $D_L(z)$ is the luminosity distance. In a spatially flat universe the latter reads,
\begin{equation}
D_L(z)=c(1+z)\int^z_{0}\frac{dz^\p}{H(z^\p)}\,,
\end{equation}
where we have momentarily kept $c$ explicitly for the sake of better understanding. Considering \eqref{eq:H2}, we can easily see that if we only use SNIa data, there is a full degeneracy between $M$ and $H_0$ in the computation of the apparent magnitudes, so it is not possible to obtain information about $\varphi$, since we cannot disentangle it from the absolute magnitude. As in GR-$\Lambda$CDM, we can only get constraints on the current fraction of matter energy in the universe, i.e. $\tilde{\Omega}_m$.
\newline
\newline
{\bf Baryon acoustic oscillations}
\newline
\newline
\noindent
The constraints obtained from the analysis in real or Fourier space of the baryon acoustic oscillations (BAO) are usually provided by galaxy surveys in terms of $D_A(z)/r_s$ and $r_s H(z)$, or in some cases by a combination of these two quantities when it is not possible to disentangle the line-of-sight and transverse information, through the so-called dilation scale $D_V$ (cf. Sec. \ref{sec:MethodData}),
\begin{equation}
\frac{D_V(z)}{r_s}=\frac{1}{r_s}\left[D_M^2(z)\frac{cz}{H(z)}\right]^{1/3}\,,
\end{equation}
$D_M=(1+z)D_{A}(z)$ being the comoving angular diameter distance, $D_A(z)=D_L(z)/(1+z)^2$ the proper angular diameter distance, and
\begin{equation}\label{eq:rs}
r_s=\int_{z_d}^{\infty}\frac{c_s(z)}{H(z)}\,dz
\end{equation}
the comoving sound horizon at the baryon drag epoch $z_d$. In the above equation, $ c_s(z)$ is  the sound speed of the photo-baryon plasma, which depends on the ratio $\rho_b^{0}/\rho^{0}_\gamma$. The current temperature of photons (and hence also its associated current energy density $\rho^{0}_\gamma$) is already known with high precision thanks to the accurate measurement of the CMB monopole \cite{Fixsen:2009ug}. Because of \eqref{eq:H1} it is obvious that we cannot extract information on $\varphi$ from BAO data when $\epsilon_{BD}=0$, since it cancels exactly in the ratio $D_A(z)/r_s$ and the product $r_s H(z)$. Thus, BAO data provide constraints on $\tilde{\Omega}_m$ and $\rho_b^{0}$, but not on $\varphi$.
\newline
\newline
{\bf Redshift-space distortions (RSD)}
\newline
\newline
\noindent
{The  LSS observable $f(z)\sigma_8(z)$, which is essentially determined from RSD measurements,  is of paramount importance to study the formation of structures in the universe.  In BD-gravity, the equation of matter field perturbations is different from that of GR and  is studied in detail in Sec.\,\ref{sec:StructureFormation}.  Here we wish to make some considerations which will help to have a rapid overview of why BD-gravity with a cosmological constant can also help to improve the description of structure formation as compared to GR. The exact differential equation for the linear density contrast of the matter perturbations,  $\delta_m=\delta\rho_m/\rho_m$,  in this context can be computed  at deep subhorizon scales and is given by (cf. Sec. \ref{sec:StructureFormation}):}
%\begin{equation}\label{eq:ExactPerturScaleFactor}
%\delta_m^\pp+\left(\frac{3}{a}+\frac{H^\prime(a)}{H(a)}\right)\delta_m^\p-\frac{4\pi G_N}{H^2(a)\varphi(a)}\frac{\rho_m(a)}{a^2}\left(\frac{4+2\omega_{BD}}{3+2\omega_{BD}}\right)\delta_m=0\,,
%\end{equation}
\begin{equation}\label{eq:ExactPerturScaleFactor}
\delta_m^\pp+\left(\frac{3}{a}+\frac{H^\prime(a)}{H(a)}\right)\delta_m^\p-\frac{4\pi \Geff(a)}{H^2(a)}\frac{\rho_m(a)}{a^2}\,\delta_m=0\,,
\end{equation}
{where for the sake of convenience we express it in terms of the scale factor and hence prime denotes here  derivative {\it w.r.t.}  such variable:  $()^\prime \equiv d()/d a$. In the above equation, $\Geff(a)$ is the effective gravitational coupling \eqref{eq:LocalGN} with $\varphi=\varphi(a)$ expressed in terms of the scale factor:}
\begin{equation}\label{eq:LocalGNa}
\Geff(a) = G(\varphi(a)) \frac{2+4\eBD}{2+3\eBD}\,.
\end{equation}
It crucially controls the  Poisson term of the perturbations equation, i.e. the last term in \eqref{eq:ExactPerturScaleFactor}.
As we can see, it is $\Geff(\varphi)$ and not just $G(\varphi)$ the coupling involved in the structure formation data since it is $\Geff(\varphi)$ the gravitational coupling felt by the test masses in BD-gravity.
It is obvious that the above Eq.\eqref{eq:ExactPerturScaleFactor} boils down to the GR form for $\eBD=0$ and $\varphi=1$.

With the help of the above equations  we  wish first to assess the bearing that $\varphi$ can have on the LSS observable $f(z)\sigma_8(z)$ at  fixed values of the current energy densities and for $\epsilon_{BD}=0$. Recall that when  $\epsilon_{BD}=0$ the BD-field cannot evolve at all, so it just remains fixed at some value.  For this consideration, we therefore set $\psi=$const. in equations \eqref{eq:Friedmannequation} and \eqref{eq:pressureequation} and of course the BD-theory becomes just a GR version with an effective coupling $\Geff=G(\varphi)=$const.  which nevertheless need not be identical to  $G_N$.  In these conditions, it is easy to verify that Eq.\,\eqref{eq:ExactPerturScaleFactor} adopts the following simpler form, in which $\Geff$ drops from the final expression:
\begin{equation}\label{eq:DC}
\delta_m^{\pp}+\frac{3}{2a} \left(1-w(a)\right) \delta^\p_m -\frac{3}{2a^2}\frac{\rho_m(a)}{\rho(a)}\delta_m=0\,.
\end{equation}
In the above expression, $w(a)=p(a)/\rho(a)$ is the equation of state (EoS) of the total cosmological fluid, hence  $\rho(a)$ and $p(a)$ stand respectively for the total density and pressure of the fluids involved (cf. Sec. \ref{sec:BDgravity}). In particular, during the epoch of structure formation the matter particles contribute a negligible contribution to the pressure and the dominant component is that of the  cosmological term: $p(a)\simeq p_\CC=-\rL$.

It is important to realize that $\varphi=$ const. does not play any role in \eqref{eq:DC}. This means that its constant value, whatever it may be,  does not affect the evolution of the density contrast, which is only determined by the fraction of matter, $\rho_m(a)/\rho(a)$,  and the  EoS of the total cosmological fluid.  The equation that rules the evolution of the density contrast is exactly the same as in the GR-$\Lambda$CDM model. Matter inhomogeneities grow in the same way regardless of the constant value $\Geff$ that we consider. Matter tends to clump more efficiently for larger values of the gravitational strength, of course, but the Hubble friction also grows in this case, since such an increase in $\Geff$ also makes the universe to expand faster.
%This can be better appreciated in \eqref{DensityConstrastLCDM}.
Surprisingly, if $\epsilon_{BD}=0$, i.e. if $\Geff=$const., both effects compensate each other. Thus, the BD growth rate $f(a)=a\delta_m^\prime(a)/\delta_m(a)$ does not change  {\it w.r.t.} the GR scenario either. But what happens with $\sigma_8(z)$? It is computed through the following expression:
\begin{equation}\label{eq:sigma8}
\sigma_8^2(z)=\frac{1}{2\pi^2}\int_0^\infty dk\,k^2\,P(k,z)\,W^2(kR_8)\,,
\end{equation}
in which  $P(k,z)$ is the power spectrum of matter fluctuations and $W(kR_8)$ is the top hat smoothing function in Fourier space, with $R_8=8h^{-1}$ Mpc. Even for $\epsilon_{BD}=0$ one would naively expect \eqref{eq:sigma8} to be sensitive to the value of $\varphi$, since some relevant features of the power spectrum clearly are. For instance, the scale associated to the matter-radiation equality reads
\begin{equation}
k_{eq}=a_{eq}H(a_{eq})=H_0\tilde{\Omega}_m\sqrt{\frac{2}{\tilde{\Omega}_r}}\,,
\end{equation}
and $H_0\propto\varphi^{-1/2}$, so the peak of $P(k,z)$ is shifted when we change $\varphi$. Also the window function itself depends on $H_0$ through $R_8$. Despite this, the integral \eqref{eq:sigma8} does not depend on the Hubble function for fixed energy densities at present, and hence neither on $\varphi$. To see this, let us  first rescale the wave number  as follows  $k=\bar{k}h$, and we obtain
\begin{equation}\label{eq:sigma8v2}
\sigma_8^2(z)=\frac{1}{2\pi^2}\int_0^\infty d\bar{k}\,\bar{k}^2\,\underbrace{P(k=\bar{k}h,z)\,h^3}_{\equiv \bar{P}(\bar{k},z)}\,W^2(\bar{k}\cdot 8\,{\rm Mpc^{-1}})\,.
\end{equation}
The only dependence on $h$ is now contained in $\bar{P}(\bar{k},z)$. We can write $P(k,z)=P_0k^{n_s}T^2(k/k_{eq})\delta^2_m(z)$, with $T(k/k_{eq})$ being the transfer function and
\begin{equation}
P_0=A_s\frac{8\pi^2}{25}\frac{k_*^{1-n_s}}{(\tilde{\Omega}_m H_0^2)^2}\,,
\end{equation}
with $A_s$ and $n_s$ being the amplitude and spectral index of the dimensionless primordial power spectrum, respectively, and $k_*$ the corresponding pivot scale. The last relation can be found using standard formulae, see e.g. \cite{GorbunovRubakovBook,Amendola:2015ksp}. Taking into account all these expressions we obtain
\begin{equation}
\bar{P}(\bar{k},z)=A_s\frac{8\pi^2}{25}\frac{\bar{k}^{1-n_s}_*\bar{k}^{n_s}}{(10^4\varsigma^2\tilde{\Omega}_m)^2} T^2(\bar{k}/\bar{k}_{eq})\delta^2_m(z)\,,
\end{equation}
where we have used $H_0=100\, h\, \varsigma$  with $\varsigma\equiv 1\,{\rm km/s/Mpc}=2.1332\times 10^{-44}$ GeV (in natural units). We see that all factors of $h$ cancel out. Now it is obvious that $\sigma_8(z)$ is not sensitive to the value of $\varphi$. We have explicitly checked this with our own modified version of the Einstein-Boltzmann system solver \texttt{CLASS} \cite{Blas:2011rf}, in which we have implemented the BD-$\Lambda$CDM model (see Sec. \ref{sec:MethodData} for details). The product $f(z)\sigma_8(z)$ does not depend on $\varphi$ when $\epsilon_{BD}=0$, so RSD data cannot constrain $\varphi$ either.
\newline
\newline
\newline
{\bf Strong-Lensing time delay distances, distance ladder determination of $H_0$, and cosmic chronometers}
\newline
\newline
\noindent
In this work, we will use the Strong-Lensing time delay angular diameter distances provided by the H0LICOW collaboration \cite{Wong:2019kwg}, see Sec. \ref{sec:MethodData}. Contrary to SNIa and BAO data, these distances are not relative, but absolute. This allows us to extract information not only on $\Om$, but on $H_0$ too.  Furthermore,  the data on $H(z)$ obtained from cosmic chronometers (CCH) give us information about these two parameters as well. Cosmic chronometers have been recently employed in the reconstruction of the expansion history of the universe using Gaussian Processes and the so-called Weighted Function Regression method \cite{Yu:2017iju,Gomez-Valent:2018hwc,Haridasu:2018gqm}, which do not rely on any particular cosmological model. The extrapolated values of the Hubble parameter found in these analyses are closer to the best-fit GR-$\Lambda$CDM value reported by Planck \cite{Aghanim:2018eyx}, around $H_0\sim (67.5-69.5)$ km/s/Mpc, but they are still compatible within  $\sim 1\sigma$ c.l. with the local determination obtained with the distance ladder technique \cite{Riess:2018uxu,Riess:2019cxk,Reid:2019tiq} and the Strong-Lensing time delay measurements by H0LICOW \cite{Wong:2019kwg}. The statistical weight of the CCH data is not as high as the one obtained from these two probes, so when combined with the latter, the resulting value for $H_0$ is still in very strong tension with Planck \cite{Gomez-Valent:2018hwc,Haridasu:2018gqm}. As mentioned before, $H_0^2\propto \rho^{0}/\varphi$ when $\epsilon_{BD}=0$. Thus, we can alleviate in principle the $H_0$-tension by keeping the same values of the current energy densities of all the species as in the best-fit GR-$\Lambda$CDM model reported by Planck \cite{Aghanim:2018eyx}, lowering the value of $\varphi$ down at cosmological scales, below $1$, and assuming some sort of screening mechanism acting on high enough density regions that allows us to evade the solar system constraints and keep unmodified the stellar physics needed to rely on CCH, SNIa, the H0LICOW data, and the local distance ladder measurement of $H_0$. By doing this we do not modify at all the SNIa, BAO and RSD observables {\it w.r.t.} the GR-$\Lambda$CDM, and we automatically improve the description of the H0LICOW data and the local determination of $H_0$, which are the observables that prefer higher values of the Hubble parameter.  Let us also mention that the fact that $\varphi<1$ throughout the cosmic history (which means $G>G_N$)  allows to have a larger value of $H$ (for similar values of the density parameters) at any time as compared to the GR-$\CC$CDM and hence a smaller value of the sound horizon distance $r_s$, Eq.\,\eqref{eq:rs}, what makes the model to keep a good  fit to the BAO data. This is confirmed by the numerical analysis presented in Tables 3-6 and 10 as compared to the conventional $\CC$CDM values, see Tables 3-5 and 7.  While the claim existing in the literature  that models which predict smaller values of $r_s$ are the preferred ones for solving the $H_0$-tension is probably correct,  we should point out that this sole fact is no guarantee of success, as one still needs in general a compensation mechanism at low energies which prevents $\sigma_8$ from increasing and hence worsening such tension. In the BD-$\CC$CDM such compensation mechanism is provided by a  negative value of $\eBD$ (as we will show later), and for this reason the two tensions can be smoothed at the same time in an harmonic way.

Overall, as we have seen from the above discussion,  according to the (long) string of supernovae, baryonic acoustic oscillations, cosmic chronometers, Strong-Lensing and local Hubble parameter data (SNIa+BAO+RSD+CCH+H0LICOW+$H_0$) it is possible to loosen the $H_0$-tension, and this is already very remarkable, but we still have to see whether this is compatible with the very precise measurements of the photon temperature fluctuations in the CMB map or not. More specifically, we have to check whether it is possible to describe these anisotropies while keeping the current energy densities close to the best-fit GR-$\Lambda$CDM model from Planck, compatible with $\varphi<1$.
\newline
\newline
%%%%%%%%%%%%%%%%%%%%%%%%%%%%%%%%%%%%%%%%%%%%%%%%%%%%%%%%%%%%%%%
\begin{figure}[t!]
\begin{center}
\label{Cls-varphi}
\includegraphics[width=6in, height=4in]{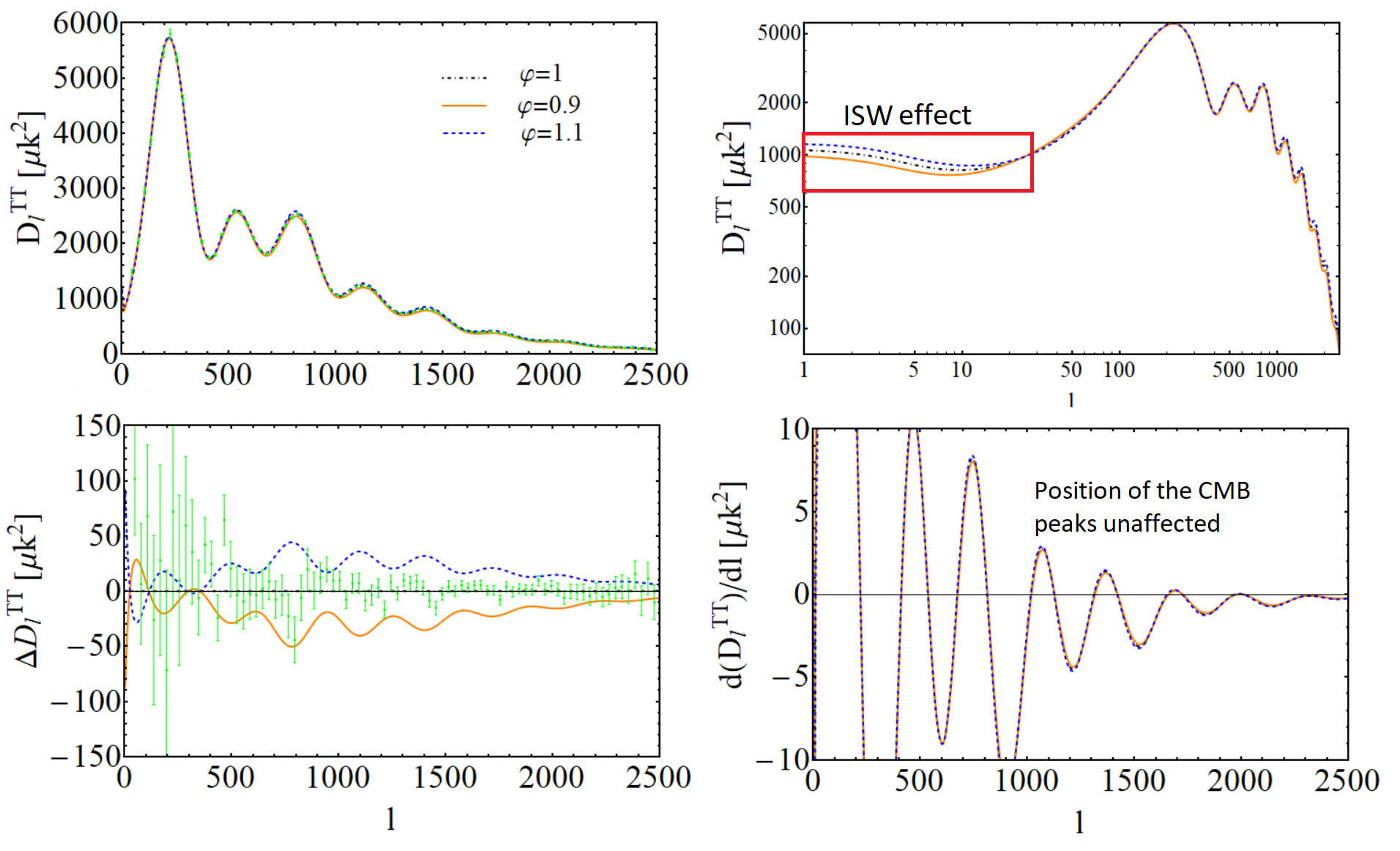}
\caption{\scriptsize {{\it Left-upper plot:} Theoretical curves of the CMB temperature anisotropies obtained by fixing the current energy densities, $\tau$ and the parameters of the primordial power spectrum to the GR-$\Lambda$CDM best-fit values obtained from the analysis of TTTEEE+lowE data by Planck 2018 \cite{Aghanim:2018eyx} (which we will refer to as the $\CC$CDM baseline configuration, denoted by a dot-dashed black line), against the BD-$\Lambda$CDM model  keeping $\epsilon_{BD}=0$ for two different constant values  $\varphi=0.9,1.1$  (orange and dashed blue lines, respectively); {\it Right-upper plot:} The same, but using a logscale in the $x$-axis to better appreciate the integrated Sachs-Wolfe effect at low multipoles; {\it Left-bottom plot:} Absolute differences of the data points and theoretical curves for $\varphi=0.9,1.1$ {\it w.r.t.} the $\varphi=1$ case; {\it Right-bottom plot:} Derivative of the functions plotted in the upper plots, to check the effect of $\varphi$ on the position of the peaks, which corresponds to the location of the zeros in this plot. No shift is observed, as expected (cf. the explanations in the main text).}}
\end{center}
\end{figure}
%%%%%%%%%%%%%%%%%%%%%%%%%%%%%%%%%%%%%%%%%%%%%%%%%%%%%%%%%%%%%%%
\newline
\newline
{\bf CMB temperature anisotropies}
\newline
\newline
\noindent We expect the peak positions of the CMB temperature (TT, in short) power spectrum to remain unaltered under changes of $\varphi$ (when $\epsilon_{BD}=0$), since they are always related with an angle, which is basically a ratio of cosmological distances ({a transverse distance to the line of sight divided by the angular diameter distance}).  If $\varphi=$const.,  such constant cancels  in the ratio, so there is no dependence on $\varphi$. In the right-bottom plot of Fig. 1 we show the derivative of the $\mathcal{D}_l^{TT}$'s for three alternative values of $\varphi$. It is clear that the location of the zeros does not depend on the value of this parameter. Hence $\varphi$ does not shift the peaks of the TT CMB spectrum when we consider it to be a constant throughout the entire cosmic history, as expected. Nevertheless, there are two things that affect its shape and both are due to the impact that $\varphi$ has on the Bardeen potentials. We have seen before that the matter density contrast is not affected by $\varphi$ when it is constant, but taking a look on the Poisson equation in the BD model (cf. Appendix D), we can convince ourselves that $\varphi$ does directly affect the value of $\Psi$ and $\Phi$, since both functions are proportional to $\rho_m\delta_m/\varphi$ at subhorizon scales, see Eqs.\,\eqref{eq:PhiplusPsi}-\eqref{eq:Poisson3}. This dependence modifies two basic CMB observables:
\begin{itemize}
\item The CMB lensing, at low scales \newtext{(large multipoles, $500\lesssim l\lesssim 2000$)}. In the left-bottom plot of Fig. 1 we show the difference of the TT CMB spectra {\it w.r.t.} the GR-$\Lambda$CDM model. A variation of $\varphi$ changes the amount of CMB lensing, which in turn modifies the shape of the spectrum mostly from the third peak onwards. \newtext{In that multipole range also the Silk damping plays an important role and leaves a signature \cite{Silk1968}.}
\item The integrated Sachs-Wolfe effect \cite{Sachs:1967er}, at large scales \newtext{(low multipoles, $l\lesssim 30$). Values of $\varphi<1$ (which, recall, lead to higher values of $H_0$) suppress the power of the $\mathcal{D}_l^{TT}$'s in that range.} This is a welcome feature, since it could help us to explain the low multipole ``anomaly'' that is found in the context of the GR-$\Lambda$CDM model (see e.g. Fig. 1 of \cite{Aghanim:2018eyx}, and \cite{Das:2013sca}). \newtext{Later on in the paper, we further discuss and confirm the alleviation  of this intriguing anomaly in light of the final fitting results, see Sec.\,\ref{sec:LowMultipoles} and Fig.\,12. This is obviously an additional bonus of the BD-$\CC$CDM framework.}
\end{itemize}

%%%%%%%%%%%%%%%%%%%%%%%%%%%%%%%%%%%%%%%%%%%%%%%%%%%%%%%%%%%%%%%
\begin{figure}[t!]
\begin{center}
\label{fig:nsVSvarphi}
\includegraphics[width=6in, height=2.4in]{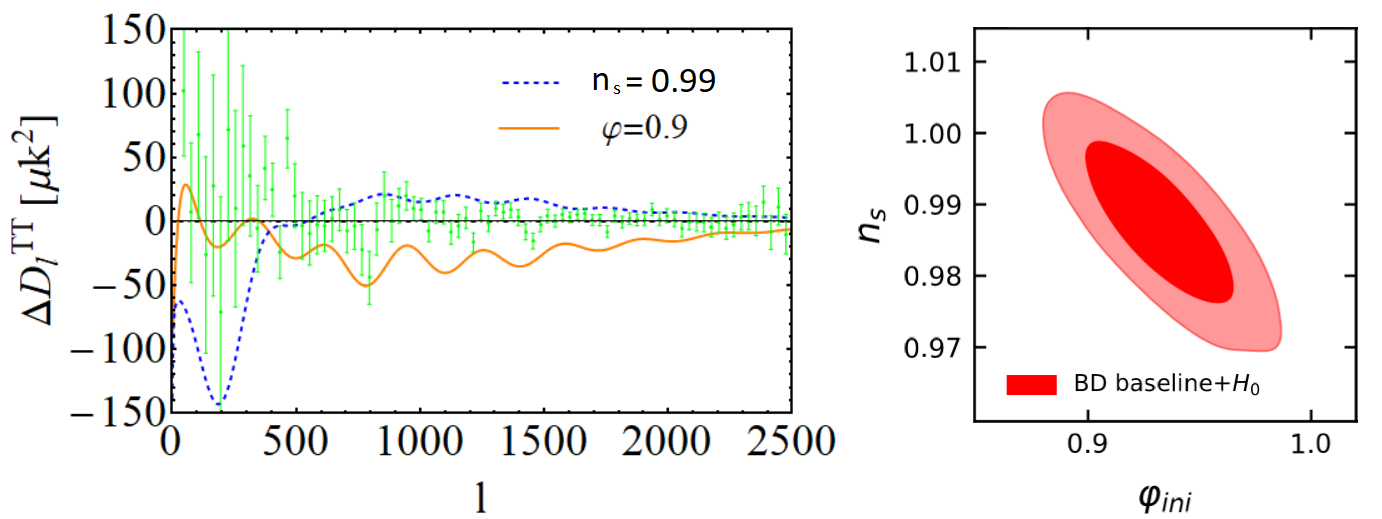}
\caption{\scriptsize {{\it Left plot:} Differences between the CMB temperature spectrum obtained using the original $\CC$CDM baseline configuration (as used in Fig. 1) and those obtained using $\varphi=0.9$ (with the baseline $n_s=0.9649$) and $n_s=0.99$ (with the baseline $\varphi=1$). This is to show that the effects induced by a lowering of $\varphi$ can be compensated by an increase of the spectral index; {\it Right plot:} We also show the $1\sigma$ and $2\sigma$ c.l. regions in the $(\varphi_{ini},n_s)$-plane, obtained using the baseline dataset, together with the Gaussian prior on $H_0$ from \cite{Reid:2019tiq}, see Sec. \ref{sec:MethodData} for details. Here one can clearly see the  anticorrelation between these two parameters.}}
\end{center}
\end{figure}
%%%%%%%%%%%%%%%%%%%%%%%%%%%%%%%%%%%%%%%%%%%%%%%%%%%%%%%%%%%%%%%

Even though in Fig. 1 we are considering large ($\sim 10\%$) relative deviations of $\varphi$ with respect to $1$, the induced deviations of the $\mathcal{D}_l^{TT}$'s are fully contained in the observational error bars at low multipoles, and they are not extremely big at large ones. The latter are only $1-2\sigma$ away from most of the data points.  We emphasize that we are only varying $\varphi$ here, so there is still plenty of room to correct these deviations by modifying the value of other parameters entering the model. To do that it would be great if we could still keep the values of the current energy densities as in the concordance model, since this would ensue the automatic fulfillment of the constraints imposed by the datasets discussed before. But is this possible? In Fig. 2 we can see that e.g. an increase of $n_s$ can compensate for the decrease of $\varphi$ pretty well. This is why in the BD model we obtain higher best-fit values of the spectral index {\it w.r.t.} the GR-$\Lambda$CDM, and a clear anti-correlation between these two parameters ({cf. Fig. 2, Tables 3-5 and 10, and Sec. \ref{sec:NumericalAnalysis} for details}). Small variations in other parameters can also help to improve the description of the data, of course, but the role of $n_s$ seems to be important. In Fig. 3 we can appreciate the change in the matter power spectrum induced by different values of $n_s$. There is a modification in the range of scales that can be observationally accessed to with the analysis of RSD, but these differences are negative at $k\lesssim 0.07\,h{\rm\,Mpc^{-1}}$ and positive at larger values of the wave number (lower scales), so there can be a compensation when $\sigma_8$ is computed through \eqref{eq:sigma8}, leaving the value of the latter stable. Moreover, we will see below that $\epsilon_{BD}$ can also help to decrease the value of $P(k)$ at $k\gtrsim 0.02\,h{\rm\,Mpc^{-1}}$, so the correct shape for the power spectrum is therefore guaranteed.

The upshot of this section is worth emphasizing:  as  it turns out,  the sole fact of considering a cosmological Newtonian coupling about $\sim 10\%$ larger than the one measured locally can allow us to fit very well all the cosmological datasets, loosening the $H_0$-tension and keeping standard values of $\sigma_8$. It has become common in the literature to divide the theoretical proposals  able to decrease  the $H_0$-tension into two different classes depending on the stage of the universe's expansion at which new physics are introduced \cite{Knox:2019rjx}: pre- and post-recombination solutions. The one we are suggesting here (see also the preceding  paper \cite{Sola:2019jek}) cannot be identified with any of these two categories, since it modifies the strength of gravity at cosmological scales not only before the decoupling of the CMB photons or the late-time universe, but during the whole cosmological history, relying on a screening mechanism able to generate $\Geff=G=G_N$ in high density ({nonrelativistic}) environments where nonlinear processes become important, as e.g. in our own solar system\footnote{The study of these screening mechanisms, see e.g. \cite{Tsujikawa:2008uc,Amendola:2015ksp,Clifton:2011jh,Li:2020uaz} and references therein, can be the subject of  future work, but here we  remark that e.g. chameleon \cite{Khoury:2003aq}, symmetron \cite{Hinterbichler:2010es} or Vainshtein mechanisms (see \cite{Kimura:2011dc} and references therein), do not screen the value of $\varphi$ during the radiation-dominated epoch. This is important to loose the $H_0$-tension in the BD-$\Lambda$CDM framework through the increase of $H(z)$ at both, the early and late universe.}. That there is indeed a  change of the gravity  strength  throughout the entire cosmological history in our study follows from the fact that $\eBD\neq0$ in the BD framework, and this is exactly the feature that we are going to exploit in the next section, a feature which adds up to the mere change of the global strength of the gravity interaction, which is still possible for $\eBD=0$ in the BD context, and that we have explored in the previous section.

%%%%%%%%%%%%%%%%%%%%%%%%%%%%%%%%%%%%%%%%%%%%%%%%%%%%%%%%%%%%%%%
\begin{figure}[t!]
\begin{center}
\label{fig:pkns}
\includegraphics[width=6.5in, height=2.3in]{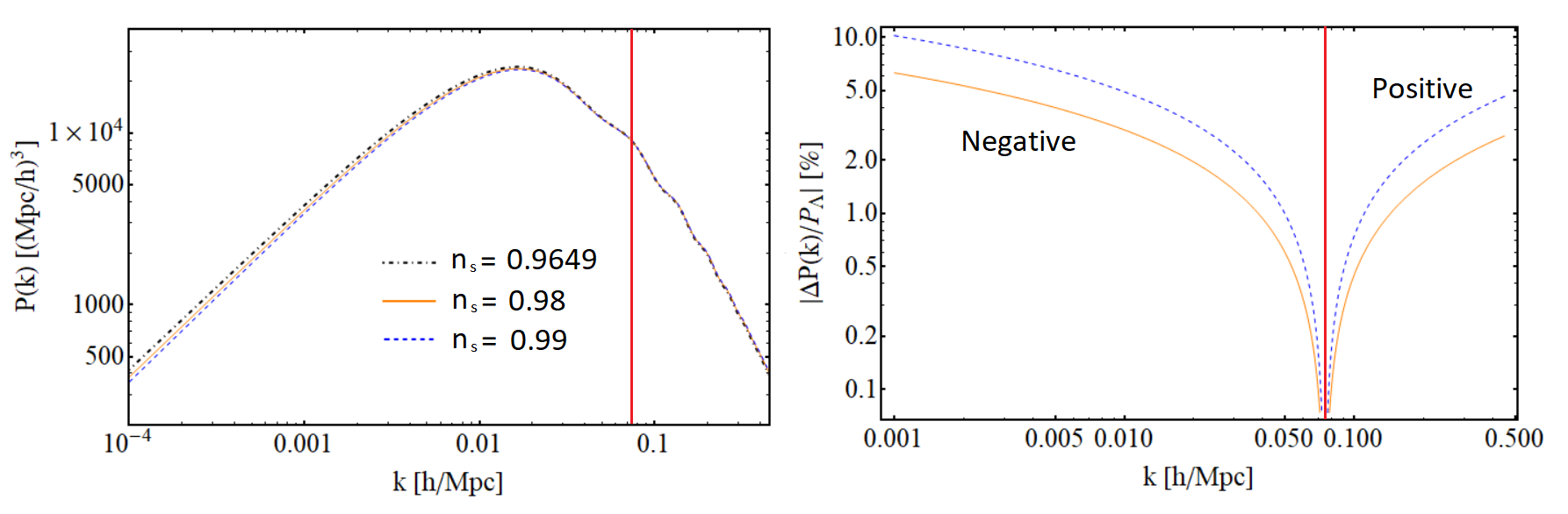}
\caption{\scriptsize {{\it Left plot:} Here we compare the linear matter power spectrum obtained for the $\CC$CDM baseline configuration (dot-dashed black line) and two alternative values of the spectral index $n_s$. The vertical red line indicates the value of the wave number at which there is the break in the right plot; {\it Right plot:} The absolute relative differences in $P(k)$ {\it w.r.t.} the baseline $\CC$CDM model. The ``positive'' (``negative'') region is the one in which $P(k)$ is larger (lower) than in the baseline setup. Both zones are delimited by a vertical red line. See the comments in the text.}}
\end{center}
\end{figure}
%%%%%%%%%%%%%%%%%%%%%%%%%%%%%%%%%%%%%%%%%%%%%%%%%%%%%%%%%%%%%%%

\subsection{Role of $\epsilon_{BD}$, and the $\sigma_8$-tension}\label{sec:eBDands8}

Next we study the effect of $\epsilon_{BD}\neq0$. We know that when we introduce the matter bispectrum information from BOSS \cite{Gil-Marin:2016wya} (which definitely prefers a lower amount of structure in the universe than the data reported in \cite{Alam:2016hwk}) in our fitting analyses, we find a stronger signal for negative values of  $\epsilon_{BD}$ (cf. Sec. \ref{sec:NumericalAnalysis}). When $\epsilon_{BD}\ne 0$ equation \eqref{eq:DC} is not valid anymore, since it was derived under the assumption of $\varphi=$const.  {We start once more from the exact perturbations equation for the matter density contrast  in the linear regime within the BD-gravity, i.e. Eq.\,\eqref{eq:ExactPerturScaleFactor}.  Let us consider its form within the approximation $|\epsilon_{BD}|\ll 1$ (as it is preferred by the data, see e.g. Table 3). We come back to the BD-field equations  \eqref{eq:Friedmannequation} and \eqref{eq:pressureequation} and apply such approximation.  In this way Eq.\,\eqref{eq:ExactPerturScaleFactor} can be expanded linearly in $\eBD$ as follows (primes are still denoting derivatives with respect to the scale factor)}:
\begin{equation}\label{eq:DC2}
\delta_m^{\pp}+\left[\frac{3}{2a}\left(1-w(a)\right)+\frac{\mathcal{F}^\p}{2}-\frac{\varphi^\p}{2\varphi}\right] \delta_m^\p-\frac{3}{2a^2}\frac{\rho_m(a)}{\rho(a)}\delta_m\left(1+\frac{\epsilon_{BD}}{2}-\mathcal{F}\right)=0\,,
\end{equation}
where we have defined
\begin{equation}\label{eq:F}
\mathcal{F}=\mathcal{F}\left(\frac{\varphi^\p}{\varphi}\right)=-a\frac{\varphi^\p}{\varphi}+\frac{\omega_{BD}}{6}a^2\left(\frac{\varphi^\p}{\varphi}\right)^2\,.
\end{equation}
In particular, we have expanded the effective gravitational coupling  \eqref{eq:LocalGNa} linearly as in \eqref{eq:LocalGN2}.
As we can show easily, the expression \eqref{eq:F} can be treated as a perturbation, since it is proportional to $\epsilon_{BD}$. To prove this, let us borrow the solution for the matter-dominated epoch (MDE) derived from the analysis of fixed points presented in Appendix B. Because the behavior of the BD-field towards the attractor at the MDE is governed by a power law of the form $\varphi\sim a^{\eBD}$ (cf. Eq.\,\eqref{eq:psiMDE}), we obtain
\begin{eqnarray}\label{eq:ximatter}
a\frac{\varphi^\p}{\varphi} &=& \epsilon_{BD}+\mathcal{O}(\epsilon^2_{BD})\,,\\
\mathcal{F}&=&-\frac{5}{6}\epsilon_{BD}+\mathcal{O}(\epsilon^2_{BD})\simeq -\frac{5}{6}\epsilon_{BD}\,;\,\ \ \ \ \ \ \ \ \mathcal{F}^\p=\mathcal{O}(\epsilon^2_{BD})\simeq 0\,.\label{eq:Fximatter}
\end{eqnarray}
This proves our contention that the function $\mathcal{F}$ in  \eqref{eq:F} is of order $\eBD$ and its effects can be treated as a perturbation to the above formulas.
Incidentally, the relative change of $\varphi$ does not depend on $\varphi$ itself.
%
%\begin{equation}
%\frac{2\epsilon_{BD}+4}{3\epsilon_{BD}+2}= 1+\frac{\epsilon_{BD}}{2}+\mathcal{O}(\epsilon_{BD}^2)\,.
%\end{equation}
%
From the definition of $\mathcal{F}$ we can now refine the old Friedmann's equation \eqref{eq:H1} or \eqref{eq:H2} (only valids for $\eBD=0$) as follows. Starting from Eq.\,\eqref{eq:Friedmannequation}, it is easy to see that it can be cast in the Friedmann-like form:
\begin{equation}\label{eq:FriedmannWithF}
H^2(a)=\frac{8\pi G_N}{3\varphi (a) (1-\mathcal{F}(a))}\,\rho(a).
\end{equation}
Despite $\mathcal{F}(a)$ evolves with the expansion, as shown by \eqref{eq:F}, it is of order $\eBD$ and evolves very slowly. In this sense, Eq.\,\eqref{eq:FriedmannWithF} behaves approximately as an ${\cal O}(\eBD)$ correction to Friedmann's equation \eqref{eq:H1}.
Setting $a=1$, the value of the current Hubble parameter satisfies
\begin{equation}\label{eq:FriedmannWithFAtPresent}
H_0^2=\frac{8\pi G_N}{3\varphi_0 (1-\mathcal{F}_0)} \rho^0,
\end{equation}
where $\varphi_0 \equiv \varphi (a=1)$ and $\mathcal{F}_0 \equiv \mathcal{F}(a=1)$. The above equation implies that $\rho^0=\rho_c^0 \varphi_0 (1-\mathcal{F}_0 )$. We may now rewrite \eqref{eq:FriedmannWithF} in the suggestive form:
%\begin{equation}\label{eq:H2withF}
%H^2(a)= H_0^2\left[1+\hat{\Omega}_{m}(a)(a^{-3}-1)+\hat{\Omega}_\gamma(a)(a^{-4}-1)+\hat{\Omega}_\nu(a)-\hat{\Omega}_\nu\right]\,,
%\end{equation}
\begin{equation}\label{eq:H2withF}
H^2(a)= H_0^2\left[\hat{\Omega}_{m}(a)a^{-3}+\hat{\Omega}_{\gamma}(a)a^{-4}+\hat{\Omega}_\nu(a)+\hat{\Omega}_\CC(a)\right]\,,
\end{equation}
provided we introduce the new `hatted' parameters $\hat{\Omega}_i(a)$, which are actually slowly varying functions of the scale factor:
%%%%%%%%%%%%%%%%%%%%%%%%%%%%%%%%%%%%%%%%%%%%%%%%%%%%%%%%%%%%%%%
\begin{figure}[t!]
\begin{center}
\label{fig:varphi}
\includegraphics[width=4.2in, height=2.8in]{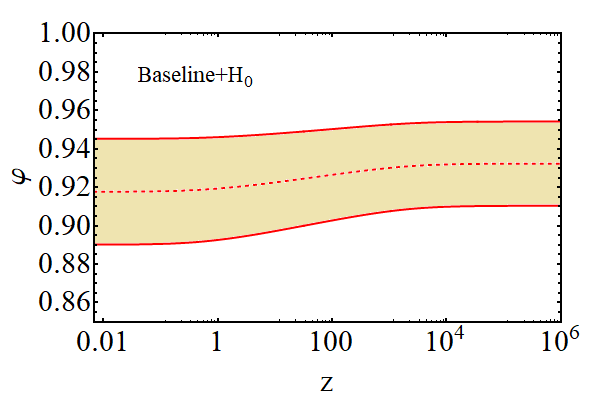}
\caption{\scriptsize{Exact numerical analysis of the evolution of $\varphi$ as a function of the redshift across the entire cosmic history, starting from the radiation-dominated epoch up to our time. We use here the values of the BD-$\CC$CDM Baseline+$H_0$ dataset indicated in the figure itself (cf. Secs. 7 and 8). In particular, $\epsilon_{BD}=-0.00199^{+0.00142}_{-0.00147}$. The band around the central (dotted) curve shows the computed $1\sigma$ uncertainty from the Markov chains of our statistical analysis. }}
\end{center}
\end{figure}
%%%%%%%%%%%%%%%%%%%%%%%%%%%%%%%%%%%%%%%%%%%%%%%%%%%%%%%%%%%%%%%
\begin{equation}\label{eq:hatOmega}
\hat{\Omega}_i (a)=\frac{\Omega_i}{\varphi (a)}\frac{1}{1-\mathcal{F}(a)}\simeq \frac{\Omega_i}{\varphi (a)}\left(1+\mathcal{F}(a)\right) \,,
\end{equation}
with $\Omega_i = \rho^0_i /\rho_c^0$ as previously. These functions also satisfy, exactly, the canonical sum rule at present: %
$\sum_i\hat{\Omega}_i(a=1)=1$.
For $\eBD=0$, the hatted parameters reduce to the old tilded ones \eqref{eq:tildeOmegues}, $\hat{\Omega}_i=\tilde{\Omega}_i$, and for typical values of $|\epsilon_{BD}|\sim \mathcal{O}(10^{-3})$ the two sets of parameter differ by $\mathcal{O}(\epsilon_{BD})$ only:
\begin{equation}\label{eq:hatOmega2}
\hat{\Omega}_i (a)=\frac{\Omega_i}{\varphi}+\mathcal{O}(\epsilon_{BD})=\tilde{\Omega}_i+\mathcal{O}(\epsilon_{BD}) \,.
\end{equation}
From \eqref{eq:DC2} we obviously recover  the previous Eq.\,\eqref{eq:DC} in the limit $\epsilon_{BD}\to 0$, and it is easy to see that for non-null values of $\epsilon_{BD}$ the density contrast acquires a dependence on the ratio $\varphi^\p/\varphi$ and its derivative, so it is sensitive to the relative change of $\varphi$ with the expansion.  Its time evolution is now possible by virtue of the third BD-field equation \eqref{eq:FieldeqPsi}, which can be expanded linearly in $\eBD$ in a similar way. After some calculations, we find
\begin{equation}\label{eq:FieldeqPsiapprox}
\varphi^\pp+\frac{1}{2a}\left(5-3w(a)\right)\varphi^\p=\frac{3\eBD}{2a^2}(1-3w(a))\varphi\,.
\end{equation}
{For $\eBD=0$ we recover the solution $\varphi=$const. In the radiation dominated epoch (RDE), $w\simeq 1/3$,  the \textit{r.h.s.} vanishes and in this case $\varphi$ need not be constant.  It is easy to see that the exact solution of this equation in that epoch is
\begin{equation}\label{eq:varphiRDE}
\varphi(a)=\varphi^{(0)}+\frac{\varphi^{(1)}}{a}\,,
\end{equation}
for arbitrary constants $\varphi^{(i)}$. The variation during the RDE is therefore very small since the dominant solution is a constant and the variation  comes only through a decaying mode $1/a\sim t^{-1/2}$ ($t$ is the cosmic time). For the MDE (for which $w=0$) there is some evolution, once more with a decaying mode but then through a sustained logarithmic term:
\begin{equation}\label{eq:varphiMDE}
\varphi(a)\sim \varphi^{(0)}\left(1+\eBD\ln a\right)+ \varphi^{(1)}a^{-3/2}\rightarrow \varphi^{(0)} \left(1+\eBD\ln a\right) \,,
\end{equation}
where coefficient $\varphi^{(0)}$ is to be adjusted from the boundary conditions between epochs\footnote{Approximate solutions to the BD-field equations for the main cosmological variables  in the different epochs are discussed in Appendix \ref{AppendixA}.}. The dynamics of $\varphi$ for $\eBD\neq0$  is actually mild in all epochs since $\eBD$  on the \textit{r.h.s.} of Eq.\,\eqref{eq:FieldeqPsiapprox} is small.
However mild it might be, the  dynamics of $\varphi$ modifies both the friction and Poisson terms in Eq. \eqref{eq:DC2}, and it is therefore  of pivotal importance to understand what are the changes that are induced by positive and negative values of $\epsilon_{BD}$ on these terms during the relevant epochs of the structure formation history.  An exact (numerical) solution is displayed in Fig. 4, where we can see that $\varphi$ remains within the approximate interval  $0.918\lesssim\varphi\lesssim 0.932$ for the entire cosmic history (starting from the RDE up to our time).  This plot has been obtained from the overall numerical fit performed to the observational data used in this analysis within one of the BD-$\CC$CDM baseline datasets considered (cf. Sec.\,\ref{sec:NumericalAnalysis}). The error band around the main curve includes the $1\sigma$-error computed from our statistical analysis.  Two very important things are to be noted at this point: on the one hand the variation of $\varphi$  is indeed small, and on the other hand $\varphi<1$, and hence $G(\varphi)>G_N$ for the whole cosmic span.}

%%%%%%%%%%%%%%%%%%%%%%%%%%%%%%%%%%%%%%%%%%%%%%%%%%%%%%%%%%%%%%%
\begin{figure}[t!]
\begin{center}
\label{fig:terms_deltaEq}
\includegraphics[width=6.2in, height=2.3in]{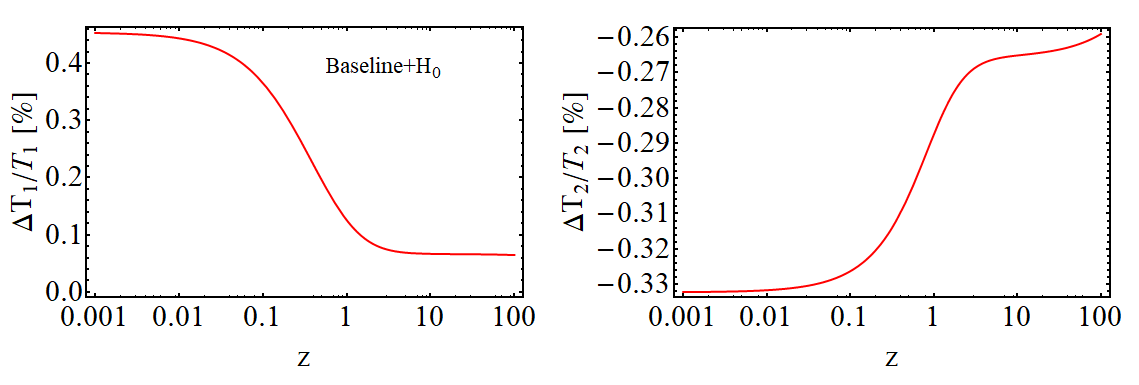}
\caption{\scriptsize{{\it Left plot:} Relative difference between the friction term $T_1$ of Eq. \eqref{eq:DC2}, $\delta_m^\pp+T_1\delta_m^\p-T_2\delta_m=0$, using the best-fit values of the BD-$\CC$CDM model obtained with the Baseline+$H_0$ dataset, i.e. with $\epsilon_{BD}=-0.00199$, and the case with $\epsilon_{BD}=0$ (cf. Table 3 for the values of the other parameters); {\it Right plot:} The same, but for the last term of Eq. \eqref{eq:DC2}, $T_2$ (Poisson term). We can clearly appreciate that negatives values of $\epsilon_{BD}$ produce a higher Hubble friction, i.e. a higher $T_1$, and a lower $T_2$, {\it w.r.t.} the $\epsilon_{BD}=0$ case. Both things lead to a decrease of $\delta_m$. The two plots have been obtained with our modified version of \texttt{CLASS}. See the main text for further details.}}
\end{center}
\end{figure}
%%%%%%%%%%%%%%%%%%%%%%%%%%%%%%%%%%%%%%%%%%%%%%%%%%%%%%%%%%%%%%%

We can start considering what is the influence of the scalar field dynamics on the perturbations during the pure MDE. Using the relations \eqref{eq:ximatter} and \eqref{eq:Fximatter} in \eqref{eq:DC2} we find ({setting $w=0$, $\rho\simeq \rho_m$ and neglecting $\rL$ in the MDE}):
\begin{equation}\label{eq:DCmatter}
\delta_m^{\pp}+\frac{\delta_m^\p}{2a}(3-\epsilon_{BD})-\frac{3}{2a^2}\delta_m\left(1+\frac{4\epsilon_{BD}}{3}\right)=0\,.
\end{equation}
{If $\epsilon_{BD}<0$ the Poisson term (the last in the above equation) decreases and, on top of that, the Hubble friction increases {\it w.r.t.} the case with $\varphi=$const. (or the GR-$\CC$CDM model, if we consider the same energy densities). Both effects help to slow down the structure formation in the universe. Of course, if $\epsilon_{BD}>0$ the opposite happens.   This is confirmed by solving explicitly Eq.\eqref{eq:DCmatter}.  Despite an exact solution to Eq.\,\eqref{eq:DCmatter} can be found, it suffices to quote it at $\mathcal{O}(\epsilon_{BD})$ and neglect the $\mathcal{O}(\epsilon^2_{BD})$ corrections.  The growing and  decaying modes at leading order read  $\delta^+_m(a)\sim a^{1+\eBD}$ and  $\delta^-_m(a)\sim a^{-\frac12(3+\eBD)}$, respectively.  The latter just fades soon into oblivion  and the former explains why negative values of $\epsilon_{BD}$ are favored by the data on RSD, since $\eBD<0$ obviously slows down the rate of structure formation and hence acts as an effective (positive) contribution to the vacuum energy density\footnote{This feature was already noticed in the preliminary treatment of Ref.\,\cite{Perez:2018qgw} for the BD theory itself, and it was actually pointed out as a general feature of the class of Running Vacuum Models (RVM),  which helps to cure the $\sigma_8$-tension\,\cite{Gomez-Valent:2018nib,Gomez-Valent:2017idt}. This is remarkable, since the RVM's  turn out  to mimic BD-gravity, as first noticed in \cite{Peracaula:2018dkg}. For a summary, see  Appendix \ref{sec:RVMconnection}.}.  The preference for negative values of $\eBD$ is especially clear when the RSD data include  the matter bispectrum information, which tends to accentuate the slowing down of the growth function, as noted repeatedly in a variety of  previous works\,\cite{Sola:2016ecz , Sola:2017jbl, Sola:2017znb,sola2017first}.  We may clearly appraise this feature also in the present study, see e.g. Tables 3-4 (with spectrum {\it and} bispectrum) and 5 (with spectrum but no bispectrum), where the $\sigma_8$ value is in general well-behaved ($\sigma_8\simeq 0.8$) in the BD-$\CC$CDM framework when $\eBD<0$, but it is clearly reduced (at a level $\sigma_8\simeq 0.78-0.79$)  in the presence of bispectrum data.  And in both cases the value of $H_0$  is in the range of $70-71$ km/s/Mpc. Most models trying to explain both tensions usually increase $\sigma_8$ substantially  ($0.82-0.85$). }

We can also study the pure vacuum-dominated epoch (VDE) in the same way. In this case $\varphi\sim a^{2\eBD}$ (cf. Appendix \ref{AppendixB}), and hence
\begin{equation}\label{eq:xivacuum}
a\frac{\varphi^\p}{\varphi} = 2\epsilon_{BD}+\mathcal{O}(\epsilon^2_{BD})\simeq 2\epsilon_{BD}\,,
\end{equation}
again with $\mathcal{F}^\p=\mathcal{O}(\epsilon^2_{BD})\simeq 0$. The Poisson term can be neglected in this case since $\rho_m\ll\rho\simeq \rL$, and hence,
\begin{equation}\label{eq:DCvacuum}
\delta_m^{\pp}+\frac{\delta_m^\p}{a}(3-\epsilon_{BD})=0\,.
\end{equation}
When the vacuum energy density rules the expansion of the universe, there is a stable constant mode solution $\delta_m=$const. and a decaying mode that decreases faster than in the GR scenario if $\epsilon_{BD}<0$, again due to the fact that the friction term is in this case larger than in the standard picture, specifically the latter reads $\delta^-_m(a)\sim a^{-2+\epsilon_{BD}}$ in the ${\cal  O}(\eBD)$ approximation.
%%%%%%%%%%%%%%%%%%%%%%%%%%%%%%%%%%%%%%%%%%%%%%%%%%%%%%%%%%%%%%%
\begin{figure}[t!]
\begin{center}
\label{fig:pkepsilon}
\includegraphics[width=6in, height=2.1in]{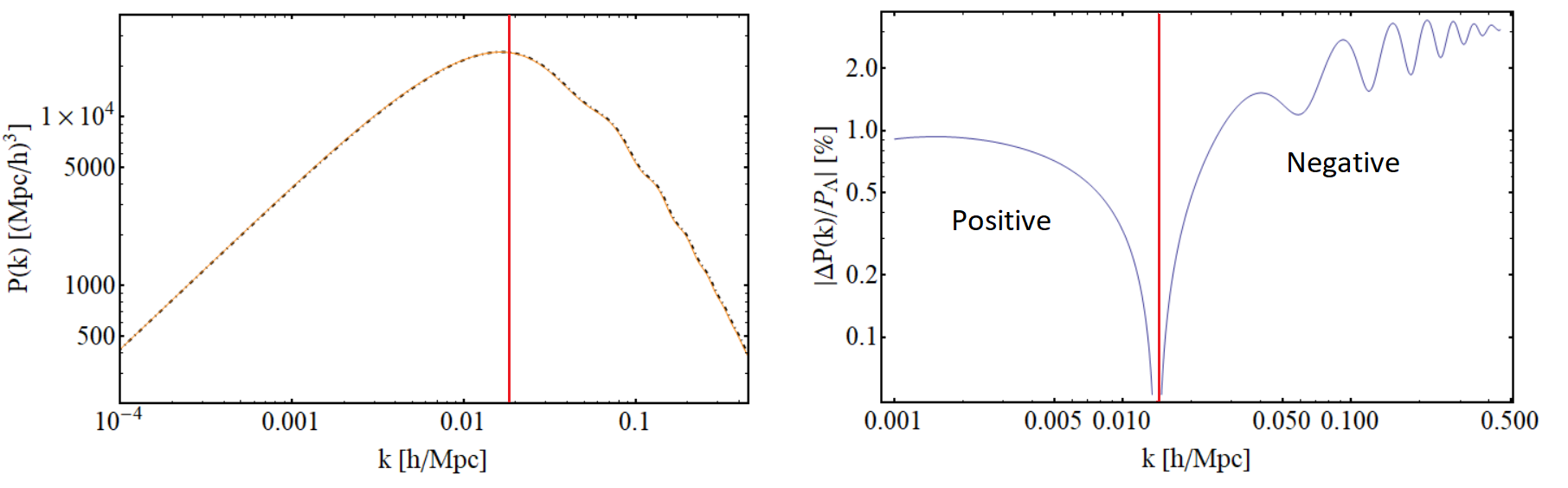}
\caption{\scriptsize{As in Fig. 3, but comparing now the $\CC$CDM baseline configuration defined there with the case $\epsilon_{BD}=-0.003$ (and $\varphi_{ini}=1$) of the BD-$\CC$CDM model.}}
\end{center}
\end{figure}
%%%%%%%%%%%%%%%%%%%%%%%%%%%%%%%%%%%%%%%%%%%%%%%%%%%%%%%%%%%%%%%
The analytical study of the transition between the matter and vacuum-dominated epochs is more difficult, but with what we have already seen it is obvious that the amount of structure generated also in this period of the cosmic expansion will be lower than in the $\varphi=$const. case if $\epsilon_{BD}$ takes a negative value. In Fig. 5 we show this explicitly.

From this analysis it should be clear that if $\epsilon_{BD}<0$ there is a decrease of the matter density contrast for fixed energy densities when compared with the GR-$\Lambda$CDM scenario, and also with the BD scenario with $\varphi=const$. In Fig. 6 we can see this feature directly in the matter power spectrum, which is  seen to be suppressed with respect to the case $\epsilon_{BD}=0$ {for those scales that are relevant for the RSD, i.e. within the range of wave numbers  $0.01 h {\rm Mpc^{-1}}\lesssim k\lesssim 0.1 h {\rm Mpc^{-1}}$  (corresponding to distance scales roughly between a few dozen to a few hundred Mpc)}. However, we still don't know whether these negative values of $\epsilon_{BD}$ can also be accommodated by the other datasets.

%%%%%%%%%%%%%%%%%%%%%%%%%%%%%%%%%%%%%%%%%%%%%%%%%%%%%%%%%%%%%%%
\begin{figure}[t!]
\begin{center}
\label{fig:Cls-epsilon}
\includegraphics[width=6in, height=4in]{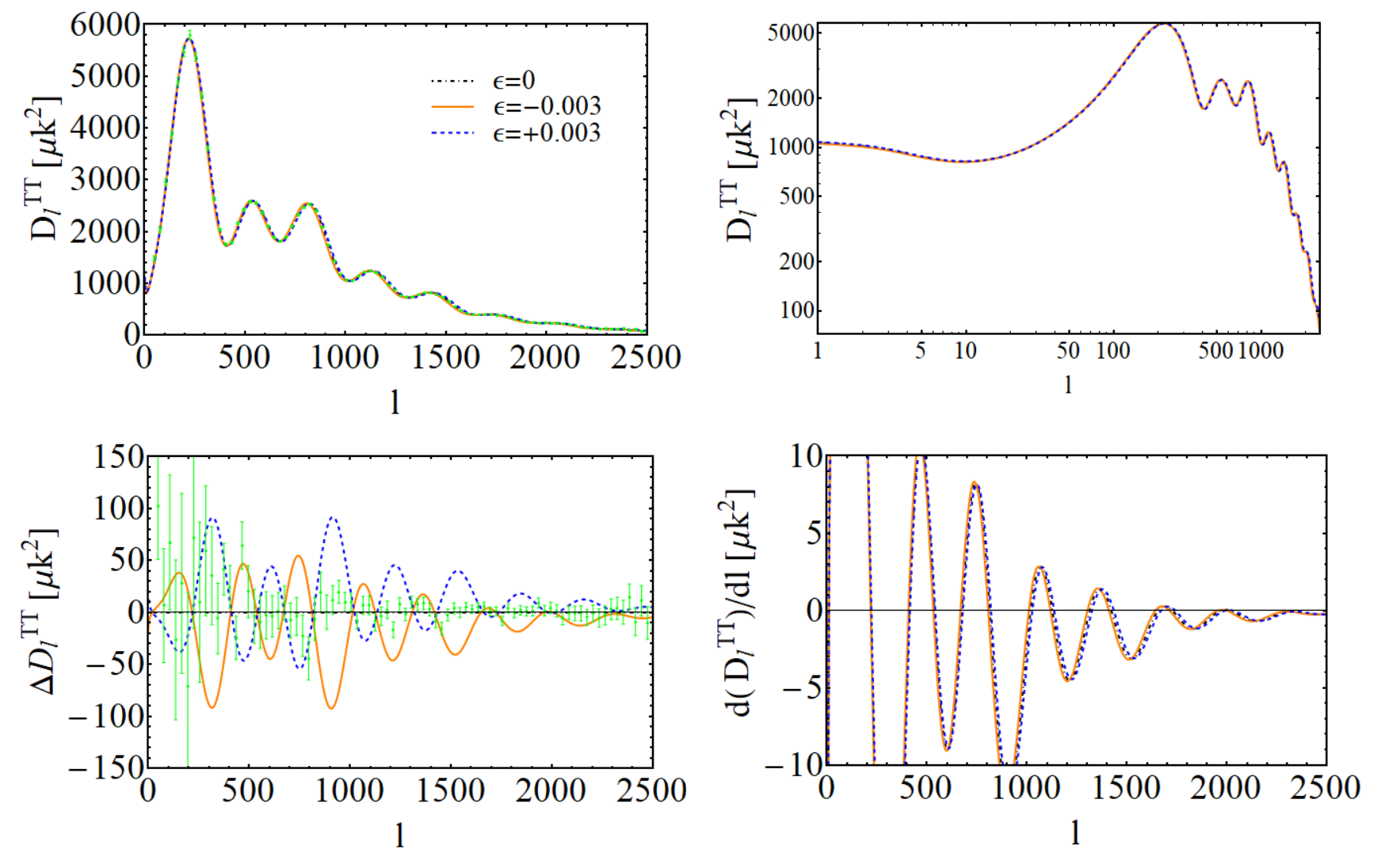}
\caption{\scriptsize{As in Fig. 1, but comparing the GR-$\CC$CDM baseline scenario (equivalent to the BD-$\CC$CDM one for  $\epsilon_{BD}=0$, $\varphi=1$) with the case $\epsilon_{BD}=\pm 0.003$, using as initial condition $\varphi_{ini}=1$.}}
\end{center}
\end{figure}
%%%%%%%%%%%%%%%%%%%%%%%%%%%%%%%%%%%%%%%%%%%%%%%%%%%%%%%%%%%%%%%

To check this, let us recall from our discussion above (see also Appendices \ref{AppendixA} and \ref{AppendixB} for more details) that the evolution of the BD-field takes place basically during the MDE.  In the RDE the scalar field is essentially frozen once the decaying mode becomes irrelevant, and although in the late-time universe $\varphi$ evolves faster than in the MDE (compare \eqref{eq:ximatter} and \eqref{eq:xivacuum}) it remains almost constant in the redshift range $z\lesssim \mathcal{O}(1)$, in which all the non-CMB data points lie, particularly the LSS data.  {Let us note that the typical values of $\eBD$  fitted from the overall set of data used in our analysis (cf. Sec. \ref{sec:MethodData})  place that parameter  in the ballpark  of $|\epsilon_{BD}|\sim \mathcal{O}(10^{-3})$ (cf. Tables 3-5, for example).  Schematically, we can think that the field takes a value $\varphi_{ini}$ during the RDE, then if $\epsilon_{BD}<0$ it decreases an amount $\Delta\varphi_{ini}<0$  with respect to its original value $\varphi_{ini}$, during the MDE, and  finally $\varphi^{0}\approx\varphi_{ini}+\Delta\varphi_{ini}$, with $\varphi(z)\approx\varphi^0$ at $z\lesssim\mathcal{O}(1)$}. Thus, Eq. \eqref{eq:H1} still applies in good approximation during the RDE and the late-time universe, but with two different values of the cosmological Newton's coupling in the two (widely separated) cosmological eras. Taking these facts into account it is easy to understand why SNIa data are not able to tell us anything about $\epsilon_{BD}$ when used alone, since again $\varphi^{0}$ is fully degenerated with the absolute magnitude parameter $M$ of the supernovae. This does not mean, though, that SNIa data cannot tell us anything about the $H_0$-tension when they are considered together with other datasets that do provide constraints on $H_0$, since the constraints that SNIa impose on $\Om$ help to break degeneracies present in the other datasets and to tighten the allowed region of parameter space. H0LICOW and CCH data, for instance, will allow us to put constraints again on $\Om$ and also on $H_0$ (hence on $\rho^{0}/\varphi^{0}$). BAO will constrain $\rho_b^{0}$, $\Om$, and now also $\Delta\varphi_{ini}/\varphi_{ini}$, which is proportional to $\epsilon_{BD}$. For instance, $r_s H(z)\propto \left(1-\frac{\Delta\varphi_{ini}}{2\varphi_{ini}}\right)$.

The BD effect caused on the temperature spectrum of CMB anisotropies is presented in the fourfold plot in Fig. 7. Due to the fact that now we have $\epsilon_{BD}\ne 0$, $\Delta\varphi\ne 0$,  a small shift in the location of the peaks is naturally generated. In the right-bottom plot of such figure one can see that negative values of $\epsilon_{BD}$ move the peaks slightly towards lower multipoles, and the other way around for positive values of this parameter. It is easy to understand why.  To start with, let us remark that we have produced all the curves of this plot using the same initial condition $\varphi_{ini}=1$ and the same $\rho^0_b$, $\rho^0_{cdm}$ and $\rho_\Lambda$ and, hence, fixing in the same way the complete evolution of the energy densities for the different plots in that figure. This means that the differences in the Hubble function can only be due to differences in the evolution of the BD scalar field. The modified expansion histories produce changes in the value of $\theta_*=r_s/D_{A,rec}$ (with $D_{A,rec}$ being the angular diameter distance to the last scattering surface), so also in the location of the peaks. If $\epsilon_{BD}<0$,  $\varphi$ decreases with the expansion, so its value at recombination and at present is lower than when $\epsilon_{BD}=0$, and correspondingly $G(\varphi)$ will be higher. Because of this, the value of the Hubble function will be larger, too, and the cosmological distances lower, so the relation between $\theta_*(\epsilon_{BD}\ne 0)$ and  $\theta_*(\epsilon_{BD}= 0)$ can be written as follows:
\begin{equation}
\theta_*(\epsilon_{BD}\ne 0)=\frac{r_s(\epsilon_{BD}\ne 0)}{D_{A,rec}(\epsilon_{BD}\ne 0)}=\frac{X\cdot r_s(\epsilon_{BD}= 0)}{Y\cdot D_{A,rec}(\epsilon_{BD}=0)}= \frac{X}{Y}\,\theta_*(\epsilon_{BD}=0)\,,
\end{equation}
{where the rescaling factors satisfy  $0<X,Y<1$ for $\epsilon_{BD}<0$. As we have already mentioned before, most of the variation of $\varphi$ occurs during the MDE, so the largest length reduction will be in the cosmic stretch from recombination to the present time, and thereby  $Y<X$}. Thus, if $\epsilon_{BD}< 0$ we find $\theta_*(\epsilon_{BD}< 0)>\theta_*(\epsilon_{BD}=0)$ and the peaks of the TT CMB spectrum shifts towards lower multipoles. Analogously, if $\epsilon_{BD}$ is positive $X,Y>1$, with $Y>X$, so $\theta_*(\epsilon_{BD}> 0)<\theta_*(\epsilon_{BD}=0)$ and the peaks move to larger multipoles. It turns out, however, that these shifts, and also the changes in the amplitude of the peaks, can be compensated by small changes in the baryon and DM energy densities, as we will show in Sec. \ref{sec:NumericalAnalysis}.

At this point we would like to recall why in the GR-$\Lambda$CDM concordance model it is not possible to reconcile the local measurements of $H_0$ with its CMB-inferred value. In the concordance model, which we assume spatially flat,  the current value of the cold dark matter density is basically fixed by the amplitude of the first peak of the CMB temperature anisotropies, and $\rho_b^0$ by the relative amplitude of the second and third peaks with respect to the first one. {As a result, even if the cosmological term plays no role in the early universe, one finds that  in order to explain the precise location of the CMB peaks the value of $\rL$ obtained from the matching of the predicted  $\theta_*$ with the measured peak positions determines $\rL$ so precisely that it leaves little margin. This causes a problem since such narrow range of values is not in the right range to explain the value of the current Hubble parameter measured with the cosmic distance ladder technique \cite{Riess:2018uxu,Riess:2019cxk,Reid:2019tiq} and the Strong-Lensing time delay angular diameter distances from H0LICOW \cite{Wong:2019kwg}. In other words,  despite the concordance model fits in a remarkably successful way the CMB and BAO, and also the SNIa data, there is an irreducible internal discordance in the parameter needs to explain with precision both the physics of the early and of the late-time universe. This is of course the very expression of the  $H_0$-tension, to which we have to add the $\sigma_8$ one.}

Cosmographic analyses based on BAO and SNIa data calibrated with the GR-$\Lambda$CDM Planck preferred value of $r_s$ also lead to low estimates of $H_0$. This is the so-called inverse cosmic distance ladder approach, adopted for instance in \cite{Aubourg:2014yra,Bernal:2016gxb,Feeney:2018mkj,Macaulay:2018fxi}. This has motivated cosmologists to look for alternative theoretical scenarios (for instance, the generic class of  EDE  proposals)  able to increase the expansion rate of the universe before the decoupling of the CMB photons and, hence, to lower $r_s$ down.  {This, in principle, demands an increase of the Hubble function at present in order not to spoil the good fit to the BAO and CMB observables.} {Nevertheless, not all the models passing the BAO and CMB constraints and predicting a larger value of $H_0$ satisfy the `golden rule' mentioned in the Introduction, since they can lead e.g. to a worsening of the $\sigma_8$-tension}.  As an example, we can mention some early DE models, e.g. those discussed in \cite{Poulin:2018cxd}. In these scenarios there is a very relevant DE component which accounts for the $\sim 7\%$ of the total matter-energy content of the universe at redshifts $\sim 3000-5000$, before recombination. This allows of course to enhance the expansion rate and reduce $r_s$. After such epoch, the DE decays into radiation. In order not to alter the position of the CMB peaks and BAO relative distances, an increase of the DM energy density is needed. According to \cite{Hill:2020osr}, this leads to an excess of density power and an increase of $\sigma_8$ which is not welcome by LSS measurements, including RSD, Weak-Lensing and galaxy clustering data.  Another example is the interesting modified gravity model analyzed in \cite{Ballesteros:2020sik}, based on changing the cosmological value of $G$ also in the pre-recombination era, thus mimicking an increase of the effective number of relativistic degrees of freedom in such epoch. The additional component gets eventually diluted at a rate faster than radiation in the MDE and it is not clear if an effect is left at present\footnote{{In stark contrast to the model of \cite{Ballesteros:2020sik}, in BD-$\Lambda$CDM cosmology the behavior of the effective $\rho_{BD}$  (acting as a kind of additional DE component during the late universe) mimics pressureless matter during the MDE epoch and modifies the effective EoS of the DE at present, see the next  Sec. \ref{sec:EffectiveEoS} for details.}}. This model also fits the CMB and BAO data well and loosens at some extent the $H_0$-tension, but violates the golden rule of the tension solver, as it spoils the structure formation owing to the very large values of  $\sigma_8\sim 0.84-0.85$ that are predicted (see the discussion in  Sec.\,\ref{sec:Discussion} for more details).

Our study shows that a value of the cosmological gravitational coupling about $\sim 10\%$ larger than $G_N$ can ameliorate in a significant way the $H_0$-tension, while keeping the values of all the current energy densities very similar to those found in the GR-$\Lambda$CDM model. If, apart from that, we also allow for a very slow running (increase) of the cosmological $G$ triggered by negative values of order $\epsilon_{BD}\sim -\mathcal{O}(10^{-3})$, we can  mitigate at the same time the $\sigma_8$-tension when only the CMB TT+lowE anisotropies are considered. When the CMB polarizations and lensing are also included in the analysis, then $\sigma_8$ is kept at the GR-$\Lambda$CDM levels, and the sign of $\epsilon_{BD}$ is not conclusive.  {In all situations we can preserve the golden rule}. We  discuss in detail the numerical  results of our analysis  in Sec. \ref{sec:NumericalAnalysis}.

\section{Effective equation of state of the dark energy in the BD-$\CC$CDM model}\label{sec:EffectiveEoS}
\noindent
{Our aim in this section is to write down the Brans-Dicke cosmological equations  \eqref{eq:Friedmannequation}-\eqref{eq:FieldeqPsi} in the context of what we may call the ``effective GR-picture''. This means to rewrite them  in such a way that they can be thought of as an effective model within the frame of GR, thus providing a parametrized departure from GR at the background level. We will see that the main outcome of this task, at least qualitatively, is that the BD-$\CC$CDM model  (despite it having a constant vacuum energy density $\rL$) appears as one in the GR class, but with a dynamical DE rather than a CC. The dynamics of such an effective form of DE is a function of the BD-field $\varphi$.  We wish to compute its effective EoS.  In order to proceed, the first step is to  rewrite Eq.\, \eqref{eq:Friedmannequation} \`a la  Friedmann:}
\begin{equation}\label{eq:BDFriedmann}
3H^2 = 8\pi{G_N}(\rho + \rvphi)\,,
\end{equation}
where $\rho$ is the total energy density as defined previously (coincident with that of the GR-$\CC$CDM model),  and  $\rvphi$ is the additional ingredient that is needed, which reads
\begin{equation}\label{rhoBD}
\rvphi\equiv \frac{3}{8\pi{G_N}}\left(H^2\dvphi - H\dot{\varphi} + \frac{\omega_{BD}}{6}\frac{\dot{\varphi}^2}{\varphi}\right).
\end{equation}
Remember the definition $\varphi(t) \equiv G_N\psi(t)$ made in \eqref{eq:definitions}, and we have now introduced
\begin{equation}\label{eq:Deltaphi}
\dvphi(t)\equiv 1-\varphi(t)\,,
\end{equation}
which tracks the small departure of $\varphi$ from one and hence of $G(\varphi)$ from $G_N$  (cf. Sec. \ref{sec:BDgravity}). Note that $\varphi=\varphi(t)$ evolves in general with the expansion, but very slowly  since $\eBD$ is presumably fairly small.

From the above  Eq.\,\eqref{rhoBD} it is pretty clear that we have absorbed all the terms beyond the $\Lambda$CDM model into the expression of $\rvphi$. While it is true that we define this quantity as if it were an energy density, it is important to bear in mind that it is not associated to any kind of particle, it is just a way to encapsulate those terms that are not present in the standard model.  This quantity, however, satisfies a local conservation law as if it were a real energy density, as we shall see in a moment.  From the generalized Friedmann equation \eqref{eq:BDFriedmann} and the explicit expression for $\rvphi$ given above we can write down the generalized cosmic sum rule verified by the BD-$\CC$CDM model in the effective GR-picture:
\begin{equation}\label{eq:SumRuleBDexact}
\Omega_m+\Omega_r+\Omega_\CC+\Omega_\varphi=1\,,
\end{equation}
where the $\Omega_i$ are the usual (current) cosmological parameters of the concordance $\CC$CDM, whereas $\Omega_\varphi$ is the additional one that parametrizes the departure of the  BD-$\CC$CDM model from the  GR-$\CC$CDM  in the context of the GR-picture, and reads
\begin{equation}\label{eq:OmegaBD}
\Omega_{\varphi}=\frac{\rho_{\varphi}^0}{\rco}.
\end{equation}
Notice that the above sum rule is exact and it is different from that in Eq.\,\eqref{eq:SumRuleBD} since the latter is only approximate for the case when $\eBD=0$ or very small.  These are two different pictures of the same BD-$\CC$CDM model. The modified cosmological parameters (\ref{eq:tildeOmegues}) depend on $\varphi$ whereas here the $\varphi$-dependence has been fully concentrated on $\Omega_\varphi$.  It is interesting to write down the exact equation \eqref{eq:SumRuleBDexact} in the form
\begin{equation}\label{eq;OMegaBD}
\Omega_m+\Omega_r+\Omega_\CC=1-\Omega_{\varphi}= 1-\Delta\varphi_0+\frac{\dot{\varphi}_0}{H_0}-\frac{\wBD}{6}\frac{\dot{\varphi}_0^2}{H_0^2\varphi_0}\,.
\end{equation}
in which $\varphi_0=\varphi(z=0)$ and $\dot{\varphi}_0=\dot{\varphi}(z=0)$. For $\eBD\simeq 0$ we know that $\varphi\simeq$const. and we can neglect the time derivative terms and then we find the approximate form $\Omega_m+\Omega_r+\Omega_\CC=1-\Omega_{\varphi}\simeq  1-\Delta\varphi_0$.
This equation suggests that a value of $\Delta\varphi_0\neq 0$ would emulate the presence of  a small  fictitious spatial curvature in the GR-picture. See e.g.\cite{Park:2019emi,Khadka:2020vlh,Cao:2020jgu} and references therein for the study of a variety models explicitly involving spatial curvature.

The second step in the process of constructing the GR-picture of the BD theory  is to express \eqref{eq:pressureequation} as in the usual pressure equation for GR, and this forces us to define a new pressure quantity $p_{\varphi}$ associated to $\rvphi$. We find
\begin{equation}
2\dot{H} + 3H^2 = -8\pi{G_N}(p + \pvphi),
\end{equation}
with
\begin{equation} \label{pBD}
p_{\varphi}\equiv \frac{1}{8\pi{G_N}}\left(-3H^2\dvphi -2\dot{H}\dvphi  + \ddot{\varphi} + 2H\dot{\varphi} + \frac{\omega_{BD}}{2}\frac{\dot{\varphi}^2}{\varphi}\right).
\end{equation}
{On the face of the above definitions \eqref{rhoBD} and \eqref{pBD}, we can now interpret the BD theory as an effective theory within the frame of General Relativity, which deviates from it an amount $\dvphi$. Indeed, for $\dvphi=0$ we have $\rvphi=\pvphi=0$ and we recover GR.  Mind that $\dvphi=0$ means not only that $\varphi=$const (hence that $\eBD=0$, equivalently $\wBD \to\infty$), but also that that constant is exactly $\varphi=1$.  In such case $\Geff$ is also constant and $\Geff=G_N$ exactly. The price that we have to pay for such a GR-like description of the BD model is the appearance of the fictitious BD-fluid with energy density $\rvphi$ and pressure $\pvphi$, which complies with the following conservation equation throughout the expansion of the universe}\,\footnote{{The new `fluid' that one has to add to GR to effectively mimic BD plays a momentous role to explain the $H_0$ and $\sigma_8$-tensions. In a way it mimics the effect of the `early DE' models mentioned in the previous section, except that the BD-fluid persists for the entire cosmic history and is instrumental both in the early as well as in the current universe so as to preserve the golden rule of the tension solver: namely, it either smoothes the two tensions of GR or improves one of them without detriment of the other. }}:
\begin{equation}
\dot{\rho}_{\varphi}+3H(\rho_{\varphi}+p_{\varphi})= 0.
\end{equation}
{One can check that this equation holds after a straightforward calculation, which makes use of the three  BD-field equations \eqref{eq:Friedmannequation}-\eqref{eq:FieldeqPsi}.}
Although at first sight the above conservation equation can be surprising actually it is not, since it is a direct consequence of the Bianchi identity. Let us now assume that the effective BD-fluid can be described by an equation of state like $\pvphi = w_{\varphi}\rvphi$, so
\begin{equation}\label{BDEoS0}
w_{\varphi}=\frac{\pvphi}{\rvphi}=\frac{-3H^2\dvphi -2\dot{H}\dvphi  + \ddot{\varphi} + 2H\dot{\varphi} + \frac{\omega_{BD}}{2}\frac{\dot{\varphi}^2}{\varphi}}{3H^2\dvphi - 3H\dot{\varphi} + \frac{\omega_{BD}}{2}\frac{\dot{\varphi}^2}{\varphi}}.
\end{equation}
The contribution from those terms containing derivatives of the BD-field are subdominant for the whole cosmic history.  We have verified this fact numerically, see Fig. 4. While the variation of $\varphi$ between the two opposite ends of the cosmic history is of $\sim 1.5\%$ and is significant for our analysis, the instantaneous variation is actually negligible. Thus,  $H\dot{\varphi}$ and  $\ddot{\varphi}$  are both much smaller than $\dot{H}\dvphi$,  and  in this limit we can approximate \eqref{BDEoS0} very accurately as
\begin{equation}\label{BDEoS}
w_{\varphi}(t) \simeq -1 - \frac{2}{3}\frac{\dot{H}}{H^2}\,\ \ \ \ \ \ ({\rm for}\ H\dot{\varphi}, \ddot{\varphi}\ll \dot{H}\dvphi)\,.
\end{equation}
This EoS turns  out to be the standard total EoS of the  $\Lambda$CDM, which boils down to the EoS corresponding to the different epochs of the cosmic evolution (i.e. $w=1/3, 0, -1$ for RDE, MDE and VDE).  This means that the EoS of the  BD-fluid mimics these epochs.   We can go a step further and define not just the BD-fluid but the combined system of the BD-fluid and the vacuum energy density represented by the density $\rL$ associated to the cosmological constant.  {We define the following effective EoS for such combined fluid:}
%%%%%%%%%%%%%%%%%%%%%%%%%%%%%%%%%%%%%%%%%%%%%%%%%%%%%%%%%%%%%%%
\begin{figure}[t!]
\begin{center}
\label{fig:weff_varphi}
\includegraphics[width=6.8in, height=2.8in]{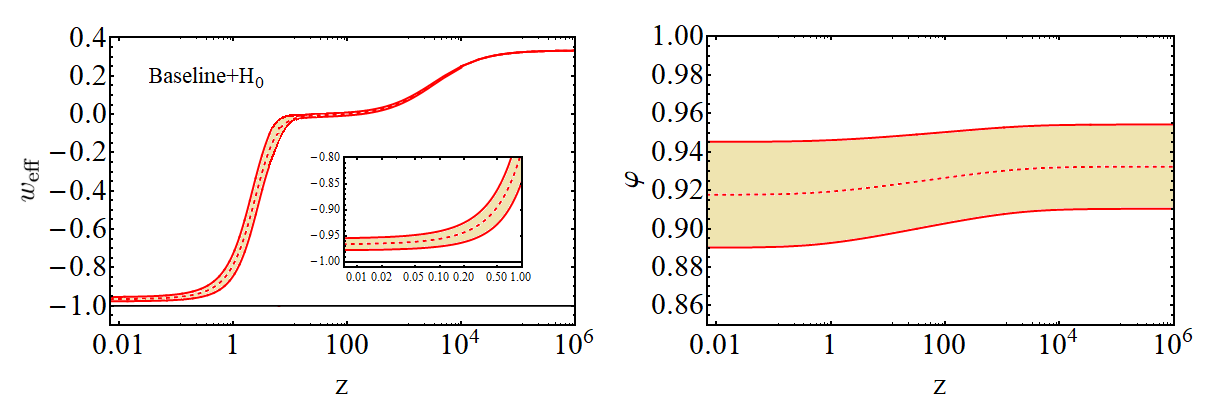}
\caption{\scriptsize{{\it Left plot:} Effective equation of state of the DE in the BD-$\CC$CDM model as a function of the redshift. The inner plot magnifies the region around our time. We can see that the BD model mimics quintessence with a significance of more than $3\sigma$; {\it Right plot:} It shows once more Fig.\, 4 in order to ease the comparison of the EoS evolution, which is associated to the evolution of the BD-field $\varphi$ -- cf. Eq.\eqref{effEoS}. The shadowed bands in these plots correspond to the 1$\sigma$ regions}.}
\end{center}
\end{figure}
%%%%%%%%%%%%%%%%%%%%%%%%%%%%%%%%%%%%%%%%%%%%%%%%%%%%%%%%%%%%%%%
%
\begin{equation}\label{effEoS}
\weff\equiv \frac{p_\Lambda + p_{\varphi}}{\rho_\Lambda + \rho_{\varphi}} = -1 + \frac{p_{\varphi} + \rho_{\varphi}}{\rho_\Lambda+ \rho_{\varphi}}=-1+\frac{-2\dot{H}\dvphi+f_1(\varphi,\dot{\varphi},\ddot{\varphi})}{\CC+ 3 H^2\,\dvphi+f_2(\varphi,\dot{\varphi})}\,,
\end{equation}
where the two functions
\begin{equation}\label{eq:f1f2}
  f_1(\varphi,\dot{\varphi},\ddot{\varphi})=\ddot{\varphi}-H\dot{\varphi}+\wBD \frac{\dot{\varphi}^2}{\varphi}\,,\ \ \ \ \ \ \  f_2(\varphi,\dot{\varphi})= -3H\dot{\varphi}+\frac{\wBD}{2}  \frac{\dot{\varphi}^2}{\varphi}
\end{equation}
{involve differentiations with respect to the slowly varying field $\varphi$ and as before they are negligible,  in absolute value,  as compared to $\dot{H}\Delta\varphi$ and  ${H^2}\Delta\varphi$. The effective EoS \eqref{effEoS} is a time-evolving quantity which mimics dynamical DE at low redshifts.  At very high redshifts $z\gg1$, well beyond the DE dominated epoch, we can neglect $\CC$ in the denominator of the EoS and the dominant term is $ 3 H^2\,\dvphi$. Similarly, in the numerator the dominant term is always $-2\dot{H}\dvphi$.  Therefore, at high redshifts the effective EoS \eqref{effEoS} behaves as \eqref{BDEoS} since $\dvphi$ cancels out: $\weff(z)\simeq w_{\varphi}(z)\,\ ( z\gg1)$, which means that it just reproduces the standard EoS of the GR-$\CC$CDM.}
The exact EoS \eqref{effEoS}, however,  must be computed numerically, and it is displayed in Fig.\,8, together with the numerical plot of $\varphi$ (which we have already shown in Fig.\,4). We have used the Baseline+$H_0$ dataset defined in Sec. \ref{sec:MethodData}.  For a semi-qualitative discussion of the combined EoS it will suffice an analytical approximation, as we did before with $w_{\varphi}$. The most relevant part of $\weff (z)$ as to the possibility of disentangling the dynamical DE effects triggered by the underlying BD model is near the present time ($z<1$). Thus, neglecting the contribution from the functions $f_{1,2}$, but now keeping the $\CC$-term in the denominator of  \eqref{effEoS} we can use the Hubble function of the concordance model and we find the following result at linear order in $\dvphi$:
\begin{equation}\label{BDEoSz0}
\weff(z)\simeq-1-\frac{2\dot{H}\dvphi}{\CC}\simeq -1+\dvphi\,\frac{\Omo}{\OLo}\,(1+z)^3\,,
\end{equation}
where $\Omo$ and $\OLo$ are the current values of the cosmological parameters, which satisfy $\Omo+\OLo=1$  for spatially flat universe.
{As has been stated before, the previous approximate formula is valid only for $z<1$, but it shows very clearly that for $\dvphi>0$ (resp. $<0$) we meet quintessence-like (resp. phantom-like) behavior.  As we have repeatedly emphasized, our analysis points to $\eBD<0$ and hence $\varphi$ decreases with the expansion, remaining smaller than one. From Eq.\,\eqref{eq:Deltaphi} this means $\dvphi>0$  and therefore we find that the effective GR behavior of the BD-$\CC$CDM is quintessence-like.}  We can be more precise at this point.
We have numerically computed the value of the exact function \eqref{effEoS} at $z=0$, taking into account the contribution from all the terms, in particular the slowly varying functions \eqref{eq:f1f2}, see Tables 3-6. The results obtained from three of the most prominent datasets defined in Sec.\,\ref{sec:MethodData} read as follows:
\begin{align}
&{\bf Baseline}:\quad &\weff(0)=& -0.983^{+0.015}_{-0.014}\\
&{\bf Baseline+H_0}:\quad &\weff(0)=& -0.966^{+0.012}_{-0.011}\\
&{\bf Baseline+H_0+SL}:\quad &\weff(0) =& -0.962\pm 0.011.
\end{align}
As can be seen, there is a non-negligible departure from the constant EoS value $-1$  of the  GR-$\Lambda$CDM, which reaches the $\sim 3\sigma$ c.l. when the prior on $H_0$ from the local distance ladder measurement by SH0ES \cite{Reid:2019tiq} is included in the analysis, and $\sim 3.5\sigma$ c.l. when also the angular diameter distances from H0LICOW \cite{Wong:2019kwg} are taken into account. The effective quintessence EoS $\weff(0)>-1$ is one of the ingredients that allows the BD-$\CC$CDM model to significantly loosen the $H_0$-tension, since it is a direct consequence of having $\varphi<1$ (or, equivalently, $G>G_N$) (cf. Sec. \ref{sec:preview} for details).
\newline
\newline
We have obtained the above results from the equations of motion once a metric was assumed; however, it is possible to obtain all the expressions listed in this section starting from the BD action \eqref{eq:BDaction} itself and then considering the FLRW metric.  {To show this, let us use the dimensionless field $\varphi=G_N\psi$ and the variable $\dvphi$ defined in \eqref{eq:Deltaphi}. }First of all we split the whole action in three pieces
\begin{equation}
S_{BD}[\varphi]=S_{EH}+S_{GR}[\varphi]+S_m,
\end{equation}
where
\begin{equation}
S_{EH}\equiv\int d^4x \sqrt{-g}\left[\frac{R}{16 \pi G_N}-\rho_\Lambda \right],
\end{equation}
is the usual Einstein-Hilbert action, whereas
\begin{equation}
S_{GR}[\varphi] \equiv \int d^4x \sqrt{-g}\frac{1}{16\pi G_N}\left[ -R\dvphi-\frac{\omega_{BD}}{\varphi}g^{\mu \nu}\partial_\nu \varphi \partial_\mu \varphi \right],
\end{equation}
is the action parametrizing the deviation of the BD theory from the  GR-picture expressed  in terms of the scalar field $\varphi$. As expected, for $\varphi=1$ that action vanishes identically. Finally,
\begin{equation}
S_m\equiv\int d^4x \sqrt{-g}\mathcal{L}_m (\chi_i,g_{\mu \nu})
\end{equation}
is the action for the matter fields. Since there is no interaction involving $\varphi$ with other components,  the BD-field $\varphi$  is covariantly conserved, as remarked in Sec. \ref{sec:BDgravity}. In order to compute the energy-momentum tensor and find out the effective density and pressure of the BD-field, we apply the usual definition of that tensor in curved spacetime:
\begin{equation} \label{energy-momentum}
T_{\mu \nu}^{BD}=-\frac{2}{\sqrt{g}}\frac{\delta S_{GR}[\varphi]}{\delta g^{\mu \nu}}\,.
\end{equation}
 After some calculations we arrive at
\begin{equation} \label{BDEMT}
\begin{split}
T_{\mu \nu}^{BD}=\frac{R_{\mu \nu}}{8\pi G_N}\dvphi-\frac{\nabla_\nu \nabla_\mu\varphi}{8\pi G_N}&+\frac{g_{\mu \nu}\Box \varphi}{8\pi G_N}+\frac{\omega_{BD}}{8\pi \varphi}\partial_\nu \varphi \partial_\mu \varphi\\
&-\frac{g_{\mu \nu}}{16\pi G_N}\left(R\dvphi+\frac{\omega_{BD}}{\varphi}g^{\alpha \beta}\partial_\alpha \varphi  \partial_\beta \varphi \right).
\end {split}
\end{equation}
Since $\varphi$ has no interactions it behaves as any free scalar field, so its energy-momentum tensor must adopt the perfect fluid form at the background level:
\begin{equation}
T_{\mu \nu}^{\rm BD}=\pvphi g_{\mu \nu}+(\rvphi+\pvphi)u_\mu u_\nu.
\end{equation}
Now we can compare this form with \eqref{BDEMT}. It is straightforward to obtain the energy density as well as the corresponding pressure, we only need to compute $\rvphi=T_{00}^{\rm BD}$ and $\pvphi = (T^{\rm BD} + \rvphi)/3$, being $T^{\rm BD} =g^{\mu\nu}T^{\rm BD}_{\mu\nu}$ the trace of the tensor. {Using at this point the spatially flat FLRW metric one can work out  the explicit result for $\rvphi$ and $\rvphi$  and reconfirm that it acquires the form previously indicated in the equations \eqref{rhoBD} and \eqref{pBD}. This  provides perhaps a more formal derivation of these formulas and serves as a cross-check of them.}

\section{Structure formation  in the linear regime. Perturbations equations}\label{sec:StructureFormation}
In order to perform a complete analysis of the model, we need to study the evolution of the perturbed cosmological quantities in the context of BD theory. For a review of the standard model perturbations equations, see e.g. \cite{Ma:1995ey,liddle_lyth_2000,Lyth:2009zz}.
We assume a FLRW metric written in conformal time, denoted by $\eta$, in which the line element is $ds^2=a^2(\eta)[-d\eta^2+(\delta_{ij}+h_{ij})dx^idx^j]$. Here $h_{ij}$ is a perturbation on the spatial part of the metric which can be expressed in momentum space as follows,
\begin{equation}\label{eq:MainhFourier}
h_{ij}(\eta,\vec{x})=\int d^3k\, e^{-i\vec{k}\cdot\vec{x}}\left[\hat{k}_i\hat{k}_j h(\eta,\vec{k})+\left(\hat{k}_i\hat{k}_j-\frac{\delta_{ij}}{3}\right)6\xi(\eta,\vec {k})\right].
\end{equation}
As we see, in momentum space the trace $h\equiv \delta^{ij}h_{ij}$ decouples from the traceless part of the perturbation, $\xi$. Now, we are going to list the perturbations equations at late stages of expansion in momentum space at deep subhorizon scales, that is, we assume $\mathcal{H}^2\ll k^2$, with  $\mathcal{H}\equiv a^\prime/a$. Although primes were used previously for derivatives with respect to the scale factor, they will henceforth stand for  derivatives with respect to the conformal time within the main text ({except in Appendix \ref{AppendixB}}): $()^\prime \equiv d()/d \eta$.  For example, it is easy to see that $\mathcal{H}= a H$.
One may work with the standard differential equation for the density contrast at deep subhorizon scales,

\begin{equation} \label{DensityConstrastLCDM}
\delta_m^{\prime \prime}+\mathcal{H}\delta_m^\prime-4\pi{G_N}\bar{\rho}_m a^2 \delta_m=0,
\end{equation}
where $\delta_m \equiv \delta \rho_m / \bar{\rho}_m$ is the density contrast, the bar over $\bar{\rho}_m$ indicates that is a background quantity and the evolution of $\mathcal{H}$ and $\bar{\rho}_m$ is the one expected by the background equations of the BD theory in Section \ref{sec:BDgravity}. This expression is just the corresponding one for the $\Lambda$CDM, completely neglecting any possible perturbation in the BD-field, namely $\delta \varphi$. However, it is possible to see that a second order differential equation for the density contrast can be written, even if the perturbation in $\varphi$ is not neglected. This is done in detail in the Appendices \ref{AppendixC} and \ref{AppendixD} for the Synchronous as well as for the Newtonian gauges, respectively. In this section, we present the main perturbations equations in the case of the Synchronous gauge and discuss the interpretation of the result.

If  $\vec{v}_m$ is the physical 3-velocity of matter  (which is much smaller than 1 and can be treated as a perturbation), then we can define its divergence, $\theta_m\equiv \vec{\nabla} \cdot (\vec{v}_m)$. At deep subhorizon scales it is possible to see that the equation governing its evolution is
\begin{equation}
\theta_m^\prime+\mathcal{H}\theta_m=0.
\end{equation}
{Since $da^{-1}/d\eta=- \mathcal{H}/a$,  we arrive to a decaying solution  $\theta_m  \propto a^{-1}$. A common assumption is to set $\theta_m \sim 0$ in the last stages of the universe, which is what we will do in our analysis. This allows us to simplify the equations. Another simplification occurs if we take into account that we are basically interested in computing the matter perturbations only at deep subhorizon scales, namely for $k^2\gg \mathcal{H}^2$, which allows us to neglect some terms as well (cf. Appendix \ref{AppendixC}).  Altogether we are led to the following set of perturbations equations in the synchronous gauge:}
\begin{equation}\label{eq:Mainsimpli0}
\delta_m^\prime=-\frac{h^\prime}{2}\,.
\end{equation}
\begin{equation}\label{eq:Mainsimpli1}
k^2\delta \varphi+\frac{h^\prime}{2}\bar{\varphi}^\prime =\frac{8 \pi G_N}{3+2\omega_{BD}}a^2 \bar{\rho}_m\delta_m\,,
\end{equation}
\begin{equation}\label{eq:Mainsimpli2}
\bar{\varphi}(\mathcal{H}h^\prime-2\xi k^2)+k^2\delta\varphi+\frac{h^\prime}{2}\bar{\varphi}^\prime=8\pi G_N a^2 \bar{\rho}_m\delta_m\,,
\end{equation}
\begin{equation}\label{eq:Mainsimpli3}
2k^2\delta \varphi+\bar{\varphi}^\prime h^\prime+\bar{\varphi}\left(h^{\prime \prime}+2h^\prime \mathcal{H}-2k^2 \xi\right)=0\,.
\end{equation}
{Combining these four equations simultaneously (cf. Appendix \ref{AppendixC} for more details) and  without doing any further approximation, one finally obtains the following compact equation for the matter density contrast of the BD theory at deep subhorizon scales:}
\begin{equation}\label{eq:ExactPerturConfTime}
\delta_m^{\prime\prime}+\mathcal{H}\delta_m^\prime-\frac{4\pi G_N a^2}{\bar{\varphi}}\bar{\rho}_m\delta_m\left(\frac{4+2\omega_{BD}}{3+2\omega_{BD}}\right)=0\,.
\end{equation}
{In other words},
\begin{equation}\label{eq:ExactPerturConfTime2}
\delta_m^{\prime\prime}+\mathcal{H}\delta_m^\prime- 4\pi \Geff(\bar{\varphi}) a^2\,\bar{\rho}_m\delta_m=0\,.
\end{equation}
The quantity
\begin{equation}\label{eq:MainGeffective}
 \Geff(\bar{\varphi})=\frac{G_N}{\bar{\varphi}}\left(\frac{4+2\omega_{BD}}{3+2\omega_{BD}}\right)=\frac{G_N}{\bar{\varphi}}\left(\frac{2+4\eBD}{2+3\eBD}\right)
\end{equation}
is precisely the effective coupling previously introduced in Eq.\,\eqref{eq:LocalGNa}; it modifies the Poisson term of the perturbations equation with respect to that of the standard model, Eq.\,\eqref{DensityConstrastLCDM}.  There is, in addition, a modification in the friction term between the two models, which is of course associated to the change in ${\cal H}$.

The argument of $\Geff$ in \eqref{eq:MainGeffective} is not $\varphi$ but the background value $\bar{\varphi}$ since the latter  is what remains in first order of perturbations from the consistent splitting of the field into the background value and the perturbation:  $\varphi=\bar{\varphi}+\delta\varphi$.  Notice that there is no dependence left on the perturbation $\delta\varphi$.  As we can see from \eqref{eq:MainGeffective}, the very same effective coupling that rules the attraction of two tests masses in BD-gravity is the coupling strength that governs the formation of structure in this theory, as it could be expected.  But this does not necessarily mean that the effective gravitational strength governing the process of structure formation is the same as for two tests masses on Earth. We shall elaborate further on this point in the next section. At the moment we note that if we compare  the above perturbations equation with the standard model one \eqref{DensityConstrastLCDM},  the former reduce to the latter in the limit $\wBD\to\infty$ (i.e. $\eBD\to 0$) \textit{and} $\bar{\varphi}=1$.

The form of \eqref{eq:ExactPerturConfTime2} in terms of the scale factor variable rather than in conformal time was given previously in Sec. \ref{sec:rolesvarphiH0}   when we considered a preview of the implications of BD-gravity on structure formation data\footnote{{Recall, however, that prime in Eq.\,\eqref{eq:ExactPerturConfTime2} stands for differentiation  {\it w.r.t.}  conformal time whereas in Eq.\,\eqref{eq:ExactPerturScaleFactor} denotes differentiation  {\it w.r.t.}  the scale factor.  These equations perfectly agree and represent the same perturbations equation for the matter density field in BD-gravity in the respective variables. They are also in accordance with the perturbations equation obeyed by the matter density field within the context of scalar-tensor theories with the general action \eqref{eq:BDactionST}\cite{Boisseau:2000pr} of which  the form (\ref{eq:BDaction2}) is a  particular case. }}. The transformation of derivatives between the two variables can be easily performed with the help of the chain rule $d/d\eta=a{\cal H} d/da$.

\section{Different BD scenarios and Mach's Principle}\label{sec:Mach}
As previously indicated, the  relation \eqref{eq:LocalGNa}, which appears now in the cosmological context in the manner \eqref{eq:MainGeffective}, follows from the computation of the gravitational field felt by two test point-like (or spherical) masses in interaction in BD-gravity within the weak-field limit\cite{BransDicke1961, dicke1962physical}, see also \cite{Fujii:2003pa} and references therein. Such relation shows in a manifest way the integration of Mach's principle within the BD context, as it postulates a link between the measured local value of the gravitational strength, $G_N$, as measured at the Earth surface, and its cosmological value, $\Geff(\varphi)$, which depends on $\varphi$ and $\wBD$. In particular, $\varphi$  may be sensitive to the mean energy densities and pressures of all the matter and energy fields that constitute the universe. {If there is no mechanism screening the BD-field on Earth, $\Geff(\bar{\varphi})(z=0)=G_N$.} {However, one can still fulfill this condition  if  Eq.\,\eqref{eq:MainGeffective} constraints the current value of the cosmological BD-field $\bar{\varphi}$ }
\begin{equation}\label{eq:constraintvarphi}
\bar{\varphi}(z=0)=\frac{4+2\omega_{BD}}{3+2\omega_{BD}}=\frac{2+4\eBD}{2+3\eBD}\simeq 1+\frac12\,\eBD+{\cal O}(\eBD^2)\,.
\end{equation}
That is to say, such constraint permits to reconcile $\Geff(\bar{\varphi})(z=0)$ with $G_N$ by still keeping  $\bar{\varphi}(z=0)\ne 1$ and $\epsilon_{BD}\ne 0$. Hence the BD-field can be dynamical and there can be a departure of $G(\bar{\varphi})$ from $G_N$ even at present. This constraint, however, is much weaker than the one following from taking the more radical approach in which $\Geff(\varphi)$ and  $G_N$  are enforced to coincide upon imposing the double condition $\eBD\to0$ (i.e. $\wBD\to\infty$) \textit{and}  $\bar{\varphi}=1$.  It is this last setup which anchors the BD theory to remain exactly (or very approximately) close to GR at all scales.

However, if we take seriously the stringent constraint imposed  by the Cassini probe on the post-Newtonian parameter $\gamma^{PN}$\cite{Bertotti:2003rm}, which leads to a very large  value of $\wBD\gtrsim 10^4$ (equivalently, a very small value of $\eBD=1/\wBD$), as we discussed in Sec.\,\ref{sec:BDgravity},  $\bar{\varphi}$ must remain almost constant throughout the cosmic expansion, thus essentially equal to $\bar{\varphi}(z=0)$. However, the Cassini limit leaves $\bar{\varphi}(z=0)$ unconstrained, so this constant  is not restricted to be in principle equal to $1$.  In this case the relation \eqref{eq:constraintvarphi} may or may not apply; there is in fact no especial reason for it {(it will depend on the effectiveness of the screening mechanism on Earth)}. If it does, i.e. if there is no screening, $\Geff$ is forced to be very close to $G_N$ {$\forall{z}$};
%even  for intermediate values of $\wBD$ (i,e,. much smaller than the Cassini's ones);
if it does not, $\bar{\varphi}$ can freely take {(almost constant)} values  which do not push $\Geff$ to stay so glued to $G_N$.  It is  interesting to see  the  extent  to which the cosmological constraints can compete with the local ones given the current status of precision they can both attain.

So, as it turns out, we find that one of the two interpretations  leads to values of $\Geff$ very close to $G_N$ {$\forall{z}$} on account of the fact that  we are  imposing  very large values of  $\oD$ {and assuming \eqref{eq:constraintvarphi},} %$\varphi(z=0)=1$ (or very close to it),
whereas the other achieves the same aim (viz. $\Geff $  can stay very close to $G_N$) for intermediate values of $\wBD$ provided they are  linked to $\bar{\varphi}(z=0)$ through the constraint  \eqref{eq:constraintvarphi}. This last option, as indicated, is not likely {since this would imply the existence of a screening at the scales probed by Cassini that may become ineffective on Earth, where the densities are higher}. Finally, we may as well have a situation where the cosmological $\Geff $  remains different from  $G_N$ even if the Cassini limit is enforced. For this to occur we need an (essentially constant) value of  $\bar{\varphi}\neq 1$ (different from the one associated to that constraint) at the cosmological level. This can still be compatible with the local constraints provided $\varphi$ is screened {on Earth} at $z=0$.

{A more open-minded and general approach, which we are going to study in this work, is to take the last mentioned option but without the Cassini limit. This means that $\eBD$ is not forced to be so small and hence  $\varphi$ can still have some appreciable dynamics. We  assume that the pure BD model applies
% at all scales (even if this is obscured at local scales), starting
from the very large cosmological domains to those at which the matter perturbations remain linear.  Equation \eqref{eq:MainGeffective} predicts the cosmological value of $\Geff$ once $\wBD$ and the initial value $\varphi_{ini}$ are fitted to the data. We can dispense with the Cassini constraint (which affects $\wBD$ only) because we assume that some kind of screening mechanism
%(alien to BD itself)
acts at very low (astrophysical) scales, namely in the nonlinear domain, without altering the pure BD model at the cosmological level.
%The BD model applies to all scales, as indicated,  but in the local neighbourhood it appears much more GR-like owing to the screening of $\varphi$.
 To construct a concrete screening mechanism would imply to specify some microscopic interaction properties of $\varphi$ with matter, but these do not affect the analysis at the cosmological level, where there are no place with high densities of material particles. But once such mechanism is constructed (even if not being the primary focus of our work) the value of the BD-field $\varphi$  is ambient dependent, so to speak, since $\varphi$  becomes sensitive to the presence of large densities of matter.   This possibility is well-known in the literature through the chameleon mechanism\,\cite{Khoury:2003aq} and in the case of the BD-field was previously considered in \cite{Avilez:2013dxa} without letting the Brans-Dicke parameter $\omega_{BD}$ to acquire negative values, and using  datasets which now can be considered a bit obsolete. Here we do, instead,  allow negative values for $\omega_{BD}$ (we have seen in Sec.\,\ref{sec:preview} the considerable advantages involved in this possibility), and moreover we are using a much more complete set of observations from all panels of data taking.  In this scenario we cannot make use of \eqref{eq:MainGeffective} to connect the locally measured value of the gravitational strength $G_N$ with the BD-field at cosmological scales. We just do not need to know how the theory exactly connects these two values. We reiterate once more: we will not focus on the screening mechanism itself here but rather on the properties of the BD-field in the universe {in the large}, i.e. at the cosmological level. As it is explained in \cite{Avilez:2013dxa} -- see also \cite{Clifton:2011jh} --  many scalar-tensor theories of gravity belonging to the Horndeski class\,\cite{Horndeski:1974wa} could lead to such kind of BD behavior at cosmological level and, hence, it deserves a dedicated and updated analysis, which is currently lacking in the literature.}

{To summarize, the following interpretations of the BD-gravity framework considered here are, in principle, possible in the light of the current observational data:}
\begin{itemize}

\item {\bf BD-Scenario I}: \textit{Rigid Scenario for both the Local and Cosmological domains}. In it, we have $\Geff(\varphi (z))\simeq G(\varphi (z))\simeq G_N$, these three couplings being so close that in practice BD-gravity is indistinguishable from GR.  In this case, the BD-gravity framework is assumed to hold on equal footing with all the scales of the universe, local and cosmological.  There are no screening effects from matter. In this context, one interprets that the limit from the Cassini probe\,\cite{Bertotti:2003rm}  leads to a very large value of $\wBD$, which enforces $\varphi$ to become essentially rigid, and one assumes that such constant value is very close to $1$ owing to the relation \eqref{eq:constraintvarphi}.  Such a rigid scenario is, however, unwarranted. It is possible, although  is not necessary since, strictly speaking,  there is no direct connection between the Cassini bound on  $\wBD$ and the value that $\varphi$ can take. Thus, in this scenario the relation \eqref{eq:constraintvarphi} is just assumed. In point of fact, bounds on  $\wBD$  can only affect the time evolution of $\varphi$, they do not constraint its value.

\item {\bf BD-Scenario II (Main)}: {\it  Locally  Constrained but Cosmologically Unconstrained Scenario}. It is our main scenario.  It assumes a constrained  situation in the local domain, caused by the presence of chameleonic forces,  but permits an unconstrained picture for the entire cosmological range. In other words,  the Cassini limit that holds for the post-Newtonian parameter $\gamma^{PN}\, $in the local astrophysical level (and hence on $\wBD$) is assumed to reflect just the presence of screening effects of matter in that nonlinear domain. These effects are acting on $\varphi$  and  produce the illusion that  $\wBD$  has a very large value (as if `dressed' or `renormalized' by the chamaleonic forces). One expects that the `intrusive' effects of matter are only possible in high density (hence nonrelativistic) environments, and in their presence we cannot actually know the real (`naked') value of $\wBD$  through local experiments alone. We assume that the screening disappears as soon as we leave the astrophysical scales and plunge into the cosmological ones; then, and only then, we can measure the naked or ``bare' value of $\wBD$  (stripped from such effects). We may assume that the screening ceases already at the LSS scales where linear structure formation occurs, \newtext{see e.g. \cite{Tsujikawa:2008uc} for examples of potentials which can help to realize this mechanism.} The bare value of $\wBD$ can then be fitted to the overall data, and in particular to the LSS formation data.  Since $\wBD$ does no longer appear that big (nor it has any a priori sign) the BD-field $\varphi$ can evolve in an appreciable way at the cosmological level: it increases with the expansion if $\wBD>0$, and decreases with the expansion if $\wBD<0$.  In this context, its  initial value, $\varphi_{ini}$, becomes a relevant cosmological parameter, which must be taken into account as a fitting parameter on equal footing with  $\wBD$ and all of the conventional parameters entering the fit.  Using the large wealth of cosmological data, these parameters can be fixed at the cosmological level without detriment of the observed physics at the local domain, provided there is a screening mechanism insuring that $\Geff(\varphi)(0)$  stays sufficiently  close  to $G_N$ in the local neighbourhood. The numerical results for this important scenario are presented in Tables 3-6 and 9.

\item  {\bf BD-Scenario III}:  {\newtext{\it Cassini-constrained Scenario}}. A more restricted version of scenario II  can appear if the `bare value' of $\wBD$ is as large as in the Cassini bound.  In such case $\wBD$ is, obviously,  perceived large in both domains,  local  and cosmological. Even so, and despite of the fact that $\varphi$ varies very slowly in this case, one can still exploit the dependence of the fit on the initial value of the BD-field, $\varphi_{ini}$, and use it as a relevant cosmological parameter.  In practice, this situation has only one additional degree of freedom as compared to Scenario I (and one less than in Scenario II), but it is worth exploring -- see our results in Table 10.  As these numerical results show, the Cassini bound still leaves considerable freedom to the BD-$\CC$CDM model for improving the $H_0$ tension (without aggravating the $\sigma_8$ one) since the value of $\varphi$ is still an active degree of freedom, despite its time evolution is now more crippled.

 \item  {\bf BD-Scenario IV}: {\it Variable-$\wBD(\varphi)$  Scenario}.  Here one admits that the parameter $\wBD$ is actually a function of the BD-field, $\wBD=\wBD(\varphi)$, which can be modeled and adapted to the constraints of the  local and cosmological domains, or even combined with the screening effects of the local universe.  We have said from the very beginning that we would assume $\wBD=$const. throughout our analysis, and in fact we shall stick to that hypothesis; so here we mention the variable $\wBD$ scenario  only for completeness.  In any case, if a function $\wBD(\varphi)$ exists such that  it takes very large values in the local universe while it takes much more moderate values in the cosmological scales,  that sort of scenario would be in the main tantamount to Scenario II insofar as concerns its cosmological implications.

\end{itemize}

{In our analysis we basically choose BD- Scenarios II and III (the latter being a particular case of the former), which represent the most tolerant  point of view within the canonical  $\wBD=$const. option. Scenario II offers the most powerful framework amenable to provide a cure for the tensions afflicting the conventional $\CC$CDM model based on  GR.  Thus, we assume that we can  measure the cosmological  value of the gravity strength in BD theory -- i.e. the value given in Eq.\,\eqref{eq:MainGeffective} -- {by using only cosmological data. We combine the information from the LSS processes involving linear structure formation with the background information obtained from low, intermediate, and very high redshift probes, including BAO, CMB, and the distance ladder measurement of $H_0$.}
%differences in the  values of  $\varphi$  which appear when we consider vastly separated epochs of the cosmic history  (such as the time of the CMB release and the current time).}
The values of  $\varphi_{ini}$ and $\wBD$ are fitted to the data, and with them we obtain not only $\varphi(z=0)$  but we determine the effective cosmological gravity strength at all epochs  from \eqref{eq:MainGeffective}. The cosmological value of the gravity coupling  can be considered as the `naked' or `bare' value of the gravitational interaction, stripped from screening effects of matter, in the same way as $\wBD$ measured at cosmological scales is the bare value freed of these effects.  Even though $\Geff$ can be different from $G_N,$  we do not object to that since it can be ascribed to screening forces caused by the huge amounts of clustered matter in the astrophysical environments.  For this reason we do not adopt the local constraints for our cosmological analysis presented in this paper, i.e. we adhere to Scenario II as our main scenario.  Remarkably enough, we shall see that Scenario III still possesses a large fraction of the potentialities inherent to Scenario II, notwithstanding the Cassini bound.  In this sense Scenarios II and III are both extremely interesting. A  smoking gun of such overarching possible picture  is the possible detection of the dynamical dark energy EoS encoded in the BD theory within the GR-picture (cf. Sec.\,\ref{sec:EffectiveEoS}), which reveals itself in the form of effective quintessence,  as well as through the large smoothing achieved of  the main tensions afflicting the conventional $\CC$CDM. From here on, we present the bulk of our analysis and detailed results after we have already discussed to a great extent their possible implications.}

\section{Data and methodology}\label{sec:MethodData}

We fit the BD-$\Lambda$CDM together with the concordance GR-$\Lambda$CDM model and the GR-XCDM (based on the XCDM parametrization of the DE \cite{Turner:1998ex}) to the wealth of cosmological data compiled from distant type Ia supernovae (SNIa), baryonic acoustic oscillations (BAO), a set of measurements of the Hubble function at different redshifts, the Large-Scale Structure (LSS) formation data encoded in $f(z_i)\sigma_8(z_i)$, and the CMB temperature and low-$l$ polarization data from the Planck satellite. The joint combination of all these individual datasets will constitute our \textbf{Baseline Data} configuration. Moreover, we also study the repercussion of some alternative data, by adding them to the aforementioned baseline setup. These additional datasets are: a prior on the value of $H_0$ (or alternatively an effective calibration prior on M) provided by the SH0ES collaboration; the CMB high-$l$ polarization and lensing data from Planck; the Strong-Lensing (SL) time delay angular diameter distances from H0LICOW; and, finally, Weak-Lensing (WL) data from KiDS. Below we provide a brief description of the datasets employed in our analyses, together with the corresponding references.
\newline
\newline
\textbf{SNIa}: We use the full Pantheon likelihood, which incorporates the information from 1048 SNIa \cite{Scolnic:2017caz}. In addition, we also include the 207 SNIa from the DES survey \cite{Abbott:2018wog}. These two SNIa samples are uncorrelated, but the correlations between the points within each sample are non-null and have been duly incorporated in our analyses through the corresponding covariance matrices.
\newline
\newline
\textbf{BAO}: We use data on both, isotropic and anisotropic BAO analyses. We provide the detailed list of data points and corresponding references in Table 1. A few comments are in order about the use of some of the BAO data points considered in this article. Regarding the Ly$\alpha$-forest data, we opt to use the auto-correlation information from \cite{Agathe:2019vsu}. Excluding the Ly$\alpha$ cross-correlation data allows us to avoid double counting issues between the latter and the eBOSS data from \cite{Gil-Marin:2018cgo}, due to the partial (although small) overlap in the list of quasars employed in these two analyses. It is also important to remark that we consider in our baseline dataset the BOSS data reported in \cite{Gil-Marin:2016wya}, which contains information from the spectrum (SP) and the bispectrum (BP). The bispectrum information could capture some details otherwise missed when only the spectrum is considered, so it is worth to use it\footnote{See also Ref.\cite{Sola:2018sjf} for additional comments on the significance of the bispectrum data as well as its potential implications on the possible detection of dynamical dark energy.}. Therefore, we study the possible significance of the bispectrum component in the data by carrying out an explicit comparison of the results obtained with the baseline configuration to those obtained by substituting the data points from \cite{Gil-Marin:2016wya} with those from \cite{Alam:2016hwk}, which only incorporate the SP information. The results are provided in Tables 3 and 5, respectively. In Tables 4 and 6-8 we use the SP+BSP combination \cite{Gil-Marin:2016wya}. In Table 10 we employ both SP and SP+BSP.
%%%%%%%%%%%%%%%%%%%%%%%%%%%%%%%%%%%%%%%%%%%%%%%%%%%%%%%%%%%%%%%%%%%%%%%%%%%%%%
%
\begin{table}[t]
\begin{center}
\resizebox{14.5cm}{!}{
%\begin{scriptsize}
\begin{tabular}{| c | c |c | c |c|}
\multicolumn{1}{c}{Survey} &  \multicolumn{1}{c}{$z$} &  \multicolumn{1}{c}{Observable} &\multicolumn{1}{c}{Measurement} & \multicolumn{1}{c}{{\small References}}
\\\hline
6dFGS+SDSS MGS & $0.122$ & $D_V(r_s/r_{s,fid})$[Mpc] & $539\pm17$[Mpc] &\cite{Carter:2018vce}
\\\hline
 WiggleZ & $0.44$ & $D_V(r_s/r_{s,fid})$[Mpc] & $1716.4\pm 83.1$[Mpc] &\cite{Kazin:2014qga} \tabularnewline
\cline{2-4} & $0.60$ & $D_V(r_s/r_{s,fid})$[Mpc] & $2220.8\pm 100.6$[Mpc]&\tabularnewline
\cline{2-4} & $0.73$ & $D_V(r_s/r_{s,fid})$[Mpc] &$2516.1\pm 86.1$[Mpc] &
\\\hline

DR12 BOSS (BSP)& $0.32$ & $Hr_s/(10^{3}km/s)$ & $11.549\pm0.385$   &\cite{Gil-Marin:2016wya}\\ \cline{3-4}
 &  & $D_A/r_s$ & $6.5986\pm0.1337$ &\tabularnewline \cline{3-4}
 \cline{2-2}& $0.57$ & $Hr_s/(10^{3}km/s)$  & $14.021\pm0.225$ &\\ \cline{3-4}
 &  & $D_A/r_s$ & $9.389\pm0.1030$ &\\\hline

DR12 BOSS (SP) & $0.38$ & $D_M(r_s/r_{s,fid})$[Mpc] & $1518\pm22$   &\cite{Alam:2016hwk} \\ \cline{3-4}
 &  & $H(r_{s,fid}/r_s)$[km/s/Mpc] & $81.5\pm1.9$ & \tabularnewline \cline{3-4}
 \cline{2-2}& $0.51$ & $D_M(r_s/r_{s,fid})$[Mpc] & $1977\pm27$ & \\ \cline{3-4}
 &  & $H(r_{s,fid}/r_s)$[km/s/Mpc] & $90.4\pm1.9$ & \\ \cline{3-4}
 \cline{2-2}& $0.61$ & $D_M(r_s/r_{s,fid})$[Mpc]  & $2283\pm32$ & \\ \cline{3-4}
 &  & $H(r_{s,fid}/r_s)$[km/s/Mpc] & $97.3\pm2.1$ & \\\hline

DES & $0.81$ & $D_A/r_s$ & $10.75\pm0.43$ &\cite{Abbott:2017wcz}
\\\hline
eBOSS DR14 & $1.19$ & $Hr_s/(10^{3}km/s)$ & $19.6782\pm1.5867  $ &\cite{Gil-Marin:2018cgo}\\ \cline{3-4}
 &  & $D_A/r_s$ & $12.6621\pm0.9876$ &\tabularnewline \cline{3-4}
 \cline{2-2}& $1.50$ & $Hr_s/(10^{3}km/s)$  & $19.8637\pm2.7187$ &\\ \cline{3-4}
 &  & $D_A/r_s$ & $12.4349\pm1.0429$ &\\ \cline{3-4}
 \cline{2-2}& $1.83$ & $Hr_s/(10^{3}km/s)$  & $26.7928\pm3.5632$ &\\ \cline{3-4}
 &  & $D_A/r_s$ & $13.1305\pm1.0465$ &\\\hline

Ly$\alpha$-F eBOSS DR14 & $2.34$ & $D_H/r_s$ & $8.86\pm0.29$   &\cite{Agathe:2019vsu}
\\ \cline{3-4} &  & $D_M/r_s$ & $37.41\pm 1.86$
&\\\hline
\end{tabular}}
% \end{scriptsize}
\caption{Published values of BAO data, see the quoted references and text in Sec. \ref{sec:MethodData}. Although we include in this table the values of $D_H/r_s=c/(r_sH)$ and $D_M/r_s$ for the Ly$\alpha$-forest auto-correlation data from \cite{Agathe:2019vsu}, we have performed the fitting analysis with the full likelihood. The fiducial values of the comoving sound horizon appearing in the table are $r_{s,fid} = 147.5$ Mpc for \cite{Carter:2018vce}, $r_{s,fid} = 148.6$ Mpc for \cite{Kazin:2014qga}, and $r_{s,fid} = 147.78$ Mpc for \cite{Alam:2016hwk}.}
\end{center}
\end{table}
%%%%%%%%%%%%%%%%%%%%%%%%%%%%%%%%%%%%%%%%%%%%%%%%%%%%%%%%%%%%%%%%%%%%%%%%%%%%%%
%
\newline
\newline
\textbf{Cosmic Chronometers}: We use the 31 data points on $H(z_i)$, at different redshifts, from \cite{Jimenez:2003iv,Simon:2004tf,Stern:2009ep,Moresco:2012jh,Zhang:2012mp,Moresco:2015cya,Moresco:2016mzx,Ratsimbazafy:2017vga}. All of them have been obtained making use of the differential age technique applied to passively evolving galaxies \cite{Jimenez:2001gg}, which provides cosmology-independent constraints on the Hubble function, but are still subject to systematics coming from the choice of the stellar population synthesis technique, and also the potential contamination of young stellar components in the quiescent galaxies \cite{Lopez-Corredoira:2017zfl,Lopez-Corredoira:2018tmn,Moresco:2018xdr}. For this reason we consider a more conservative dataset that takes into account these additional uncertainties. To be concrete, we use the processed sample presented in Table 2 of \cite{Gomez-Valent:2018gvm}. See therein for further details.
\newline
\newline
\textbf{CMB}: In our baseline dataset we consider the full Planck 2018 TT+lowE likelihood \cite{Aghanim:2018eyx}. In order to study the influence of the CMB high-$l$ polarizations and lensing we consider two alternative (non-baseline) datasets, in which we substitute the TT+lowE likelihood by: (i) the TTTEEE+lowE likelihood, which incorporates the information of high multipole polarizations; (ii) the full TTTEEE+lowE+lensing likelihood, in which we also incorporate the Planck 2018 lensing data. In Tables 6 and 7 these scenarios are denoted as B+$H_0$+pol and B+$H_0$+pol+lens, respectively.
%
%%%%%%%%%%%%%%%%%%%%%%%%%%%%%%%%%%%%%%%%%%%%%%%%%%%%%%%%%%%%%%%%%%%%%%%%%%%%%%
%
\begin{table}[t]
\begin{center}
\resizebox{10cm}{!}{
%\begin{scriptsize}
\begin{tabular}{| c | c |c | c |}
\multicolumn{1}{c}{Survey} &  \multicolumn{1}{c}{$z$} &  \multicolumn{1}{c}{$f(z)\sigma_8(z)$} & \multicolumn{1}{c}{{\small References}}
\\\hline
6dFGS+2MTF & $0.03$ & $0.404^{+0.082}_{-0.081}$ & \cite{Qin:2019axr}
\\\hline
SDSS-DR7 & $0.10$ & $0.376\pm 0.038$ & \cite{Shi:2017qpr}
\\\hline
GAMA & $0.18$ & $0.29\pm 0.10$ & \cite{Simpson:2015yfa}
\\ \cline{2-4}& $0.38$ & $0.44\pm0.06$ & \cite{Blake:2013nif}
\\\hline
DR12 BOSS (BSP)& $0.32$ & $0.427\pm 0.056$  & \cite{Gil-Marin:2016wya}\\ \cline{2-3}
 & $0.57$ & $0.426\pm 0.029$ & \\\hline
 WiggleZ & $0.22$ & $0.42\pm 0.07$ & \cite{Blake:2011rj} \tabularnewline
\cline{2-3} & $0.41$ & $0.45\pm0.04$ & \tabularnewline
\cline{2-3} & $0.60$ & $0.43\pm0.04$ & \tabularnewline
\cline{2-3} & $0.78$ & $0.38\pm0.04$ &
\\\hline

DR12 BOSS (SP) & $0.38$ & $0.497\pm 0.045$ & \cite{Alam:2016hwk}\tabularnewline
\cline{2-3} & $0.51$ & $0.458\pm0.038$ & \tabularnewline
\cline{2-3} & $0.61$ & $0.436\pm0.034$ &
\\\hline

VIPERS & $0.60$ & $0.49\pm 0.12$ & \cite{Mohammad:2018mdy}
\\ \cline{2-3}& $0.86$ & $0.46\pm0.09$ &
\\\hline
VVDS & $0.77$ & $0.49\pm0.18$ & \cite{Guzzo:2008ac},\cite{Song:2008qt}
\\\hline
FastSound & $1.36$ & $0.482\pm0.116$ & \cite{Okumura:2015lvp}
\\\hline
eBOSS DR14 & $1.19$ & $0.4736\pm 0.0992$ & \cite{Gil-Marin:2018cgo} \tabularnewline
\cline{2-3} & $1.50$ & $0.3436\pm0.1104$ & \tabularnewline
\cline{2-3} & $1.83$ & $0.4998\pm0.1111$ &

\\\hline
 \end{tabular}}
% \end{scriptsize}
\caption{Published values of $f(z)\sigma_8(z)$, see the quoted references and text in Sec. \ref{sec:MethodData}.}
\end{center}
\end{table}
%%%%%%%%%%%%%%%%%%%%%%%%%%%%%%%%%%%%%%%%%%%%%%%%%%%%%%%%%%%%%%%%%%%%%%%%%%%%%%
%
\newline
\newline
\textbf{LSS}: In this paper the LSS dataset contains the data points on the product of the ordinary growth rate $f(z_i)$ with $\sigma_8(z_i)$ at different effective redshifts. They are all listed in Table 2, together with the references of interest. In order to correct the potential bias introduced by the particular choice of a fiducial model in the original observational analyses we apply the rescaling correction explained in \cite{Macaulay:2013swa}. See also Sec. II.2 of \cite{Nesseris:2017vor}. The internal correlations between the BAO and RSD data from \cite{Gil-Marin:2016wya}, \cite{Alam:2016hwk} and \cite{Gil-Marin:2018cgo} have been duly taken into account through the corresponding covariance matrices provided in these three references.
\newline
\newline
\textbf{Prior on $H_0$}: We include as a prior in almost all the non-baseline datasets the value of the Hubble parameter measured by the SH0ES collaboration, $H_0= (73.5\pm 1.4)$ km/s/Mpc \cite{Reid:2019tiq}. It is obtained with the cosmic distance ladder method using an improved calibration of the Cepheid period-luminosity relation. It is based on distances obtained from detached eclipsing binaries located in the Large Magellanic Cloud, masers in the galaxy NGC $4258$ and Milky Way parallaxes. This measurement is in $4.1\sigma$ tension with the value obtained by the Planck team under the TTTEEE+lowE+lensing dataset, and using the GR-$\Lambda$CDM model, $H_0= (67.36\pm 0.54)$ km/s/Mpc \cite{Aghanim:2018eyx}. In only one alternative dataset we opt to use instead the SH0ES effective calibration prior on the absolute magnitude  $M$ of the SNIa, as provided in \cite{Camarena:2019rmj}:  $M=-19.2191\pm 0.0405$. This case is denoted ``$M$'' in our tables. It is obtained from the calibration of `nearby' SNIa (at $z\lesssim 0.01$) with Cepheids\,\cite{Riess:2019cxk}.
%Mira variables and TRGBs located in their host galaxies.
It has been recently argued in \cite{Camarena:2019moy,Camarena:2019rmj} (and later on also in \cite{Dhawan:2020xmp,Benevento:2020fev}) that in cosmological studies it is better to use this SH0ES constraint rather than the direct prior on $H_0$ when combined with low-redshift SNIa data to avoid double counting issues. We show that the results obtained with these two methods are compatible and lead to completely consistent results (see the discussion in Sec. \ref{sec:NumericalAnalysis}, and also Tables 3 and 6).
\newline
\newline
\textbf{SL}: In one of the non-baseline datasets we use in combination with the SH0ES prior on $H_0$ the data extracted from the six gravitational lensed quasars of variable luminosity reported by the H0LICOW team. They measure the time delay produced by the deflection of the light rays due to the presence of an intervening lensing mass. After modeling the gravitational lens it is possible to compute the so-called time delay distance $D_{\Delta{t}}$ (cf. \cite{Wong:2019kwg} and references therein). As explained in Sec. \ref{sec:preview}, the fact of being absolute distances (instead of relative, as in the SNIa and BAO datasets) {allows them to directly constrain the Hubble parameter in the context of the GR-$\Lambda$CDM as follows:  $H_0=73.3^{+1.7}_{-1.8}$ km/s/Mpc} . It turns out that for the three sources B1608+656, RX51131-1231 and HE0435-1223, the posterior distribution of $D_{\Delta t}$ can be well approximated by the analytical expression of the skewed log-normal distribution,
\begin{equation}\label{eq:skewed}
\mathcal{L}(D_{\Delta t}) = \frac{1}{\sqrt{2\pi}(D_{\Delta t} - \lambda_D)\sigma_D}\,{\rm exp}\left[-\frac{(\ln(D_{\Delta t} -\lambda_D) - \mu_D)^2}{2\sigma^2_D}\right]\,,
\end{equation}
where the corresponding values for the fitted parameters $\mu_D$, $\sigma_D$ and $\lambda_D$ are reported in Table 3 of \cite{Wong:2019kwg}. On the other hand, the former procedure cannot be applied to the three remaining lenses, i.e. SDSS 1206+4332, WFI 2033-4723 and PG 1115+080. From the corresponding Markov chains provided by H0LICOW\footnote{http://shsuyu.github.io/H0LiCOW/site/} we have instead constructed the associated analytical posterior distributions of the time delay angular diameter distances for each of them. Taking advantage of the fact that the number of points in each bin is proportional to $\mathcal{L}(D_{\Delta t})$ evaluated at the average $D_{\Delta t}$ for each bin, we can get the values for $-\ln(\mathcal{L}/\mathcal{L}_{max})$ and fit the output to obtain its analytical expression as a function of $D_{\Delta t}$. The fitting function can be as accurate as wanted, e.g. using a polynomial of order as high as needed. The outcome of this procedure is used instead of \eqref{eq:skewed} for the three aforesaid lenses.

Interestingly, the non-detection of Strong-Lensing time delay variations can be used to put an upper bound on $\dot{G}/G$ at the redshift and location of the lens \cite{Giani:2020fpz}. These constraints, though, cannot be applied to our model since we assume that the field is screened in these dense regions (see Scenarios II and III in Sec. \ref{sec:Mach}). Moreover, they are still too weak -- $\dot{G}/G\lesssim 10^{-2}\,yr^{-1}$ \cite{Giani:2020fpz} -- to have an impact on our results.
\newline
\newline
\textbf{WL}: The Kilo-Degree Survey (KiDS) has measured the Weak-Lensing statistical distortion of angles and shapes of galaxy images caused by the presence of inhomogeneities in the low-redshift universe \cite{Hildebrandt:2016iqg,Joudaki:2017zdt,Kohlinger:2018sxx,Wright:2020ppw}. The two-point correlation functions of these angle distortions are related to the power spectrum of matter density fluctuations, and from it it is possible to obtain constraints on the parameter combination $S_8=\sigma_8\sqrt{\Omega_m/0.3}$. However, we have not included WL data in our baseline setup. At the computational level, nonlinear effects for small angular scales are calculated with the Halofit module \cite{Takahashi:2012em}, which only works fine for the GR-$\Lambda$CDM and minimal extensions of it, as the aforementioned XCDM \cite{Turner:1998ex} and also for the CPL parametrization of the DE EoS parameter\, \cite{Chevallier:2000qy,Linder:2002et}. Thus, it is not able to model accurately the potential low-scale particularities of the BD-$\Lambda$CDM model. We have tried to remove these low angular scales from our analysis in order to avoid the operation of Halofit, but in that case the loss of information is very big, and $S_8$ remains basically unconstrained. In Tables 6 and 7 we show the results obtained using the full KiDS likelihood \cite{Hildebrandt:2016iqg,Kohlinger:2018sxx} \footnote{See http://kids.strw.leidenuniv.nl/sciencedata.php for more details.}, i.e. including all the scales (also the small ones). The results, though, should be interpreted with caution.
\newline
\newline
We study the performance of the BD-$\CC$CDM, GR-$\CC$CDM and GR-XCDM models under different datasets. In the following we briefly summarize the composition of each of them:

\begin{itemize}

\item {\bf Baseline (B)}: Here we include the Planck 2018 TT+lowE CMB data, together with SNIa +BAO+$H(z_i)$+LSS (see Table 3 and 8). It is important to remark that for the BOSS BAO+LSS data we consider \cite{Gil-Marin:2016wya}, which includes the information from the spectrum (SP) as well as from the bispectrum (BSP). We construct some other datasets using this baseline configuration as the main building block. See the other items, below.

\item {\bf Baseline{\boldmath+$H_0$}} ({\bf B{\boldmath+$H_0$}}): Here we add the SH0ES prior on the $H_0$ parameter from \cite{Reid:2019tiq} to the baseline dataset (see again Tables 3 and 8).

\item {\bf Baseline{\boldmath+$H_0$+}SL}: The inclusion of the Strong-Lensing (SL) data from \cite{Wong:2019kwg} exacerbates  more the $H_0$-tension in the context of the GR-$\CC$CDM model (see e.g. \cite{Verde:2019ivm} and Sec. \ref{sec:Discussion}), so it is interesting to also study the ability of the BD-$\CC$CDM to fit the SL data when they are combined with the previous B+$H_0$ dataset, and compare the results with those obtained with the GR-$\CC$CDM. The corresponding fitting results are displayed in Table 4.

\item  {\bf Spectrum}: In this dataset we replace the SP+BSP data from \cite{Gil-Marin:2016wya} used in the Baseline dataset (see the first item, above) by the data from \cite{Alam:2016hwk}, which only contains the spectrum (SP) information (i.e. the usual matter power spectrum).

\item {\bf Spectrum{\boldmath+$H_0$}}: As in the preceding item, but including the $H_0$ prior from SH0ES \cite{Reid:2019tiq}.

\end{itemize}

The aforementioned datasets are all based on the BD-Scenario II (cf. Sec. \ref{sec:Mach}) and can be considered as the main ones (cf. Tables 3-5, and 8), nevertheless we also consider a bunch of alternative datasets (also based on the BD-Scenario II). We present the corresponding results in Table 6 and the first five columns of Table 7 for the BD-$\CC$-CDM and GR-$\CC$CDM models, respectively.

\begin{itemize}

\item {\bf B{\boldmath+$M$}}: In this scenario we replace the prior on $H_0$ \cite{Reid:2019tiq} employed in the B+$H_0$ dataset with the effective SH0ES calibration prior on the absolute magnitude of SNIa $M$ provided in \cite{Camarena:2019rmj}.

\item {\bf B{\boldmath+$H_0$+}pol}: Here we add the CMB high-$l$ polarization data from Planck 2018 \cite{Aghanim:2018eyx} to the B+$H_0$ dataset described before, i.e. we consider the Planck 2018 TTTEEE+lowE likelihood for the CMB.

\item {\bf B{\boldmath+$H_0$+}pol+lens}: In addition to the datasets considered in the above case we also include the CMB lensing data from Planck 2018 \cite{Aghanim:2018eyx}, i.e. we use the Planck 2018 TTTEEE+lowE+ lensing likelihood.

\item {\bf B{\boldmath+$H_0$+}WL}: In this alternative case we consider the Weak-Lensing (WL) data from KiDS \cite{Hildebrandt:2016iqg,Kohlinger:2018sxx}, together with the B+$H_0$ dataset.

\item  {\bf CMB+BAO+SNIa}: By considering only this data combination, we study the performance of the the BD-$\CC$CDM and the GR-$\CC$CDM models under a more limited dataset, obtained upon the removal of the data that trigger the $H_0$ and $\sigma_8$ tensions, i.e. the prior on $H_0$ from SH0ES and the LSS data. The use of the BAO+SNIa data helps to break the strong degeneracies found in parameter space when only the CMB is considered. Here we use the TT+lowE Planck 2018 likelihood \cite{Aghanim:2018eyx}.

\end{itemize}

Finally, in Table 10 we present the results obtained for the BD-$\CC$CDM in the context of the Cassini-constrained scenario, or Scenario III (see Sec. \ref{sec:Mach} for the details). The corresponding results for the GR-$\CC$CDM are shown in the third column of Table 3, and the last three rows of Table 7. In all these datasets we include the Cassini bound\,\cite{Bertotti:2003rm} (see Sec. \ref{sec:BDgravity} for details). The main purpose of this scenario is to test the ability of the BD-$\CC$CDM to fit the observational data with $\eBD\simeq 0$ {\it and} $\varphi\neq 1$.

\begin{itemize}

\item {\bf B{\boldmath+$H_0$+}Cassini}: It contains the very same datasets as in the Baseline+$H_0$ case, but here we also include the Cassini constraint.

\item {\bf B{\boldmath+$H_0$+}Cassini (No LSS)}: Here we study the impact of the LSS data in the context of Scenario III, by removing them from the previous B+$H_0$+Cassini dataset.

\item {\bf Dataset \cite{Ballesteros:2020sik}}: To ease the comparison with the results obtained in \cite{Ballesteros:2020sik}, here we use exactly the same dataset as in that reference, namely: the Planck 2018 TTTEEE+lowE+lensing likelihood \cite{Aghanim:2018eyx}, the BAO data from \cite{Beutler:2011hx,Ross:2014qpa,Alam:2016hwk}, and the SH0ES prior from \cite{Riess:2019cxk}, $H_0=74.03\pm 1.42$ km/s/Mpc.

\item {\bf Dataset\cite{Ballesteros:2020sik}+LSS}: Here we consider an extension of the previous scenario by adding the LSS data on top of the data from \cite{Ballesteros:2020sik}.

\end{itemize}

We believe that all the datasets and scenarios studied in this work cover a wide range of possibilities and show in great detail which is the phenomenological performance of the BD-$\CC$CDM, GR-$\CC$CDM and GR-XCDM models.

The speed of gravitational waves at $z\approx 0$, $c_{gw}$, has been recently constrained to be extremely close to the speed of light, $|c_{gw}/c-1|\lesssim 5\cdot 10^{-16}$ \cite{TheLIGOScientific:2017qsa}. In the BD-$\CC$CDM model $c_{gw}=c\,$ $\forall{z}$, so this constraint is automatically fulfilled, see Appendix C.6 and references \cite{Creminelli:2017sry,Ezquiaga:2017ekz} for further details. We have also checked that the BD-$\CC$CDM respects the bound on $G(\varphi)$ at the Big Bang Nucleosynthesis (BBN) epoch, $|G(\varphi_{\rm BBN})/G_N-1|\lesssim 0.1$ \cite{Uzan:2010pm}, since $G(\varphi_{\rm BBN})\simeq G(\varphi_{ini})$ and our best-fit values satisfy $G(\varphi_{ini})>0.9 G_N$ regardless of the dataset under consideration, see the fitting results in Tables 3-6 and 10.

To obtain the posterior distributions and corresponding constraints for the various dataset combinations described above we have run the Monte Carlo sampler \texttt{Montepython}\footnote{http://baudren.github.io/montepython.html} \cite{Audren:2012wb} together with the Einstein-Boltzmann system solver \texttt{CLASS}\footnote{http://lesgourg.github.io/class\_public/class.html} \cite{Blas:2011rf}. We have duly modified the latter to implement the background and linear perturbations equations of the BD-$\Lambda$CDM model. {We use adiabatic initial conditions for all matter species. Let us note that the initial perturbation of the BD-field and its time derivative can be set to zero, as the DM velocity divergence when the synchronous gauge is employed, see Appendix C.5 for a brief discussion.} To get the contour plots and one-dimensional posterior distributions of the parameters entering the models we have used the \texttt{Python} package \texttt{GetDist}\footnote{https://getdist.readthedocs.io/en/latest/} \cite{Lewis:2019xzd}, and to compute the full Bayesian evidences for the different models and dataset combinations, we have employed the code \texttt{MCEvidence}\footnote{https://github.com/yabebalFantaye/MCEvidence} \cite{Heavens:2017afc}. The Deviance Information Criterion (DIC) \cite{DIC} has been computed with our own numerical code. The results are displayed in Tables 3-10, and also in Figs. 9-11. They are discussed in the next section.

\section{Numerical analysis. Results}\label{sec:NumericalAnalysis}

In the following we put the models under consideration to the test, using the various datasets described in Sec. \ref{sec:MethodData}. We perform the statistical analysis of the models in terms of a joint likelihood function, which is the product of the individual likelihoods for each data source and includes the corresponding covariance matrices. For a fairer comparison with the GR-$\CC$CDM  we use standard information criteria in which the presence of extra parameters in a given model is conveniently penalized so as to achieve a balanced comparison with the model having less parameters. More concretely, we employ the full Bayesian evidence to duly quantify the fitting ability of the BD-$\CC$CDM model as compared to its GR analogue. Given a dataset $\mathcal{D}$, the probability of a certain model $M_i$ to be the best one among a given set of models $\{M\}$ reads,
%%%%%%%%%%%%%%%%%%%%%%%%%%%%%%%%%%%%%%%%%%%%%%%%%%%%%%%%%%%%%%%
\begin{figure}[t!]
\begin{center}
\label{fig:BayesRatio}
\includegraphics[width=4.5in, height=3.5in]{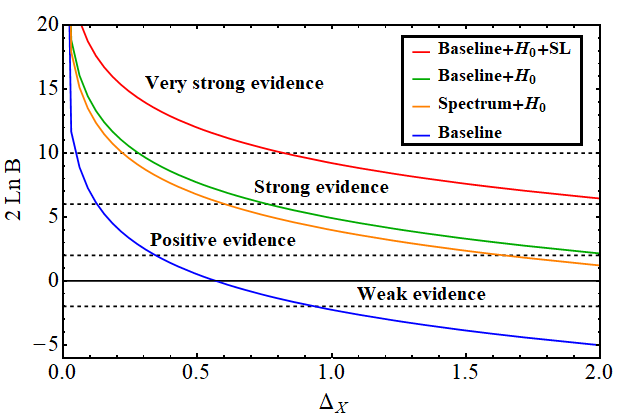}
\caption{\scriptsize{The full Bayesian evidence curves for the BD-$\CC$CDM as compared to the GR-$\CC$CDM, using different datasets and as a function of $\Delta_X=\Delta\epsilon_{BD}/10^{-2}=\Delta\varphi_{ini}/0.2$, with $\Delta\epsilon_{BD}$ and $\Delta\varphi_{ini}$ being the (flat) prior ranges for $\epsilon_{BD}$ and $\varphi_{ini}$, respectively. The curves are computed using the exact evidence formula, Eq. \eqref{eq:evidence}. The marked evidence ranges conform with the conventional Jeffreys' scale, see the main text in Sec. \ref{sec:NumericalAnalysis}.}}
\end{center}
\end{figure}
%%%%%%%%%%%%%%%%%%%%%%%%%%%%%%%%%%%%%%%%%%%%%%%%%%%%%%%%%%%%%%%
%
\begin{equation}\label{eq:BayesTheorem}
P(M_i|\mathcal{D})=\frac{P(M_i)\mathcal{E}(\mathcal{D}|M_i)}{P(\mathcal{D})}\,,
\end{equation}
{where $P(M_i)$ is the prior probability of the model $M_i$, $P(\mathcal{D})$ the probability of having the dataset $\mathcal{D}$, and the normalization condition $\sum_{j\in\{M\}}P(M_j)=1$ is assumed. The quantity $\mathcal{E}(\mathcal{D}|M_i)$ is the so-called marginal likelihood or evidence\,\cite{Amendola:2015ksp}. If the model $M_i$ has $n$ parameters contained in the vector $\vec{p}^{M_i}=(p^{M_i}_1, p^{M_i}_2,...,p^{M_i}_n)$, the evidence takes the following form:}
%%%%%%%%%%%%%%%%%%%%%%%%%%%%%%%%%%%%%%%%%%%%%%%%%%%%%%%%%%%%%%%%%%
%%%%%%%%%%%%%%%%%%%%%%%%%%%%%%%%%%%%%%%%%%%%%%%%%%%%%%%%%%%%%%%%%
%%%%%%%%%%%%%%%%%%%%%%%%%%%%%%%%%%%%%%%%%%%%%%%%%%%%%%%%%%%%%%%%%

\renewcommand{\arraystretch}{1.1}
\begin{table}[t!]
\begin{center}
\resizebox{1\textwidth}{!}{

\begin{tabular}{|c  |c | c |  c | c | c  |}
 \multicolumn{1}{c}{} & \multicolumn{2}{c}{Baseline} & \multicolumn{2}{c}{Baseline+$H_0$}
\\\hline
{\scriptsize Parameter} & {\scriptsize GR-$\Lambda$CDM}  & {\scriptsize BD-$\Lambda$CDM} & {\scriptsize GR-$\Lambda$CDM}  &  {\scriptsize BD-$\Lambda$CDM}
\\\hline
{\scriptsize $H_0$ (km/s/Mpc)}  & {\scriptsize $68.20^{+0.41}_{-0.40}$} & {\scriptsize $68.86^{+1.15}_{-1.24}$} & {\scriptsize $68.57^{+0.36}_{-0.42}$}  & {\scriptsize $70.83^{+0.92}_{-0.95}$}
\\\hline
{\scriptsize$\omega_b$} & {\scriptsize $0.02227^{+0.00019}_{-0.00018}$}  & {\scriptsize $0.02251^{+0.00026}_{-0.00027}$} & {\scriptsize $0.02238\pm 0.00019$}  &  {\scriptsize $0.02275^{+0.00024}_{-0.00026}$}
\\\hline
{\scriptsize$\omega_{cdm}$} & {\scriptsize $0.11763^{+0.00090}_{-0.00092}$}  & {\scriptsize $0.11598^{+0.00159}_{-0.00152}$} & {\scriptsize $0.11699^{+0.00092}_{-0.00083}$}  &  {\scriptsize $0.11574^{+0.00164}_{-0.00158}$}
\\\hline
{\scriptsize$\tau$} & {{\scriptsize$0.050^{+0.004}_{-0.008}$}} & {{\scriptsize$0.052^{+0.006}_{-0.008}$}} & {{\scriptsize$0.051^{+0.005}_{-0.008}$}}  &   {{\scriptsize$0.053^{+0.006}_{-0.008}$}}
\\\hline
{\scriptsize$n_s$} & {{\scriptsize$0.9683^{+0.0039}_{-0.0038}$}}  & {{\scriptsize$0.9775^{+0.0084}_{-0.0086}$}} & {{\scriptsize$0.9703^{+0.0038}_{-0.0036}$}} &   {{\scriptsize$0.9873^{+0.0076}_{-0.0075}$}}
\\\hline
{\scriptsize$\sigma_8$}  & {{\scriptsize$0.797^{+0.005}_{-0.006}$}}  & {{\scriptsize$0.785\pm 0.013$}} & {{\scriptsize$0.796^{+0.006}_{-0.007}$}}  &   {{\scriptsize$0.789\pm 0.013$}}
\\\hline
{\scriptsize$r_s$ (Mpc)}  & {{\scriptsize$147.83^{+0.29}_{-0.30}$}}  & {{\scriptsize$145.89^{+2.26}_{-2.49}$}} & {{\scriptsize$147.88\pm 0.31$}}  &   {{\scriptsize$142.46^{+1.84}_{-1.86}$}}
\\\hline
{\scriptsize $\epsilon_{BD}$} & - & {{\scriptsize $-0.00184^{+0.00140}_{-0.00142}$}} & - &   {{\scriptsize $-0.00199^{+0.00142}_{-0.00147}$}}
\\\hline
{\scriptsize$\varphi_{ini}$} & - & {{\scriptsize $0.974^{+0.027}_{-0.031}$}} & - &    {{\scriptsize $0.932^{+0.022}_{-0.023}$}}
\\\hline
{\scriptsize$\varphi(0)$} & - & {{\scriptsize $0.960^{+0.032}_{-0.037}$}} & - &    {{\scriptsize $0.918^{+0.027}_{-0.029}$}}
\\\hline
{\scriptsize$\weff(0)$} & - & {{\scriptsize $-0.983^{+0.015}_{-0.014}$}} & - &    {{\scriptsize $-0.966^{+0.012}_{-0.011}$}}
\\\hline
{\tiny $\dot{G}(0)/G(0) (10^{-13}yr^{-1})$} & - & {{\scriptsize $2.022^{+1.585}_{-1.518}$}} & - &    {{\scriptsize $2.256^{+1.658}_{-1.621}$}}
\\\hline
{\scriptsize$\chi^2_{min}$} & {\scriptsize 2271.98}  & {\scriptsize 2271.82} & {\scriptsize 2285.50}  &  {\scriptsize 2276.04}
\\\hline
{\scriptsize$2\ln B$} & {\scriptsize -}  & {\scriptsize -2.26} & {\scriptsize -}  &  {\scriptsize +4.92}
\\\hline
{\scriptsize$\Delta {\rm DIC}$} & {\scriptsize -}  & {\scriptsize -0.54} & {\scriptsize -}  &  {\scriptsize +4.90}
\\\hline
\end{tabular}}
\end{center}
\label{tableFit3}
\caption{\scriptsize The mean fit values and $68.3\%$ confidence limits for the considered models using our baseline dataset in the first block, i.e. SNIa+$H(z)$+BAO+LSS+CMB TT data, and baseline+$H_0$ in the second one (cf. Sec. \ref{sec:MethodData} for details). These results have been obtained within our main BD scenario (Scenario II of Sec.\,\ref{sec:Mach}). In all cases a massive neutrino of $0.06$ eV has been included. First we display the fitting results for the six conventional parameters, namely: $H_0$, the reduced density parameters for baryons ($\omega_{b}=\Omega_b h^2$) and CDM ($\omega_{cdm}=\Omega_{cdm} h^2$), the reionization optical depth $\tau$, the spectral index $n_s$ of the primordial power-law power spectrum, and, for convenience, instead of the amplitude $A_s$ of such spectrum we list the values of $\sigma_8$. We also include the sound horizon at the baryon drag epoch, $r_s$, obtained as a derived parameter. Right after we list the values of the free parameters that characterize the BD model: $\epsilon_{BD}$ \eqref{eq:definitions} and the initial condition for the BD-field, $\varphi_{ini}$. We also include the values of the BD-field, the (exact) effective EoS parameter \eqref{effEoS}, and the ratio between the derivative and the value of Newton's coupling, all computed at $z=0$. Finally, we report the values of the minimum of the $\chi^2$-function, $\chi^2_{min}$, the exact Bayes ratios (computed under the conditions explained in the main text of Sec. 8), and the DIC. It is also worth to remark that the baseline dataset employed here includes the contribution not only of the spectrum, but also the bispectrum information from BOSS \cite{Gil-Marin:2016wya}, see Sec. 7 for details.}
\end{table}
\begin{equation}\label{eq:evidence}
\mathcal{E}(\mathcal{D}|M_i)=\int \mathcal{L}(\mathcal{D}|\vec{p}^{M_i},M_i)\pi(\vec{p}^{M_i}) d^np^{M_i}\,,
\end{equation}
{with $\mathcal{L}(\mathcal{D}|\vec{p}^{M_i},M_i)$ the likelihood and $\pi(\vec{p}^{M_i})$ the prior of the parameters entering the model $M_i$. The evidence is larger for those models that have more overlapping volume between the likelihood and the prior distributions, but penalizes the use of additional parameters having a non-null impact on the likelihood. Hence, the evidence constitutes a good way of quantifying the performance of the model by implementing in practice the Occam razor principle. We can compare the fitting performance of BD-$\CC$CDM and GR-$\Lambda$CDM models by assuming equal prior probability for both of them, i.e. $P({\rm BD-}\CC{\rm CDM})=P({\rm GR-}\CC{\rm CDM})$ (``Principle of Insufficient Reason''). The ratio of their associated probabilities can then be directly written as the ratio of their corresponding evidences, i.e.}
%%%%%%%%%%%%%%%%%%%%%%%%%%%%%%%%%%%%%%%%%%%%%%%%%%%%%%%%%%%%%%%%%%
%%%%%%%%%%%%%%%%%%%%%%%%%%%%%%%%%%%%%%%%%%%%%%%%%%%%%%%%%%%%%%%%%
%%%%%%%%%%%%%%%%%%%%%%%%%%%%%%%%%%%%%%%%%%%%%%%%%%%%%%%%%%%%%%%%%

\renewcommand{\arraystretch}{1.1}
\begin{table}[t!]
\begin{center}
\resizebox{0.65\textwidth}{!}{

\begin{tabular}{|c  |c | c |  c |}
 \multicolumn{1}{c}{} & \multicolumn{2}{c}{Baseline+$H_0$+SL}
\\\hline
{\scriptsize Parameter} & {\scriptsize GR-$\CC$CDM}  & {\scriptsize BD-$\CC$CDM}
\\\hline
{\scriptsize $H_0$ (km/s/Mpc)}  & {\scriptsize $68.74^{+0.37}_{-0.40}$} & {\scriptsize $71.30^{+0.80}_{-0.84}$}
\\\hline
{\scriptsize$\omega_b$} & {\scriptsize $0.02242^{+0.00018}_{-0.00019}$}  & {\scriptsize $0.02281\pm 0.00025$}
\\\hline
{\scriptsize$\omega_{cdm}$} & {\scriptsize $0.11666^{+0.00087}_{-0.00086}$}  & {\scriptsize $0.11560^{+0.00158}_{-0.00169}$}
\\\hline
{\scriptsize$\tau$} & {{\scriptsize$0.051^{+0.005}_{-0.008}$}} & {{\scriptsize$0.053^{+0.006}_{-0.008}$}}
\\\hline
{\scriptsize$n_s$} & {{\scriptsize$0.9708^{+0.0036}_{-0.0038}$}}  & {{\scriptsize$0.9901^{+0.0075}_{-0.0070}$}}
\\\hline
{\scriptsize$\sigma_8$}  & {{\scriptsize$0.795^{+0.006}_{-0.007}$}}  & {{\scriptsize$0.789\pm 0.013$}}
\\\hline
{\scriptsize$r_s$ (Mpc)}  & {{\scriptsize$147.93^{+0.30}_{-0.31}$}}  & {{\scriptsize$141.68^{+1.69}_{-1.73}$}}
\\\hline
{\scriptsize $\epsilon_{BD}$}  & {\scriptsize -}   & {{\scriptsize$-0.00208^{+0.00151}_{-0.00140}$}}
\\\hline
{\scriptsize$\varphi_{ini}$}  & {\scriptsize -}   & {{\scriptsize$0.923^{+0.019}_{-0.021}$}}
\\\hline
{\scriptsize$\varphi(0)$}  & {\scriptsize -}   & {{\scriptsize$0.908^{+0.026}_{-0.028}$}}
\\\hline
{\scriptsize$\weff(0)$}  & {\scriptsize -}   & {{\scriptsize$-0.962\pm0.011$}}
\\\hline
{\tiny $\dot{G}(0)/G(0) (10^{-13}yr^{-1})$} & {\scriptsize -}   & {{\scriptsize$2.375^{+1.612}_{-1.721}$}}
\\\hline
{\scriptsize$\chi^2_{min}$} & {\scriptsize 2320.40}  & {\scriptsize 2305.80}
\\\hline
{\scriptsize$2\ln B$} & {\scriptsize -}  & {\scriptsize $+9.22$}
\\\hline
{\scriptsize$\Delta {\rm DIC}$} & {\scriptsize -}  & {\scriptsize $+9.15$}
\\\hline
\end{tabular}}
\end{center}
\label{tableFit4}
\caption{\scriptsize Fitting results as in Table 3, but adding the Strong-Lensing data, i.e. we use here the dataset Baseline+$H_0$+SL, for both the GR-$\CC$CDM and the BD-$\CC$CDM.}
\end{table}

%%%%%%%%%%%%%%%%%%%%%%%%%%%%%%%%%%%%%%%%%%%%%%%%%%%%%%%%%%%%%%%%%%
%%%%%%%%%%%%%%%%%%%%%%%%%%%%%%%%%%%%%%%%%%%%%%%%%%%%%%%%%%%%%%%%%
%%%%%%%%%%%%%%%%%%%%%%%%%%%%%%%%%%%%%%%%%%%%%%%%%%%%%%%%%%%%%%%%%

\renewcommand{\arraystretch}{1.1}
\begin{table}[t!]
\begin{center}
\resizebox{1\textwidth}{!}{

\begin{tabular}{|c  |c | c |  c | c | c  |}
 \multicolumn{1}{c}{} & \multicolumn{2}{c}{Spectrum} & \multicolumn{2}{c}{Spectrum+$H_0$}
\\\hline
{\scriptsize Parameter} & {\scriptsize GR-$\Lambda$CDM}  & {\scriptsize BD-$\Lambda$CDM} & {\scriptsize GR-$\Lambda$CDM}  &  {\scriptsize BD-$\Lambda$CDM}
\\\hline
{\scriptsize $H_0$ (km/s/Mpc)}  & {\scriptsize $68.00^{+0.47}_{-0.48}$} & {\scriptsize $68.86^{+1.27}_{-1.34}$} & {\scriptsize $68.61^{+0.46}_{-0.49}$}  & {\scriptsize $70.94^{+1.00}_{-0.98}$}
\\\hline
{\scriptsize$\omega_b$} & {\scriptsize $0.02223^{+0.00020}_{-0.00021}$}  & {\scriptsize $0.02241\pm 0.00027$} & {\scriptsize $0.02239^{+0.00020}_{-0.00019}$}  &  {\scriptsize $0.02264^{+0.00026}_{-0.00025}$}
\\\hline
{\scriptsize$\omega_{cdm}$} & {\scriptsize $0.11809^{+0.00112}_{-0.00095}$}  & {\scriptsize $0.11743^{+0.00168}_{-0.00170}$} & {\scriptsize $0.11695\pm 0.00104$}  &  {\scriptsize $0.11702^{+0.00169}_{-0.00167}$}
\\\hline
{\scriptsize$\tau$} & {{\scriptsize$0.051^{+0.005}_{-0.008}$}} & {{\scriptsize$0.053^{+0.006}_{-0.008}$}} & {{\scriptsize$0.053^{+0.006}_{-0.008}$}}  &   {{\scriptsize$0.054^{+0.007}_{-0.008}$}}
\\\hline
{\scriptsize$n_s$} & {{\scriptsize$0.9673^{+0.0039}_{-0.0044}$}}  & {{\scriptsize$0.9742^{+0.0086}_{-0.0090}$}} & {{\scriptsize$0.9705^{+0.0040}_{-0.0041}$}} &   {{\scriptsize$0.9845^{+0.0076}_{-0.0077}$}}
\\\hline
{\scriptsize$\sigma_8$}  & {{\scriptsize$0.800^{+0.006}_{-0.007}$}}  & {{\scriptsize$0.798\pm 0.014$}} & {{\scriptsize$0.798\pm 0.007$}}  &   {{\scriptsize$0.801^{+0.015}_{-0.014}$}}
\\\hline
{\scriptsize$r_s$ (Mpc)}  & {{\scriptsize$147.75^{+0.31}_{-0.35}$}}  & {{\scriptsize$145.82^{+2.33}_{-2.52}$}} & {{\scriptsize$147.88^{+0.33}_{-0.32}$}}  &   {{\scriptsize$142.55^{+1.71}_{-1.96}$}}
\\\hline
{\scriptsize $\epsilon_{BD}$} & - & {{\scriptsize $-0.00079^{+0.00158}_{-0.00157}$}} & - &   {{\scriptsize $-0.00081^{+0.00162}_{-0.00165}$}}
\\\hline
{\scriptsize$\varphi_{ini}$} & - & {{\scriptsize $0.976^{+0.028}_{-0.032}$}} & - &    {{\scriptsize $0.935^{+0.020}_{-0.024}$}}
\\\hline
{\scriptsize$\varphi(0)$} & - & {{\scriptsize $0.970^{+0.034}_{-0.038}$}} & - &    {{\scriptsize $0.929^{+0.028}_{-0.030}$}}
\\\hline
{\scriptsize$\weff(0)$} & - & {{\scriptsize $-0.987^{+0.016}_{-0.014}$}} & - &    {{\scriptsize $-0.971^{+0.013}_{-0.011}$}}
\\\hline
{\tiny $\dot{G}(0)/G(0) (10^{-13}yr^{-1})$} & - & {{\scriptsize $0.864^{+1.711}_{-1.734}$}} & - &    {{\scriptsize $0.913^{+1.895}_{-1.791}$}}
\\\hline
{\scriptsize$\chi^2_{min}$} & {\scriptsize 2269.04}  & {\scriptsize 2268.28} & {\scriptsize 2283.66}  &  {\scriptsize 2274.64}
\\\hline
{\scriptsize$2\ln B$} & {\scriptsize -}  & {\scriptsize $-2.94$} & {\scriptsize -}  &  {\scriptsize $+3.98$}
\\\hline
{\scriptsize$\Delta {\rm DIC}$} & {\scriptsize -}  & {\scriptsize $-3.36$} & {\scriptsize -}  &  {\scriptsize $+4.76$}
\\\hline
\end{tabular}}
\end{center}
\label{tableFit5}
\caption{\scriptsize  As in Table 3, but replacing the BOSS BAO+LSS data from \cite{Gil-Marin:2016wya} with those from \cite{Alam:2016hwk}, which only includes the spectrum information. See the discussion of the results in Sec. \ref{sec:NumericalAnalysis}.}
\end{table}

%%%%%%%%%%%%%%%%%%%%%%%%%%%%%%%%%%%%%%%%%%%%%%%%%%%%%%%%%%%%%%%%%%%%%%%%%%%%%%%%%%%%%%%%%

%%%%%%%%%%%%%%%%%%%%%%%%%%%%%%%%%%%%%%%%%%%%%%%%%%%%%%%%%%%%%%%%%%%%%%%%%%%%%%%%%%%%%%%%%
%
\begin{equation}\label{eq:BayesRatio}
\frac{P({\rm BD-}\CC{\rm CDM}|\mathcal{D})}{P({\rm GR-}\CC{\rm CDM}|\mathcal{D})}=\frac{\mathcal{E}(\mathcal{D}|{\rm BD-}\CC{\rm CDM})}{\mathcal{E}(\mathcal{D}|{\rm GR-}\CC{\rm CDM})} \equiv B\,,
\end{equation}
where $B$ is the so-called Bayes ratio (or Bayes factor) and is the quantity we are interested in. Notice that when $B>1$ this means that data prefer the BD-$\CC$CDM model over the GR version, but of course depending on how much larger than 1 it is we will have different levels of statistical significance for such preference. It is common to adopt in the literature the so-called Jeffreys' scale to categorize the level of evidence that one can infer from the computed value of the Bayes ratio. Jeffrey's scale actually is usually written not directly in terms of $B$, but in terms of $2\ln B$. The latter is sometimes estimated with a simple Schwarz (or Bayesian) information criterion $\Delta$BIC \cite{Schwarz1978,KassRaftery1995}, although   $2\ln B$ is a much more rigorous, sophisticated (and difficult to compute) statistics than just the usual  $\Delta$BIC estimates based on using the minimum value of $\chi^2$, the number of points and the number  of independent fitting parameters. If   $2\ln B$ lies below $2$ in absolute value, then we can conclude that the evidence in favor of BD-$\CC$CDM (against GR-$\CC$CDM)  is at most only {\it weak}, and in all cases {\it not conclusive}; if $2<2\ln B<6$ the evidence is said to be {\it positive}; if, instead, $6<2\ln B<10$, then it is considered to be {\it strong}, whereas if $2\ln B>10$ one is entitled to speak of {\it very strong} evidence in favor of the BD-$\CC$CDM over the GR-$\CC$CDM model. For more technical details related with the evidence and the Bayes ratio we refer the reader to \cite{KassRaftery1995,Burnham2002,Amendola:2015ksp}. Notice that the computation of \eqref{eq:BayesRatio} is not easy in general; in fact, it can be rather cumbersome since we usually work with models with a high number of parameters, so the multiple integrals that we need to compute become quite demanding from the computational point of view. We have calculated the evidences numerically, of course, processing the Markov chains obtained from the Monte Carlo analyses carried out with \texttt{CLASS}+\texttt{MontePython} \cite{Blas:2011rf,Audren:2012wb} with the numerical code \texttt{MCEvidence} \cite{Heavens:2017afc}, which is publicly available (cf. Sec. \ref{sec:MethodData}). We report the values obtained for $2\ln B$ \eqref{eq:BayesRatio} in Tables 3-6, 8, and 10. Table 3 contains the fitting results for the BD- and GR$-\CC$CDM models obtained with the Baseline and Baseline+$H_0$ datasets. In Table 4 we present the results for the Baseline+$H_0$+SL dataset. In Table 5 we show the output of the fitting analysis for the same models and using the same data as in Table 3, but changing the BOSS data from \cite{Gil-Marin:2016wya}, which contain both  the mater spectrum and bispectrum information, by the BOSS data from \cite{Alam:2016hwk}, which only incorporate the spectrum part (i.e. the usual matter power spectrum). In Table 6 we plug the results obtained for the BD-$\CC$CDM with the alternative datasets described in Sec. \ref{sec:MethodData}, and in Table 7 we show the corresponding results for the GR-$\CC$CDM. Next, in Table 8 we present the results with the Baseline and Baseline+$H_0$ data configurations obtained using the GR-XCDM parametrization. In Table 9 we display the values of the parameters $\sigma_8$ and $S_8$ for the GR- and BD-$\CC$CDM models, as well as the parameter $\tilde{S}_8 = S_8/\sqrt{\varphi(0)}$ for the BD. Finally, in Table 10 we present the fitting results for the BD-$\CC$CDM model, considering the (Cassini-constrained) Scenario III described in Sec. \ref{sec:Mach}.

\renewcommand{\arraystretch}{1.1}
\begin{table}[t!]
\begin{center}
\resizebox{1\textwidth}{!}{
\begin{tabular}{|c  |c | c |  c | c | c | c | c | c| c|}
 \multicolumn{1}{c}{} & \multicolumn{2}{c}{} & \multicolumn{1}{c}{} & \multicolumn{1}{c}{} & \multicolumn{1}{c}{} & \multicolumn{1}{c}{} & \multicolumn{1}{c}{}
\\\hline
Datasets & {\small $H_0$}  & {\small $\omega_m$} & {\small $\sigma_8$ } & {\small $r_s$ (Mpc) }  & {\small $\epsilon_{BD}\cdot 10^{3}$ }  & {\small $\weff(0)$}  & {\small $2\ln B$}
\\\hline
{\small B+$M$}  & {\small $71.19^{+0.92}_{-1.02}$} & {\small $0.1390^{+0.0014}_{-0.0015}$} & {\small $0.788^{+0.012}_{-0.013}$} & {\small $141.87^{+2.06}_{-1.81}$} & {\small $-2.16^{+1.42}_{-1.36}$} & {\small $-0.963^{+0.012}_{-0.011}$} & {\small $+10.38$}
\\\hline
{\small B+$H_0$+pol}  & {\small $69.85^{+0.81}_{-0.85}$} & {\small $0.1409^{+0.0012}_{-0.0011}$} & {\small $0.801\pm 0.011$} & {\small $144.72^{+1.51}_{-1.83}$} & {\small $-0.30^{+1.20}_{-1.23}$} &  {\small $-0.985^{+0.012}_{-0.009}$} & {\small $-1.44$ }
\\\hline
{\small B+$H_0$+pol+lens}  & {\small $69.74^{+0.82}_{-0.77}$} & {\small $0.1416^{+0.0011}_{-0.0010}$} & {\small $0.808\pm 0.009$} & {\small $144.66^{+1.56}_{-1.61}$} & {\small $0.00^{+1.11}_{-1.07} $} &  {\small $-0.986\pm 0.010$} & {\small $-1.98$ }
\\\hline
{\small B+$H_0$+WL}  & {\small $70.69^{+0.91}_{-0.90}$} & {\small $0.1398^{+0.0015}_{-0.0013}$} & {\small $0.794^{+0.011}_{-0.012}$} & {\small $142.76^{+1.79}_{-1.86}$} & {\small $-1.42^{+1.29}_{-1.37}$} & {\small $-0.970^{+0.011}_{-0.010}$} & {\small $+4.34$ }
\\\hline
{\small CMB+BAO+SNIa}  & {\small $68.63^{+1.44}_{-1.50}$} & {\small $0.1425\pm 0.0019$} & {\small $0.818^{+0.017}_{-0.018}$} & {\small $146.62^{+2.92}_{-2.93}$} & {\small $1.14^{+1.84}_{-1.68}$} &  {\small $-0.999^{+0.020}_{-0.017}$} & {\small $-3.00$ }
\\\hline
\end{tabular}}
\end{center}
\label{tableFit6}
\caption{\scriptsize Fitting results for the BD-$\Lambda$CDM model obtained with some alternative datasets and in all cases within the main BD Scenario II. Due to the lack of space, we employ some abbreviations, namely: {\bf B} for the Baseline dataset described in Sec. \ref{sec:MethodData}; {\bf pol} for the Planck 2018 (TE+EE) high-$l$  polarization data; and {\bf lens} for the CMB lensing. {The $\omega_m$ parameter contains the contribution of baryons and dark matter.  In the last row, CMB refers to the TT+lowE Planck 2018 likelihood (cf.  Sec. \ref{sec:MethodData} for details on the data).   $H_0$ is given in km/s/Mpc. For a discussion of the results, see Sec. \ref{sec:NumericalAnalysis}}.}
\end{table}

%%%%%%%%%%%%%%%%%%%%%%%%%%%%%%%%%%%%%%%%%%%%%%%%%%%%%%%%%%%%%%%%%%%%%%%%%%%%%%%%%%%%%%%%%%%%%%%%%%%

The evidence \eqref{eq:evidence} clearly depends on the priors for the parameters entering the model. For the 6 parameters in common in the BD- and GR-$\Lambda$CDM models, namely, $(\omega_b=\Omega_b h^2,\omega_{cdm}=\Omega_{cdm}h^2,H_0,\tau,A_s,n_s)$, if we use the same flat priors in both models they cancel exactly in the computation of the Bayes ratio \eqref{eq:BayesRatio}. Thus, the latter does not depend on the priors for these parameters if their ranges are big enough so as to not alter the shape of the likelihood severely. The Bayes ratio is, though, sensitive to the priors for the two additional parameters introduced in the BD-$\CC$CDM model in our Scenario II (cf. Sec. \ref{sec:Mach}), i.e. $\varphi_{ini}$ and $\epsilon_{BD}$, since they are not canceled in \eqref{eq:BayesRatio}. We study the dependence of the evidence on these priors in Fig.\,9, where we plot $2\ln B$ obtained for the BD-$\CC$CDM model from different datasets, and as a function of a quantity that we call $\Delta_X$, defined as
\begin{equation}\label{eq:DeltaX}
\Delta_X\equiv \frac{\Delta\epsilon_{BD}}{10^{-2}}=\frac{\Delta\varphi_{ini}}{0.2}\,,
\end{equation}
with $\Delta\epsilon_{BD}$ and $\Delta\varphi_{ini}$ being the flat prior ranges of $\epsilon_{BD}$ and $\varphi_{ini}$, centered at $\epsilon_{BD}=0$ and $\varphi_{ini}=1$, respectively. $\Delta_X$ will be equal to one when $\Delta\epsilon_{BD}=10^{-2}$ and $\Delta\varphi_{ini}=0.2$, which are natural values for these prior ranges. The former implies $\oD>100$ and the latter could be associated to the range $0.9<\varphi_{ini}<1.1$. We do not expect $\oD\lesssim 100$ since this would imply an exceedingly large departure from GR, even at cosmological scales, where this lower bound was already set using the first releases of WMAP CMB data almost twenty years ago, see e.g. \cite{Nagata:2003qn,Acquaviva:2004ti}. Regarding the prior range $0.9<\varphi_{ini}<1.1$, it is also quite natural, since this is necessary to satisfy the BBN bounds \cite{Uzan:2010pm}. In all tables we report the values of $2\ln B$ obtained by setting the natural value $\Delta_X=1$, and in Fig.\,9, as mentioned before, we also show how this quantity changes with the prior width, in terms of the variable $\Delta_X$ \eqref{eq:DeltaX}.

In the Cassini-constrained Scenario III (cf. again Sec. \ref{sec:Mach}), we also allow variations of $\varphi_{ini}$ and $\epsilon_{BD}$ in our Monte Carlo runs, of course, but the natural prior range for $\eBD$ is now much smaller than in Scenario II, since now we expect it to be more constrained by the local observations. It is more natural to take in this case a prior range $\Delta\eBD=5\cdot 10^{-5}$ (still larger than Cassini's bound), and this is what we do in all the analyses of this scenario. See the comments in Sec. \ref{sec:Discussion}, and Table 10.

In Tables 3-5 apart from the Bayes ratio, we also include the Deviance Information Criterion \cite{DIC}, which is strictly speaking an approximation of the exact Bayesian approach that works fine when the posterior distributions are sufficiently Gaussian. The DIC is defined as
\begin{equation}
{\rm DIC}=\chi^2(\hat{\theta})+2p_D\,.
\end{equation}
Here $p_D=\overline{\chi^2}-\chi^2(\hat{\theta})$ is the effective number of parameters of the model and $\overline{\chi^2}$ the mean of the overall $\chi^2$ distribution. DIC is particularly suitable for us, since we can easily compute all the quantities involved directly from the Markov chains generated with \texttt{MontePython}. To compare the ability of the BD- and GR-$\Lambda$CDM models to fit the data, one has to compute the respective differences of DIC values between the first and second models. They are denoted $\Delta$DIC in our tables, and this quantity is the analogous to $2\ln B$.

 \subsection{Comparing with the  XCDM parametrization}

\renewcommand{\arraystretch}{1.1}
\begin{table}[t!]
\begin{center}
\resizebox{1\textwidth}{!}{
\begin{tabular}{|c  |c | c |  c | c | c |}
 \multicolumn{1}{c}{} & \multicolumn{2}{c}{} & \multicolumn{1}{c}{} & \multicolumn{1}{c}{}
\\\hline
{\scriptsize Datasets} & {\small $H_0$}  & {\small $\omega_m$} & {\small $\sigma_8$ } & {\small $r_s$ (Mpc) }
\\\hline
{\small B+$M$} & {\small $68.64^{+0.39}_{-0.38}$} & {\small $0.1398\pm 0.0009$} & {\small $0.796^{+0.005}_{-0.006}$} & {\small $147.87^{+0.29}_{-0.30}$}
\\\hline
{\small B+$H_0$+pol}  & {\small $68.50^{+0.33}_{-0.36}$} & {\small $0.1408^{+0.0007}_{-0.0008}$} & {\small $0.799\pm 0.006$} & {\small $147.53\pm 0.022$}
\\\hline
{\small B+$H_0$+pol+lens}  & {\small $68.38^{+0.35}_{-0.33}$} & {\small $0.1411\pm 0.0007$} & {\small $0.803\pm0.006$} & {\small $147.44^{+0.22}_{-0.21}$}
\\\hline
{\small B+$H_0$+WL}  & {\small $68.64\pm 0.37$} & {\small $0.1399^{+0.0008}_{-0.0009}$} & {\small $0.795^{+0.005}_{-0.007}$} & {\small $147.89^{+0.31}_{-0.29}$}
\\\hline
{\small CMB+BAO+SNIa}  & {\small $67.91^{+0.39}_{-0.41}$} & {\small $0.1413\pm 0.0009$} & {\small $0.805^{+0.006}_{-0.007}$} & {\small $147.66^{+0.31}_{-0.29}$}
\\\hline\hline
{\small B+$H_0$ (No LSS)}  & {\small $68.38^{+0.42}_{-0.37}$} & {\small $0.1405\pm 0.0009$} & {\small $0.802^{+0.007}_{-0.008}$} & {\small $147.76\pm 0.30$}
\\\hline
{\small Dataset \cite{Ballesteros:2020sik}}  & {\small $68.17^{+0.43}_{-0.44}$} & {\small $0.1416\pm 0.0009$} & {\small $0.810^{+0.006}_{-0.007}$} & {\small $147.35^{+0.22}_{-0.24}$}
\\\hline
{\small Dataset \cite{Ballesteros:2020sik} + LSS (SP)}  & {\small $68.36\pm 0.42$}  & {\small $0.1412\pm 0.0009$} & {\small $0.806\pm 0.006$} & {\small $147.43\pm $0.23}
\\\hline
\end{tabular}}
\end{center}
\label{tableFit7}
\caption{\scriptsize Different fitting results for the GR-$\Lambda$CDM model. The first five rows correspond to the different non-baseline datasets explored for the  BD-$\CC$CDM  in Table 6. The last three rows correspond  to other scenarios tested with the BD-$\CC$CDM model in Table 10, see  Sec.\,\ref{sec:Discussion} for more details. $H_0$ is given in km/s/Mpc.}
\end{table}

As mentioned, in our numerical analysis of the data we also wish to consider the effect of a simple but powerful DDE parametrization, which is the traditional XCDM\cite{Turner:1998ex}. In this very simple framework, the DE is self-conserved and is associated to some unspecified entity or fluid (called X) which exists together with ordinary baryonic and cold dark matter, but it does not have any interaction with them.  The energy density of X is simply given by $\rho_X(a) = \rho_{X0}a^{-3(1+w_0)}$,   $\rho_{X0}=\rho_\Lambda$ being the current DE density value and $w_0$ the (constant) EoS parameter of such fluid. More complex parametrizations of the EoS can be considered, for instance the CPL one\,\cite{Chevallier:2000qy,Linder:2002et}, in which there is a  time evolution of the EoS itself. However, we have previously shown its incapability to improve the XCDM performance in solving the two tensions, see\,\,\cite{Sola:2018sjf}. Thus, in this work we prefer to stay as closer as possible to the standard cosmological model and we will limit ourselves to analyze the XCDM only. By setting $w_0=-1$ we retrieve the $\Lambda$CDM model with constant $\rho_\Lambda$. For $w_0 \gtrsim-1$ the XCDM mimics quintessence, whereas for $w_0 \lesssim -1$ it mimics phantom DE. The fitting results generated from the XCDM on our datasets are used in our analysis as a figure of merit or benchmark to compare with the corresponding fitting efficiency of both the BD-$\CC$CDM and the GR-$\CC$CDM models. In the next section, we comment on the comparison.
%

%%%%%%%%%%%%%%%%%%%%%%%%%%%%%%%%%%%%%%%%%%%%%%%%%%%%%%%%%%%%%%%%%%%%%%%%%%%%%%%

%%%%%%%%%%%%%%%%%%%%%%%%%%%%%%%%%%%%%%%%%%%%%%%%%%%%%%%%%%%%%%%%%%
%%%%%%%%%%%%%%%%%%%%%%%%%%%%%%%%%%%%%%%%%%%%%%%%%%%%%%%%%%%%%%%%%
%%%%%%%%%%%%%%%%%%%%%%%%%%%%%%%%%%%%%%%%%%%%%%%%%%%%%%%%%%%%%%%%%

\setlength{\tabcolsep}{0.7em}
\renewcommand{\arraystretch}{1.1}
\begin{table}[t!]
\begin{center}
\resizebox{0.7\textwidth}{!}{

\begin{tabular}{|c  |c | c |  c |}
 \multicolumn{1}{c}{} & \multicolumn{2}{c}{GR-XCDM}
\\\hline
{\normalsize Parameter} & {\normalsize Baseline}  & {\normalsize Baseline+$H_0$}
\\\hline
{\normalsize  $H_0$ (km/s/Mpc) }  & {\normalsize  $67.34^{+0.63}_{-0.66}$ } & {\normalsize $68.40^{+0.60}_{-0.62}$}
\\\hline
{\normalsize$\omega_b$} & {\normalsize   $0.02235^{+0.00021}_{-0.00020}$}  & {\normalsize $0.02239^{+0.00019}_{-0.00020}$}
\\\hline
{\normalsize$\omega_{cdm}$} & {\normalsize $0.11649^{+0.00108}_{-0.00111}$}  & {\normalsize $0.11671^{+0.00117}_{-0.00109}$}
\\\hline
{\normalsize$\tau$} & {{\normalsize$0.053^{+0.006}_{-0.008}$}} & {{\normalsize$0.051^{+0.005}_{-0.008}$}}
\\\hline
{\normalsize$n_s$} & {{\normalsize$0.9709\pm 0.0043$}}  & {{\normalsize$0.9707^{+0.0042}_{-0.0043}$}}
\\\hline
{\normalsize$\sigma_8$}  & {{\normalsize$0.782^{+0.011}_{-0.010}$}}  & {{\normalsize$0.792\pm 0.011$}}
\\\hline
{\normalsize$r_s$ (Mpc)}  & {{\normalsize$148.05^{+0.32}_{-0.34}$}}  & {{\normalsize$147.95^{+0.33}_{-0.34}$}}
\\\hline
{\normalsize$w_0$}  & {{\normalsize$-0.956\pm0.026$}}  & {{\normalsize$-0.991^{+0.026}_{-0.024}$}}
\\\hline
{\normalsize$\chi^2_{min}$} & {\normalsize 2269.88}  & {\normalsize 2285.22}
\\\hline
{\normalsize$2\ln B$} & {\normalsize $-2.23$}  & {\normalsize $-5.21$}
\\\hline
\end{tabular}}
\end{center}
\label{tableFit8}
\caption{\scriptsize As in Table 3, but for the XCDM parametrization (within GR). Motivated by previous works (see e.g. \cite{Sola:2018sjf}), we have used the (flat) prior $-1.1<w_0<-0.9$.}
\end{table}

\vspace{0.3cm}

\section{Discussion and extended considerations}\label{sec:Discussion}

In this work, we have dealt with Brans-Dicke  (BD) theory in  extenso. Our main goal was to assess if BD-gravity can help to smooth out the main two tensions besetting the usual concordance $\CC$CDM model (based on GR): i) the $H_0$-tension (the most acute existing discordance at present), and ii) the $\sigma_8$-tension, which despite not being so sharp it often occurs that the (many) models in the market dealing with the former tend to seriously  aggravate the latter.  As we have explained in the Introduction,  the  `golden rule' to be preserved by the tension solver should be to find a clue on how to tackle the main discrepancy (on the local $H_0$ parameter)  while at the same time to curb the $\sigma_8$ one, or at least not to worsen it. We have found that BD-gravity could be a key paradigm capable of such achievement.  Specifically, we have considered the original BD model with the only addition of a cosmological constant (CC), and we have performed a comprehensive analysis  in the light of a rich and updated set of observations. These involve a large variety of experimental inputs of various kinds, such as  the long chain SNIa+$H(z)$+BAO+LSS+CMB of data sources, which we have considered  at different levels and combinations; and tested with the inclusion of other potentially important factors such as the influence of the bispectrum component in the structure formation data (apart from the ordinary power spectrum); and also assessed the impact of  gravitational lensing data of different sorts (Weak and Strong-Lensing).

Although  BD-gravity is fundamentally different from GR, we have found very useful to try to pick out possible measurable signs of the new gravitational paradigm by considering the two frameworks in the (spatially flat) FLRW metric and compare the  versions of the $\CC$CDM model resulting in each case, which we have called BD-$\CC$CDM and GR-$\CC$CDM, respectively.  We have parametrized the departure of the former from the latter at the background level  (cf. Sec.\,\ref{sec:EffectiveEoS}) and we have seen that BD-$\CC$CDM can appear in the form of a dynamical dark energy (DDE) version of the GR-$\CC$CDM, in which the vacuum energy density is evolving through a non-trivial EoS  (cf. Fig.\,8).  We have called it  the `GR-picture' of the BD theory.  The resulting effective behavior is  $\CC$CDM-like with, however, a mild time-evolving  quasi-vacuum component. In fact, such behavior is not `pure vacuum' -- which is why we call it quasi-vacuum --  despite of the fact that the original BD-$\CC$CDM theory possesses a rigid cosmological constant.  Specifically, using the numerical fitting results of our analysis we find that such EoS  shows up in effective quintessence-like form at more than $3\sigma$ c.l. (this is perfectly appreciable at naked eye in Fig. 8). Our fit to the data demonstrates that such an effective  representation of BD-gravity can be competitive with the concordance model with a rigid $\CC$-term. It may actually create the fiction that the DE is dynamical when viewed within the GR framework, whilst it is  just a rigid CC in the underlying  BD action.  The practical outcome is that the BD approach with a CC definitely  helps to smooth out some of the tensions afflicting the $\CC$CDM in a manner very similar to the Running Vacuum Model, see e.g.\,\cite{Sola:2016hnq,Sola:2016ecz,Sola:2017znb,Sola:2017jbl,Gomez-Valent:2017idt,Gomez-Valent:2018nib}, and this success might ultimately reveal the signs of the  BD theory. We conclude that finding  traces of vacuum dynamics, accompanied with apparent deviations from the standard matter conservation law\,\cite{Fritzsch:2012qc} could be the `smoking gun' pointing to the possibility that the gravity theory sitting behind these effects is not GR but BD.

\renewcommand{\arraystretch}{1.1}
\begin{table}[t!]
\begin{center}
\resizebox{1\textwidth}{!}{

\begin{tabular}{|c  |c | c |  c |c|c|c|}
 \multicolumn{1}{c}{} & \multicolumn{2}{c}{}
\\\hline
{\normalsize Scenarios} & {\normalsize ${\sigma}_{8(GR)}$} & {\normalsize ${\sigma}_{8(BD)}$} & {\normalsize ${S}_{8(GR)}$} & {\normalsize ${S}_{8(BD)}$}  & {\normalsize $\tilde{S}_{8(BD)}$}
\\\hline
{\normalsize Baseline } & {\normalsize $0.797^{+0.005}_{-0.006}$} & {\normalsize $0.785\pm 0.013$} & {\normalsize $0.800^{+0.010}_{-0.011}$} & {\normalsize $0.777^{+0.019}_{-0.020}$} & {\normalsize $0.793\pm 0.012$}
\\\hline
{\normalsize Baseline+$H_0$} & {\normalsize $0.796^{+0.006}_{-0.007}$} & {\normalsize $0.789\pm 0.013$} & {\normalsize $0.793^{+0.011}_{-0.010}$} & {\normalsize $0.758\pm 0.018$} & {\normalsize $0.792\pm 0.013$}
\\\hline
{\normalsize Baseline+$H_0$+SL} & {\normalsize $0.795^{+0.006}_{-0.007}$} & {\normalsize $0.789\pm 0.013$}& {{\normalsize$0.789^{+0.011}_{-0.010}$}} & {\normalsize $0.753^{+0.017}_{-0.018}$}  & {{\normalsize$0.791^{+0.012}_{-0.013}$}}
\\\hline
{\normalsize Spectrum}  & {\normalsize $0.800^{+0.006}_{-0.007}$}& {\normalsize $0.798\pm 0.014$} &{\normalsize $0.807\pm 0.013$} & {\normalsize $0.793^{+0.021}_{-0.022}$} & {\normalsize $0.805\pm 0.014$}
\\\hline
{\normalsize Spectrum+$H_0$} & {\normalsize $0.798\pm 0.007$} & {\normalsize $0.801^{+0.015}_{-0.014}$} & {{\normalsize$0.794^{+0.012}_{-0.013}$}} & {\normalsize $0.773^{+0.019}_{-0.021}$} & {{\normalsize$0.802^{+0.013}_{-0.014}$}}
\\\hline
\end{tabular}}
\end{center}
\label{tableFit9}
\caption{\scriptsize Fitted values for the $\sigma_{8(M)}$ (here M stands for GR or BD) obtained under different dataset configurations. We also include the derived values of $S_{8(M)}$ = $\sigma_8\sqrt{\Omega_m/0.3}$ and the renormalized $\tilde{S}_8$ = $\sigma_8\sqrt{\tilde{\Omega}_m/0.3}$, with $\tilde{\Omega}_m$ defined as in Sec.\,\ref{sec:rolesvarphiH0}. These results correspond to the main Scenario II of the BD-$\CC$CDM model.}
\end{table}

%%%%%%%%%%%%%%%%%%%%%%%%%%%%%%%%%%%%%%%%%%%%%%%%%%%%%%%%%%%%%%%%%%
%%%%%%%%%%%%%%%%%%%%%%%%%%%%%%%%%%%%%%%%%%%%%%%%%%%%%%%%%%%%%%%%%
%%%%%%%%%%%%%%%%%%%%%%%%%%%%%%%%%%%%%%%%%%%%%%%%%%%%%%%%%%%%%%%%%

\subsection{Alleviating the $H_0$-tension}\label{sec:H0-tension}

Our analysis of the BD theory with a CC, taken at face value,  suggests that the reason for the enhancement of $H_0$ in the BD model is because the effective gravitational coupling acting at cosmological scales, $\Geff\sim G_N/\varphi$, is higher than the one measured on Earth (see Fig.\,4). This possibility allows the best-fit current energy densities of all the species to remain compatible at $\lesssim 1\sigma$ c.l. with the ones obtained in the GR-$\CC$CDM model (cf. e.g. Tables 3-5).  %In fact, within the BD-$\CC$CDM we may keep the values of the tilded parameters $\tilde{\Omega}_i=\rho_i/\rho$ --  which we have  defined in Eq. (\ref{eq:tildeOmegues}) --  stable {\it w.r.t.} its GR analogue, and at the same time augment the strength of the  gravitational coupling $G(\varphi)$ at cosmological scales.
Thus, since $\varphi<1$  we find that the increase of the Hubble parameter is basically due to the increase of  the effective $G$, and there is no need for a strong modification of the energy densities of the various species filling the universe. This is a welcome feature since  the measurable cosmological mass parameter in the BD-$\CC$CDM model is, for sufficiently small $\eBD$,  not the usual $\Omega_m$, but precisely the tilded one, related to it through $\tilde{\Omega}_m=\Omega_m/\varphi$.  The latter   is about   $\sim  8-9\%$  bigger than its standard model counterpart  ($\tilde{\Omega}_m> \Omega_m$)   as it follows from  Fig. \,4, where we can read off the current value  $\varphi(z=0)\simeq 0.918$.  Now, because at the background level  it is possible to write an approximate  Friedmann's equation \eqref{eq:H2}  in terms of the tilded parameters, these are indeed the ones that are actually  measured from SNIa and BAO observations in the BD context\footnote{Recall that for $\eBD\neq0$ the tilded  parameters $\tilde{\Omega}_i$  (which were originally defined for $\eBD=0$) receive a correction and become the hatted parameters $\hat{\Omega}_i$ introduced in Eq.\,\eqref{eq:hatOmega}.  However, the difference between them is of ${\cal O}(\eBD)$, see Eq.\,\eqref{eq:hatOmega2}, and since $|\epsilon_{BD}|\sim \mathcal{O}(10^{-3})$ it can be ignored.}. The differences, however, as we have just pointed out, are not to be attributed to a change in the physical energy content of matter but to the fact that $\varphi<1$  throughout the entire cosmic history, as clearly shown in Fig.\, 4.  Obviously, the measurement of the parameters $\tilde{\Omega}_i$  can be performed  through the very same data  and procedures well accounted for  in the context of the GR-$\CC$CDM framework.  This explanation is perfectly consistent with the fact that when the Friedmann's law is expressed in terms of the effetive $G$, as indicated  in Eq.\,\eqref{eq:H1}, the local value of the Hubble parameter $H_0$ becomes bigger owing to $\Geff=G_N/\varphi$ being bigger than $G_N$.  Thus,  when we compare the early and local measurements of $H_0$ we do not meet  any anomaly in this approach.

 We also recall at this point that there is no correction from $\wBD$  on the effective coupling $\Geff$, Eq.\,\eqref{eq:MainGeffective},  in the local domain. This is because in our context  $\wBD$ appears as being very large owing to the assumed screening of the BD-field caused by the clustered matter (cf. Sec.\,\ref{sec:Mach}).  From Fig.\,4 and Table 3 we find that the BD model leads to a value of  $H_0$ a factor  $\Geff^{1/2}(z=0)/G_N^{1/2}\sim 1/\varphi^{1/2}(z=0)=1/\sqrt{0.918}$, i.e. $\sim 4.5\%$, bigger than the one inferred from the CMB in the GR-$\CC$CDM model, in which $\Geff=G_N\,(\forall{z})$. It is reassuring to realize  that such a `renormalization factor'  can  enhance the low Planck 2018  CMB measurement of the  Hubble parameter (viz. $H_0=67.4\pm 0.5$ km/s/Mpc\,\cite{Aghanim:2018eyx})  up to the range of  $70-71$km/s/Mpc (cf. e.g. Tables 3-6), hence  much closer to the local measurements. For example,   SH0ES  yields $ H_0= (73.5\pm 1.4)$ km/s/Mpc\,\cite{Reid:2019tiq}; and when the latter is combined  with Strong-Lensing data from the H$0$LICOW collab.\cite{Wong:2019kwg}  it leads to  $ H_0= (73.42\pm 1.09)$ km/s/Mpc.  This combined  value  is $5\sigma$ at odds with the Planck 2018 measurement, a serious tension.

\renewcommand{\arraystretch}{1.1}
\begin{table}[t!]
\begin{center}
\resizebox{1\textwidth}{!}{

\begin{tabular}{|c  |c | c |  c | c | c  |}
 \multicolumn{1}{c}{} & \multicolumn{4}{c}{BD-$\Lambda$CDM (Scenario III: Cassini-constrained)}
\\\hline
{\scriptsize Parameter} & {\scriptsize B+$H_0$ (No LSS)}  & {\scriptsize B+$H_0$ } & {\scriptsize Dataset \cite{Ballesteros:2020sik}}  &  {\scriptsize Dataset \cite{Ballesteros:2020sik} + LSS (SP)}
\\\hline
{\scriptsize $H_0$ (km/s/Mpc)}  & {\scriptsize $70.99^{+0.94}_{-0.97}$} & {\scriptsize $70.80^{+0.81}_{-0.91}$} & {\scriptsize $70.01^{+0.86}_{-0.92}$}  & {\scriptsize $70.03^{+0.90}_{-0.88}$}
\\\hline
{\scriptsize$\omega_b$} & {\scriptsize $0.02257\pm 0.00021$}  & {\scriptsize $0.02256^{+0.00019}_{-0.00020}$} & {\scriptsize $0.02271\pm 0.00016$}  &  {\scriptsize $0.02272^{+0.00015}_{-0.00016}$}
\\\hline
{\scriptsize$\omega_{cdm}$} & {\scriptsize $0.11839^{+0.00093}_{-0.00094}$}  & {\scriptsize $0.11748\pm 0.00089$} & {\scriptsize $0.11885^{+0.00092}_{-0.00095}$}  &  {\scriptsize $0.11827^{+0.00089}_{-0.00093}$}
\\\hline
{\scriptsize$\tau$} & {\scriptsize $0.057^{+0.007}_{-0.008}$} & {\scriptsize$0.050^{+0.004}_{-0.008}$} & {\scriptsize$0.061^{+0.007}_{-0.008}$}  &   {\scriptsize$0.058^{+0.006}_{-0.008}$}
\\\hline
{\scriptsize$n_s$} & {\scriptsize $0.9824^{+0.0057}_{-0.0058}$}  & {\scriptsize$0.9811^{+0.0051}_{-0.0052}$} & {\scriptsize$0.9783^{+0.0052}_{-0.0059}$} &   {\scriptsize$0.9701^{+0.0056}_{-0.0054}$}
\\\hline
{\scriptsize$\sigma_8$}  & {\scriptsize $0.815^{+0.008}_{-0.009}$}  & {\scriptsize$0.804^{+0.006}_{-0.007}$} & {\scriptsize$0.817\pm 0.007$}  &   {\scriptsize$0.812^{+0.006}_{-0.007}$}
\\\hline
{\scriptsize$r_s$ (Mpc)}  & {\scriptsize $142.14^{+1.91}_{-1.72}$}  & {\scriptsize$143.31^{+1.72}_{-1.63}$} & {\scriptsize$143.58^{+1.62}_{-1.55}$}  &   {\scriptsize$144.10^{+1.62}_{-1.52}$}
\\\hline
{\scriptsize $\epsilon_{BD}$} & {\scriptsize $-0.00002\pm 0.00002$} & {\scriptsize $-0.00002\pm 0.00002$} & {\scriptsize $-0.00002\pm 0.00002$} &  {\scriptsize $-0.00002\pm 0.00002$}
\\\hline
{\scriptsize$\varphi_{ini}$} & {\scriptsize $0.933\pm 0.021$} & {\scriptsize $0.944\pm 0.020$} &{\scriptsize $0.955^{+0.018}_{-0.019}$}&    {\scriptsize $0.960^{+0.020}_{-0.018}$}
\\\hline
{\scriptsize$\varphi(0)$} & {\scriptsize $0.933^{+0.020}_{-0.021}$} & {\scriptsize $0.944^{+0.019}_{-0.020}$} &{\scriptsize $0.955^{+0.018}_{-0.019}$}     &{\scriptsize $0.960^{+0.020}_{-0.017}$}
\\\hline
{\scriptsize$\weff(0)$} & {\scriptsize $-0.972\pm 0.009$} & {\scriptsize $-0.977\pm 0.008$} & {\scriptsize $-0.981^{+0.008}_{-0.007}$} &    {\scriptsize $-0.983\pm 0.008$}
\\\hline
{\tiny $\dot{G}(0)/G(0) (10^{-13}yr^{-1})$} & {\scriptsize $0.025^{+0.025}_{-0.026}$} & {\scriptsize $0.026^{+0.027}_{-0.028}$} & {\scriptsize $0.023^{+0.026}_{-0.027}$} &   {\scriptsize $0.020\pm 0.026$}
\\\hline
{\scriptsize$\chi^2_{min}$} & {\scriptsize 2256.14}  & {\scriptsize 2278.34} & {\scriptsize 2797.44}  &  {\scriptsize 2812.68}
\\\hline
{\scriptsize$2\ln B$} & {\scriptsize +9.03}  & {\scriptsize $+5.21$} & {\scriptsize +3.45}  &  {\scriptsize +2.21}
\\\hline
\end{tabular}}
\end{center}
\label{tableFit10}
\caption{\scriptsize Fitting results for the BD-$\CC$CDM, in the context of the BD-Scenario III explained in Sec. \ref{sec:Mach} under different datasets.  As characteristic of Scenario III, in all of these datasets the Cassini constraint on the post-Newtonian parameter $\gamma^{PN}$ has been imposed \cite{Bertotti:2003rm}. In the first two fitting columns we use the Baseline+$H_0$ dataset described in Sec. \ref{sec:MethodData}. However, we exclude the LSS data in the first column while it is kept in the second. In the third and fourth fitting columns we report on the results obtained using the very same dataset as in Ref.\,\cite{Ballesteros:2020sik}, just to ease the comparison between the BD-$\CC$CDM and the variable $G$ model studied in that reference (cf. their Table 1). This dataset includes the Planck 2018 TTTEEE+lensing likelihood \cite{Aghanim:2018eyx}, BAO data from \cite{Beutler:2011hx,Ross:2014qpa,Alam:2016hwk} and the SH0ES prior from \cite{Riess:2019cxk}. In the last fitting column, however, we add the LSS data to the previous set but with no bispectrum (cf. Table 2 and Sec. \ref{sec:MethodData}). The corresponding results for the  GR-$\CC$CDM can be found in Tables 3 and 7.}
\end{table}

%%%%%%%%%%%%%%%%%%%%%%%%%%%%%%%%%%%%%%%%%%%%%%%%%%%%%%%%%%%%%%%%%%%%%%%%%%%%%%%%%%%%%%%%%

%%%%%%%%%%%%%%%%%%%%%%%%%%%%%%%%%%%%%%%%%%%%%%%%%%%%%%%%%%%%%%%
\begin{figure}[t!]
\begin{center}
\label{fig:triangular}
\includegraphics[width=6.5in, height=6.5in]{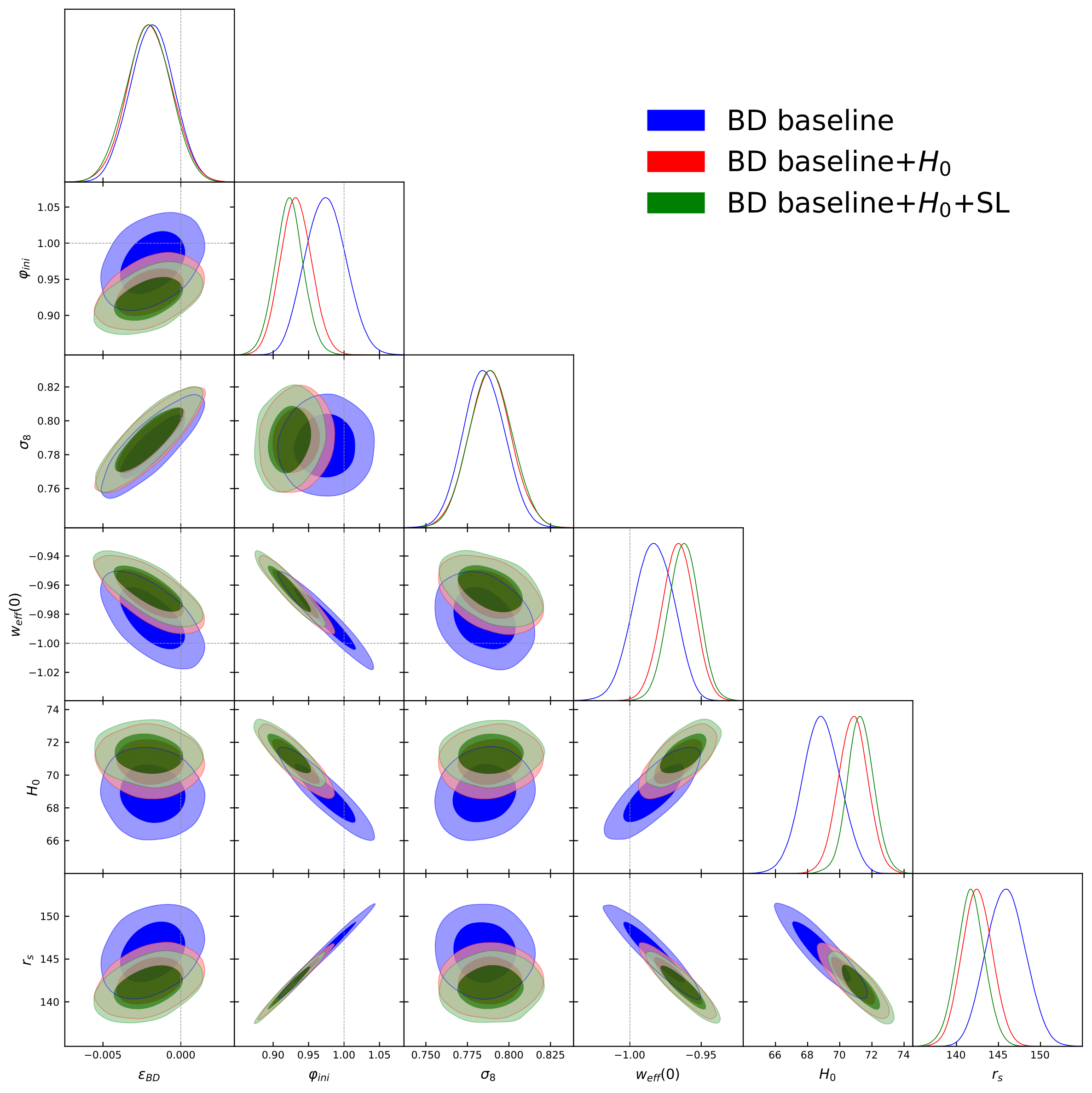}
\caption{\scriptsize{Triangular matrix containing the two-dimensional marginalized distributions for some relevant combinations of parameters in the BD-$\CC$CDM model (at $1\sigma$ and $2\sigma$ c.l.), together with the corresponding one-dimensional marginalized likelihoods for each of them. $H_0$ is expressed in km/s/Mpc, and $r_s$ in Mpc. See Tables 3 and 4 for the numerical fitting results.}}
\end{center}
\end{figure}
%%%%%%%%%%%%%%%%%%%%%%%%%%%%%%%%%%%%%%%%%%%%%%%%%%%%%%%%%%%%%%%

On the other hand, if we compare e.g. the fitting value predicted within the BD-$\CC$CDM model from Table 3 (namely $H_0=70.83^{+0.92}_{-0.95}$ km/s/Mpc) with the aforementioned  SH0ES determination, we can see that the difference is of only $1.58\sigma$. If we next compare our fitting result from Table\,4  ($H_0= 71.30^{+0.80}_{-0.84}$ km/s/Mpc), which incorporates the H$0$LICOW Strong-Lensing data in the fit as well, with the combined SH0ES  and H$0$LICOW result (viz. the one which is in $5\sigma$ tension with the CMB value)  we obtain once more an inconspicuous tension of only  $1.55\sigma$.  In either case  it is far away from any perturbing discrepancy.  In fact, no discrepancy which is not reaching a significance of at least  $3\sigma$ can be considered sufficiently worrisome.

\subsection{Alleviating the $\sigma_8$-tension}\label{sec:s8-tension}

Furthermore, the smoothing of the tension applies to the $\sigma_8$ parameter as well, with the result that it essentially disappears within a similar level of inconspicuousness. In fact, values such as $\sigma_8=0.789\pm 0.013$ and $\tilde{S}_8=0.792\pm 0.013$ (obtained within the Baseline$+H_0$ dataset, see Table 9) are in good agreement with weak gravitational lensing observations derived from shear data (cf. the WL data block mentioned in Sec. \ref{sec:MethodData}). Let us take  the value by Joudaki et al. 2018 of the combined observable $S_8 = 0.742\pm 0.035$, for example,  obtained by KiDS-450, 2dFLenS and BOSS collaborations from a joint analysis of weak gravitational lensing tomography and overlapping redshift-space galaxy clustering\,\cite{Joudaki:2017zdt}.  These observations can be compared  with our prediction for ${S}_8=\sigma_8\sqrt{\Omega_m/0.3}$  and  with the `renormalized' form of that quantity within the BD-$\CC$CDM model, namely $\tilde{S}_8=\sigma_8\sqrt{\tilde{\Omega}_m/0.3}$,  which depends on the modified cosmological parameter $\tilde{\Omega}_m$, which is slightly higher,  recall  our Eq.\,\eqref{eq:tildeOmegues}\footnote{Although we could use the hatted parameter $\hat{S}_8=\sigma_8\sqrt{\hat{\Omega}_m/0.3}$, instead of  $\tilde{S}_8$, we have already pointed out  that the difference between $\hat{\Omega}_m$ and $\tilde{\Omega}_m$ is negligible for $|\epsilon_{BD}|\sim \mathcal{O}(10^{-3})$, and so is between $\hat{S}_8$ and  $\tilde{S}_8$.}. Both ${S}_8$ and $\tilde{S}_8$ are displayed together in Table 9 for the main scenarios, also in company with $\sigma_8$ values for the GR and BD models.  Differences of the mentioned experimental measurements with respect to e.g. our prediction for the Baseline$+H_0$ dataset,   are at the level of  $0.5\sigma-1.3\sigma$ depending on whether we use $S_8$ or $\tilde{S}_8$, whence statistically irrelevant in any case. More details can be appraised on some of  these observables and their correlation with $H_0$ in Figs. 10 and 11, on which we shall further comment later on.

%%%%%%%%%%%%%%%%%%%%%%%%%%%%%%%%%%%%%%%%%%%%%%%%%%%%%%%%%%%%%%%
\begin{figure}[t!]
\begin{center}
\label{fig:H0S8}
\includegraphics[width=4.3in, height=3.8in]{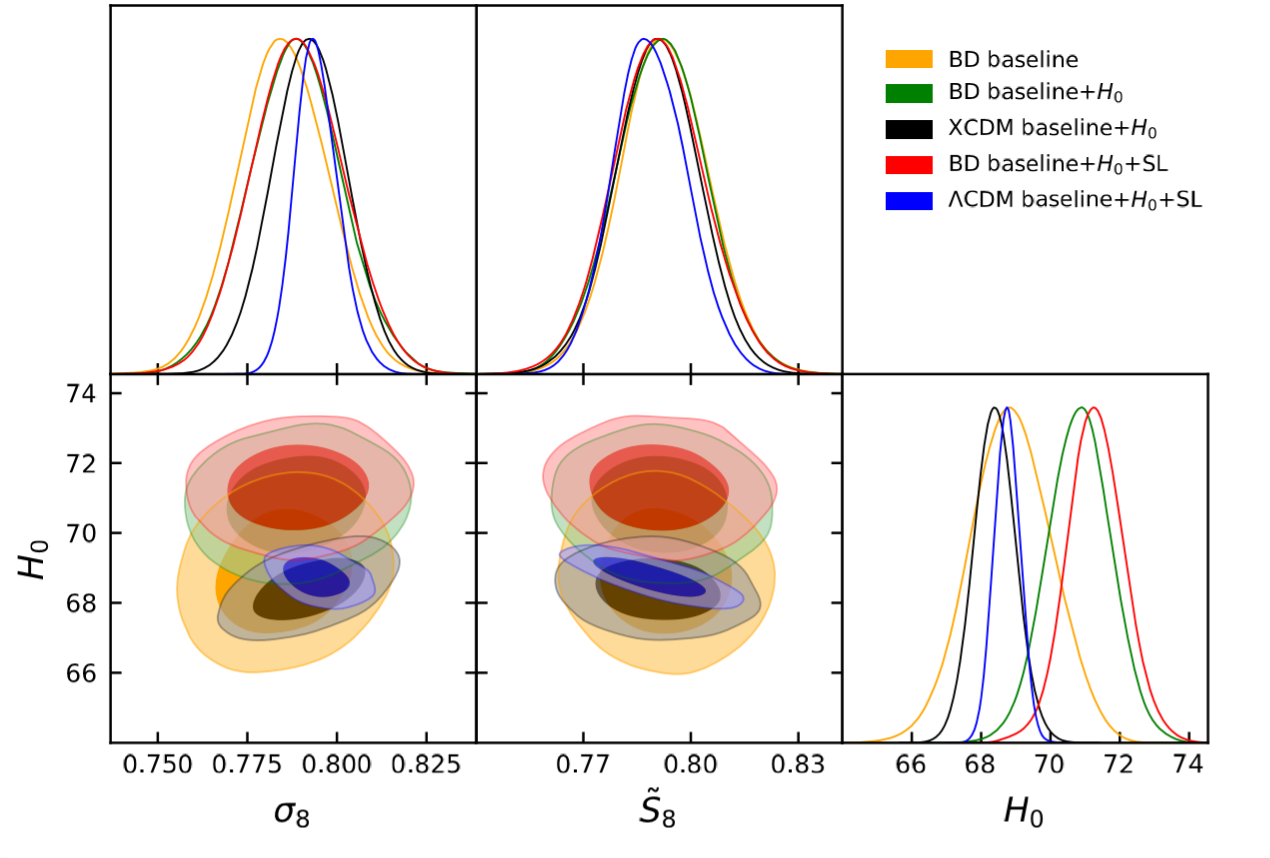}
\caption{\scriptsize {Constraints obtained for $\sigma_8$ and $\tilde{S}_8$ versus $H_0$ (in km/s/Mpc) from the fitting analyses of the GR- and BD-$\CC$CDM models, and the GR-XCDM parametrization. We show both, the contour lines in the corresponding planes of parameter space, and the associated marginalized one-dimensional posteriors. The centering of the parameters in the ranges $\sigma_8<0.80$ and $H_0\gtrsim71$ km/s/Mpc is a clear sign of the smoothening of the $\sigma_8$-tension and, more conspicuously, of the $H_0$ one within the BD-$\CC$CDM model. We can also see that while a simple XCDM parametrization for the DE can help to diminish $\sigma_8$ as compared to the concordance model, it is however unable to improve the $H_0$ tension, which is kept at a similar level as within the concordance model.}}
\end{center}
\end{figure}
%%%%%%%%%%%%%%%%%%%%%%%%%%%%%%%%%%%%%%%%%%%%%%%%%%%%%%%%%%%%%%%

\subsection{Comparing different scenarios}\label{sec:Others}

We have also tested the performance of the BD- and GR-$\CC$CDM models when we include the CMB high-$l$ (TE+EE) polarization data from Planck 2018, with and without the CMB lensing, in combination with the baseline dataset and the SH0ES prior on $H_0$ (cf. Tables 6 and 7). The values of the Hubble parameter  in these cases are a little bit lower than when we consider only the temperature and low-$l$ polarization (TT+lowE) CMB data, but the tension is nevertheless significantly reduced, being now of only $\sim 2.2\sigma$ c.l., whereas in the GR-$\CC$CDM model it is kept at the $\sim 3.5\sigma$ level. The values of $\sigma_8$ are still low, $\sim 0.80-0.81$. The information criteria in these cases, though, have no preference for any of the two models, they are not conclusive.

We also examine the results that are obtained when we do not include in our fitting analyses any of the data sources that trigger the tensions. We consider here the CMB, BAO and SNIa datasets (denoted as CMB+BAO+SNIa in Tables 6 and 7), but exclude the use of the SH0ES prior on $H_0$ and the LSS data.  As expected,  the evidence for the BD model decreases since now we do not give to it the chance of showing its power.  Even though the description of the data improves,  it is not enough to compensate for the penalty received owing to the use of the two additional parameters $(\epsilon_{BD},\varphi_{ini})$,  and in this case we read  $2\ln B=-3.00$ (from Table 6). Thus, there is here a marginal preference for the GR scenario, but as previously mentioned, this is completely normal, since we are removing precisely the data sources whose correct description demands for new physics. Even so, the $H_0$-tension is again remarkably reduced from  $3.8\sigma$  in  GR-$\CC$CDM to only  $2.4\sigma$  in the BD-$\CC$CDM. The respective values of $\sigma_8$ remain compatible with $0.80$  within  $\sim 1\sigma$.

It is also interesting to compare the results that we have obtained within the BD framework  with other approaches in the literature,  in which the variation of $G$ is dealt with as a small departure from GR, namely in a context where the action still contains a large mass scale $M$ near the Planck mass  $m_{\rm Pl}$, together with some scalar field which parametrizes the deviations from it. This is of course fundamentally different from the BD paradigm but it bears relation owing to the variation of the effective $G$, and it has also been used to try to smooth out the tensions. However, as already mentioned in the Introduction, and also in Sec. \ref{sec:preview}, it is not easy at all for a given model to fulfill the `golden rule', i.e. to loosen the two tensions at a time, or just to alleviate one of them without worsening the other.  Different proposals have appeared in the market trying to curb the $H_0$-tension, e.g. the so-called early dark energy models\,\cite{Poulin:2018cxd,Chudaykin:2020acu,Braglia:2020bym}, and the model with variable $G$ recently considered in \cite{Ballesteros:2020sik,Braglia:2020iik,Ballardini:2020iws}. Although the physical mechanism of the EDE and the aforementioned models with variable $G$ is of course very different, their aim is pretty similar. They reduce the sound horizon $r_s$ at recombination in order to force the increase of the Hubble function in the late universe. This allows them to generate larger values of $H_0$ in order to keep a good fit to the CMB and BAO data, but this happens only at the expense of increasing the tension in $\sigma_8$, since they do not have any compensation mechanism able to keep the structure formation in the late universe at low enough levels.  Some of these models  appear not to be particularly disfavored notwithstanding. But this is simply because they did not use LSS data in their fits, i.e. they did not put their models to the test of structure formation and for this reason they have more margin to adjust the remaining  observables without getting any statistical punishment.  So the fact that the significant increase of $\sigma_8$  that they find is not statistically penalized is precisely because they do not use  LSS data,  such as e.g. those on the observable $f(z)\sigma_8(z)$ displayed in our Table 2.  In this respect, EDE cosmologies are an example; they seem to be unable to alleviate the $H_0$-tension when LSS data are taken into account, as shown in \cite{Hill:2020osr}.

\subsection{Imposing the Cassini constraint}\label{sec:Cassini}

To further illustrate the capability of the BD-$\CC$CDM model to fit the data under more severe conditions, in  Table 10 we consider four possible settings to fit our Cassini-constrained  BD-Scenario III defined in Sec.\,\ref{sec:Mach}. Recall that this BD scenario involves the stringent Cassini bound on the post-Newtonian parameter $\gamma^{PN}$ \cite{Bertotti:2003rm}, which we have discussed in Sec.\,\ref{sec:BDgravity}. The first two fitting columns of Table 10 correspond to our usual Baseline$+H_0$ dataset,  in one case (first fitting column in that table) we omit the LSS data, whereas in the second column we restore it. In this way we can check the effect of the structure formation data on the goodness of the fit. The comparison between the results presented in these two columns shows, first and foremost, that the Cassini bound does not have a drastically nullifying effect, namely it does not render the BD-$\CC$CDM model irrelevant to the extent of  making it indistinguishable from GR-$\CC$CDM, not at all, since the quality of the fits is still fairly high (confer the Bayes factors in the last row). In truth, the fit quality is still comparable to that of the Baseline$+H_0$ scenario (cf. Table 3). However, the description of the LSS data is naturally poorer since $\eBD$ is smaller and the model cannot handle so well  the features of the structure formation epoch, thus yielding slightly higher values of $\sigma_8$. Second, the fact that the scenario without LSS furnishes a higher Bayes factor  just exemplifies the aforementioned circumstance that when cosmological models are tested without using this kind of data the results may in fact not be sufficiently reliable. When the LSS data enter the fit (third fitting colum in that table), we observe, interestingly enough, that the BD-$\CC$CDM model is still able to keep $H_0$  in the safe range, it does improve the value of $\sigma_8$ as well  (i.e. it becomes lower) and, on top of that, it carries a (`smoking gun')  signal of almost $2.9\sigma$ c.l. --  encoded in the value of $\weff$ --  pointing to quintessence-like behavior. Overall it is quite encouraging since it shows that the Cassini bound does not exceedingly hamper the BD-$\CC$CDM model capabilities. Such bound constraints the time evolution of $\varphi$ (because $|\eBD|$ is forced to be much smaller) but it does not preclude $\varphi$ from choosing a suitable value in compensation (cf. BD-Scenario III in Sec.\,\ref{sec:Mach}).

In the last two columns of Table 10, we can further  check what are the changes in the previous fitting results when we use a  more restricted dataset,  e.g. the one used in Ref.\,\cite{Ballesteros:2020sik}, in which the Cassini bound is also implemented. These authors study a model which represents a modification of GR through an effective $G\sim 1/M^2$, with $M$ a mass very near  the Planck mass, $\mPl$, which is allowed to change slowly through a scalar field $\phi$ as follows: $M^2\to M^2+\beta\phi ^2$, where $\beta$ is a small (dimensionless) parameter. The authors assume that the Cassini bound on the post-Newtonian parameter $\gamma^{PN}$ \cite{Bertotti:2003rm} is in force (see Sec. \ref{sec:MethodData} for details). However, they do not consider LSS data (only CMB and BAO). We may compare the results they obtain within that variable $G$ model (cf. their Table 1) with those we obtain within the BD-Scenario III under the very same dataset as these authors. The results are displayed in the third fitting column of Table 10.  We obtain  $H_0=(70.01^{+0.86}_{-0.92})$ km/s/Mpc and $\sigma_8=0.817\pm 0.007$, whereas they obtain $H_0=(69.2^{+0.62}_{-0.75})$ km/s/Mpc and $\sigma_8=0.843^{+0.015}_{-0.024}$. Clearly, the BD-$\CC$CDM is able to produce larger central values of $H_0$ and lower values of $\sigma_8$, even under the Cassini bound, although the differences are  within errors. The value of $2\ln\,B$ lies around $+3.5$ and hence points to a mild positive evidence in favor of the BD model. This is consistent with the associated deviation we find  of $G(0)$ from $G_N$ at $2.43\sigma$ c.l., and with a signal of effective quintessence at $2.53\sigma$ c.l. within the GR-picture.

Let us  now consider what is obtained if we add up the LSS data to this same BD-Scenario III, still with the restricted dataset od Ref.\,\cite{Ballesteros:2020sik}. As expected, the inclusion of the structure formation data
%, which remember contains the important information on SP+BSP,
pushes the value of $\sigma_8=0.812^{+0.006}_{-0.007}$ down as compared to their absence ($\sigma_8=0.817\pm 0.007$). This is the most remarkable difference between the two cases, as one cannot appreciate significant changes in the other parameters. Something very similar happens when we compare the values of $\sigma_8$ of the first two fitting columns of Table 10, in which we consider the B+$H_0$ dataset without LSS data (first fitting column) and with LSS data (second fitting column). The relative improvement {\it w.r.t.} the model of \cite{Ballesteros:2020sik} is therefore greater in the presence of LSS data, whose use has been omitted in that reference. In that variable $G$ model one finds a larger value of $H(z)$ at recombination thanks to the larger values of $G$ in that epoch, but at present $G$ is forced to be almost equal to $G_N$ (being the differences not relevant for cosmology). This means that: (i) in that model $G$ decreases with time, which leads to a kind of effective $\epsilon_{BD}>0$; (ii) the model cannot increase $H_0$ with a large value of $G(z=0)$. Both facts do limit significantly  the effectiveness of the model in loosening the tensions. In the BD-$\CC$CDM model under consideration, instead, we find that $G$ has to be $\sim 8-9\%$ larger than $G_N$ not only in the pre-recombination universe, but also at present, and this allows to reduce significally the $H_0$-tension. Moreover, we find that a mild increase of the cosmological $G$ with the expansion leads also to an alleviation of the $\sigma_8$-tension.

Under all of the datasets studied in Table 10 we obtain central values of $|\epsilon_{BD}|\sim \mathcal{O}(10^{-5})$, which are compatible with 0 at $1\sigma$. Notice that this value is just of order of the Cassini bound on $\eBD$, as could be expected. Notwithstanding, and remarkably enough, the stringent bound imposed by the Cassini constraint, which enforces $\epsilon_{BD}$ to remain two orders of magnitude lower than in the main Baseline  scenarios, is nevertheless insufficient to wipe out the  positive effects from the BD-$\CC$CDM model. They are still capable to emerge with a sizeable part of the genuine BD signal. This is, as anticipated in Sec.\,\ref{sec:Mach}, mainly due to the fact that the Cassini bound cannot restrict the value of the BD field  $\varphi$, only its time evolution.

\subsection{More on alleviating simultaneously the two tensions}\label{sec:TwoTensions}

A few additional comments on our results concerning the parameters $H_0$  and $\sigma_8$ are now in order. Their overall impact can be better assessed by examining  the triangular matrix of fitted contours involving all the main parameters, as shown in Fig.\,10,
in which we offer the numerical results of several superimposed analyses based on different datasets, all of them within Baseline scenarios. We project the contour lines in the corresponding planes of parameter space, and  show the associated marginalized one-dimensional posteriors.   The fitted value of the EoS parameter at $z=0$  shown there, $\weff(0)$, is to be understood, of course, as a derived parameter from the prime ones of the fit, but we include this information along with the other parameters in order to further display the significance of the obtained signal:  $\gtrsim 3\sigma$  quintessence-like behavior.  Such signal, therefore, mimics  `GR+ DDE' and  hints at something beyond pure GR-$\CC$CDM. What we find in our study is that such time-evolving DE behavior is actually of quasi-vacuum type and appears as a kind of  signature of the underlying BD theory\footnote{As we recall in subsection \ref{sec:RVMconnection} of Appendix \ref{AppendixA}, such  kind of  dynamical behavior of the vacuum is characteristic of the Running Vacuum Model (RVM),  a version of the   $\CC$CDM in which the vacuum energy density is not just a constant  but involves also a dynamical term $\sim H^2$.  The description of the BD-$\CC$CDM model in the GR-picture mimics a  behavior of this sort \,\cite{Peracaula:2018dkg,Perez:2018qgw}.}.

Figure\,10 provides  a truly panoramic and graphical view of our main fitting results,  and  from where we can comfortably  judge the impact of the BD framework for describing the overall cosmological data. It is fair to say that it appears at a level highly competitive with GR --  in fact, superior to it.  In all datasets involving the local $H_0$ input in the fit analyses, the improvement is substantial and manifest. Let us stand out only three of the entries in that graphical matrix:  i) for the parameters $(\sigma_8,H_0)$,  all the contours in the main dataset scenarios are centered around values of $\sigma_8<0.80$ and $H_0\gtrsim71$ km/s/Mpc, which are the coveted ranges for every model aiming at smoothing the two tensions at a time;  ii) for the pair $(H_0,r_s)$, the contours are centered around the same range of relevant  $H_0$ values as before, and also around values of the comoving sound horizon (at the baryon drag epoch) $r_s\lesssim142$ Mpc, these being significantly smaller  than those of the concordance $\CC$CDM  (cf. Table 7) and hence consistent with larger values of the expansion rate at that epoch;  iii) and for  $(\weff(0), H_0)$,  the relevant  range $H_0\gtrsim71$ km/s/Mpc is once more picked out, together with an effective quintessence signal $\weff(0)>-1$ at more than $3\sigma$ (specifically $3.45\sigma$ for the scenario of Table 4, in which strong lensing data are included in the fit).

It is also interesting to focus once more our attention on  Fig.\, 11, where we provide devoted contours involving  both  $\tilde{S}_8$ and $\sigma_8$  versus $H_0$.  On top of the observations we have previously  made on these observables, we  can compare here our basic dataset scenarios for the BD-$\CC$CDM model with the yield of a simple  XCDM parametrization of the DDE. In previous studies we had already shown that such parametrization can help to deal with the $\sigma_8$ tension\,\cite{Sola:2018sjf}. Nonetheless, as we can see here, it proves completely impotent for solving  or minimally helping to alleviate the $H_0$-tension since the values predicted for this parameter stay as low as in the concordance model. This shows, once more,  that in order to address a possible solution to the two tensions simultaneously, it is not enough to have just some form of dynamics in the DE sector; one really needs a truly specific one,  e.g. the one provided (in an effective way) by the BD-$\CC$CDM model.

\subsection{Predicted relative variation of the effective gravitational strength}\label{sec:VariationG}

In the context of the BD framework it is imperative, in fact mandatory,  to discuss the current values of the relative variation of the effective gravitational strength, viz. of $\dot{G}(0)/G(0)$, which follow from our fitting analyses (see the main Tables 3-5). The possible time evolution of that quantity hinges directly on $\eBD$, of course, since the latter is the parameter that controls the (cosmological) evolution of the gravitational coupling in the BD theory. It is easy to see from Eq.\,\eqref{eq:definitions} that $\dot{G}(0)/G(0)=-\dot{\varphi}(0)/\varphi(0)\simeq -\eBD H_0$, where we use the fact that $\varphi\sim a^{\eBD}$ in the matter-dominated epoch (cf. Appendix \ref{AppendixB}). Recalling that $H_0\simeq 7\times 10^{-11}$ yr$^{-1}$ (for $h\simeq 0.70$), we find  $\dot{G}(0)/G(0)\simeq -\eBD\cdot 10^{-10}\,yr^{-1}$. Under our main BD-Scenario II (cf.\,Sec.\,\ref{sec:Mach}) we obtain values for $\dot{G}(0)/G(0)$ of order $\mathcal{O}(10^{-13})\,yr^{-1}$ (and positive), just because $\epsilon_{BD}\sim \mathcal{O}(10^{-3})$ (and negative). Being  $\dot{G}(0)/G(0)>0$ it means that the effective gravitational coupling obtained by our global cosmological fit increases with the expansion, and hence it was smaller in the past. This suggests that the sign $\eBD<0$, which is directly picked out by the data, prefers a kind of asymptotically free behavior for the gravitational coupling since the epochs in the past are more energetic,  in fact characterized by larger values of $H$ (with natural dimension of energy).  The central values show a mild time variation at present, at a level of
 $1.3\sigma$,  both for $\eBD$ and $\dot{G}(0)/G(0)$, when the bispectrum data from BOSS is also included in the analysis (cf. Tables 3 and 4). Such departure goes below $1\sigma$ level when only the spectrum is considered (see Table 5). In the context of the BD-Scenario III, in which $\eBD$ is very tightly constrained by the Cassini bound \cite{Bertotti:2003rm}, namely at a level of $\mathcal{O}(10^{-5})$, we find $\dot{G}(0)/G(0)\sim 10^{-15}\,yr^{-1}$, which is compatible with 0 at $1\sigma$. All that said, we should emphasize once more  that the fitting values that we obtain for $\dot{G}(0)/G(0)$ refer to the cosmological time variation of $G$ and, therefore, cannot be directly compared with constraints existing in the literature based on strict local gravity measurements, such as e.g. those from the lunar laser ranging experiment -- $\dot{G}(0)/G(0)=(2\pm 7)\cdot 10^{-13}yr^{-1}$ \cite{Muller:2007zzb} -- (see e.g. the review \cite{Uzan:2010pm} for a detailed presentation of many other local constraints on $\dot{G}(0)/G(0)$). Even though this bound turns out to be preserved within our analysis, it is not in force at the cosmological level provided  an screening mechanism acting at these scales is assumed, as in our case. Thus, the local measurements have no bearing a priori on the BD-$\CC$CDM cosmology. The opposite may not be true, for despite  the fact that the values reported in our tables are model-dependent, they prove to be quite efficient and show that the cosmological observations can compete in precision with the local measurements.

 %%%%%%%%%%%%%%%%%%%%%%%%%%%%%%%%%%%%%%%%%%%%%%%%%%%%%%%%%%%%%%%%%%%%%%%%%%%%%%%%%%%%%
\begin{figure}[t!]
\begin{center}
\label{fig:Cls-ISW}
\includegraphics[width=4.in, height=3in]{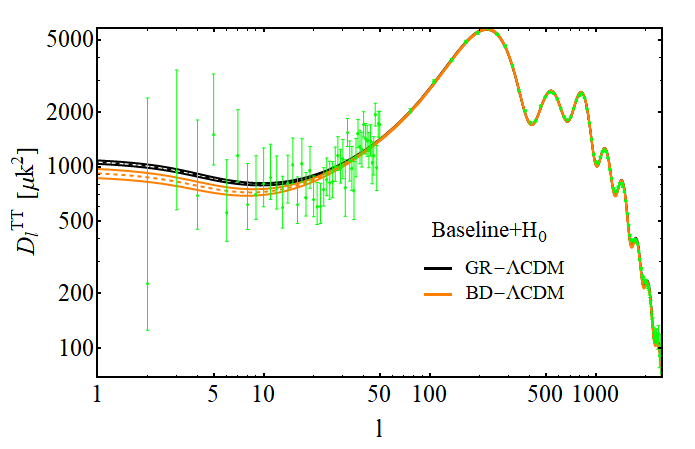}
\caption{\scriptsize{CMB temperature power spectrum for the GR-$\CC$CDM (in black) and BD-$\CC$CDM (in orange), obtained from the fitting results within the Baseline+$H_0$ dataset (cf. Table 3 and Sec. \ref{sec:MethodData}). We plot the central curves together with the corresponding $1\sigma$ bands. In the inner plot we zoom in the multipole range $l\in [0,30]$ and include the Planck 2018 \cite{Aghanim:2018eyx} data error bars (in green). At low multipoles ($l\lesssim 30$) the BD-$\CC$CDM model produces less power than the concordance GR-$\CC$CDM model owing to the suppression of the Integrated Sachs-Wolfe effect that we have discussed in Sec. \ref{sec:preview}. The differences are $\gtrsim 1\sigma$, and allow to soften a well-known low-multipole CMB anomaly. See the main text for further discussion.}}
\end{center}
\end{figure}
%%%%%%%%%%%%%%%%%%%%%%%%%%%%%%%%%%%%%%%%%%%%%%%%%%%%%%%%%%%%%%%

 %%%%%%%%%%%%%%%%%%%%%%%%%%%%%%%%%%%%%%%%%%%%%%%%%%%%%%%%%%%%%%%%%%%%%%%%%%%%%%%%%%%%%%
 \begin{table}[h!]
\begin{center}
\begin{tabular}{|c  |c | c |  c | c | c  |}
 \multicolumn{1}{c}{} & \multicolumn{2}{c}{Baseline} & \multicolumn{2}{c}{Baseline+$H_0$}
\\\hline
{\scriptsize Parameter} & {\scriptsize GR-$\Lambda$CDM}  & {\scriptsize BD-$\Lambda$CDM} & {\scriptsize GR-$\Lambda$CDM}  &  {\scriptsize BD-$\Lambda$CDM}
\\\hline
{\scriptsize $H_0$ (km/s/Mpc)}  & {\scriptsize $67.75^{+0.46}_{-0.48}$} & {\scriptsize $68.86^{+1.26}_{-1.22}$} & {\scriptsize $68.35^{+0.49}_{-0.46}$}  & {\scriptsize $70.81^{+0.95}_{-0.92}$}
\\\hline
{\scriptsize$\omega_b$} & {\scriptsize $0.02231^{+0.00018}_{-0.00020}$}  & {\scriptsize $0.02214\pm 0.00031$} & {\scriptsize $0.02239\pm 0.00019$}  &  {\scriptsize $0.02239^{+0.00029}_{-0.00032}$}
\\\hline
{\scriptsize$\omega_{cdm}$} & {\scriptsize $0.11655^{+0.00110}_{-0.00111}$}  & {\scriptsize $0.11860^{+0.00194}_{-0.00197}$} & {\scriptsize $0.11625^{+0.00099}_{-0.00106}$}  &  {\scriptsize $0.11821^{+0.00214}_{-0.00200}$}
\\\hline
{\scriptsize$\tau$} & {{\scriptsize$0.053^{+0.006}_{-0.008}$}} & {{\scriptsize$0.054^{+0.006}_{-0.008}$}} & {{\scriptsize$0.053^{+0.006}_{-0.008}$}}  &   {{\scriptsize$0.054^{+0.006}_{-0.009}$}}
\\\hline
{\scriptsize$n_s$} & {{\scriptsize$0.9706^{+0.0041}_{-0.0040}$}}  & {{\scriptsize$0.9691^{+0.0104}_{-0.0093}$}} & {{\scriptsize$0.9717^{+0.0041}_{-0.0042}$}} &   {{\scriptsize$0.9791^{+0.0097}_{-0.0090}$}}
\\\hline
{\scriptsize$\sigma_8$}  & {{\scriptsize$0.770^{+0.016}_{-0.018}$}}  & {{\scriptsize$0.759^{+0.0184}_{-0.0163}$}} & {{\scriptsize$0.780^{+0.018}_{-0.015}$}}  &   {{\scriptsize$0.762^{+0.021}_{-0.018}$}}
\\\hline
{\scriptsize$r_s$ (Mpc)}  & {{\scriptsize$148.03\pm0.33$}}  & {{\scriptsize$146.16^{+2.40}_{-2.58}$}} & {{\scriptsize$148.04^{+0.32}_{-0.34}$}}  &   {{\scriptsize$142.81^{+1.91}_{-2.03}$}}
\\\hline
{\scriptsize$m_{\nu}$ (eV)}  & {{\scriptsize$0.161^{+0.058}_{-0.059}$}}  & {{\scriptsize$0.409^{+0.139}_{-0.198}$}} & {{\scriptsize$0.118^{+0.053}_{-0.058}$}}  &   {{\scriptsize$0.409^{+0.183}_{-0.228}$}}
\\\hline
{\scriptsize $\epsilon_{BD}$} & - & {{\scriptsize $0.00459^{+0.00316}_{-0.00319}$}} & - &   {{\scriptsize $0.00433^{+0.00395}_{-0.00336}$}}
\\\hline
{\scriptsize$\varphi_{ini}$} & - & {{\scriptsize $0.979^{+0.028}_{-0.032}$}} & - &    {{\scriptsize $0.938^{+0.023}_{-0.024}$}}
\\\hline
{\scriptsize$\varphi(0)$} & - & {{\scriptsize $1.016^{+0.041}_{-0.049}$}} & - &    {{\scriptsize $0.972^{+0.042}_{-0.043}$}}
\\\hline
{\scriptsize$\omega_{eff}(0)$} & - & {{\scriptsize $-1.005^{+0.021}_{-0.017}$}} & - &    {{\scriptsize $-0.986^{+0.017}_{-0.016}$}}
\\\hline
{\tiny $\dot{G}(0)/G(0) (10^{-13}yr^{-1})$} & - & {{\scriptsize $-5.048^{+3.485}_{-3.434}$}} & - &    {{\scriptsize $-4.915^{+3.749}_{-4.501}$}}
\\\hline
{\scriptsize$\chi^2_{min}$} & {\scriptsize 2270.54}  & {\scriptsize 2268.44} & {\scriptsize 2285.02}  &  {\scriptsize 2274.76}
\\\hline
{\scriptsize$2\ln B$} & {\scriptsize -}  & {\scriptsize -0.12} & {\scriptsize -}  &  {\scriptsize +9.34}
\\\hline
{\scriptsize$\Delta {\rm DIC}$} & {\scriptsize -}  & {\scriptsize -0.81} & {\scriptsize -}  &  {\scriptsize +8.22}
\\\hline
\end{tabular}
\caption{\newtext{As in Table 3, but here we allow the variation of the mass of the massive neutrino ($m_\nu$) in the Monte Carlo routine, instead of setting it to $0.06$ eV. The other two neutrinos remain massless. We have used the same conservative prior range for $m_\nu\in [0,1]$ eV in both, the GR- and BD-$\Lambda$CDM models.}}
\end{center}
\label{tableFit3}
\end{table}

 \subsection{One more bonus: suppressing the power at low multipoles }\label{sec:LowMultipoles}

An additional bonus from the  BD cosmology is worth mentioning before we close this lengthy study. It is found in the description of the CMB temperature anisotropies. As we have discussed in Sec. \ref{sec:preview}, the BD-$\CC$CDM model is, in principle, able to suppress the power at low multipoles ($l\lesssim 30$), thereby softening one of the so-called CMB anomalies that are encountered in the context of the GR-$\CC$CDM model. This is basically due to the low values of $\varphi<1$ preferred by the data, which in turn produce a suppression  of the Integrated Sachs-Wolfe (ISW) effect\,\cite{Sachs:1967er,Das:2013sca}. We have confirmed that this suppression actually occurs for the best-fit values of the parameters in our analysis, cf. Fig. 12. The aforementioned anomaly is not very severe, since the  power at low multipoles is affected by a large cosmic variance and cannot be measured very precisely. Nevertheless, it is a subtle anomaly which has been there unaccounted for a long time and could not be improved in a consistent way: that is to say, usually models ameliorating the low tail of the spectrum do spoil the high part of it. However, here  the suppression of power with respect to the GR-$\CC$CDM at $l\lesssim 30$ is fully consistent and is another very welcome feature of the BD-$\CC$CDM model, which is not easy at all to attain.  It is interesting to mention that the ISW effect can be probed by cross correlating the CMB temperature maps with the LSS data, e.g. with foreground galaxies number counts, especially if using  future surveys which should have much smaller uncertainties.  This can be a useful probe for DE and a possible distinctive signature for DDE theories\,\cite{Pogosian:2004wa,Zucca:2019ohv}\,\footnote{We thank L. Pogosian for interesting comments along these lines.}.  The upshot of our investigation is that once more  the `golden rule' mentioned in the Introduction is preserved here and the curing effects from the BD-$\CC$CDM stay aligned: the  three tensions of the  GR-$\CC$CDM ($H_0$, $\sigma_8$ and  the exceeding CMB power at low multipoles) can be improved at a time.

\subsection{The effect of massive neutrinos }\label{sec:MassiveNeutrino}

\newtext{Finally, it is worth assessing the impact of leaving the sum of the three neutrino masses ($M_\nu\equiv\sum m_\nu$) as a free parameter, rather than fixing it. We model this scenario in two different ways:}

\begin{itemize}
    \item \newtext{Considering one massive ($m_\nu$) and two massless neutrinos, so $M_\nu=m_\nu$. Here we try to mimic the physics encountered when the neutrino masses follow the normal \newtext{hierarchy}, in which one of the neutrinos is much heavier than the other two, for reasonable values of $M_\nu$.}
    \item \newtext{Considering three massive neutrinos of equal mass $m_\nu$, such that $M_\nu=3m_\nu$. This is the degenerated case, utilized also in the analysis by Planck 2018 \cite{Aghanim:2018eyx} (cf. Sec. 7.5.1 therein).}
\end{itemize}

\newtext{In both cases, the BD-$\CC$CDM has one additional parameter {\it w.r.t.} the same model with fixed mass $M_\nu=m_\nu=0.06$ eV that we have discussed in the previous sections, and three more than the vanilla $\Lambda$CDM model: $(\eBD,\varphi_{ini}, M_\nu)$. Upon taking into account the constraints obtained from neutrino oscillation experiments on the mass-squared splittings through the corresponding likelihoods we would perhaps find more precise bounds on the sum of the neutrino masses\,\cite{Loureiro:2018pdz}, but here we want to carry out a more qualitative analysis to study how massive neutrinos may impact on our results, considering only cosmological data. The results of the first mass scenario are shown in Table 11. Those for the second scenario are not tabulated, but are commented below.}

\newtext{As can be seen from Table 11, the effect of having a neutrino with an adjustable mass picked out by the fitting process is non-negligible. It produces a significant lowering of the value of $\sigma_8$ while preserving the value of $H_0$ at a level comparable to the previous tables in which the neutrino had a fixed mass of $0.06$ eV, therefore preserving what we have called the `golden rule'.  Let us note, however, that the sign of $\eBD$ has changed now with respect to the situation with a fixed light mass, it is no longer negative but positive and implies a value of the BD parameter of $\wBD\simeq230$.  Last but not least, the fitted value of the neutrino mass is $m_\nu=0.409$ eV, which is significantly higher than the upper bound placed by Planck 2018 under the combination TTTEEE+lowE+lensing+BAO: $M_\nu\equiv\sum m_\nu<0.120$ eV ($95\%$ c.l.)\,\cite{Aghanim:2018eyx}. }

\newtext{Similar results are obtained with the second mass scenario mentioned above.  Let us summarize them. For the Baseline+$H_0$ dataset, the common fitted mass value obtained for the three neutrinos is $m_\nu=0.120^{+0.054}_{-0.068}$ eV, which means that $M_\nu\simeq 0.360$ eV, slightly lower than in the first scenario. The corresponding BD parameter reads comparable, $\wBD\simeq 261$, and the values for $\sigma_8$ and $H_0$ are also very close to the previous case, so the two scenarios share similar advantages. The fact that the Planck 2018 upper limit for $M_\nu$ is overshooted in both neutrino mass scenarios does not necessarily exclude them, as the limits on  $M_\nu$ are model-dependent, see e.g. \cite{Loureiro:2018pdz} and \cite{Ballardini:2020iws}. In particular, Planck 2018 obviously used  GR-$\CC$CDM. We can check in  Table 11 that the neutrino mass values for BD-$\CC$CDM are substantially different. The results that we have obtain for GR-$\CC$CDM are fully compatible with those from Planck 2018.}
%in particular the result we obtain for GR-$\CC$CDM in the Baseline case is essentially coincident with the Planck 2018 result. }
\newtext{We conclude that the influence of neutrino masses on the BD-$\CC$CDM fitting results is potentially significant, but it cannot be settled at this point. The subject obviously deserves further consideration in the future. }

\section{Conclusions}\label{sec:Conclusions}

To summarize, we have presented a rather comprehensive work on the current status of the Brans-Dicke theory with a cosmological constant in the light of the modern observations. Such framework constitutes a new version of the concordance $\CC$CDM model in the context of a distinct gravity paradigm, in which the gravitational constant is no longer a fundamental constant of Nature but a dynamical field. We have called this framework BD-$\CC$CDM model to distinguish it from the conventional one, the GR-$\CC$CDM model, based on General Relativity.  Our work is a highly expanded and fully updated analysis of our previous and much shorter presentation\,\cite{Sola:2019jek}, in which we have  replaced the Planck 2015 data by the Planck 2018 data,  and we have included additional sets of modern cosmological observations. We reconfirm the results of \,\cite{Sola:2019jek} and provide now a bunch of new results which fit in with the conclusions of our previous work and reinforce its theses.  To wit:  in the light of the figures and tables that we have presented in the current study we may assert that the BD-$\CC$CDM model fares better not only as compared to the GR-$\CC$CDM  with a rigid cosmological constant (cf. Tables 3-7 and 10) but also when the CC term is replaced with a dynamical parametrization of the DE, such as the traditional XCDM, which acts as a benchmark (cf. Table 8).  We find that the GR-XCDM is completely unable to enhance the value of $H_0$ beyond that of the concordance model.  In particular, in Tables 3-4  we can see that the information criteria (Bayes factor and Deviance Information Criterion) do favor  significantly and consistently the BD-$\CC$CDM model as compared to GR-$\CC$CDM and  GR-XCDM. There is a very good resonance between the Bayesian evidence criterion and the DIC differences, which definitely uphold the BD framework at a level of $+5$ units for the Baseline+$H_0$ dataset scenario, meaning that the degree of support of BD versus GR is in between \textit{positive} to \textit{strong} (cf. Sec.\,\ref{sec:NumericalAnalysis}). This support is further enhanced up to more than $+9$ units, hence  in between \textit{strong} to \textit{very strong}, for the case  when we include the Strong-Lensing data in the fit (see Table 4). The exact Bayesian evidence curves computed in Fig.\,9 reconfirm these results in a graphical way. The pure baseline dataset scenario (in which the local $H_0$ value is not included) shows weak evidence; however, as soon as the local $H_0$ value is fitted along with the remaining parameters the evidence increases rapidly and steadily,  reaching the status of positive, strong and almost very strong depending on the datasets.

\newtext{Another dataset scenario which is particularly favored in our analysis} is the one based on considering the effective calibration prior on the absolute magnitude $M$ of the nearer SNIa data in the distance ladder (as defined in Sec.\,\ref{sec:MethodData}), instead of the local value $H_0$ from SH0ES. The results for the Baseline dataset in combination with $M$ (denoted B+$M$) can be read off from Table 6 (first row).  We can see it yields a tantalizing overall output, with values of $H_0$ and  $\sigma_8$ in the correct ranges for solving the two tensions, and fully compatible with the results obtained using the prior on $H_0$, as expected. In addition, the corresponding Bayes factor for this scenario points to a remarkably high value $2\ln B>+10$, thereby carrying a very strong Bayesian evidence,  in fact comparable to the Baseline+$H_0$+SL scenario of Table 4.  In all of the mentioned cases in our summary the information criteria  definitely endorse the BD-cosmology versus the GR one.

\newtext{Finally, we have also assessed the influence of the neutrino masses in the context of the BD-$\CC$CDM model, see Table 11. We have found that massive neutrinos can help to further reduce the predicted value of $\sigma_8$ to the level of what is precisely needed to describe the weak gravitational lensing observations derived from direct shear data. This would completely dissolve the $\sigma_8$-tension without detriment of the positive results obtained to loosen the $H_0$-tension, i.e. by preserving the `golden rule'.  Such a conclusion is, however, provisional as it requires a devoted study of the neutrino sector (extending the analysis of Sec.\,\ref{sec:MassiveNeutrino}) which is beyond the scope of the current paper.}

 Overall, the statistical support in favor of the BD-$\CC$CDM model against the concordance  GR-$\CC$CDM model is rather significant. It is not only that the $H_0$ and $\sigma_8$ tensions are simultaneously dwarfed to a level where they are both rendered inessential ($\lesssim 1.5\sigma$), but also that all tested BD scenarios involving the local $H_0$ value provide a much better global fit than the concordance model on the basis  of a rich and updated set of modern observations from all the main cosmological data sources available at present. If we take into account that the BD-$\CC$CDM framework is not just some \textit{ad hoc} phenomenological toy-model, or some last-minute smart parametrization just concocted to solve or mitigate the two tensions, but the next-to-leading fundamental theory candidate directly competing with GR, it may give us a sense of the potential significance of these results.

\vspace{0.5cm}

{\bf Acknowledgements}
JSP, JdCP and CMP are partially supported by projects  FPA2016-76005-C2-1-P (MINECO), 2017-SGR-929 (Generalitat de Catalunya) and MDM-2014-0369 (ICCUB). JdCP is also supported by a  FPI fellowship associated to the project FPA2016-76005-C2-1-P. CMP is partially supported by the fellowship 2019 FI-B 00351. AGV is funded by the Deutsche Forschungsgemeinschaft (DFG) - Project number 415335479.  This work is also partially supported by the COST Association Action CA18108  ``Quantum Gravity Phenomenology in the Multimessenger Approach  (QG-MM)''.  We thank A.  Notari and L. Pogosian for discussions.

\vspace{0.5cm}

%\newpage

{\bf Special acknowledgement dedicated to  Roberto D. Peccei}
\vspace{0.2cm}

We would like to dedicate this comprehensive work to the memory of Professor Roberto D. Peccei.   Those of us (JSP)  who had the  immense fortune  to know him, and to work with him, at DESY (Hamburg) and UCLA (Los Angeles), will never forget such an extraordinary privilege in our lives.  We  learned not only (a lot of)  physics from him but also the very meaning of doing Physics and Science `in capital letters', which consists not just in doing research for the sake of pure knowledge  but for the sake of a better  Humanity and a better World.   We have lost a great man, a great scientist and a most beloved mentor and friend.   Thanks for everything Roberto!  I/we will never forget you.

\vspace{1cm}
%\newpage

\appendix

\section{Semi-analytical solutions in different epochs}\label{AppendixA}

Our aim in this section is to find semi-analytical solutions for the BD equations in the various epochs of the cosmic history, by using a perturbative approach. We express the BD-field and the scale factor up to linear order in $\epsilon_{BD}$ as follows,
\begin{align}
&\varphi(t)=\varphi^{(0)}+\epsilon_{BD}\varphi^{(1)} (t)+\mathcal{O}(\epsilon_{BD}^2)\label{ExpansionOfvarphi} \\
& a(t)=a^{(0)}(t)+\epsilon_{BD} a^{(1)}(t)+\mathcal{O}(\epsilon_{BD}^2)\,, \label{ExpansionOfa}
\end{align}
with the functions with superscript $(0)$ denoting the solutions of the background equations in standard GR with a constant Newtonian coupling that can be in general different from $G_N$, and the functions with superscript $(1)$ denoting the first-order corrections induced by a non-null $\epsilon_{BD}$. Neglecting the higher-order terms is a very good approximation for all the relevant epochs of the expansion history due to the small values of $\epsilon_{BD}$ allowed by the data. Plugging these expressions in Eqs. \eqref{eq:Friedmannequation}-\eqref{eq:FullConservationLaw} we can solve the system and obtain the dominant energy density at each epoch and the Hubble function, which of course can also be written as
\begin{align}
&\rho_N(t) = \rho_N^{(0)}(t)+\epsilon_{BD} \rho_N^{(1)}(t)+\mathcal{O}(\epsilon_{BD}^2)\,\label{ExpansionOfrho}\\
& H(t) = H^{(0)}(t)+\epsilon_{BD} H^{(1)}(t)   +\mathcal{O}(\epsilon_{BD}^2)\,,\label{ExpansionOfH}
\end{align}
respectively, where $N=R,M,\Lambda$ denotes the solution at the radiation, matter, and $\Lambda$-dominated epochs\footnote{For the sake of simplicity and to ease the obtention of analytical expressions, in this appendix we consider three massless neutrinos.}. We make use of the following relations,
\begin{equation}
\frac{1}{2\omega_{BD}+3}=\frac{\epsilon_{BD}}{2}+\mathcal{O}(\epsilon_{BD}^2)\,, \qquad
\frac{\dot{\varphi}}{\varphi}=\epsilon_{BD} \frac{\dot{\varphi}^{(1)}}{\varphi^{(0)}}+\mathcal{O}(\epsilon_{BD}^2)\,,\qquad
\frac{\omega_{BD}}{2}\left(\frac{\dot{\varphi}}{\varphi} \right)^2=\frac{\epsilon_{BD}}{2}\left(\frac{\dot{\varphi}^{(1)}}{\varphi^{(0)}}\right)^2+\mathcal{O}(\epsilon_{BD}^2)\,.
\end{equation}
The Klein-Gordon equation is already of first order in $\epsilon_{BD}$,
\begin{equation}
\ddot{\varphi}^{(1)}+3H^{(0)}\dot{\varphi}^{(1)}= 4\pi G_N(\rho^{(0)}_N-3p^{(0)}_N)\,, \label{eq:KGFirstOrder}
\end{equation}
and this allows us to find $\varphi^{(1)}$ without knowing $H^{(1)}$ nor the linear corrections for the energy densities and pressures. The Friedmann equation leads to
\begin{equation}
H^{(0)}=\left(\frac{8\pi{G_N}}{3\varphi^{(0)}}\rho_N^{(0)}\right)^{1/2}\,,\label{eq:FriedmannZeroOrder}
\end{equation}
at zeroth order, and
\begin{equation}
6H^{(0)}H^{(1)}+3H^{(0)}\frac{\dot{\varphi}^{(1)}}{\varphi^{(0)}}-\frac{1}{2}\left(\frac{\dot{\varphi}^{(1)}}{\varphi^{(0)}}\right)^2=3\left(H^{(0)}\right)^2 \left(\frac{\rho^{(1)}_N}{\rho_N^{(0)}} - \frac{\varphi^{(1)}}{\varphi^{(0)}}\right)\,, \label{eq:FriedmannFirstOrder}
\end{equation}
at first order, which let us to compute $a^{(1)}(t)$ once we get $\rho^{(1)}_N(t,a^{(1)}(t))$ from the conservation equation and substitute it in the {\it r.h.s}. $H^{(1)}$ is then trivially obtained using the computed scale factor. Alternatively, one can also combine \eqref{eq:FriedmannFirstOrder} with the pressure equation to obtain a differential equation for $H^{(1)}$ and directly solve it. Proceeding in this way one obtains, though, an additional integration constant that must be fixed using the Friedmann and conservation equations.

\subsection{Radiation dominated epoch (RDE)}
In the RDE the trace of the energy-momentum tensor is negligible when compared with the total energy density in the universe, so we can set the right-hand side of \eqref{eq:KGFirstOrder} to zero. The leading order of the scale factor and the Hubble function take the following form, respectively: $a^{(0)}(t)=A t^{1/2}$,  $H^{(0)}(t)=1/(2t)$, with $A \equiv (32\pi{G_N}\rho^{0}_r/3\varphi^{(0)})^{1/2}$ and $\rho^{0}_{r}$ the current value of the radiation energy density. Using these relations we can find the BD-field as well as the scale factor at first order. They read,
\begin{equation}
\varphi (t) =\varphi^{(0)}+\epsilon_{BD}\left(C_{R1} + C_{R2}\,t^{-1/2}\right)+\mathcal{O}(\epsilon_{BD}^2)\,,
\end{equation}
and
\begin{equation}
a(t)= At^{1/2}\left( 1  + \epsilon_{BD}\left[\frac{C_{R2}^2 \ln(t)}{24(\varphi^{(0)})^2{t}}-\frac{C_{R1}}{4\varphi^{(0)}}+\frac{C_{R3}}{t} \right] +\mathcal{O}(\epsilon_{BD}^2) \right)\,,
\end{equation}
where $C_{R1}$, $C_{R2}$ and $C_{R3}$ are integration constants. {Let us note that in the RDE the evolution of the BD-field $\varphi$ is essentially frozen, since there is no growing mode. The term evolving as $\sim t^{-1/2}$ is the decaying mode, and after some time we are eventually left with a constant contribution}.

 Finally, it is easy to find the corresponding Hubble function
\begin{equation}
H(t) = \frac{1}{2t}\left(1  + \epsilon_{BD}\left[ -\frac{C_{R2}^2}{12(\varphi^{(0)})^2}\frac{\ln(t)}{t}+\frac{1}{t}\left(\frac{C_{R2}^2}{12(\varphi^{(0)})^2}-2C_{R3} \right) \right] +\mathcal{O}(\epsilon_{BD}^2)\right)\,.
\end{equation}
At late enough times it is natural to consider that decaying mode is already negligible, and this allows us to simplify a lot the expressions, by setting $C_{R2}=0$. The scalar field remains then constant in very good approximation when radiation rules the expansion of the universe, and the other cosmological functions are { the same as in GR ($a\sim t^{1/2}$,  $H\sim \frac{1}{2t}$), but with an effective gravitational coupling $G=G_N/\varphi$.}

\subsection{Matter dominated epoch (MDE)}

When nonrelativistic matter is dominant in the universe the scalar field evolves as
\begin{equation}\label{eq:varphiMDEApp}
\varphi(t) = \varphi^{(0)} + \epsilon_{BD}\left[\frac{C_{M1}}{t}+\frac{2\varphi^{(0)}}{3}\ln(t) + C_{M2}\right] +\mathcal{O}(\epsilon_{BD}^2)\,,
\end{equation}
where $C_{M1}$ and $C_{M2}$ are integration constants. At leading order the scale factor and the Hubble function take the following form, $a^{(0)}(t) = Bt^{2/3}$ and $H^{(0)}(t) = 2/3t$ respectively, with $B \equiv \left(6\pi{G_N}\rho^0_m/\varphi^{(0)}\right)^{1/3}$ and $\rho^0_m$ the current value of the matter energy density. If we neglect the decaying mode in \eqref{eq:varphiMDEApp} we find $\varphi(a)=\varphi^{(0)}(1+\epsilon_{BD}\ln a+\mathcal{O}(\epsilon^2_{BD}))$. This solution is also found from the analysis of fixed points of Sec. \ref{AppendixB}, at leading order in $\epsilon_{BD}$. The scale factor reads in this case,
\begin{equation}
a(t) = Bt^{2/3}\left(1 + \epsilon_{BD}\left[ -\left(\frac{C_{M1}}{\varphi^{(0)}}\right)^2\frac{1}{8t^2}-\frac{C_{M2}}{3\varphi^{(0)}}-\frac{1}{18}-\frac{2}{9}\ln(t)+\frac{C_{M3}}{t} \right] +\mathcal{O}(\epsilon_{BD}^2) \right)\,,
\end{equation}
where $C_{M3}$ is another integration factor, and the Hubble function,
\begin{equation}
H(t) = \frac{2}{3t}\left(1 +\epsilon_{BD}\left[\left(\frac{C_{M1}}{\varphi^{(0)}}\right)^2\frac{3}{8t^2}-\frac{1}{3}-\frac{3C_{M3}}{2t} \right]+\mathcal{O}(\epsilon_{BD}^2)\right)\,.
\end{equation}
{Once more, after we neglect the contribution from the decaying modes, the usual cosmological functions are as is GR in the MDE ($a\sim t^{2/3}$,  $H\sim \frac{2}{3t}$). However, in contrast to the RDE there is some mild evolution (a logarithmic one with the cosmic time) of the BD-field. Since $\varphi^{(0)}$ is obviously positive, it follows from  Eq.\,\eqref{eq:varphiMDEApp} that the sign of such evolution (implying growing or decreasing behavior) is entirely defined by the sign of the BD-parameter $\eBD$.  Our fit to the overall data clearly shows that $\eBD<0$ (cf. Sec. \ref{sec:NumericalAnalysis})  and hence $\varphi$ decreases with the expansion during the MDE, which means that the effective gravitation coupling $G=G_N/\varphi$ increases with the expansion.}

\subsection{$\Lambda$-dominated  or VDE}
Here we assume that the energy density of the universe is completely dominated by a vacuum fluid, with constant energy density $\rho_\Lambda$.  {This period occurs in the very early universe during inflation, and in the very late one when matter has diluted significantly and a new period of inflation occurs}. The usual solution for the Hubble function in that epoch is $H^{(0)}(t)=H_\Lambda$, where $H_\Lambda$ is a constant, fixed by \eqref{eq:FriedmannZeroOrder}. Taking this into account we find
\begin{equation}
\varphi(t)=\varphi^{(0)}+\epsilon_{BD}\left(2H_\Lambda \varphi^{(0)} t+C_{\Lambda 1} e^{-3H_\Lambda t}+C_{\Lambda 2} \right)+\mathcal{O}(\epsilon_{BD}^2), \label{eq:varphi(t)Inflation}
\end{equation}
and for the scale factor
\begin{equation}
\begin{split}
&a(t)=C_{\Lambda 4}e^{H_\Lambda t}\left(1+\epsilon_{BD}\left[-\frac{H_\Lambda^2 t^2}{2}-\left(\frac{C_{\Lambda 1}}{\varphi^{(0)}}\right)^2\frac{e^{-6H_\Lambda t}}{8}+C_{\Lambda 3}H_\Lambda t\left(\frac{C_{\Lambda 2}}{2\varphi^{(0)}}+\frac{2}{3} \right) \right]\right)+\mathcal{O}(\epsilon_{BD}^2),
\end{split}
\end{equation}
where $C_{\Lambda 1},$ $C_{\Lambda 2}$, $C_{\Lambda 3}$, and $C_{\Lambda 4}$ are integration constants. Note that by performing the limit $\epsilon_{BD} \rightarrow 0$ we recover the usual GR expressions, as in all the previous formulas. The Hubble function at first order takes the form
\begin{equation}
\begin{split}
&H(t)=H_\Lambda\left(1 +\epsilon_{BD}\left[-\frac{2}3+\frac{3}{4}\left(\frac{C_{\Lambda 1}}{\varphi^{(0)}}\right)^2e^{-6H_\Lambda t}-H_\Lambda t-\frac{C_{\Lambda 2}}{2\varphi^{(0)}} \right]+\mathcal{O}(\epsilon_{BD}^2)\right).
\end{split}
\label{H(t)Inflation}
\end{equation}
The term accompanied by the constant $C_{\Lambda 1}$ in \eqref{eq:varphi(t)Inflation} is a decaying mode, which we considered to be negligible already in the RDE. Thus, we can safely remove it, and $C_{\Lambda 2}$ can be fixed by the condition $\varphi(t_*)=\varphi_*$, at some $t_*$ deeply in the VDE. The scalar field evolves then as $\varphi(t)\sim 2\epsilon_{BD}H_\Lambda t\sim 2\epsilon_{BD}\ln a$, which is the behavior we obtain also from the analysis of fixed points (cf. appendix \ref{AppendixB}).  {We can also see that during this epoch the BD-field decreases with the expansion because $\eBD<0$, as indicated before.}

\subsection{Mixture of matter and vacuum energy}

In a universe with a non-negligible amount of vacuum and matter energy densities, it is also possible to obtain an analytical expression for the BD scalar field at leading order in $\epsilon_{BD}$. Unfortunately, this is not the case for the scale factor and the Hubble function, so we will present here only the formula for $\varphi$. In the $\Lambda$CDM the scale factor is given by
\begin{equation}
a^{(0)}(t)=\left(\frac{\tilde{\Omega}_{m}}{\tilde{\Omega}_{\Lambda}}\right)^{1/3} \sinh^{2/3} \left(\frac{3}{2}\sqrt{\tilde{\Omega}_{\Lambda}}H_0 t \right)\,,
\end{equation}
so
\begin{equation}
H^{(0)}(t)=H_0\sqrt{\tilde{\Omega}_\CC} \coth \left(\frac{3}{2}\sqrt{\tilde{\Omega}_{\Lambda}}H_0 t\right)\,.
\end{equation}
Solving the Klein-Gordon equation, we obtain
\begin{equation}\label{eq:varphi1VDE}
\varphi^{(1)}(t)=\sqrt{\tilde{\Omega}_\Lambda}H_0 t  \coth \left(\frac{3}{2}\sqrt{\tilde{\Omega}_\Lambda}H_0 t \right)+\frac{2}{3}\ln\left( \sinh \left(\frac{3}{2}\sqrt{\tilde{\Omega}_\Lambda}H_0 t \right) \right)\,.
\end{equation}
One can easily check that in the limits $H_0t\ll 1$ and $H_0 t\gg 1$ we recover the behavior that we have found in previous sections for the matter and $\Lambda$-dominated universes, respectively. {When we substitute the previous expression in Eq.\,\eqref{ExpansionOfvarphi}, we confirm once more that $\varphi$ decreases with the expansion (since $\eBD<0$ and $\dot{\varphi}^{(1)}(t)>0$ $\forall{t}$). This is the period when the universe is composed of a  mixture of matter and vacuum energy at comparable proportions, and corresponds to the current universe.  Thus  $G=G_N/\varphi$ increases with the expansion in the present universe as it was also the case in the preceding MDE period, which is of course an important feature that helps to solve the $H_0$-tension, as explained in different parts of the paper, and in particular in the preview Sec. \ref{sec:preview}.}

%%%%%%%%%%%%%%%%%%%%%%%%%%%%%%%%%%%%%%%%%%%%%%%%%%%%%%%%%%%%%%%%%%%%%%
%%%%%%%%%%%%%%%%%%%%%%%%%%%%%%%%%%%%%%%%%%%%%%%%%%%%%%%%%%%%%%%%%%%%%%

\subsection{Connection of the BD-$\CC$CDM model with the  Running Vacuum Model}\label{sec:RVMconnection}

Analytical solutions to the system  \eqref{eq:Friedmannequation}-\eqref{eq:FieldeqPsi}  are not known and for this reason our actual analysis proceeds numerically.  However, as we have seen in the previous sections it is possible to search for  approximate solutions in the different epochs, which can help to better understand the numerical results and the qualitative behavior of the BD-$\CC$CDM model.  Actually, a first attempt in this direction trying to show that BD-$\CC$CDM  can mimic the Running Vacuum Model (RVM) was done in \cite{Peracaula:2018dkg,Perez:2018qgw} and we refer the reader to these references for details. Here we just summarize the results and adapt them to the current notation.  It is based on searching for  solutions in the MDE in the form of a power-law ansatz in which the BD-field $\varphi$ evolves very slowly:
\begin{equation}
\varphi(a) = \varphi_0\,a^{-\epsilon}\  \qquad (|\epsilon|\ll1)\,.\label{powerlaw}
\end{equation}
Obviously $\epsilon$ must be a very small parameter in absolute value since $G(a)\equiv G(\varphi(a))$ cannot depart too much from $G_N$.  On comparing with the analysis of fixed points given in Appendix \ref{AppendixB} -- cf. Eq.\,\eqref{eq:psiMDE} -- we can anticipate that $\epsilon\propto -\eBD$, although we do not expect perfect identification since \eqref{powerlaw} is a mere ansatz solution in the MDE whereas \eqref{eq:psiMDE} is an exact phase trajectory in that epoch.  For $\epsilon>0$ (hence $\eBD<0$), the effective  coupling increases with the expansion and hence is asymptotically free since  $G(a)$  is smaller in the past, which is the epoch when the
Hubble rate (with natural dimension of energy) is bigger. For  $\epsilon<0$ ($\eBD>0$), instead, $G(a)$  decreases with the expansion.

Using the power-law ansatz (\ref{powerlaw}) we find
\begin{equation}
 \frac{\dot\varphi}{\varphi}= -\epsilon {H}\,,\ \ \ \ \ \ \ \ \ \ \ \
 \frac{\ddot\varphi}{\varphi} = -\epsilon\dot{H} + \epsilon^2{H^2}. \label{derivatives}
\end{equation}
Plugging these  relations into the system of equations  \eqref{eq:Friedmannequation}-\eqref{eq:FieldeqPsi}  and after some calculation it is possible to arrive at the following pair of Friedmann-like equations to ${\cal O}(\epsilon)$\cite{Peracaula:2018dkg,Perez:2018qgw}:
\begin{equation}\label{eq:effective Friedmann}
   H^2=\frac{8\pi G}{3}\left(\rho^0_{m} a^{-3+\epsilon}+\rDE(H)\right)
\end{equation}
and
\begin{equation}\label{eq:currentacceleration}
\frac{\ddot{a}}{a}=-\frac{4\pi G}{3}\,\left(\rho_m^0 a^{-3+\epsilon}+\rDE(H)+3p_{\Lambda}\right)\,,
\end{equation}
with $G=G_N/\varphi_0$. The first equation emulates an effective Friedmann's equation with time-evolving cosmological term, in which the DE appears as dynamical:
\begin{equation}\label{eq:rLeff}
  \rDE(H)=\rL+\frac{3\,\nu_{\rm eff}}{8\pi G} H^2\,.
\end{equation}
Here
\begin{equation}
\ \nu_{\rm eff}\equiv\epsilon\left(1+\frac16\,\oD\epsilon\right) \label{nueff}
\end{equation}
is the coefficient controlling the dynamical character of the dark energy \eqref{eq:rLeff}. The structure of this dynamical dark energy  (DDE) is reminiscent of the Running Vacuum Model (RVM), see \cite{Sola:2013gha,sola2015cosmology,Gomez-Valent:2017tkh} and references therein.  In the language used in this paper, the analogue of the above RVM form of the Friedmann equation can be derived from Eq.\,\eqref{eq:FriedmannWithF} upon taking into account that the function ${\cal F}$ is of ${\cal O}(\eBD)$. In fact,  ${\cal F}$ is the precise analogue of $\nu_{\rm eff}$, for if we set $\epsilon\to -\eBD$  in \eqref{nueff} it boils down to the value quoted in  Eq.\eqref{eq:Fximatter}. The two languages are similar, but not identical, for the reasons explained above.

Notice from \eqref{eq:effective Friedmann} that,  to ${\cal O}(\epsilon)$:
\begin{equation}\label{eq:SumRule}
\Omo +\OLo =1-\nu_{\rm eff}\,,
\end{equation}
so the usual sum rule of GR is slightly violated by the BD model when parametrized as a deviation with respect to GR.%-- \joantext{compare it with Eq.\,\eqref{eq:SumRuleBD2}}.
Only for $\epsilon=0$ we have $\nu_{\rm eff}=0$ and then  we recover the usual cosmic sum rule\footnote{It is interesting to note that the presence of $\nueff\neq0$ emulates a fictitious spatial curvature. This is  the analog, in RVM language, of the similar situation  noted  in Sec. \ref{sec:EffectiveEoS} when we defined the GR-picture of the BD model. A persistent irreducible value of this parameter in future observations might serve also as a hint of the underlying BD physics.}.
The parameter $\nu_{\rm eff}$ becomes associated to the dynamics  of the DE. Worth noticing, the above expression adopts  the form of the RVM, see  \cite{Sola:2013gha,sola2015cosmology,Gomez-Valent:2017tkh} and references therein, in particular \cite{Shapiro:2000dz,Sola:2007sv,Shapiro:2009dh} -- where the running parameter is usually denoted $\nu$ and is associated to the $\beta$-function of the running vacuum.  Recently, the parameter $\nu$ (and in general the structure of the RVM) has been elucidated from direct calculations in QFT in curved spacetime within GR\,\cite{Moreno-Pulido:2020anb}. For additional discussions on the running of the CC term, see e.g. \cite{Babic:2004ev,Ward:2010qs,Antipin:2017pbt}. The RVM has been shown to be phenomenologically promising to alleviate some of the existing tensions within the $\CC$CDM, particularly the  $\sigma_8$-tension\,\,\cite{Sola:2016ecz , Sola:2017jbl , Gomez-Valent:2018nib ,Gomez-Valent:2017idt,Sola:2017znb , sola2017first , Sola:2016hnq , Sola:2015wwa , Geng:2017apd,Rezaei:2019xwo,Geng:2020mga}. It is therefore not surprising that the mimicking of the RVM by the BD-$\CC$CDM model enjoys of the same virtues. In actual fact, the particular RVM form obtained in BD-gravity (we may call it ``BD-RVM'' for short)  is even more successful since it can cure both tensions, the $H_0$ and $\sigma_8$ one.  The reason why the BD-RVM can cure also the $H_0$-tension is because we need the evolution of the effective gravitational coupling $\Geff$ to achieve that, as we have seen in the preview Sec.\,\ref{sec:preview}, whereas the $\sigma_8$-tension can be cured with $\nueff$, which is associated to $\epsilon\propto-\eBD$  (the second ingredient characteristic of BD-gravity), and hence the two key elements are there to make a successful phenomenological performance.

On the other hand, from \eqref{eq:currentacceleration} it follows that the EoS for the effective DDE is
\begin{equation}\label{eq:EffEoS}
\weff(z)=\frac{p_{\Lambda}}{\rDE(H)}\simeq -1+\frac{3\nueff}{8\pi G \rL}\,H^2(z)=-1+\frac{\nueff}{\Omega_\CC}\,\frac{H^2(z)}{H_0^2}\,,
%\simeq -1+\nueff\,\frac{\Omo}{\OLo}\,(1+z)^3\,,
\end{equation}
where use has been made of (\ref{eq:rLeff}). It follows that the BD-RVM, in contrast to the original RVM, does not describe a  DE of pure vacuum form ($p_{\Lambda}=-\rL$) but a DE whose EoS departs slightly from the pure vacuum. In fact, for $\epsilon>0\ (\epsilon<0) $  we have $\nu_{\rm eff}>0\  (\nu_{\rm eff}<0)$ and the effective DDE behaves quintessence (phantom)-like.  For $\epsilon\to 0$ (hence  $\nu_{\rm eff}\to 0$) we have  $\weff\to -1$ ($\CC$CDM).   As could be expected, Eq.\eqref{eq:EffEoS} is the BD-RVM version of the effective EoS that we obtained in Sec.\,\ref{sec:EffectiveEoS}  -- see Eq.\,\eqref{BDEoSz0}.  The two languages are consistent. Indeed, by comparison we see that $\nueff$ here plays the role of $\dvphi$ there. We know that $\dvphi=1-\varphi>0$, i.e. $\varphi<1$,  for $\eBD<0$, as we have shown previously, which is consistent with the fact that $\nueff\propto\epsilon\propto -\eBD>0$.  Finally, since $\weff$  approaches $-1$ from above (cf. Fig. 8) it corresponds to an effective quintessence behavior, which is more pronounced the more we explore the EoS into our past.

%%%%%%%%%%%%%%%%%%%%%%%%%%%%%%%%%%%%%%%%%%%%%%%%%%%%%%%%%%%%%%%%%%%%%%
%%%%%%%%%%%%%%%%%%%%%%%%%%%%%%%%%%%%%%%%%%%%%%%%%%%%%%%%%%%%%%%%%%%%%%

\section{Fixed Points in BD-$\CC$CDM  cosmology}\label{AppendixB}

In order to study the fixed points of this system of differential equations we must define new variables such that the system becomes of first order in the derivatives when it is rewritten in terms of the new variables. It is useful though to firstly carry out the change $t\to N\equiv \ln(a)$. This preliminary step will help us to identify in an easier way how we must define the new variables. When written in terms of $N$ the system takes the following form\footnote{{Primes in this appendix stand for derivatives {\it w.r.t.}  to the variable $N=\ln a$, i.e.  $()^\prime \equiv d()/d N$. We consider, as in Appendix A, three massless neutrinos.}}:

\begin{equation}
\frac{\psi^{\pp}}{\psi}+\frac{H^\p}{H}\frac{\psi^\p}{\psi}+3\frac{\psi^\p}{\psi}=\frac{8\pi}{3+2\w}\left(\frac{\rho_m+4\rho_\Lambda}{\psi H^2}\right)\,,
\end{equation}

\begin{equation}
3+3\frac{\psi^\p}{\psi}-\frac{\w}{2}\left(\frac{\psi^\p}{\psi}\right)^2=\frac{8\pi}{\psi H^2}(\rho_r+\rho_m+\rho_\Lambda)\,,
\end{equation}

\begin{equation}
3+\frac{H^\p}{H}\left(2+\frac{\psi^\p}{\psi}\right)+\frac{\psi^{\pp}}{\psi}+2\frac{\psi^\p}{\psi}+\frac{\w}{2}\left(\frac{\psi^\p}{\psi}\right)^2=\frac{8\pi}{\psi H^2}\left(\rho_\Lambda-\frac{\rho_r}{3}\right)\,,
\end{equation}

\begin{equation}
\rho^\p_r+4\rho_r=0\,,
\end{equation}

\begin{equation}
\rho^\p_m+3\rho_m=0\,.
\end{equation}

\noindent
Now one can define the following quantities:

\begin{equation}
\xp\equiv \frac{\psi^\p}{\psi}\quad ; \quad x_i^2\equiv\frac{8\pi \rho_i}{H^2\psi}\,,
\end{equation}

\noindent
where $i=r,m,\Lambda$. In terms of these variables the system of equations can be easily written as follows:

\begin{equation}\label{eq:eq1}
\xp^\p+\xp^2+\frac{H^\p}{H}\xp+3\xp=\frac{\xm^2+4\xl^2}{3+2\w}\,,
\end{equation}

\begin{equation}\label{eq:eq2}
3+3\xp-\frac{\w}{2}\xp^2=\xr^2+\xm^2+\xl^2\,,
\end{equation}

\begin{equation}\label{eq:eq3}
3+\frac{H^\p}{H}\left(2+\xp\right)+\xp^\p+\xp^2+2\xp+\frac{\w}{2}\xp^2=\xl^2-\frac{\xr^2}{3}\,,
\end{equation}

\begin{equation}\label{eq:eq4}
\xr^\p=-\xr\left(\frac{H^\p}{H}+2+\frac{\xp}{2}\right)\,,
\end{equation}

\begin{equation}\label{eq:eq5}
\xm^\p=-\xm\left(\frac{H^\p}{H}+\frac{3}{2}+\frac{\xp}{2}\right)\,.
\end{equation}

\noindent
This system is of first order, as wanted. We have five equations and five unknowns, namely: $\xr,\xm,\xp,\xl,H^\p/H$. We can reduce significantly the complexity of the system if we just isolate $H^\p/H$ from \eqref{eq:eq1} and $\xl$ from \eqref{eq:eq2},

\begin{equation}
\frac{H^\p}{H}=\frac{1}{\xp}\left[\frac{\xm^2+4\xl^2}{3+2\w}-\xp^\p-3\xp-\xp^2\right]\,,
\end{equation}

\begin{equation}
\xl^2=3+3\xp-\frac{\w}{2}\xp^2-\xr^2-\xm^2\,,
\end{equation}

\noindent
and substitute the resulting expressions in the other equations, i.e. in \eqref{eq:eq3}, \eqref{eq:eq4}, and \eqref{eq:eq5}. Doing this, and after a little bit of algebra, one finally obtains three equations written only in terms of $\xr,\xm,\xp$:

\begin{equation}\label{DynSysPsi}
\xp^\p=-\xp\left[3+3\xp-\frac{1}{2}\w\xp^2-\frac{2}{3}\xr^2-\frac{\xm^2}{2}-\left(\frac{1}{\xp}+\frac{1}{2}\right)\left(\frac{12+12\xp-2\w\xp^2-4\xr^2-3\xm^2}{3+2\w}\right)\right]\,,
\end{equation}

\begin{equation}\label{DynSysR}
\xr^\p=-\xr\left[2+\frac{5}{2}\xp-\frac{\w}{2}\xp^2-\frac{2}{3}\xr^2-\frac{\xm^2}{2}-\frac{1}{2}\left(\frac{12+12\xp-2\w\xp^2-4\xr^2-3\xm^2}{3+2\w}\right)\right]\,,
\end{equation}

\begin{equation}\label{DynSysM}
\xm^\p=-\xm\left[\frac{3}{2}+\frac{5}{2}\xp-\frac{\w}{2}\xp^2-\frac{2}{3}\xr^2-\frac{\xm^2}{2}-\frac{1}{2}\left(\frac{12+12\xp-2\w\xp^2-4\xr^2-3\xm^2}{3+2\w}\right)\right]\,.
\end{equation}

\noindent
They allow us to search for the fixed points of the system. There is an important restriction produced by \eqref{eq:eq4} and \eqref{eq:eq5}. Supposing that $x_r \neq 0$ and $x_m\neq 0$, we see that the mentioned equations impose that
\begin{equation}
\frac{x_r^\prime}{x_r}=-\frac{x_m^\prime}{x_m}+\frac{1}{2}.
\end{equation}
This equation is not compatible with the conditions of fixed point, so that we should assume that $x_r=0$, $x_m=0$ or both conditions at the same time. The fixed points are:
\newline
\newline
\textbf{RDE}

\begin{equation}
(\xr,\xm,\xl,\xp)_{\rm RD} = \left(\sqrt{3},0,0,0\right)\,,
\end{equation}
the Jacobian of the nonlinear system \eqref{DynSysPsi}, \eqref{DynSysR}, \eqref{DynSysM} has eigenvalues $\lambda^{RD}_1=-1$, $\lambda^{RD}_2 =1/2$, $\lambda^{RD}_3=4$ so that is an unstable point.
\newline
\newline
\textbf{MDE}

\begin{equation}
(\xr,\xm,\xl,\xp)_{\rm MD} = \left(0,\frac{\sqrt{12+17\w+6\w^2}}{\sqrt{2}|1+\w|},0,\frac{1}{1+\w}\right)\,,
\end{equation}
the Jacobian of the nonlinear system \eqref{DynSysPsi}, \eqref{DynSysR}, \eqref{DynSysM} has eigenvalues $\lambda^{MD}_1=-1/2,$ $\lambda^{MD}_2\approx-3/2,$ $\lambda^{MD}_3\approx 3$ so that is also an unstable point.
\newline
\newline
\textbf{$\Lambda$-dominated or VDE}

\begin{equation}
(\xr,\xm,\xl,\xp)_{\rm \Lambda D} = \left(0,0,\frac{\sqrt{15+28\w+12\w^2}}{|1+2\w|},\frac{4}{1+2\w}\right)\,
\end{equation}
the Jacobian of the nonlinear system \eqref{DynSysPsi}, \eqref{DynSysR}, \eqref{DynSysM} has eigenvalues $\lambda^{\Lambda}_1=-2,$ $\lambda^{\Lambda}_2 =-3/2,$ $\lambda^{\Lambda}_3 \approx -3$ so that is a stable point.

The first and second fixed points are unstable, whereas the latter is stable. We have assumed that $|\w|\gg1$ for approximating the eigenvalues.

The first one is very well-known, since regardless of the initial conditions for the BD scalar field we already know that the velocity of the scalar field decays during the RDE, and the solution tends to the attractor with $\psi^\p=0$ ($\xp=0$) and full domination of radiation, i.e. $\xr=\sqrt{3}$. The Hubble function and BD scalar field are:

\begin{equation}
\psi_{RD}(a)=\psi_r\,,
\end{equation}

\begin{equation}
H^2_{RD}(a)=\frac{8\pi}{3\psi_r}\rho^{0}_{r}a^{-4}\,,
\end{equation}
where $\psi_r$ is an arbitrary constant and $\rho^{0}_{r}$ is the value of the radiation energy density at present.
This fixed point is unstable because at some moment nonrelativistic matter starts to dominate the expansion. When this happens the solution starts to look for the new attractor, the one of the MDE. We stress  that this is an exact solution. The  BD scalar field and the Hubble function during the MDE take the following form:

\begin{equation}\label{eq:psiMDE}
\psi_{MD}(a)=Ca^{\frac{1}{1+\w}}\,,
\end{equation}

\begin{equation}
H^2_{MD}(a)=\frac{16\pi \rho_m^{0}(1+\w)^2a^{-\left(\frac{4+3\w}{1+\w}\right)}}{C(12+17\w+6\w^2)}\,,
\end{equation}

\noindent
where $C$ is an arbitrary constant and $\rho_m^{0}$ is the value of the matter energy density at present. The MD fixed point is, again, unstable, because the MDE finishes and the VDE starts. The solution searches now for the last fixed point, which is stable. In this last case, we have obtained the following Hubble function and BD-field,

\begin{equation}
\psi_\Lambda (a)= D a^{4/(1+2\w)}\,,
\end{equation}

\begin{equation}
H^2_\Lambda (a)=\frac{8\pi \rho_\Lambda }{D}\frac{(1+2\w)^2}{15+28\w +12 \w^2}a^{-4/(1+2\w)}\,,
\end{equation}
where $D$ is an arbitrary constant and $\rho_\Lambda$ is the constant value of vacuum energy.

One can easily check that the solutions computed in Appendix A, of first order in $\epsilon_{\rm BD}=1/\w$, coincide (once the decaying modes become irrelevant) with the ones presented here when the latter are Taylor-expanded up to first order in this parameter as well. The results we have found here are consistent with previous studies on fixed points in cosmological dynamical systems, see e.g. \cite{Bahamonde:2017ize} and references therein.

\section{Cosmological perturbations in the synchronous gauge} \label{AppendixC}

In this section we explicitly derive the set of perturbed equations for the Brans-Dicke model in the synchronous gauge, and up to linear order in the perturbed quantities. The perturbed FLRW metric written in conformal time $\eta$ reads
\begin{equation}\label{eq:LineElementSyn}
ds^2=a^2(\eta)[-d\eta^2+(\delta_{ij}+h_{ij})dx^idx^j]\,,
\end{equation}
where {$\eta$ is the conformal time, $\vec{x}$ the spatial comoving coordinates, and} $h_{ij}$ is the metric perturbation. We apply in this work the usual metric formalism and therefore we assume that the Christoffel symbols are unequivocally determined by the metric through the Levi-Civita connection. Thus, the perturbed Chrystoffel symbols together with the perturbed Riemann and Ricci tensors, and the perturbed Ricci scalar, can be easily written in terms of $h_{ij}$, its trace $h\equiv \delta^{ij}h_{ij}$, and their spacetime derivatives. Although the corresponding expressions can be already found in the literature, we have opted to provide them together with the perturbed energy-momentum tensor in Sec. \ref{sec:PertGeoEM} for completeness. In Secs. \ref{sec:PertEqPos} and \ref{sec:PertEqMom} we write the main perturbed equations in position and momentum space, respectively, and in Sec. \ref{sec:PertEqLowScales} we derive and discuss the equation that rules the growth of matter perturbations in the matter and $\Lambda$-dominated universe at deep subhorizon scales. {In Sec. \ref{sec:IC} we provide a brief note on the initial conditions for all the perturbed quantities. Finally, in Sec. \ref{sec:GWs} we discuss tensor perturbations in the BD-$\Lambda$CDM cosmology.}

%%%%%%%%%%%%%%%%%%%%%%%%%%%%%%%%%%%%%%%%%%%
%%%%%%%%%%%%%%%%%%%%%%%%%%%%%%%%%%%%%%%%%%%
%%%%%%%%%%%%%%%%%%%%%%%%%%%%%%%%%%%%%%%%%%%

\subsection{Perturbed geometric quantities, energy-momentum tensor, and other relevant terms appearing in the field equations}\label{sec:PertGeoEM}

The elements of the metric tensor and its inverse can be straightfordwardly obtained from \eqref{eq:LineElementSyn}. They read as follows,
\begin{equation}
\begin{split}
&g_{00}=-a^2\qquad g_{ij}=a^2(\delta_{ij}+h_{ij})\quad ; \quad g^{00}=-1/a^2\qquad g^{ij}=\frac{1}{a^2}(\delta_{ij}-h_{ij}).
\end{split}
\end{equation}
Plugging them into the formula of the Levi-Civita connection one gets the following expressions for the Christoffel symbols:
\begin{equation}
\begin{split}
&\Gamma^{0}_{00}=\mathcal{H}\,,\qquad \Gamma^{0}_{i0}=\Gamma^{i}_{00}=0\,,\qquad \Gamma^{0}_{ij}=\mathcal{H}(\delta_{ij}+h_{ij})+\frac{h^\prime_{ij}}{2}\,,\\
&\Gamma^{i}_{j0}=\mathcal{H}\delta_{ij}+\frac{h^\prime_{ij}}{2}\,,\qquad \Gamma^{i}_{jl}=\frac{1}{2}\left(h_{ij,l}+h_{il,j}-h_{jl,i}\right)\,,\\
\end{split}
\end{equation}
where each prime denotes a derivative with respect to the conformal time, i.e.\,$d/d\eta$,  the lower commas partial derivatives with respect to spatial (comoving) coordinates, and $\mathcal{H}\equiv a^\prime/a$. The contributions of all the second and higher order terms in perturbation theory have been neglected since we are not interested here in analyzing nonlinear structure formation processes, {in which also the details of the screening mechanism acting in the nonlinear regime could become important}. The Ricci tensor components can be computed making use of the above formulas,
\begin{equation}
\begin{split}
R_{00}=&-3\mathcal{H}^\prime-\frac{h^{\prime \prime}}{2}-\frac{\mathcal{H}}{2}h^\prime,\qquad R_{0i}=\frac{1}{2}\left(\partial_jh^\prime_{ij}-\partial_{i}h^\prime\right)\,,\\
R_{ij}=&(\delta_{ij}+h_{ij})(\mathcal{H}^\prime+2\mathcal{H}^2)+\frac{h^{\prime\prime}_{ij}}{2}+\frac{\mathcal{H}}{2}h^\prime\delta_{ij}+\mathcal{H}h^\prime_{ij}+\frac{1}{2}(h_{li,jl}+h_{lj,il}-h_{ij,ll}-h_{,ij})\,.
\end{split}
\end{equation}
Notice that we have applied Einstein's summation convention. Contracting the Ricci tensor with the metric we finally obtain the scalar curvature,
\begin{equation}
\begin{split}
a^2R=6(\mathcal{H}^\prime+\mathcal{H}^2)+h^{\prime\prime}+3\mathcal{H}h^\prime+h_{li,li}-h_{,ll}\,.
\end{split}
\end{equation}
Equipped with these tools we can proceed to compute the components of the Einstein tensor, which read
\begin{equation}\label{eq:PerturbEinstTensor}
\begin{split}
G_{00}=&3\mathcal{H}^2+\mathcal{H}h^\prime+\frac{1}{2}\left(h_{li,li}-h_{,ll}\right)\,,\\
G_{i0}=&\frac{1}{2}\left(\partial_jh^\prime_{ij}-\partial_{i}h^\prime\right)\,,\\
G_{ij}=&-(\delta_{ij}+h_{ij})(2\mathcal{H}^\prime+\mathcal{H}^2)+\frac{h^{\prime\prime}_{ij}}{2}-\frac{h^{\prime\prime}}{2}\delta_{ij}-\mathcal{H}h^\prime\delta_{ij}+\mathcal{H}h^\prime_{ij}\\
&+\frac{1}{2}(h_{li,jl}+h_{lj,il}-h_{ij,ll}-h_{,ij}-h_{lt,lt}\delta_{ij}+h_{,ll}\delta_{ij})\,.
\end{split}
\end{equation}
It is also convenient to obtain the trace of $G_{ij}$, since it will be employed in subsequent calculations,
\begin{equation}
G_{ii}=-(3+h)(2\mathcal{H}^\prime+\mathcal{H}^2)-h^{\prime\prime}-2\mathcal{H}h^\prime+\frac{1}{2}(h_{,ll}-h_{li,li})\,.
\end{equation}
As in Ref. \cite{Ma:1995ey}, we can express $h_{ij}$ as follows,
\begin{equation}\label{eq:hFourier}
h_{ij}(\eta,\vec{x})=\int d^3k\, e^{-i\vec{k}\cdot\vec{x}}\left[\hat{k}_i\hat{k}_j h(\eta,\vec{k})+\left(\hat{k}_i\hat{k}_j-\frac{\delta_{ij}}{3}\right)6\xi(\eta,\vec{k})\right]\,,
\end{equation}
where $\hat{k}_i=k_i/k$ with $k=|\vec{k}|$, and $h(\eta,\vec{k})$ the Fourier transform of the trace of $h_{ij}(\eta,\vec{x})$. When we work in Fourier space we will denote it $h$, like in position space, without specifying its dependence on the wave number $\vec{k}$ explicitly. Plugging \eqref{eq:hFourier} into the perturbed part of \eqref{eq:PerturbEinstTensor} we obtain the elements of the perturbed Einstein tensor in Fourier space. We will employ them later on. They read,
\begin{equation}\label{eq:PerturbEinstTensor2}
\begin{split}
\delta G_{00}=&\mathcal{H}h^\prime-2\xi k^2\,,\\
\delta G_{i0}=&-ik_i2\xi^\prime\,,\\
\delta G_{ii}=&-h(2\mathcal{H}^\prime+\mathcal{H}^2)-h^{\prime\prime}-2\mathcal{H}h^\prime+2\xi k^2\,.
\end{split}
\end{equation}
We do not write here the Fourier transform of $\delta G_{ij}$ because we will not use it later.

In order to compute the perturbed energy-momentum tensor of the perfect fluids that fill the universe with Eq. \eqref{eq:EMT} we must know which is the form of their perturbed 4-velocities. It is easy to show that they are just given by
\begin{equation}
u^\mu=\frac{1}{a}(1,v^i)\qquad ;\qquad u_\mu=a (-1,v^i)\,,
\end{equation}
with $v^i=\frac{dx^i}{d\eta}$. Using this in Eq. \eqref{eq:EMT} and splitting the total energy density and pressure in a background and a perturbed parts, i.e. considering $\rho(\eta,\vec{x})=\bar{\rho}(\eta)+\delta\rho(\eta,\vec{x})$ and $p(\eta,\vec{x})=\bar{p}(\eta)+\delta p(\eta,\vec{x})$, we obtain the following elements of the perturbed energy-momentum tensor,
\begin{equation}
\begin{split}
T_{00} =&\, a^2(\bar{\rho}+\delta\rho)\,,\\
T_{ij} =&\, a^2\bar{p}(\delta_{ij}+h_{ij})+a^2\delta_{ij}\delta p\,,\\
T_{0i} =&\, -a^2(\bar{p}+\bar{\rho})v^i\,,\\
T \equiv &\, g^{\mu\nu}T_{\mu\nu} = 3(\bar{p}+\delta p)-\bar{\rho}-\delta\rho\,,
\end{split}
\end{equation}
where a sum over all the species in the universe is taken for granted. The following quantities will also be useful in subsequent calculations.
\begin{equation}
\begin{split}
\partial_\alpha\varphi \partial^\alpha\varphi=&\,-\frac{1}{a^2}\left[(\bar{\varphi}^\prime)^2+2\bar{\varphi}^\prime\delta\varphi^\prime\right] \,,\\
a^2\Box\varphi =&\, -\bar{\varphi}^{\prime\prime}-2\mathcal{H}\bar{\varphi}^\prime-\delta\varphi^{\prime\prime}+\nabla^2\delta\varphi-2\mathcal{H}\delta\varphi^\prime-\frac{h^\prime}{2}\bar{\varphi}^\prime\,,\\
\nabla_0\nabla_0\varphi=&\,\bar{\varphi}^{\prime\prime}+\delta\varphi^{\prime\prime}-\mathcal{H}\bar{\varphi}^\prime-\mathcal{H}\delta\varphi^\prime\,,\\
\nabla_i\nabla_0\varphi =&\,\partial_i\delta\varphi^\prime-\mathcal{H}\partial_i\delta\varphi\,,\\
\nabla_i\nabla_j =& \partial_i\partial_j\delta\varphi-\bar{\varphi}^\prime\left[\mathcal{H}(\delta_{ij}+h_{ij})+\frac{h_{ij}^\prime}{2}\right]-\delta_{ij}\mathcal{H}\delta\varphi^\prime\,.
\end{split}
\end{equation}
Here we have split the BD-field as the sum of the mean (background) field $\bar{\varphi}$ and its corresponding perturbation $\delta\varphi$, i.e. $\varphi(\eta,\vec{x})=\bar{\varphi}(\eta)+\delta\varphi(\eta,\vec{x})$.

%%%%%%%%%%%%%%%%%%%%%%%%%%%%%%%%%%%%%%%%%%%
%%%%%%%%%%%%%%%%%%%%%%%%%%%%%%%%%%%%%%%%%%%
%%%%%%%%%%%%%%%%%%%%%%%%%%%%%%%%%%%%%%%%%%%

\subsection{Perturbation equations in position space}\label{sec:PertEqPos}

We apply now the machinery derived in the previous subsection to perturb the modified Einstein's Eqs. \eqref{eq:BDFieldEquation1}, the covariant conservation of the energy-momentum tensor and the Klein-Gordon equation \eqref{eq:BDFieldEquation2}. The last one reads,
\begin{equation}\label{eq:pertKG}
-\delta \varphi^{\prime \prime}-2\mathcal{H}\delta \varphi^\prime+\nabla^2 \delta \varphi-\frac{h^\prime}{2}\bar{\varphi}^\prime =\frac{8\pi G_N}{3+2\omega_{BD}}a^2(3\delta p-\delta\rho)\, ,
\end{equation}
where $\nabla^2 \equiv \sum\limits_{i=1}^{3}\partial_i^2$. The perturbed $00$, $0i$, and $ij$ components of Einstein's equation lead, respectively, to
\begin{equation}\label{eq:pertT00}
\begin{split}
\bar{\varphi}\left(\mathcal{H}h^\prime+\frac{h_{li,li}-h_{,ll}}{2}\right)+& 3\mathcal{H}^2\delta\varphi-\nabla^2\delta\varphi+3\mathcal{H}\delta\varphi^\prime+\frac{h^\prime}{2}\bar{\varphi}^\prime\\
&+\frac{\omega_{BD}}{2\bar{\varphi}}\left[\frac{\delta\varphi}{\bar{\varphi}}(\bar{\varphi}^\prime)^2-2\bar{\varphi}^\prime\delta\varphi^\prime\right] = 8\pi G_N a^2\delta \rho\,,
\end{split}
\end{equation}
\begin{equation}\label{eq:pertT0i}
\begin{split}
\bar{\varphi}\left(\frac{\partial_{j}h^\prime_{ij}-\partial_ih^\prime}{2}\right)-\partial_i\delta\varphi^\prime+\mathcal{H}\partial_i\delta\varphi-\frac{\omega_{BD}}{\bar{\varphi}}\bar{\varphi}^\prime\partial_i\delta\varphi=-8\pi G_N a^2(\bar{\rho}+\bar{p})v^i\,,
\end{split}
\end{equation}
\begin{equation}\label{eq:pertTij}
\begin{split}
\delta_{ij} & \left[  -\delta\varphi(2\mathcal{H}^\prime+\mathcal{H}^2)-\bar{\varphi}\left(\frac{h^{\prime\prime}}{2}+\mathcal{H}h^\prime\right)-\mathcal{H}\delta\varphi^\prime-\delta\varphi^{\prime\prime}+\nabla^2\delta\varphi-\frac{h^\prime}{2}\bar{\varphi}^\prime+\bar{\varphi}\left(\frac{h_{,ll}-h_{lt,lt}}{2}\right)\right.\\
&\left. +\frac{\omega_{BD}}{2\bar{\varphi}}\left(\frac{\delta\varphi}{\bar{\varphi}}(\bar{\varphi}^\prime)^2-2\bar{\varphi}^\prime\delta\varphi^\prime\right)\right]+h_{ij}^\prime\left(\mathcal{H}\bar{\varphi}+\frac{\bar{\varphi}^\prime}{2}\right)-\partial_i\partial_j\delta\varphi+\frac{h_{ij}^{\prime\prime}}{2}\bar{\varphi}\\
&-h_{ij}\left[\bar{\varphi}^{\prime\prime}+\bar{\varphi}(2\mathcal{H}^\prime+\mathcal{H}^2) +\mathcal{H}\bar{\varphi}^\prime+\frac{\omega_{BD}}{2\bar{\varphi}}(\bar{\varphi}^\prime)^2\right]+\frac{\bar{\varphi}}{2}\left(h_{li,jl}+h_{lj,il}-h_{ij,ll}-h_{,ij}\right)=8\pi G_N \delta T_{ij}\,.
\end{split}
\end{equation}
Finally, the covariant conservation of the energy-momentum tensor leads to the following pair of extra equations, i.e. $\nabla^\mu T_{\mu \nu}=0$, for $\nu=0$ and $\nu=i$, respectively,
\begin{equation}\label{eq:covariant0}
\sum_{j} \bar{\rho}_{(j)}\left[\delta^\prime_{(j)}+3\mathcal{H}\left(\frac{\delta p_{(j)}}{\delta \rho_{(j)}}-{w}_{(j)}\right)\delta_{(j)}+(1+{w}_{(j)})\left(\theta_{(j)}+\frac{h^\prime}{2}\right)\right]=0\,,
\end{equation}
\begin{equation}\label{eq:covarianti}
\sum_{j}\bar{\rho}_{(j)}(1+{w}_{(j)})\left[\theta^\prime_{(j)}+\left( \mathcal{H}(1-3{w}_{(j)})+\frac{{w}_{(j)}^\prime}{1+{w}_{(j)}} \right)\theta_{(j)}+\frac{\nabla^2\delta p_{(j)}}{(1+{w}_{(j)})\bar{\rho}_{(j)}}+\frac{{w}_{(j)}}{1+{w}_{(j)}}\partial_i\partial_l h_{il}\right]=0\,,
\end{equation}
where $\theta_{(j)}\equiv \partial_i v^i_{(j)}$, $\delta_{(j)}\equiv \delta \rho_{(j)}/\bar{\rho}_{(j)}$, ${w}_{(j)}\equiv \bar{p}_{(j)}/\bar{\rho}_{(j)}$, and the sums run over all the species $j$ that fill the universe.

%%%%%%%%%%%%%%%%%%%%%%%%%%%%%%%%%%%%%%%%%%%
%%%%%%%%%%%%%%%%%%%%%%%%%%%%%%%%%%%%%%%%%%%
%%%%%%%%%%%%%%%%%%%%%%%%%%%%%%%%%%%%%%%%%%%

\subsection{Perturbation equations in momentum space}\label{sec:PertEqMom}

As it is well-known, working in momentum space simplifies a lot the treatment of the cosmological perturbations, basically because at linear order in perturbation theory the different modes of the perturbed quantities do not couple to each other, i.e. there is no mixture of wave numbers and we can safely omit the subscript $k$ for the modes. Here we limit ourselves to just write the expressions provided in the previous subsection, but in momentum space. The calculations are straightforward and no further details are thus needed.
\newline
\newline
\noindent Equation \eqref{eq:pertKG}
\begin{equation}\label{eq:pertKGk}
\delta \varphi^{\prime \prime}+2\mathcal{H}\delta \varphi^\prime+k^2\delta \varphi+\frac{h^\prime}{2}\bar{\varphi}^\prime =\frac{8 \pi G_N}{3+2\omega_{BD}}a^2(\delta \rho-3\delta p)\,.
\end{equation}
Equation \eqref{eq:pertT00}
\begin{equation}\label{eq:pertT00k}
\begin{split}
\bar{\varphi}(\mathcal{H}h^\prime-2\xi k^2)+(3\mathcal{H}^2+&k^2)\delta\varphi+3\mathcal{H}\delta\bar{\varphi}^\prime+\frac{h^\prime}{2}\bar{\varphi}^\prime\\
&+\frac{\omega_{BD}}{2\bar{\varphi}}\left[\frac{\delta\varphi}{\bar{\varphi}}(\bar{\varphi}^\prime)^2-2\bar{\varphi}^\prime\delta\varphi^\prime\right]=8\pi G_N a^2 \delta \rho\,.
\end{split}
\end{equation}
Equation \eqref{eq:pertT0i}
\begin{equation}
-2\bar{\varphi}k^2\xi^\prime+k^2\delta\varphi^\prime-\mathcal{H}k^2\delta \varphi+\omega_{BD}k^2\delta \varphi\left(\frac{\bar{\varphi}^\prime}{\bar{\varphi}}\right)=-8\pi G_N a^2 (\bar{\rho}+\bar{p})\theta\,.
\end{equation}
Trace of equation \eqref{eq:pertTij}, and after making use of the pressure equation \eqref{eq:pressureequation}
\begin{equation}
\begin{split}
-\delta \varphi(6\mathcal{H}^\prime+3\mathcal{H}^2+2k^2)-3\delta \varphi^{\prime \prime}-3\mathcal{H}\delta \varphi^\prime&-\bar{\varphi}^\prime h^\prime+\frac{3\omega_{BD}\bar{\varphi}^\prime}{2\bar{\varphi}}\left(\frac{\bar{\varphi}^\prime}{\bar{\varphi}}\delta\varphi-2\delta \varphi^\prime\right) \\
&+\bar{\varphi}\left( -h^{\prime \prime}-2h^\prime \mathcal{H}+2k^2 \xi \right)=24\pi G_N a^2 \delta p\,.
\end{split}
\end{equation}
Equation \eqref{eq:covariant0}
\begin{equation}\label{eq:covariant0k}
\sum_{j} \bar{\rho}_{(j)}\left[\delta^\prime_{(j)}+3\mathcal{H}\left(\frac{\delta p_{(j)}}{\delta \rho_{(j)}}-{w}_{(j)}\right)\delta_{(j)}+(1+{w}_{(j)})\left(\theta_{(j)}+\frac{h^\prime}{2}\right)\right]=0\,.
\end{equation}
Equation \eqref{eq:covarianti}
\begin{equation}\label{eq:covariantiK}
\sum_{j}\bar{\rho}_{(j)}(1+{w}_{(j)})\left[\theta^\prime_{(j)}+\left( \mathcal{H}(1-3{w}_{(j)})+\frac{{w}_{(j)}^\prime}{1+{w}_{(j)}} \right)\theta_{(j)}-\frac{k^2\delta_{(j)}}{(1+{w}_{(j)})}\frac{\delta p_{(j)}}{\delta\rho_{(j)}}-\frac{k^2{w}_{(j)}}{1+{w}_{(j)}}(h+4\xi)\right]=0\,.
\end{equation}
Another useful and compact relation can be obtained from the $\hat{k}_i\hat{k}_j$ part of Eq. \eqref{eq:pertTij} in momentum space. The result, after using again the pressure equation \eqref{eq:pressureequation}, reads
\begin{equation}\label{eq:kikjk}
\begin{split}
&h^{\prime \prime}+6\xi^{\prime \prime}+(h^\prime +6\xi^\prime)\left(2\mathcal{H}+\frac{\bar{\varphi}^\prime}{\bar{\varphi}}\right)+2k^2\left(\frac{\delta\varphi}{\varphi}-\xi\right)=0\,.
\end{split}
\end{equation}
%

%%%%%%%%%%%%%%%%%%%%%%%%%%%%%%%%%%%%%%%%%%%
%%%%%%%%%%%%%%%%%%%%%%%%%%%%%%%%%%%%%%%%%%%
%%%%%%%%%%%%%%%%%%%%%%%%%%%%%%%%%%%%%%%%%%%

\subsection{Matter density contrast equation at deep subhorizon scales}\label{sec:PertEqLowScales}

Let us restrict us now to the matter and $\Lambda$-dominated epochs and see what is the evolution of matter perturbations in the late stages of the universe's expansion and deeply inside the horizon. Using the fact that vacuum does not cluster when it is described by a cosmological constant and matter is covariantly conserved, and also taking into account that radiation has only very mild impact on the Large-Scale Structure formation processes we want to study, we can obtain the following relation from \eqref{eq:covariantiK},
\begin{equation}
\theta^\prime_m=-\mathcal{H}\theta_m\,.
\end{equation}
This leads to a decaying solution for the velocity potential gradient, $\theta_m=\theta_m^{0}/a$, and in practice we can take $\theta_m\sim 0$. By doing this in \eqref{eq:covariant0k} we find
\begin{equation}\label{eq:simpli0}
\delta_m^\prime=-\frac{h^\prime}{2}\,.
\end{equation}
At low scales, Eqs. \eqref{eq:pertKGk}, \eqref{eq:pertT00k} and \eqref{eq:kikjk} simplify, after neglecting some terms which are clearly subdominant at large $k$'s, giving rise to
\begin{equation}\label{eq:simpli1}
k^2\delta \varphi+\frac{h^\prime}{2}\bar{\varphi}^\prime =\frac{8 \pi G_N}{3+2\omega_{BD}}a^2 \bar{\rho}_m\delta_m\,,
\end{equation}
\begin{equation}\label{eq:simpli2}
\bar{\varphi}(\mathcal{H}h^\prime-2\xi k^2)+k^2\delta\varphi+\frac{h^\prime}{2}\bar{\varphi}^\prime=8\pi G_N a^2 \bar{\rho}_m\delta_m\,,
\end{equation}
\begin{equation}\label{eq:simpli3}
2k^2\delta \varphi+\bar{\varphi}^\prime h^\prime+\bar{\varphi}\left(h^{\prime \prime}+2h^\prime \mathcal{H}-2k^2 \xi\right)=0\,.
\end{equation}
Using \eqref{eq:simpli0} in \eqref{eq:simpli1} one can isolate $\delta\varphi=\delta\varphi(\delta_m,\delta_m^\prime)$, and doing the same in \eqref{eq:simpli2} one gets $\xi=\xi(\delta_m,\delta_m^\prime)$. Introducing these expressions in \eqref{eq:simpli3}, and after making use again of \eqref{eq:simpli0}, one finally obtains the equation for the matter density contrast at deep subhorizon scales,
\begin{equation}
\delta_m^{\prime\prime}+\mathcal{H}\delta_m^\prime-\frac{4\pi G_N a^2}{\bar{\varphi}}\bar{\rho}_m\delta_m\left(\frac{4+2\omega_{BD}}{3+2\omega_{BD}}\right)=0\,.
\end{equation}
 The expression in terms of the scale factor is given in the main text, Eq.\,\eqref{eq:ExactPerturScaleFactor}, where we recall that primes there mean $d/da$ whereas here $d/d\eta$. A quick comparison of the last term of this equation with the one that is obtained in the GR-$\Lambda$CDM allows us to note that the effective value of the gravitational constant that is controlling the formation of linear structures at subhorizon scales is
\begin{equation}\label{eq:Geffective}
G_{{\rm eff}}(\bar{\varphi})=\frac{G_N}{\bar{\varphi}}\left(\frac{4+2\omega_{BD}}{3+2\omega_{BD}}\right)\,.
\end{equation}
{More details are provided in the main body of the paper, see Sec. \ref{sec:StructureFormation}.}

%%%%%%%%%%%%%%%%%%%%%%%%%%%%%%%%%%%%%%%%%

\subsection{Brief note on the initial conditions}\label{sec:IC}

We consider adiabatic initial conditions for the various species filling the universe. For the DM velocity divergence, we use the usual synchronous condition $\theta_{cdm,ini}=0$.  We would like to point out here that the initial perturbation of the BD-field and its time derivative can also be set to zero. This is because the modes of interest were superhorizon modes during the radiation-dominated epoch, and in that period of the universe's expansion Eq. \eqref{eq:pertKG} reduces to
\begin{equation}
\delta\varphi^{\pp}+\frac{2}{\eta}\delta\varphi^\p+k^2\delta\varphi=0\,,
\end{equation}
where we have used $\mathcal{H}=\eta^{-1}$. The solution of this equation reads,
\begin{equation}
\delta\varphi(k,\eta)=\frac{A(k)}{k\eta}\cos(k\eta+\beta(k))\,,
\end{equation}
with $A(k)$ and $\beta(k)$ being an amplitude and a phase, respectively. This solution corresponds to a damped oscillation, which is decaying fastly and can be naturally set to zero \cite{Chen:1999qh}. The initial conditions are thus equal to the ones in the GR-$\Lambda$CDM scenario \cite{Ma:1995ey}, but substituting $G_N$ by $G(\bar{\varphi}_{ini})$.

%%%%%%%%%%%%%%%%%%%%%%%%%%%%%%%%%%%%%%%%%

\subsection{Gravitational waves in BD-$\Lambda$CDM cosmology}\label{sec:GWs}

Gravitational waves (GWs) are given by the traceless and transverse part of the metric fluctuations, $h^T_{ij}$, which contains two degrees of freedom (corresponding to the two polarization states, usually denoted as $\times$ and $+$). Hence, they satisfy $h^T=0$ and $\partial_i h^T_{ij}=0$. As scalar, vector and tensor cosmological perturbations decouple from each other at linear order, we can consider the line element
\begin{equation}
ds^2=a^2(\eta)[-d\eta^2+(\delta_{ij}+h^T_{ij})dx^i dx^j]
\end{equation}
and study the ij component of Einstein's equations. In order to do so we can directly take the traceless and transverse part of equation \eqref{eq:pertTij}. We obtain,
\begin{equation}\label{eq:Gij}
-h_{ij}\left[\bar{\varphi}(2\mathcal{H}^\prime+\mathcal{H}^2)+\bar{\varphi}^\pp+\mathcal{H}\bar{\varphi}^\p +\frac{\omega_{BD}}{2\bar{\varphi}}(\bar{\varphi}^\p)^2\right]+h_{ij}^\p\left(\mathcal{H}\bar{\varphi}+\frac{\bar{\varphi}^\p}{2}\right)+\frac{\bar{\varphi}}{2}h^\pp_{ij}+\frac{\bar{\varphi}}{2} k^2 h_{ij}=8\pi G_N a^2\bar{p}\,h_{ij}\,.
\end{equation}
Notice that we have omitted the superscript $T$ for simplicity, doing $h^T_{ij}\rightarrow h_{ij}$. This equation can be reduced by using the background pressure equation \eqref{eq:pressureequation}, yielding
\begin{equation}
h^\pp_{ij}+h^\p_{ij}\left(2\mathcal{H}+\frac{\bar{\varphi}^\p}{\bar{\varphi}}\right)+k^2 h_{ij}=0\,.
\end{equation}
For a general scalar-tensor theory of gravity one has
\begin{equation}
h^\pp_{ij}+\mathcal{H}h^\p_{ij}\left(2+\alpha_M\right) + k^2(1+\alpha_T)h_{ij} = 0\,,
\end{equation}
where $\alpha_M$ and $\alpha_T$  are functions that parametrize the deviations from standard GR. The former modifies the friction term, and is basically the running of the effective Planck mass, whereas the latter is directly related with the speed of propagation of the GWs,  $c_{gw}$, since $\alpha_T=c^2_{gw}-1$. In BD, $\alpha_M=d\ln(\bar{\varphi})/d\ln(a)$ (e.g. at leading order in $\epsilon_{BD}$, we have $\alpha_M=\epsilon_{BD}$ in the pure MDE and $\alpha_M=2\epsilon_{BD}$ in the VDE, cf. Appendix B) and $\alpha_T=0$, so $c_{gw}=1$. This function, $\alpha_T$, has been recently constrained to be $|\alpha_T(z\approx 0)|\lesssim 10^{-15}$ using the measurement of the gravitational wave event GW170817 and the accompanying electromagnetic counterpart GRB170817A \cite{TheLIGOScientific:2017qsa}, located both at a distance of $40^{+8}_{-14}$ Mpc from us. BD theory automatically satisfies this constraint \cite{Creminelli:2017sry,Ezquiaga:2017ekz}, since GWs propagate exactly at the speed of light.

%%%%%%%%%%%%%%%%%%%%%%%%%%%%%%%%%%%%%%%%%%%
%%%%%%%%%%%%%%%%%%%%%%%%%%%%%%%%%%%%%%%%%%%

\section{Pertubation theory in Newtonian gauge}\label{AppendixD}
So far, we have been working with the synchronous gauge in BD linear perturbation theory. For completeness we are going to provide the equations in the conformal Newtonian (or longitudinal) gauge. {This appendix has the same structure as the previous one (we only skip the recomputation of tensor perturbations in this gauge, and the discussion of the initial conditions)}. In particular, we will show that the same density contrast differential equation for subhorizon scales that we found in Sec. \ref{sec:PertEqLowScales} arises also for the Newtonian gauge, as expected. In the last section we provide the transformation equations of the perturbed quantities from one gauge to another in analogy to \cite{Ma:1995ey}.

%%%%%%%%%%%%%%%%%%%%%%%%%%%%%%%%%%%%%%%%%%%
%%%%%%%%%%%%%%%%%%%%%%%%%%%%%%%%%%%%%%%%%%%
%%%%%%%%%%%%%%%%%%%%%%%%%%%%%%%%%%%%%%%%%%%

\subsection{Perturbed geometric quantities, energy-momentum tensor, and other relevant terms appearing in the field equations}
The square of the line element in the perturbed flat FLRW universe in the Newtonian Gauge reads as follows,
\begin{equation}
ds^2=a^2[-(1+2\Phi)d\eta^2+(1+2\Psi)\delta_{ij}dx^idx^j]\,,
\end{equation}
where the pair $\Phi$ and $\Psi$ the so-called Bardeen potentials, which are functions of conformal time and space. In the following we provide the perturbed expressions (at linear order) for the various geometrical quantities that will be used later. As before, the primes denote a derivative with respect to the conformal time, and $\mathcal{H}\equiv a^\prime/a$.
Thus, the metric elements are
\begin{equation}
\begin{split}
g_{00}=-a^2 (1+2\Phi),\qquad g_{ij}=a^2 (1+2\Psi)\delta_{ij} \quad; \quad g^{00}=-\frac{1}{a^2}(1-2\Phi), \qquad g^{ij}=\frac{1}{a^2}(1-2\Psi)\delta_{ij}.
\end{split}
\end{equation}
We can compute the Christoffel symbols associated to the metric in the usual way:
\begin{equation}
\Gamma^0_{00}=\mathcal{H}+\Phi^\prime, \qquad \Gamma^0_{0i}=\Gamma^i_{00}=\partial_i\Phi, \qquad \Gamma^0_{ij}=\delta_{ij}[\mathcal{H}(1+2\Psi-2\Phi)+\Psi^\prime],
\end{equation}
$$\qquad \Gamma^i_{j0}=\delta^i_j (\mathcal{H}+\Psi^\prime) ,\qquad \Gamma^i_{jl}=\delta^i_j\partial_l\Psi+\delta^i_l\partial_j\Psi-\delta_{jl}\partial_i\Psi .$$
The components of the Ricci tensor are
\begin{equation}
R_{00}=-3\mathcal{H}^\prime+\nabla^2\Phi-3\Psi^{\prime\prime}+3\mathcal{H}(\Phi^\prime-\Psi^\prime), \qquad R_{0i}=-2\partial_i\Psi^\prime+2\mathcal{H}\partial_i\Phi,
\end{equation}
$$R_{ij}=-\partial_i\partial_j(\Psi+\Phi)+\delta_{ij}\left[(2\mathcal{H}^2+\mathcal{H}^\prime)(1+2\Psi-2\Phi)-\nabla^2\Psi+\Psi^{\prime\prime}+5\mathcal{H}\Psi^\prime-\mathcal{H}\Phi^\prime\right].$$
Contracting the indices of the previous tensor we are able to compute the Ricci scalar
\begin{equation}
Ra^2=6(\mathcal{H}^2+\mathcal{H}^\prime)(1-2\Phi)-2\nabla^2(\Phi+2\Psi)+6\Psi^{\prime\prime}-6\mathcal{H}\Phi^\prime+18\mathcal{H}\Psi^\prime.
\end{equation}
The components of the Einstein tensor entering Einstein's equations are
\begin{equation}
G_{00}=3\mathcal{H}^2+6\mathcal{H}\Psi^\prime-2\nabla^2\Psi,
\end{equation}
$$G_{ij}=-\partial_i\partial_j(\Psi+\Phi)+\delta_{ij}\left[-(\mathcal{H}^2+2\mathcal{H}^\prime)(1+2\Psi-2\Phi)+2\mathcal{H}(\Phi^\prime-2\Psi^\prime)+\nabla^2(\Psi+\Phi)-2\Psi^{\prime\prime}\right],$$
$$G_{0i}=-2\partial_i\Psi^\prime+2\mathcal{H}\partial_i\Phi.$$
As we have done for the synchronous gauge, we consider the energy-momentum tensor of a perfect fluid, Eq. \eqref{eq:EMT}, and split the energy densities and pressures as before. The perturbed 4-velocity $u^\mu$ and its covariant form $u_\mu$ read now, respectively,
\begin{equation}
u^\mu=\frac{1}{a}(1-\Phi,v^i)\qquad u_\mu=a(-[1+\Phi],v^i)\,,
\end{equation}
where $\vec{v}$ is the physical 3-velocity of the fluid, whose modulus is much lower than 1, so we can treat it as a linear perturbation. Taking all this into account one can compute the perturbed elements of $T_{\mu\nu}$ and its trace:
$$T_{00}=a^2[(1+2\Phi)\bar{\rho}+\delta\rho]\,,$$
\begin{equation}
T_{ij}=a^2\delta_{ij}[\bar{p}(1+2\Psi)+\delta p]\,,
\end{equation}
$$T_{0i}=-a^2 v^i(\bar{\rho}+\bar{p})\,,$$
$$T=3(\bar{p}+\delta p)-\bar{\rho}-\delta\rho\,.$$
Now, we provide the formulas of some other perturbed expressions depending on $\varphi$ that will be also useful in subsequent computations:
\begin{equation}
a^2\Box\varphi = -\bar{\varphi}^{\prime\prime}-2\mathcal{H}\bar{\varphi}^\prime-\delta\varphi^{\prime\prime}+2\bar{\varphi}^{\prime\prime}\Phi+\nabla^2\delta\varphi-2\mathcal{H}\delta\varphi^\prime+\bar{\varphi}^\prime(\Phi^\prime-3\Psi^\prime)+4\mathcal{H}\Phi\bar{\varphi}^\prime ,
\end{equation}
\begin{equation}
\partial_\alpha\varphi\partial^\alpha\varphi=-\frac{(\bar{\varphi}^{\prime})^2}{a^2}(1-2\Phi)-\frac{2}{a^2}\bar{\varphi}^\prime\delta\varphi^\prime ,
\end{equation}
\begin{equation}
\nabla_0\nabla_0\varphi = \bar{\varphi}^{\prime\prime}+\delta\varphi^{\prime\prime}-\bar{\varphi}^{\prime}(\mathcal{H}+\Phi^\prime)-\mathcal{H}\delta\varphi^\prime ,
\end{equation}
\begin{equation}
\nabla_i\nabla_j\varphi=\partial_i\partial_j\delta\varphi-\bar{\varphi}^\prime\delta_{ij}[\mathcal{H}(1+2\Psi-2\Phi)+\Psi^\prime]-\mathcal{H}\delta\varphi^\prime\delta_{ij} ,
\end{equation}
\begin{equation}
\nabla_i\nabla_0\varphi=\partial_i(\delta\varphi^\prime-\bar{\varphi}^\prime\Phi-\mathcal{H}\delta\varphi) .
\end{equation}

%%%%%%%%%%%%%%%%%%%%%%%%%%%%%%%%%%%%%%%%%%%
%%%%%%%%%%%%%%%%%%%%%%%%%%%%%%%%%%%%%%%%%%%
%%%%%%%%%%%%%%%%%%%%%%%%%%%%%%%%%%%%%%%%%%%

\subsection{Perturbation equations in position space}

The perturbed Einstein equations read as follows,
\newline
\newline
\noindent
For $\mu=i,\nu=j$, $i\neq j$:
\begin{equation}\label{eq:gFieldij1}
\Psi+\Phi=-\frac{\delta\varphi}{\bar{\varphi}}\,.
\end{equation}
As we can see, the presence of the perturbation of the BD-field, $\delta\varphi\neq0$ induces anisotropic stress since it  violates the usual $\Phi=-\Psi$ setting of  GR-$\CC$CDM, which holds good  only in the absence of anisotropic stress (induced e.g. by massive neutrinos).

\noindent For $i=j$:
\begin{equation}
\begin{split}
&(\Phi^\prime-2\Psi^\prime)(2\mathcal{H}\bar{\varphi}+\bar{\varphi}^\prime)+(\Phi-\Psi)\left[2\bar{\varphi}(\mathcal{H}^2+2\mathcal{H}^\prime)+2\bar{\varphi}^{\prime\prime}+2\mathcal{H}\bar{\varphi}^\prime+\frac{\omega_{BD}}{\bar{\varphi}}(\bar{\varphi}^\prime)^2\right]\\
  {}&-2\Psi^{\prime\prime}\bar{\varphi}-\delta\varphi^{\prime\prime}-\mathcal{H}\delta\varphi^\prime-\delta\varphi(\mathcal{H}^2+2\mathcal{H}^\prime)+\frac{\omega_{BD}}{2\bar{\varphi}}\left[\frac{\delta\varphi}{\bar{\varphi}}(\bar{\varphi}^\prime)^2-2\bar{\varphi}^\prime\delta\varphi^\prime\right]= 8\pi G_N a^2(2\bar{p}\Psi+\delta p)\,. \label{eq:gFieldij2}
\end{split}
\end{equation}
For $\mu=0,\nu=0$:
\begin{equation}
\begin{split}
\bar{\varphi}(6\mathcal{H}\Psi^\prime-2\nabla^2\Psi)+3\mathcal{H}^2\delta\varphi-&\nabla^2\delta\varphi+3\mathcal{H}\delta\varphi^\prime+3\Psi^\prime\bar{\varphi}^\prime\\
  {}&+\frac{\omega_{BD}}{2\bar{\varphi}}\left[\frac{\delta\varphi}{\bar{\varphi}}(\bar{\varphi}^\prime)^2-2\bar{\varphi}^\prime\delta\varphi^\prime\right]=8\pi G_N a^2(2\Phi\bar{\rho}+\delta\rho)\,.\label{eq:gField00}
\end{split}
\end{equation}
The perturbed covariant conservation equations leads to:
\begin{equation}\label{eq:consEq0}
\sum_j \bar{\rho}_{(j)}\left[\delta^\prime_{(j)}+\delta_{(j)}\frac{\rho_{(j)}^\prime}{\rho_{(j)}}+3\mathcal{H}\delta_{(j)}\left(1+\frac{\delta p_{(j)}}{\delta \rho_{(j)}} \right)+(1+{w}_{(j)})(\theta_{(j)}+3\Psi^\prime) \right]=0\,,
\end{equation}
\begin{equation}
\sum_j \left[4\mathcal{H}\theta_{(j)}\bar{\rho}_{(j)}(1+{w}_{(j)})+\frac{d}{d\eta}(\theta_{(j)}\bar{\rho}_{(j)}(1+{w}_{(j)})) +\bar{\rho}_{(j)}(1+{w}_{(j)})\nabla^2\Phi+\nabla^2 \delta p_{(j)}\right]=0,
\end{equation}
where again, as in Appendix C, $\theta_{(j)}\equiv \partial_i v^i_{(j)}$, $\delta_{(j)}\equiv \delta \rho_{(j)}/\bar{\rho}_{(j)}$, ${w}_{(j)}\equiv \bar{p}_{(j)}/\bar{\rho}_{(j)}$, and the sums run over all the species $j$ that fill the universe.

On the other hand, the perturbed part of the Klein-Gordon equation can be written as
\begin{equation}\label{eq:BDfield2}
-\delta\varphi^{\prime\prime}+2\bar{\varphi}^{\prime\prime}\Phi+\nabla^2\delta\varphi-2\mathcal{H}\delta\varphi^\prime+\bar{\varphi}^\prime(\Phi^\prime-3\Psi^\prime)+4\mathcal{H}\Phi\bar{\varphi}^\prime=\frac{8\pi G_N}{3+2\omega_{BD}}a^2(3\delta p-\delta\rho).
\end{equation}
%%%%%%%%%%%%%%%%%%%%%%%%%%%%%%%%%%%%%%%%%%%
%%%%%%%%%%%%%%%%%%%%%%%%%%%%%%%%%%%%%%%%%%%
%%%%%%%%%%%%%%%%%%%%%%%%%%%%%%%%%%%%%%%%%%%

\subsection{Perturbation equations in momentum space}
From now on we will work in momentum space. Let us show first the Einstein equations.
For $\mu=i,\nu=j$, $i\neq j$:
\begin{equation}\label{eq:gFieldij1Mom}
\Psi+\Phi=-\frac{\delta\varphi}{\bar{\varphi}}\,.
\end{equation}
For $i=j$:
\begin{equation}
\begin{split}
&(\Phi^\prime-2\Psi^\prime)(2\mathcal{H}\bar{\varphi}+\bar{\varphi}^\prime)+(\Phi-\Psi)\left[2\bar{\varphi}(\mathcal{H}^2+2\mathcal{H}^\prime)+2\bar{\varphi}^{\prime\prime}+2\mathcal{H}\bar{\varphi}^\prime+\frac{\omega_{BD}}{\bar{\varphi}}(\bar{\varphi}^\prime)^2\right] \\
  {}&-2\Psi^{\prime\prime}\bar{\varphi}-\delta\varphi^{\prime\prime}-\mathcal{H}\delta\varphi^\prime-\delta\varphi(\mathcal{H}^2+2\mathcal{H}^\prime)+\frac{\omega_{BD}}{2\bar{\varphi}}\left[\frac{\delta\varphi}{\bar{\varphi}}(\bar{\varphi}^\prime)^2-2\bar{\varphi}^\prime\delta\varphi^\prime\right]= 8\pi G_N a^2(2\bar{p}\Psi+\delta p)\,. \label{eq:gFieldij2Mom}
\end{split}
\end{equation}
Finally for $\mu=0,\nu=0$:
\begin{equation}
\begin{split}
\bar{\varphi}(6\mathcal{H}\Psi^\prime+2k^2\Psi)+3\mathcal{H}^2\delta\varphi+&k^2\delta\varphi+3\mathcal{H}\delta\varphi^\prime+3\Psi^\prime\bar{\varphi}^\prime\\
  {}&+\frac{\omega_{BD}}{2\bar{\varphi}}\left[\frac{\delta\varphi}{\bar{\varphi}}(\bar{\varphi}^\prime)^2-2\bar{\varphi}^\prime\delta\varphi^\prime\right]=8\pi G_N a^2(2\Phi\bar{\rho}+\delta\rho)\,.\label{eq:gField00Mom}
\end{split}
\end{equation}
Notice that the previous equation yields the usual  perturbed Poisson equation for $\delta\varphi=0$  --  which implies $\Phi=-\Psi$, according to \eqref{eq:gFieldij1Mom}. Doing also $\bar{\varphi}=1$, at deep subhorizon scales it boils down to the expected simpler form   $k^2\Psi=-k^2\Phi=-4\pi G_N a^2\delta\rho$,  since $k^2\gg a^2 G_N\rho\sim  a^2 H^2= \mathcal{H}^2$.

The perturbation of the covariant conservation equation, with $\nu=0$, gives
\begin{equation}\label{eq:consEq0Mom}
\sum_j \bar{\rho}_{(j)}\left[\delta^\prime_{(j)}+3\mathcal{H}\delta_{(j)}\left(\frac{\delta p_{(j)}}{\delta \rho_{(j)}}-{w}_{(j)}\right)+(1+{w}_{(j)})(\theta_{(j)}+3\Psi^\prime) \right]=0.
\end{equation}
And for $\nu=i$, we obtain:
\begin{equation}\label{eq:consEqiMom}
\sum_j \left[4\mathcal{H}\theta_{(j)}\bar{\rho}_{(j)}(1+{w}_{(j)})+\frac{d}{d\eta}(\theta_{(j)}\bar{\rho}_{(j)}(1+{w}_{(j)})) -\bar{\rho}_{(j)}(1+{w}_{(j)})k^2\Phi-k^2 \delta p_{(j)}\right]=0.
\end{equation}
So far, these conservation equations take the same form as in the GR-$\Lambda$CDM. On the other hand, the perturbed Klein-Gordon equation reads
\begin{equation}\label{eq:BDfield2Mom}
-\delta\varphi^{\prime\prime}+2\bar{\varphi}^{\prime\prime}\Phi-k^2\delta\varphi-2\mathcal{H}\delta\varphi^\prime+\bar{\varphi}^\prime(\Phi^\prime-3\Psi^\prime)+4\mathcal{H}\Phi\bar{\varphi}^\prime=\frac{8\pi G_N}{3+2\omega_{BD}}a^2(3\delta p-\delta\rho).
\end{equation}
%
%%%%%%%%%%%%%%%%%%%%%%%%%%%%%%%%%%%%%%%%%%%
%%%%%%%%%%%%%%%%%%%%%%%%%%%%%%%%%%%%%%%%%%%
%%%%%%%%%%%%%%%%%%%%%%%%%%%%%%%%%%%%%%%%%%%

\subsection{Matter density contrast equation at deep subhorizon scales}\label{sec:PertEqLowScalesNewtonian}

As done in Appendix C for the synchronous gauge, we study now the evolution of matter perturbations at deep subhorizon scales, i.e. at those scales at which $k^2\gg \mathcal{H}^2$ (deep subhorizon scales). In this limit, Eq. \eqref{eq:consEq0Mom} boils down to
\begin{equation}
\delta_m^\prime+\theta_m+3\Psi^\prime=0\,,
\end{equation}
and \eqref{eq:consEqiMom} can be written as
\begin{equation}
\theta^\prime_m +\mathcal{H} \theta_m-k^2\Phi=0.
\end{equation}
These equations can be easily combined to make disappear the dependence on $\theta_m$. If we do that, we obtain the following approximate second order differential equation for the matter density contrast,
\begin{equation}\label{eq:perturbk2Phi}
\delta_m^{\prime \prime}+\mathcal{H}\delta_m^\prime+k^2 \Phi=0.
\end{equation}
The problem is now reduced to find an expression for $k^2\Phi$ in terms of background quantities and $\delta_m$. Collecting \eqref{eq:gFieldij1Mom}, \eqref{eq:gField00Mom} and \eqref{eq:BDfield2Mom},
\begin{equation}\label{eq:PhiplusPsi}
-\frac{\delta \varphi}{\bar{\varphi}}=\Psi+\Phi,
\end{equation}
\begin{equation}\label{eq:Poisson1}
2k^2 \Psi+k^2 \frac{\delta \varphi}{\bar{\varphi}}=\frac{8\pi G_N a^2}{\bar{\varphi}} \bar{\rho}_m \delta_m,
\end{equation}
\begin{equation}\label{eq:Poisson2}
k^2 \delta \varphi=\frac{8\pi G_N a^2}{3+2\omega_{BD}}\bar{\rho}_m \delta_m,
\end{equation}
respectively. We can see, as expected, that for $\wBD\to\infty$ \textit{and} $\bar{\varphi}=1$ the first equation above gives $\Phi=-\Psi$ (no anisotropy stress) and the third one renders a trivial equality ($0=0$), whereas the second equation yields Poisson equation $k^2\Psi=-k^2\Phi=4\pi G_N a^2\delta\rho$.
In the general case, by combining the above equations one finds:
\begin{equation}\label{eq:Poisson3}
k^2\Phi=-\frac{4\pi G_N a^2 \bar{\rho}_m \delta_m}{\bar{\varphi}}\left( \frac{4+2\omega_{BD}}{3+2\omega_{BD}} \right)\,.
\end{equation}
So, finally, inserting the previous relation in \eqref{eq:perturbk2Phi} we are led to the desired equation for the density contrast at deep subhorizon scales:
\begin{equation}
\delta_m^{\prime\prime}+\mathcal{H}\delta_m^\prime-\frac{4\pi G_N}{\bar{\varphi}} a^2\bar{\rho}_m\delta_m \left(\frac{4+2\omega_{BD}}{3+2\omega_{BD}}\right)=0\,,
\end{equation}
or, alternatively, in terms of the cosmic time $t$,
\begin{equation}\label{eq:dcEq}
\ddot{\delta}_m+2H\dot{\delta}_m-\frac{4\pi G_N}{\bar{\varphi}}\bar{\rho}_m\delta_m\left(\frac{4+2\omega_{BD}}{3+2\omega_{BD}}\right)=0\,,
\end{equation}
with the dots denoting derivatives with respect to $t$.  As expected, these equations coincide with the density contrast equation at deep subhorizon scales for the synchronous gauge and we can recover the standard $\Lambda$CDM result for  $\omega_{BD}\to \infty$ and $\bar{\varphi}\to 1$.

As already mentioned above, because of \eqref{eq:PhiplusPsi} non-null scalar field perturbations induce a deviation of the anisotropic stress, $-\Psi/\Phi$, from $1$, i.e. the GR-$\Lambda$CDM value. At scales well below the horizon,
\begin{equation}
-\frac{\Psi}{\Phi} = \frac{1+\oD}{2+\oD}=1-\epsilon_{BD}+\mathcal{O}(\epsilon^2_{BD})\,,
\end{equation}
so constraints on the anisotropic stress directly translate into constraints on $\epsilon_{BD}$, and a deviation of this quantity from $1$ at the linear regime would be a clear signature of non-standard gravitational physics. A model-independent reconstruction of the anisotropic stress from observations has been recently done in \cite{Pinho:2018unz}. Unfortunately, the error bars are still of order $\mathcal{O}(1)$, so these model-independent results cannot put tight constraints on $\epsilon_{BD}$ (yet).

%%%%%%%%%%%%%%%%%%%%%%%%%%%%%%%%%%%%%%%%%%%
%%%%%%%%%%%%%%%%%%%%%%%%%%%%%%%%%%%%%%%%%%%
%%%%%%%%%%%%%%%%%%%%%%%%%%%%%%%%%%%%%%%%%%%

\subsection{Transformations between Gauges}\label{sec:TransformationGauges}
It is possible to establish a set of equations relating the different perturbation quantities in both gauges at the same coordinates in momentum space. For the potentials, we have the well known relations of \cite{Ma:1995ey},
\begin{equation}
\Phi ( \vec{k} , \eta)=\frac{1}{2k^2}\left\{ h^{\prime \prime}( \vec{k},\eta)+6\xi^{\prime \prime}(\vec{k},\eta)+\mathcal{H}\left[ h^{\prime }(\vec{k},\eta)+6\xi^{\prime }(\vec{k},\eta)\right] \right\}\,,
\end{equation}
\begin{equation}
\Psi (\vec{k},\eta)= -\xi(\vec{k},\eta)+\frac{1}{2k^3}\mathcal{H}\left[h^\prime (\vec{k},\eta)+6\xi^\prime(\vec{k},\eta)\right]\,,
\end{equation}
and for the other perturbed quantities we find,
\begin{equation}
\begin{split}
\delta_S &= \delta_N-\alpha \frac{\bar{\rho}^\prime}{\bar{\rho}}\,,\\
\theta_S &= \theta_N- \alpha k^2\,,\\
\delta p_S &= \delta p_N-\alpha \bar{p}^\prime\,,\\
\delta \varphi_S &=\delta \varphi_N-\alpha \bar{\varphi}^\prime\,.
\end{split}
\end{equation}
Here, the subscripts $N$ and $S$ mean {\it newtonian} and {\it synchronous}, respectively, and
\begin{equation}
\alpha \equiv \frac{1}{2k^2}\left[h^\prime+6\xi^\prime \right]\,.
\end{equation}

%%%%%%%%%%%%%%%%%%%%%%%%%%%%%%%%%%%%%%%%%%%
%%%%%%%%%%%%%%%%%%%%%%%%%%%%%%%%%%%%%%%%%%%
%%%%%%%%%%%%%%%%%%%%%%%%%%%%%%%%%%%%%%%%%%%

%%%\bibliographystyle{ieeetr}
%\bibliography{referencesBD2}
%\bibliography{BD_Final_Version}
%%%\bibliography{BD_Revised11Sept}

\begin{thebibliography}{100}

\bibitem{Riess:1998cb}
A.~G. Riess {\em et~al.}, ``Observational evidence from supernovae for an
  accelerating universe and a cosmological constant,'' {\em Astron. J.},
  vol.~116, pp.~1009--1038, 1998.
\newblock arXiv:astro-ph/9805201.

\bibitem{Perlmutter:1998np}
S.~Perlmutter {\em et~al.}, ``{Measurements of $\Omega$ and $\Lambda$ from 42
  high redshift supernovae},'' {\em Astrophys. J.}, vol.~517, pp.~565--586,
  1999.
\newblock arXiv:astro-ph/9812133.

\bibitem{Aghanim:2018eyx}
N.~Aghanim {\em et~al.}, ``{Planck 2018 results. VI. Cosmological
  parameters},'' 2018.
\newblock arXiv:1807.06209.

\bibitem{peebles:1993}
P.~J.~E. Peebles, {\em {Principles of physical cosmology}}.
\newblock Princeton University Press, 1993.

\bibitem{Weinberg:1988cp}
S.~Weinberg, ``{The Cosmological Constant Problem},'' {\em Rev. Mod. Phys.},
  vol.~61, pp.~1--23, 1989.

\bibitem{Sola:2013gha}
J.~Sol{\`{a}}, ``{Cosmological constant and vacuum energy: old and new
  ideas},'' {\em J. Phys. Conf. Ser.}, vol.~453, p.~012015, 2013.
\newblock arXiv:1306.1527.

\bibitem{Sahni:1999gb}
V.~Sahni and A.~A. Starobinsky, ``{The Case for a positive cosmological Lambda
  term},'' {\em Int.\ J.\ Mod.\ Phys.\ D}, vol.~9, pp.~373--444, 2000.
\newblock arXiv:astro-ph/9904398.

\bibitem{Peebles:2002gy}
P.~J.~E. Peebles and B.~Ratra, ``{The Cosmological Constant and Dark Energy},''
  {\em Rev. Mod. Phys.}, vol.~75, pp.~559--606, 2003.
\newblock arXiv:astro-ph/0207347.

\bibitem{Padmanabhan:2002ji}
T.~Padmanabhan, ``{Cosmological constant: The Weight of the vacuum},'' {\em
  Phys. Rept.}, vol.~380, pp.~235--320, 2003.
\newblock arXiv:hep-th/0212290.

\bibitem{Copeland:2006wr}
E.~J. Copeland, M.~Sami, and S.~Tsujikawa, ``{Dynamics of dark energy},'' {\em
  Int.\ J.\ Mod.\ Phys.\ D}, vol.~15, pp.~1753--1936, 2006.
\newblock arXiv:hep-th/0603057.

\bibitem{Amendola:2015ksp}
L.~Amendola and S.~Tsujikawa, {\em {Dark Energy}: {Theory and Observations}}.
\newblock Cambridge University Press, 2015.

\bibitem{gibbons1985very}
A.D. Dolgov, in: {\it The very Early Universe}, Ed. G. Gibbons, S.W. Hawking,
  S.T. Tiklos, Cambridge University Press, 1982.

\bibitem{Abbott:1984qf}
L.~F. Abbott, ``{A Mechanism for Reducing the Value of the Cosmological
  Constant},'' {\em Phys. Lett.}, vol.~150B, pp.~427--430, 1985.

\bibitem{Peccei:1987mm}
R.~D. Peccei, J.~Sol{\`{a}}, and C.~Wetterich, ``{Adjusting the Cosmological
  Constant Dynamically: Cosmons and a New Force Weaker Than Gravity},'' {\em
  Phys. Lett.}, vol.~B195, pp.~183--190, 1987.

\bibitem{Barr:1986ya}
S.~M. Barr, ``{An Attempt at a Classical Cancellation of the Cosmological
  Constant},'' {\em Phys. Rev.}, vol.~D36, p.~1691, 1987.

\bibitem{Ford:1987de}
L.~Ford, ``{Cosmological constant damping by unstable scalar fields},'' {\em
  Phys. Rev. D}, vol.~35, p.~2339, 1987.

\bibitem{Overduin:1998zv}
J.~Overduin and F.~Cooperstock, ``{Evolution of the scale factor with a
  variable cosmological term},'' {\em Phys.\ Rev.\ D}, vol.~58, p.~043506,
  1998.
\newblock arXiv:astro-ph/9805260.

\bibitem{Ozer:1985ws}
M.~Ozer and M.~Taha, ``{A Solution to the Main Cosmological Problems},'' {\em
  Phys.\ Lett.\ B}, vol.~171, pp.~363--365, 1986.

\bibitem{Ozer:1985wr}
M.~Ozer and M.~Taha, ``{A Model of the Universe with Time Dependent
  Cosmological Constant Free of Cosmological Problems},'' {\em Nucl.\ Phys.\
  B}, vol.~287, pp.~776--796, 1987.

\bibitem{Bertolami:1986bg}
O.~Bertolami, ``{Time dependent cosmological term},'' {\em Nuovo Cim.\ B},
  vol.~93, pp.~36--42, 1986.

\bibitem{Freese:1986dd}
K.~Freese, F.~C. Adams, J.~A. Frieman, and E.~Mottola, ``{Cosmology with
  Decaying Vacuum Energy},'' {\em Nucl.\ Phys.\ B}, vol.~287, pp.~797--814,
  1987.

\bibitem{Carvalho:1991ut}
J.~Carvalho, J.~Lima, and I.~Waga, ``{On the cosmological consequences of a
  time dependent lambda term},'' {\em Phys.\ Rev.\ D}, vol.~46, pp.~2404--2407,
  1992.

\bibitem{1997cpp..conf..123S}
P.~J. {Steinhardt}, {\em {Critical Problems in Physics. Cosmological Challenges
  for the 21st Century}}.
\newblock Princeton University Press, 1997.

\bibitem{Steinhardt:2003st}
P.~J. Steinhardt, ``{A quintessential introduction to dark energy},'' {\em
  Phil.\ Trans.\ Roy.\ Soc.\ Lond.\ A}, vol.~361, pp.~2497--2513, 2003.

\bibitem{Verde:2019ivm}
L.~Verde, T.~Treu, and A.~G. Riess, ``{Tensions between the early and late
  Universe},'' {\em Nature Astronomy}, vol.~3, pp.~891--895, 2019.
\newblock arXiv:1907.10625.

\bibitem{Macaulay:2013swa}
E.~Macaulay, I.~K. Wehus, and H.~K. Eriksen, ``{Lower Growth Rate from Recent
  Redshift Space Distortion Measurements than Expected from Planck},'' {\em
  Phys. Rev. Lett.}, vol.~111, no.~16, p.~161301, 2013.
\newblock arXiv:1303.6583.

\bibitem{Nesseris:2017vor}
S.~Nesseris, G.~Pantazis, and L.~Perivolaropoulos, ``{Tension and constraints
  on modified gravity parametrizations of $G_{\textrm{eff}}(z)$ from growth
  rate and Planck data},'' {\em Phys. Rev.}, vol.~D96, no.~2, p.~023542, 2017.
\newblock arXiv:1703.10538.

\bibitem{DiValentino:2020zio}
E.~Di~Valentino {\em et~al.}, ``{Cosmology Intertwined II: The Hubble Constant
  Tension},'' 2020.
\newblock arXiv:2008.11284.

\bibitem{DiValentino:2020vvd}
E.~Di~Valentino {\em et~al.}, ``{Cosmology Intertwined III: $f \sigma_8$ and
  $S_8$},'' 2020.
\newblock arXiv:2008.11285.

\bibitem{Riess:2016jrr}
A.~G. Riess {\em et~al.}, ``{A 2.4\% Determination of the Local Value of the
  Hubble Constant},'' {\em Astrophys. J.}, vol.~826, no.~1, p.~56, 2016.
\newblock arXiv:1604.01424.

\bibitem{Riess:2018uxu}
A.~G. Riess {\em et~al.}, ``{New Parallaxes of Galactic Cepheids from Spatially
  Scanning the Hubble Space Telescope: Implications for the Hubble Constant},''
  {\em Astrophys. J.}, vol.~855, no.~2, p.~136, 2018.
\newblock arXiv:1801.01120.

\bibitem{Riess:2019cxk}
A.~G. Riess, S.~Casertano, W.~Yuan, L.~M. Macri, and D.~Scolnic, ``{Large
  Magellanic Cloud Cepheid Standards Provide a 1\% Foundation for the
  Determination of the Hubble Constant and Stronger Evidence for Physics beyond
  $\Lambda$CDM},'' {\em Astrophys. J.}, vol.~876, no.~1, p.~85, 2019.
\newblock arXiv:1903.07603.

\bibitem{Reid:2019tiq}
M.~J. Reid, D.~W. Pesce, and A.~G. Riess, ``{An Improved Distance to NGC 4258
  and its Implications for the Hubble Constant},'' {\em Astrophys. J.},
  vol.~886, no.~2, p.~L27, 2019.
\newblock arXiv:1908.05625.

\bibitem{Wong:2019kwg}
K.~C. Wong {\em et~al.}, ``{H0LiCOW XIII. A 2.4\% measurement of $H_{0}$ from
  lensed quasars: $5.3\sigma$ tension between early and late-Universe
  probes},'' 2019.
\newblock arXiv:1907.04869.

\bibitem{Kazantzidis:2018rnb}
L.~Kazantzidis and L.~Perivolaropoulos, ``{Evolution of the $f\sigma_8$ tension
  with the Planck15/$\Lambda$CDM determination and implications for modified
  gravity theories},'' {\em Phys. Rev.}, vol.~D97, no.~10, p.~103503, 2018.
\newblock arXiv:1803.01337.

\bibitem{DiValentino:2018gcu}
E.~Di~Valentino and S.~Bridle, ``{Exploring the Tension between Current Cosmic
  Microwave Background and Cosmic Shear Data},'' {\em Symmetry}, vol.~10,
  no.~11, p.~585, 2018.

\bibitem{SolaPeracaula:2018xsi}
J.~Sol{\`{a}}~Peracaula, ``{Tensions in the $\Lambda$CDM and vacuum
  dynamics},'' {\em Int. J. Mod. Phys.}, vol.~A33, no.~31, p.~1844009, 2018.

\bibitem{Skara:2019usd}
F.~Skara and L.~Perivolaropoulos, ``{Tension of the $E_G$ statistic and
  redshift space distortion data with the Planck - $\Lambda CDM$ model and
  implications for weakening gravity},'' {\em Phys. Rev.}, vol.~D101, no.~6,
  p.~063521, 2020.
\newblock arXiv:1911.10609.

\bibitem{Gil-Marin:2016wya}
H.~Gil-Mar{\'{i}}n, W.~J. Percival, L.~Verde, J.~R. Brownstein, C.-H. Chuang,
  F.-S. Kitaura, S.~A. Rodr{\'{i}}guez-Torres, and M.~D. Olmstead, ``{The
  clustering of galaxies in the SDSS-III Baryon Oscillation Spectroscopic
  Survey: RSD measurement from the power spectrum and bispectrum of the DR12
  BOSS galaxies},'' {\em Mon. Not. Roy. Astron. Soc.}, vol.~465, no.~2,
  pp.~1757--1788, 2017.
\newblock arXiv:1606.00439.

\bibitem{Hildebrandt:2016iqg}
H.~Hildebrandt {\em et~al.}, ``{KiDS-450: Cosmological parameter constraints
  from tomographic weak gravitational lensing},'' {\em Mon. Not. Roy. Astron.
  Soc.}, vol.~465, p.~1454, 2017.
\newblock arXiv:1606.05338.

\bibitem{Joudaki:2017zdt}
S.~Joudaki {\em et~al.}, ``{KiDS-450 + 2dFLenS: Cosmological parameter
  constraints from weak gravitational lensing tomography and overlapping
  redshift-space galaxy clustering},'' {\em Mon. Not. Roy. Astron. Soc.},
  vol.~474, no.~4, pp.~4894--4924, 2018.
\newblock arXiv:1707.06627.

\bibitem{Kohlinger:2018sxx}
F.~Köhlinger, B.~Joachimi, M.~Asgari, M.~Viola, S.~Joudaki, and T.~Tröster,
  ``{A Bayesian quantification of consistency in correlated data sets},'' {\em
  Mon. Not. Roy. Astron. Soc.}, vol.~484, no.~3, pp.~3126--3153, 2019.
\newblock arXiv:1809.01406.

\bibitem{Wright:2020ppw}
A.~H. Wright, H.~Hildebrandt, J.~L. v.~d. Busch, C.~Heymans, B.~Joachimi,
  A.~Kannawadi, and K.~Kuijken, ``{KiDS+VIKING-450: Improved cosmological
  parameter constraints from redshift calibration with self-organising maps},''
  {\em Astron. Astrophys.}, vol.~640, p.~L14, 2020.
\newblock arXiv:2005.04207.

\bibitem{mccrea_1935}
E.~A. Milne, ``Relativity, gravitation and world-structure,'' {\em The
  Mathematical Gazette}, vol.~19, no.~235, pp.~299--301, 1935.

\bibitem{dirac1937cosmological}
P.~A.~M. Dirac, ``The cosmological constants,'' {\em Nature}, vol.~139,
  no.~3512, p.~323, 1937.

\bibitem{dirac1938cosmological}
P.~A.~M. Dirac, ``{A New Basis for Cosmology},'' {\em Proceedings of the Royal
  Society of London Series A}, vol.~165, no.~921, pp.~199--208, 1938.

\bibitem{jordan1937naturwiss}
P.~{Jordan}, ``{Die physikalischen Weltkonstanten},'' {\em
  Naturwissenschaften}, vol.~25, no.~32, pp.~513--517, 1937.

\bibitem{jordan1939ZPhys}
P.~Jordan, ``{{\"U}ber die kosmologische Konstanz der
  Feinstrukturkonstanten},'' {\em Zeitschrift f\"{u}r Physik}, vol.~113,
  no.~9-10, pp.~660--662, 1939.

\bibitem{jordan1955schwerkraft}
P.~Jordan, {\em Schwerkraft und Weltall}, vol.~107.
\newblock Vieweg, 1955.

\bibitem{Fierz:1956zz}
M.~Fierz, ``{On the physical interpretation of P. Jordan's extended theory of
  gravitation},'' {\em Helv. Phys. Acta}, vol.~29, pp.~128--134, 1956.

\bibitem{BransDicke1961}
C.~Brans and R.~Dicke, ``Mach's principle and a relativistic theory of
  gravitation,'' {\em Phys. Rev}, vol.~124, p.~925, 1961.

\bibitem{brans1962mach}
C.~Brans, ``Mach's principle and a relativistic theory of gravitation.
  \uppercase{II},'' {\em Phys. Rev.}, vol.~125, no.~6, p.~2194, 1962.

\bibitem{dicke1962physical}
R.~Dicke, ``Mach's principle and invariance under transformation of units,''
  {\em Phys. Rev.}, vol.~125, no.~6, p.~2163, 1962.

\bibitem{Bergmann:1968ve}
P.~G. Bergmann, ``{Comments on the scalar-tensor theory},'' {\em Int. J. Theor.
  Phys.}, vol.~1, pp.~25--36, 1968.

\bibitem{Nordtvedt:1970uv}
K.~Nordtvedt, ``{PostNewtonian metric for a general class of scalar-tensor
  gravitational theories and observational consequences},'' {\em Astrophys.
  J.}, vol.~161, pp.~1059--1067, 1970.

\bibitem{Wagoner:1970vr}
R.~V. Wagoner, ``Scalar-tensor theory and gravitational waves,'' {\em Physical
  Review D}, vol.~1, no.~12, p.~3209, 1970.

\bibitem{Horndeski:1974wa}
G.~W. Horndeski, ``{Second-order scalar-tensor field equations in a
  four-dimensional space},'' {\em Int. J. Theor. Phys.}, vol.~10, pp.~363--384,
  1974.

\bibitem{Fujii:2003pa}
Y.~Fujii and K.-i. Maeda, {\em {The scalar-tensor theory of gravitation}}.
\newblock Cambridge Monographs on Mathematical Physics, Cambridge University
  Press, 2007.

\bibitem{Sotiriou:2008rp}
T.~P. Sotiriou and V.~Faraoni, ``{f(R) Theories Of Gravity},'' {\em Rev. Mod.
  Phys.}, vol.~82, pp.~451--497, 2010.
\newblock arXiv:0805.1726.

\bibitem{Capozziello:2011et}
S.~Capozziello and M.~De~Laurentis, ``{Extended Theories of Gravity},'' {\em
  Phys. Rept.}, vol.~509, pp.~167--321, 2011.
\newblock arXiv:1108.6266.

\bibitem{Clifton:2011jh}
T.~Clifton, P.~G. Ferreira, A.~Padilla, and C.~Skordis, ``{Modified Gravity and
  Cosmology},'' {\em Phys. Rept.}, vol.~513, pp.~1--189, 2012.
\newblock arXiv:1106.2476.

\bibitem{Will:2014kxa}
C.~M. Will, ``{The Confrontation between General Relativity and Experiment},''
  {\em Living Rev. Rel.}, vol.~17, p.~4, 2014.
\newblock arXiv:1403.7377.

\bibitem{Sola:2018sjf}
J.~Sol{\`{a}}~Peracaula, A.~G{\'{o}}mez-Valent, and J.~de~Cruz~P{\'{e}}rez,
  ``{Signs of Dynamical Dark Energy in Current Observations},'' {\em Phys. Dark
  Univ.}, vol.~25, p.~100311, 2019.
\newblock arXiv:1811.03505.

\bibitem{Gomez-Valent:2020mqn}
A.~Gómez-Valent, V.~Pettorino, and L.~Amendola, ``{Update on coupled dark
  energy and the $H_0$ tension},'' {\em Phys. Rev. D}, vol.~101, no.~12,
  p.~123513, 2020.
\newblock arXiv:2004.00610.

\bibitem{sola2015cosmology}
J.~Solà and A.~Gómez-Valent, ``{The $\bar{\Lambda}{\rm CDM}$ cosmology: From
  inflation to dark energy through running $\Lambda$},'' {\em Int. J. Mod.
  Phys. D}, vol.~24, p.~1541003, 2015.
\newblock arXiv:1501.03832.

\bibitem{Gomez-Valent:2017tkh}
A.~Gómez-Valent, {\em {Vacuum energy in Quantum Field Theory and Cosmology}}.
\newblock PhD thesis, ICC, Barcelona U., 2017.
\newblock arXiv:1710.01978.

\bibitem{lima2013expansion}
J.~Lima, S.~Basilakos, and J.~Sola, ``{Expansion History with Decaying Vacuum:
  A Complete Cosmological Scenario},'' {\em Mon. Not. Roy. Astron. Soc.},
  vol.~431, pp.~923--929, 2013.
\newblock arXiv:1209.2802.

\bibitem{perico2013complete}
E.~Perico, J.~Lima, S.~Basilakos, and J.~Sol\`a, ``{Complete cosmic history
  with a dynamical $\Lambda$= $\Lambda$ (H) term},'' {\em Physical Review D},
  vol.~88, no.~6, p.~063531, 2013.

\bibitem{peracaula2020particle}
J.~Solà~Peracaula and H.~Yu, ``{Particle and entropy production in the Running
  Vacuum Universe},'' {\em Gen. Rel. Grav.}, vol.~52, no.~2, p.~17, 2020.
\newblock arXiv:1910.01638.

\bibitem{Basilakos:2019acj}
S.~Basilakos, N.~E. Mavromatos, and J.~Solà~Peracaula, ``{Gravitational and
  Chiral Anomalies in the Running Vacuum Universe and Matter-Antimatter
  Asymmetry},'' {\em Phys. Rev.}, vol.~D101, no.~4, p.~045001, 2020.
\newblock arXiv:1907.04890.

\bibitem{Basilakos:2020qmu}
S.~Basilakos, N.~E. Mavromatos, and J.~Solà~Peracaula, ``{Quantum Anomalies in
  String-Inspired Running Vacuum Universe: Inflation and Axion Dark Matter},''
  {\em Phys. Lett.}, vol.~B803, p.~135342, 2020.
\newblock arXiv:2001.03465.

\bibitem{Moreno-Pulido:2020anb}
C.~Moreno-Pulido and J.~Sola, ``{Running vacuum in quantum field theory in
  curved spacetime: renormalizing $\rho_{vac}$ without $\sim m^4$ terms},''
  {\em Eur. Phys. J. C}, vol.~80, no.~8, p.~692, 2020.
\newblock arXiv:2005.03164.

\bibitem{Sola:2016ecz}
J.~Sol{\`{a}}~Peracaula, J.~de~Cruz~P{\'{e}}rez, and A.~G{\'{o}}mez-Valent,
  ``{Dynamical dark energy vs. $\Lambda$ = const in light of observations},''
  {\em EPL}, vol.~121, no.~3, p.~39001, 2018.
\newblock arXiv:1606.00450.

\bibitem{Sola:2017jbl}
J.~Sol{\`{a}}~Peracaula, J.~de~Cruz~P{\'{e}}rez, and A.~G{\'{o}}mez-Valent,
  ``{Possible signals of vacuum dynamics in the Universe},'' {\em Mon. Not.
  Roy. Astron. Soc.}, vol.~478, no.~4, pp.~4357--4373, 2018.
\newblock arXiv:1703.08218.

\bibitem{Gomez-Valent:2018nib}
A.~G{\'{o}}mez-Valent and J.~Sol{\`{a}}~Peracaula, ``{Density perturbations for
  running vacuum: a successful approach to structure formation and to the
  $\sigma_8$-tension},'' {\em Mon. Not. Roy. Astron. Soc.}, vol.~478, no.~1,
  pp.~126--145, 2018.
\newblock arXiv:1801.08501.

\bibitem{Gomez-Valent:2017idt}
A.~G\'omez-Valent and J.~Sol\`a, ``{Relaxing the $\sigma_8$-tension through
  running vacuum in the Universe},'' {\em EPL}, vol.~120, no.~3, p.~39001,
  2017.
\newblock arXiv:1711.00692.

\bibitem{Sola:2017znb}
J.~Sol{\`{a}}, A.~G{\'{o}}mez-Valent, and J.~de~Cruz~P{\'{e}}rez, ``{The $H_0$
  tension in light of vacuum dynamics in the Universe},'' {\em Phys. Lett.},
  vol.~B774, pp.~317--324, 2017.
\newblock arXiv:1705.06723.

\bibitem{sola2017first}
J.~Solà, A.~Gómez-Valent, and J.~de~Cruz~Pérez, ``{First evidence of running
  cosmic vacuum: challenging the concordance model},'' {\em Astrophys. J.},
  vol.~836, no.~1, p.~43, 2017.
\newblock arXiv:1602.02103.

\bibitem{Sola:2016hnq}
J.~Sol{\`{a}}, A.~G{\'{o}}mez-Valent, and J.~de~Cruz~P{\'{e}}rez, ``{Dynamical
  dark energy: scalar fields and running vacuum},'' {\em Mod. Phys. Lett.},
  vol.~A32, no.~9, p.~1750054, 2017.
\newblock arXiv:1610.08965.

\bibitem{Sola:2015wwa}
J.~Sol\`a, A.~G\'omez-Valent, and J.~de~Cruz~P\'erez, ``{Hints of dynamical
  vacuum energy in the expanding Universe},'' {\em Astrophys. J.}, vol.~811,
  p.~L14, 2015.
\newblock arXiv:1506.05793.

\bibitem{Geng:2017apd}
C.-Q. Geng, C.-C. Lee, and L.~Yin, ``{Constraints on running vacuum model with
  $H(z)$ and $f \sigma_8$},'' {\em JCAP}, vol.~1708, p.~032, 2017.
\newblock arXiv:1704.02136.

\bibitem{Rezaei:2019xwo}
M.~Rezaei, M.~Malekjani, and J.~Sol\`a, ``{Can dark energy be expressed as a
  power series of the Hubble parameter?},'' {\em Phys. Rev. D}, vol.~100,
  no.~2, p.~023539, 2019.
\newblock arXiv:1905.00100.

\bibitem{Geng:2020mga}
C.-Q. Geng, C.-C. Lee, and L.~Yin, ``{Constraints on a special running vacuum
  model},'' {\em Eur. Phys. J.}, vol.~C80, no.~1, p.~69, 2020.
\newblock arXiv:2001.05092.

\bibitem{Gomez-Valent:2014rxa}
A.~Gómez-Valent, J.~Solà, and S.~Basilakos, ``{Dynamical vacuum energy in the
  expanding Universe confronted with observations: a dedicated study},'' {\em
  JCAP}, vol.~01, p.~004, 2015.
\newblock arXiv:1409.7048.

\bibitem{Gomez-Valent:2014fda}
A.~G\'omez-Valent and J.~Sol\`a, ``{Vacuum models with a linear and a quadratic
  term in H: structure formation and number counts analysis},'' {\em Mon. Not.
  Roy. Astron. Soc.}, vol.~448, pp.~2810--2821, 2015.
\newblock arXiv:1412.3785.

\bibitem{Grande:2011xf}
J.~Grande, J.~Sol\`a, S.~Basilakos, and M.~Plionis, ``{Hubble expansion and
  structure formation in the running FLRW model of the cosmic evolution},''
  {\em JCAP}, vol.~08, p.~007, 2011.
\newblock arXiv:1103.4632.

\bibitem{Basilakos:2009wi}
S.~Basilakos, M.~Plionis, and J.~Solà, ``{Hubble expansion \& Structure
  Formation in Time Varying Vacuum Models},'' {\em Phys. Rev. D}, vol.~80,
  p.~083511, 2009.
\newblock arXiv:0907.4555.

\bibitem{Peracaula:2018dkg}
J.~Solà~Peracaula, ``{Brans–Dicke gravity: From Higgs physics to (dynamical)
  dark energy},'' {\em Int. J. Mod. Phys.}, vol.~D27, no.~14, p.~1847029, 2018.
\newblock arXiv:1805.09810.

\bibitem{Perez:2018qgw}
J.~de~Cruz~P{\'{e}}rez and J.~Sol{\`{a}}~Peracaula, ``{Brans–Dicke cosmology
  mimicking running vacuum},'' {\em Mod. Phys. Lett.}, vol.~A33, no.~38,
  p.~1850228, 2018.
\newblock arXiv:1809.03329.

\bibitem{Sola:2019jek}
J.~Sol{\`{a}}~Peracaula, A.~G{\'{o}}mez-Valent, J.~de~Cruz~P{\'{e}}rez, and
  C.~Moreno-Pulido, ``{Brans–Dicke Gravity with a Cosmological Constant
  Smoothes Out $\Lambda$CDM Tensions},'' {\em Astrophys. J.}, vol.~886, no.~1,
  p.~L6, 2019.
\newblock arXiv:1909.02554.

\bibitem{Zhao:2017cud}
G.-B. Zhao {\em et~al.}, ``{Dynamical dark energy in light of the latest
  observations},'' {\em Nat. Astron.}, vol.~1, no.~9, pp.~627--632, 2017.
\newblock arXiv:1701.08165.

\bibitem{DiValentino:2019dzu}
E.~Di~Valentino, A.~Melchiorri, and J.~Silk, ``{Cosmological constraints in
  extended parameter space from the Planck 2018 Legacy release},'' {\em JCAP},
  vol.~2001, no.~01, p.~013, 2020.
\newblock arXiv:1908.01391.

\bibitem{DiValentino:2017rcr}
E.~Di~Valentino, E.~V. Linder, and A.~Melchiorri, ``{Vacuum phase transition
  solves the $H_0$ tension},'' {\em Phys. Rev.}, vol.~D97, no.~4, p.~043528,
  2018.
\newblock arXiv:1710.02153.

\bibitem{DiValentino:2017iww}
E.~Di~Valentino, A.~Melchiorri, and O.~Mena, ``{Can interacting dark energy
  solve the $H_0$ tension?},'' {\em Phys. Rev.}, vol.~D96, no.~4, p.~043503,
  2017.
\newblock arXiv:1704.08342.

\bibitem{DiValentino:2017zyq}
E.~Di~Valentino, A.~Melchiorri, E.~V. Linder, and J.~Silk, ``{Constraining Dark
  Energy Dynamics in Extended Parameter Space},'' {\em Phys. Rev.}, vol.~D96,
  no.~2, p.~023523, 2017.
\newblock arXiv:1704.00762.

\bibitem{DiValentino:2016hlg}
E.~Di~Valentino, A.~Melchiorri, and J.~Silk, ``{Reconciling Planck with the
  local value of $H_0$ in extended parameter space},'' {\em Phys. Lett.},
  vol.~B761, pp.~242--246, 2016.
\newblock arXiv:1606.00634.

\bibitem{Martinelli:2019dau}
M.~Martinelli, N.~B. Hogg, S.~Peirone, M.~Bruni, and D.~Wands, ``{Constraints
  on the interacting vacuum–geodesic CDM scenario},'' {\em Mon. Not. Roy.
  Astron. Soc.}, vol.~488, no.~3, pp.~3423--3438, 2019.
\newblock arXiv:1902.10694.

\bibitem{Salvatelli:2014zta}
V.~Salvatelli, N.~Said, M.~Bruni, A.~Melchiorri, and D.~Wands, ``{Indications
  of a late-time interaction in the dark sector},'' {\em Phys. Rev. Lett.},
  vol.~113, no.~18, p.~181301, 2014.
\newblock arXiv:1406.7297.

\bibitem{Costa:2016tpb}
A.~A. Costa, X.-D. Xu, B.~Wang, and E.~Abdalla, ``{Constraints on interacting
  dark energy models from Planck 2015 and redshift-space distortion data},''
  {\em JCAP}, vol.~1701, no.~01, p.~028, 2017.
\newblock arXiv:1605.04138.

\bibitem{Anand:2017wsj}
S.~Anand, P.~Chaubal, A.~Mazumdar, and S.~Mohanty, ``{Cosmic viscosity as a
  remedy for tension between PLANCK and LSS data},'' {\em JCAP}, vol.~1711,
  p.~005, 2017.
\newblock arXiv:1708.07030.

\bibitem{An:2017crg}
R.~An, C.~Feng, and B.~Wang, ``{Relieving the Tension between Weak Lensing and
  Cosmic Microwave Background with Interacting Dark Matter and Dark Energy
  Models},'' {\em JCAP}, vol.~1802, p.~038, 2018.
\newblock arXiv:1711.06799.

\bibitem{Li:2015vla}
Y.-H. Li, J.-F. Zhang, and X.~Zhang, ``{Testing models of vacuum energy
  interacting with cold dark matter},'' {\em Phys. Rev.}, vol.~D93, no.~2,
  p.~023002, 2016.
\newblock arXiv:1506.06349.

\bibitem{Li:2014cee}
Y.-H. Li, J.-F. Zhang, and X.~Zhang, ``{Exploring the full parameter space for
  an interacting dark energy model with recent observations including
  redshift-space distortions: Application of the parametrized post-Friedmann
  approach},'' {\em Phys. Rev.}, vol.~D90, no.~12, p.~123007, 2014.
\newblock arXiv:1409.7205.

\bibitem{Li:2014eha}
Y.-H. Li, J.-F. Zhang, and X.~Zhang, ``{Parametrized Post-Friedmann Framework
  for Interacting Dark Energy},'' {\em Phys. Rev.}, vol.~D90, no.~6, p.~063005,
  2014.
\newblock arXiv:1404.5220.

\bibitem{Hazra:2018opk}
D.~K. Hazra, A.~Shafieloo, and T.~Souradeep, ``{Parameter discordance in Planck
  CMB and low-redshift measurements: projection in the primordial power
  spectrum},'' {\em JCAP}, vol.~1904, p.~036, 2019.
\newblock arXiv:1810.08101.

\bibitem{Yan:2019gbw}
S.-F. Yan, P.~Zhang, J.-W. Chen, X.-Z. Zhang, Y.-F. Cai, and E.~N. Saridakis,
  ``{Interpreting cosmological tensions from the effective field theory of
  torsional gravity},'' {\em Phys. Rev. D}, vol.~101, no.~12, p.~121301, 2020.
\newblock arXiv:1909.06388.

\bibitem{Liao:2020zko}
K.~Liao, A.~Shafieloo, R.~E. Keeley, and E.~V. Linder, ``{Determining $H_0$
  Model-Independently and Consistency Tests},'' {\em Astrophys. J.}, vol.~895,
  no.~2, p.~L29, 2020.
\newblock arXiv:2002.10605.

\bibitem{WangChen2020}
K.~Wang and L.~Chen, ``{Constraints on Newton's constant from cosmological
  observations},'' {\em Eur. Phys. J. C}, vol.~80, no.~6, p.~570, 2020.
\newblock arXiv:2004.13976.

\bibitem{Jedamzik:2020krr}
K.~Jedamzik and L.~Pogosian, ``{Relieving the Hubble tension with primordial
  magnetic fields},'' 2020.
\newblock arXiv:2004.09487.

\bibitem{Vagnozzi:2019ezj}
S.~Vagnozzi, ``{New physics in light of the $H_0$ tension: An alternative
  view},'' {\em Phys. Rev. D}, vol.~102, no.~2, p.~023518, 2020.
\newblock arXiv:1907.07569.

\bibitem{Calderon:2020hoc}
R.~Calderón, R.~Gannouji, B.~L'Huillier, and D.~Polarski, ``{A negative
  cosmological constant in the dark sector?},'' 2020.
\newblock arXiv:2008.10237.

\bibitem{DiValentino:2019jae}
E.~Di~Valentino, A.~Melchiorri, O.~Mena, and S.~Vagnozzi, ``{Nonminimal dark
  sector physics and cosmological tensions},'' {\em Phys. Rev.}, vol.~D101,
  no.~6, p.~063502, 2020.
\newblock arXiv:1910.09853.

\bibitem{Alestas:2020mvb}
G.~Alestas, L.~Kazantzidis, and L.~Perivolaropoulos, ``{$H_0$ tension, phantom
  dark energy, and cosmological parameter degeneracies},'' {\em Phys. Rev. D},
  vol.~101, no.~12, p.~123516, 2020.
\newblock arXiv:2004.08363.

\bibitem{Poulin:2018cxd}
V.~Poulin, T.~L. Smith, T.~Karwal, and M.~Kamionkowski, ``{Early Dark Energy
  Can Resolve The Hubble Tension},'' {\em Phys. Rev. Lett.}, vol.~122, no.~22,
  p.~221301, 2019.
\newblock arXiv:1811.04083.

\bibitem{Hill:2020osr}
J.~C. Hill, E.~McDonough, M.~W. Toomey, and S.~Alexander, ``{Early dark energy
  does not restore cosmological concordance},'' {\em Phys. Rev. D}, vol.~102,
  no.~4, p.~043507, 2020.
\newblock arXiv:2003.07355.

\bibitem{Chudaykin:2020acu}
A.~Chudaykin, D.~Gorbunov, and N.~Nedelko, ``{Combined analysis of Planck and
  SPTPol data favors the early dark energy models},'' {\em JCAP}, vol.~2008,
  p.~013, 2020.
\newblock arXiv:2004.13046.

\bibitem{Braglia:2020bym}
M.~Braglia, W.~T. Emond, F.~Finelli, A.~E. Gumrukcuoglu, and K.~Koyama,
  ``{Unified framework for Early Dark Energy from $\alpha$-attractors},'' 2020.
\newblock arXiv:2005.14053.

\bibitem{Umilta:2015cta}
C.~Umilt{\`{a}}, M.~Ballardini, F.~Finelli, and D.~Paoletti, ``{CMB and BAO
  constraints for an induced gravity dark energy model with a quartic
  potential},'' {\em JCAP}, vol.~1508, p.~017, 2015.
\newblock arXiv:1507.00718.

\bibitem{Ballardini:2016cvy}
M.~Ballardini, F.~Finelli, C.~Umilt{\`{a}}, and D.~Paoletti, ``{Cosmological
  constraints on induced gravity dark energy models},'' {\em JCAP}, vol.~1605,
  p.~067, 2016.
\newblock arXiv:1601.03387.

\bibitem{Rossi:2019lgt}
M.~Rossi, M.~Ballardini, M.~Braglia, F.~Finelli, D.~Paoletti, A.~A.
  Starobinsky, and C.~Umilt{\`{a}}, ``{Cosmological constraints on
  post-Newtonian parameters in effectively massless scalar-tensor theories of
  gravity},'' {\em Phys. Rev.}, vol.~D100, no.~10, p.~103524, 2019.
\newblock arXiv:1906.10218.

\bibitem{Ballesteros:2020sik}
G.~Ballesteros, A.~Notari, and F.~Rompineve, ``{The $H_0$ tension: $\Delta G_N$
  vs. $\Delta N_{\rm eff}$},'' 2020.
\newblock arXiv:2004.05049.

\bibitem{Braglia:2020iik}
M.~Braglia, M.~Ballardini, W.~T. Emond, F.~Finelli, A.~E. Gumrukcuoglu,
  K.~Koyama, and D.~Paoletti, ``{Larger value for $H_0$ by an evolving
  gravitational constant},'' {\em Phys. Rev. D}, vol.~102, no.~2, p.~023529,
  2020.
\newblock arXiv:2004.11161.

\bibitem{Ballardini:2020iws}
M.~Ballardini, M.~Braglia, F.~Finelli, D.~Paoletti, A.~A. Starobinsky, and
  C.~Umiltà, ``{Scalar-tensor theories of gravity, neutrino physics, and the
  $H_0$ tension},'' 2020.
\newblock arXiv:2004.14349.

%\bibitem{Ballardini1793412}
%M.~Ballardini, M.~Braglia, F.~Finelli, D.~Paoletti, A.~A. Starobinsky, and
%  C.~Umiltà, ``{Scalar-tensor theories of gravity, neutrino physics, and the
%  $H_0$ tension},'' 2020.
%\newblock arXiv:2004.14349.

\bibitem{Bertini:2019xws}
N.~R. Bertini, W.~S. Hipólito-Ricaldi, F.~de~Melo-Santos, and D.~C. Rodrigues,
  ``{Cosmological framework for renormalization group extended gravity at the
  action level},'' {\em Eur. Phys. J. C}, vol.~80, no.~5, p.~479, 2020.
\newblock arXiv:1908.03960.

\bibitem{Rodrigues:2015hba}
D.~C. Rodrigues, B.~Chauvineau, and O.~F. Piattella, ``{Scalar-Tensor gravity
  with system-dependent potential and its relation with Renormalization Group
  extended General Relativity},'' {\em JCAP}, vol.~1509, p.~009, 2015.
\newblock arXiv:1504.05119.

\bibitem{MTW:1974}
C.~W. Misner, K.~S. Thorn, and J.~A. Wheeler, {\em Gravitation}.
\newblock San Francisco: Freeman, 1974.

\bibitem{Sola:2016our}
J.~Sol\`a, E.~Karimkhani, and A.~Khodam-Mohammadi, ``{Higgs potential from
  extended Brans–Dicke theory and the time-evolution of the fundamental
  constants},'' {\em Class. Quant. Grav.}, vol.~34, no.~2, p.~025006, 2017.
\newblock arXiv:1609.00350.

\bibitem{Faraoni:1998yq}
V.~Faraoni, ``{The $\omega\to\infty$ limit of Brans Dicke theory},'' {\em Phys.
  Lett.}, vol.~A245, pp.~26--30, 1998.
\newblock arXiv:gr-qc/9805057.

\bibitem{Faraoni:1999yp}
V.~Faraoni, ``{Illusions of general relativity in Brans-Dicke gravity},'' {\em
  Phys. Rev.}, vol.~D59, p.~084021, 1999.
\newblock arXiv:gr-qc/9902083.

\bibitem{mathiazhagan1984inflationary}
C.~Mathiazhagan and V.~Johri, ``An inflationary universe in {Brans-Dicke}
  theory: a hopeful sign of theoretical estimation of the gravitational
  constant,'' {\em Classical and Quantum Gravity}, vol.~1, no.~2, p.~L29, 1984.

\bibitem{La1989}
D.~La and P.~J. Steinhardt, ``Extended inflationary cosmology,'' {\em Phys.
  Rev. Lett.}, vol.~62, pp.~376--378, Jan 1989.

\bibitem{Weinberg1989}
E.~J. Weinberg, ``Some problems with extended inflation,'' {\em Phys. Rev. D},
  vol.~40, pp.~3950--3959, Dec 1989.

\bibitem{barrow1990extended}
J.~D. Barrow and K.-i. Maeda, ``Extended inflationary universes,'' {\em Nuclear
  Physics B}, vol.~341, no.~1, pp.~294--308, 1990.

\bibitem{Nariai:1969vh}
H.~Nariai, ``{On the Brans solution in the scalar-tensor theory of
  gravitation},'' {\em Prog. Theor. Phys.}, vol.~42, pp.~742--744, 1969.

\bibitem{endo1977cosmological}
M.~Endo and T.~Fukui, ``{The cosmological term and a modified Brans-Dicke
  cosmology},'' {\em General Relativity and Gravitation}, vol.~8, no.~10,
  pp.~833--839, 1977.

\bibitem{Uehara1982}
K.~Uehara and C.~W. Kim, ``{Brans-Dicke cosmology with the cosmological
  constant},'' {\em Phys. Rev. D}, vol.~26, pp.~2575--2579, Nov 1982.

\bibitem{lorenz1984exact}
D.~Lorenz-Petzold, ``Exact brans-dicke cosmologies with a cosmological
  constant,'' {\em Astrophysics and space science}, vol.~100, no.~1-2,
  pp.~461--465, 1984.

\bibitem{romero1992brans}
C.~Romero and A.~Barros, ``Brans-dicke cosmology and the cosmological constant:
  The spectrum of vacuum solutions,'' {\em Astrophysics and space science},
  vol.~192, no.~2, pp.~263--274, 1992.

\bibitem{Tretyakova2012}
D.~A. Tretyakova, A.~A. Shatskiy, I.~D. Novikov, and S.~Alexeyev, ``Nonsingular
  brans-dicke-$\ensuremath{\Lambda}$ cosmology,'' {\em Phys. Rev. D}, vol.~85,
  p.~124059, Jun 2012.
\newblock arXiv:1112.3770.

\bibitem{esposito2001scalar}
G.~Esposito-Farese and D.~Polarski, ``Scalar-tensor gravity in an accelerating
  universe,'' {\em Physical Review D}, vol.~63, no.~6, p.~063504, 2001.
\newblock arXiv:gr-qc/0009034.

\bibitem{Alsing:2011er}
J.~Alsing, E.~Berti, C.~M. Will, and H.~Zaglauer, ``{Gravitational radiation
  from compact binary systems in the massive Brans-Dicke theory of gravity},''
  {\em Phys. Rev. D}, vol.~85, p.~064041, 2012.
\newblock arXiv:1112.4903.

\bibitem{Ozer:2017oik}
{\"Ozer, Hatice and Delice, Özgür}, ``{Linearized modified gravity theories
  with a cosmological term: advance of perihelion and deflection of light},''
  {\em Class. Quant. Grav.}, vol.~35, no.~6, p.~065002, 2018.
\newblock arXiv:1708.05900.

\bibitem{Faraoni:2004pi}
V.~Faraoni, {\em {Cosmology in scalar tensor gravity}}.
\newblock Springer, {2004}.

\bibitem{Boisseau:2000pr}
B.~Boisseau, G.~Esposito-Farese, D.~Polarski, and A.~A. Starobinsky,
  ``{Reconstruction of a scalar-tensor theory of gravity in an accelerating
  universe},'' {\em Phys. Rev. Lett.}, vol.~85, p.~2236, 2000.
\newblock arXiv:gr-qc/0001066.

\bibitem{Bertotti:2003rm}
B.~Bertotti, L.~Iess, and P.~Tortora, ``{A test of general relativity using
  radio links with the Cassini spacecraft},'' {\em Nature}, vol.~425,
  pp.~374--376, 2003.

\bibitem{Avilez:2013dxa}
A.~Avilez and C.~Skordis, ``{Cosmological constraints on Brans-Dicke theory},''
  {\em Phys. Rev. Lett.}, vol.~113, no.~1, p.~011101, 2014.
\newblock arXiv:1303.4330.

\bibitem{Yadav:2019fjx}
H.~Amirhashchi and A.~K. Yadav, ``{Constraining An Exact Brans-Dicke gravity
  theory with Recent Observations},'' {\em Phys. Dark Univ.}, vol.~30,
  p.~100711, 2020.
\newblock arXiv:1908.04735.

\bibitem{Fixsen:2009ug}
D.~J. Fixsen, ``{The Temperature of the Cosmic Microwave Background},'' {\em
  Astrophys. J.}, vol.~707, pp.~916--920, 2009.
\newblock arXiv:0911.1955.

\bibitem{GorbunovRubakovBook}
D.~S. Gorbunov and V.~A. Rubakov, {\em Introduction to the Theory of the Early
  Universe. Cosmological perturbations and inflationary theory}.
\newblock Singapore: World Scientific, 2011.

\bibitem{Blas:2011rf}
D.~Blas, J.~Lesgourgues, and T.~Tram, ``{The Cosmic Linear Anisotropy Solving
  System (CLASS) II: Approximation schemes},'' {\em JCAP}, vol.~1107, p.~034,
  2011.
\newblock arXiv:1104.2933.

\bibitem{Yu:2017iju}
H.~Yu, B.~Ratra, and F.-Y. Wang, ``{Hubble Parameter and Baryon Acoustic
  Oscillation Measurement Constraints on the Hubble Constant, the Deviation
  from the Spatially Flat $\Lambda$CDM Model, the Deceleration–Acceleration
  Transition Redshift, and Spatial Curvature},'' {\em Astrophys. J.}, vol.~856,
  no.~1, p.~3, 2018.
\newblock arXiv:1711.03437.

\bibitem{Gomez-Valent:2018hwc}
A.~Gómez-Valent and L.~Amendola, ``{$H_0$ from cosmic chronometers and Type Ia
  supernovae, with Gaussian Processes and the novel Weighted Polynomial
  Regression method},'' {\em JCAP}, vol.~04, p.~051, 2018.
\newblock arXiv:1802.01505.

\bibitem{Haridasu:2018gqm}
B.~S. Haridasu, V.~V. Luković, M.~Moresco, and N.~Vittorio, ``{An improved
  model-independent assessment of the late-time cosmic expansion},'' {\em
  JCAP}, vol.~1810, no.~10, p.~015, 2018.
\newblock arXiv:1805.03595.

\bibitem{Silk1968}
J.~{Silk}, ``{Cosmic Black-Body Radiation and Galaxy Formation},'' {\em
  Astrophys. J.}, vol.~151, p.~459, Feb. 1968.

\bibitem{Sachs:1967er}
R.~K. Sachs and A.~M. Wolfe, ``{Perturbations of a cosmological model and
  angular variations of the microwave background},'' {\em Astrophys. J.},
  vol.~147, pp.~73--90, 1967.
\newblock [Gen. Rel. Grav.39,1929(2007)].

\bibitem{Das:2013sca}
S.~Das and T.~Souradeep, ``{Suppressing CMB low multipoles with ISW effect},''
  {\em JCAP}, vol.~02, p.~002, 2014.
\newblock arXiv:1312.0025.

\bibitem{Knox:2019rjx}
L.~Knox and M.~Millea, ``{Hubble constant hunter’s guide},'' {\em Phys.
  Rev.}, vol.~D101, no.~4, p.~043533, 2020.
\newblock arXiv:1908.03663.

\bibitem{Tsujikawa:2008uc}
S.~Tsujikawa, K.~Uddin, S.~Mizuno, R.~Tavakol, and J.~Yokoyama, ``{Constraints
  on scalar-tensor models of dark energy from observational and local gravity
  tests},'' {\em Phys. Rev. D}, vol.~77, p.~103009, 2008.
\newblock arXiv:0803.1106.

\bibitem{Li:2020uaz}
B.~Li and K.~Koyama, {\em {Modified Gravity}}.
\newblock WSP, 2020.

\bibitem{Khoury:2003aq}
J.~Khoury and A.~Weltman, ``{Chameleon fields: Awaiting surprises for tests of
  gravity in space},'' {\em Phys. Rev. Lett.}, vol.~93, p.~171104, 2004.
\newblock arXiv:astro-ph/0309300.

\bibitem{Hinterbichler:2010es}
K.~Hinterbichler and J.~Khoury, ``{Symmetron Fields: Screening Long-Range
  Forces Through Local Symmetry Restoration},'' {\em Phys. Rev. Lett.},
  vol.~104, p.~231301, 2010.
\newblock arXiv:1001.4525.

\bibitem{Kimura:2011dc}
R.~Kimura, T.~Kobayashi, and K.~Yamamoto, ``{Vainshtein screening in a
  cosmological background in the most general second-order scalar-tensor
  theory},'' {\em Phys. Rev. D}, vol.~85, p.~024023, 2012.
\newblock arXiv:1111.6749.

\bibitem{Alam:2016hwk}
S.~Alam {\em et~al.}, ``{The clustering of galaxies in the completed SDSS-III
  Baryon Oscillation Spectroscopic Survey: cosmological analysis of the DR12
  galaxy sample},'' {\em Mon. Not. Roy. Astron. Soc.}, vol.~470, no.~3,
  pp.~2617--2652, 2017.
\newblock arXiv:1607.03155.

\bibitem{Aubourg:2014yra}
E.~Aubourg {\em et~al.}, ``{Cosmological implications of baryon acoustic
  oscillation measurements},'' {\em Phys. Rev.}, vol.~D92, no.~12, p.~123516,
  2015.
\newblock arXiv:1411.1074.

\bibitem{Bernal:2016gxb}
J.~L. Bernal, L.~Verde, and A.~G. Riess, ``{The trouble with $H_0$},'' {\em
  JCAP}, vol.~1610, p.~019, 2016.
\newblock arXiv:1607.05617.

\bibitem{Feeney:2018mkj}
S.~M. Feeney, H.~V. Peiris, A.~R. Williamson, S.~M. Nissanke, D.~J. Mortlock,
  J.~Alsing, and D.~Scolnic, ``{Prospects for resolving the Hubble constant
  tension with standard sirens},'' {\em Phys. Rev. Lett.}, vol.~122, no.~6,
  p.~061105, 2019.
\newblock arXiv:1802.03404.

\bibitem{Macaulay:2018fxi}
E.~Macaulay {\em et~al.}, ``{First Cosmological Results using Type Ia
  Supernovae from the Dark Energy Survey: Measurement of the Hubble
  Constant},'' {\em Mon. Not. Roy. Astron. Soc.}, vol.~486, no.~2,
  pp.~2184--2196, 2019.
\newblock arXiv:1811.02376.

\bibitem{Park:2019emi}
C.-G. Park and B.~Ratra, ``{Using SPT polarization, $Planck$ 2015, and non-CMB
  data to constrain tilted spatially-flat and untilted nonflat $\Lambda$CDM ,
  XCDM, and $\phi$CDM dark energy inflation cosmologies},'' {\em Phys. Rev. D},
  vol.~101, no.~8, p.~083508, 2020.

\bibitem{Khadka:2020vlh}
N.~Khadka and B.~Ratra, ``{Using quasar X-ray and UV flux measurements to
  constrain cosmological model parameters},'' {\em Mon. Not. Roy. Astron.
  Soc.}, vol.~497, no.~1, pp.~263--278, 2020.
\newblock arXiv:2004.09979.

\bibitem{Cao:2020jgu}
S.~Cao, J.~Ryan, and B.~Ratra, ``{Cosmological constraints from HII starburst
  galaxy apparent magnitude and other cosmological measurements},'' {\em Mon.
  Not. Roy. Astron. Soc.}, vol.~497, pp.~3191--3203, 2020.
\newblock arXiv:2005.12617.

\bibitem{Ma:1995ey}
C.-P. Ma and E.~Bertschinger, ``{Cosmological perturbation theory in the
  synchronous and conformal Newtonian gauges},'' {\em Astrophys. J.}, vol.~455,
  pp.~7--25, 1995.
\newblock arXiv:astro-ph/9506072.

\bibitem{liddle_lyth_2000}
A.~R. Liddle and D.~H. Lyth, {\em Cosmological Inflation and Large-Scale
  Structure}.
\newblock Cambridge University Press, 2000.

\bibitem{Lyth:2009zz}
D.~H. Lyth and A.~R. Liddle, {\em {The primordial density perturbation:
  Cosmology, inflation and the origin of structure}}.
\newblock Cambridge University Press, 2009.

\bibitem{Turner:1998ex}
M.~S. Turner and M.~J. White, ``{CDM models with a smooth component},'' {\em
  Phys. Rev.}, vol.~D56, no.~8, p.~R4439, 1997.
\newblock arXiv:astro-ph/9701138.

\bibitem{Scolnic:2017caz}
D.~M. Scolnic {\em et~al.}, ``{The Complete Light-curve Sample of
  Spectroscopically Confirmed SNe Ia from Pan-STARRS1 and Cosmological
  Constraints from the Combined Pantheon Sample},'' {\em Astrophys. J.},
  vol.~859, no.~2, p.~101, 2018.
\newblock arXiv:1710.00845.

\bibitem{Abbott:2018wog}
T.~M.~C. Abbott {\em et~al.}, ``{First Cosmology Results using Type Ia
  Supernovae from the Dark Energy Survey: Constraints on Cosmological
  Parameters},'' {\em Astrophys. J.}, vol.~872, no.~2, p.~L30, 2019.
\newblock arXiv:1811.02374.

\bibitem{Agathe:2019vsu}
V.~de~Sainte~Agathe {\em et~al.}, ``{Baryon acoustic oscillations at z = 2.34
  from the correlations of Ly$\alpha$ absorption in eBOSS DR14},'' {\em Astron.
  Astrophys.}, vol.~629, p.~A85, 2019.
\newblock arXiv:1904.03400.

\bibitem{Gil-Marin:2018cgo}
H.~Gil-Mar{\'{i}}n {\em et~al.}, ``{The clustering of the SDSS-IV extended
  Baryon Oscillation Spectroscopic Survey DR14 quasar sample: structure growth
  rate measurement from the anisotropic quasar power spectrum in the redshift
  range $0.8 < z < 2.2$},'' {\em Mon. Not. Roy. Astron. Soc.}, vol.~477, no.~2,
  pp.~1604--1638, 2018.
\newblock arXiv:1801.02689.

\bibitem{Carter:2018vce}
P.~Carter, F.~Beutler, W.~J. Percival, C.~Blake, J.~Koda, and A.~J. Ross,
  ``{Low Redshift Baryon Acoustic Oscillation Measurement from the
  Reconstructed 6-degree Field Galaxy Survey},'' {\em Mon. Not. Roy. Astron.
  Soc.}, vol.~481, no.~2, pp.~2371--2383, 2018.
\newblock arXiv:1803.01746.

\bibitem{Kazin:2014qga}
E.~A. Kazin {\em et~al.}, ``{The WiggleZ Dark Energy Survey: improved distance
  measurements to z = 1 with reconstruction of the baryonic acoustic
  feature},'' {\em Mon. Not. Roy. Astron. Soc.}, vol.~441, no.~4,
  pp.~3524--3542, 2014.
\newblock arXiv:1401.0358.

\bibitem{Abbott:2017wcz}
T.~M.~C. Abbott {\em et~al.}, ``{Dark Energy Survey Year 1 Results: Measurement
  of the Baryon Acoustic Oscillation scale in the distribution of galaxies to
  redshift 1},'' {\em Mon. Not. Roy. Astron. Soc.}, vol.~483, no.~4,
  pp.~4866--4883, 2019.
\newblock arXiv:1712.06209.

\bibitem{Jimenez:2003iv}
R.~Jim{\'{e}}nez, L.~Verde, T.~Treu, and D.~Stern, ``{Constraints on the
  equation of state of dark energy and the Hubble constant from stellar ages
  and the CMB},'' {\em Astrophys. J.}, vol.~593, pp.~622--629, 2003.
\newblock arXiv:astro-ph/0302560.

\bibitem{Simon:2004tf}
J.~Simon, L.~Verde, and R.~Jimenez, ``{Constraints on the redshift dependence
  of the dark energy potential},'' {\em Phys. Rev.}, vol.~D71, p.~123001, 2005.
\newblock arXiv:astro-ph/0412269.

\bibitem{Stern:2009ep}
D.~Stern, R.~Jim{\'{e}}nez, L.~Verde, M.~Kamionkowski, and S.~A. Stanford,
  ``{Cosmic Chronometers: Constraining the Equation of State of Dark Energy. I:
  H(z) Measurements},'' {\em JCAP}, vol.~1002, p.~008, 2010.
\newblock arXiv:0907.3149.

\bibitem{Moresco:2012jh}
M.~Moresco {\em et~al.}, ``{Improved constraints on the expansion rate of the
  Universe up to z~1.1 from the spectroscopic evolution of cosmic
  chronometers},'' {\em JCAP}, vol.~1208, p.~006, 2012.
\newblock arXiv:1201.3609.

\bibitem{Zhang:2012mp}
C.~Zhang, H.~Zhang, S.~Yuan, T.-J. Zhang, and Y.-C. Sun, ``{Four new
  observational $H(z)$ data from luminous red galaxies in the Sloan Digital Sky
  Survey data release seven},'' {\em Res. Astron. Astrophys.}, vol.~14, no.~10,
  pp.~1221--1233, 2014.
\newblock arXiv:1207.4541.

\bibitem{Moresco:2015cya}
M.~Moresco, ``{Raising the bar: new constraints on the Hubble parameter with
  cosmic chronometers at $z \sim 2$},'' {\em Mon. Not. Roy. Astron. Soc.},
  vol.~450, no.~1, pp.~L16--L20, 2015.
\newblock arXiv:1503.01116.

\bibitem{Moresco:2016mzx}
M.~Moresco, L.~Pozzetti, A.~Cimatti, R.~Jim{\'{e}}nez, C.~Maraston, L.~Verde,
  D.~Thomas, A.~Citro, R.~Tojeiro, and D.~Wilkinson, ``{A 6\% measurement of
  the Hubble parameter at $z\sim0.45$: direct evidence of the epoch of cosmic
  re-acceleration},'' {\em JCAP}, vol.~1605, no.~05, p.~014, 2016.
\newblock arXiv:1601.01701.

\bibitem{Ratsimbazafy:2017vga}
A.~L. Ratsimbazafy, S.~I. Loubser, S.~M. Crawford, C.~M. Cress, B.~A. Bassett,
  R.~C. Nichol, and P.~V{\"{a}}is{\"{a}}nen, ``{Age-dating Luminous Red
  Galaxies observed with the Southern African Large Telescope},'' {\em Mon.
  Not. Roy. Astron. Soc.}, vol.~467, no.~3, pp.~3239--3254, 2017.
\newblock arXiv:1702.00418.

\bibitem{Jimenez:2001gg}
R.~Jim{\'{e}}nez and A.~Loeb, ``{Constraining cosmological parameters based on
  relative galaxy ages},'' {\em Astrophys. J.}, vol.~573, pp.~37--42, 2002.
\newblock arXiv:astro-ph/0106145.

\bibitem{Lopez-Corredoira:2017zfl}
M.~L{\'{o}}pez-Corredoira, A.~Vazdekis, C.~M. Guti{\'{e}}rrez, and
  N.~Castro-Rodr{\'{i}}guez, ``{Stellar content of extremely red quiescent
  galaxies at $z>2$},'' {\em Astron. Astrophys.}, vol.~600, p.~A91, 2017.
\newblock arXiv:1702.00380.

\bibitem{Lopez-Corredoira:2018tmn}
M.~L{\'{o}}pez-Corredoira and A.~Vazdekis, ``{Impact of young stellar
  components on quiescent galaxies: deconstructing cosmic chronometers},'' {\em
  Astron. Astrophys.}, vol.~614, p.~A127, 2018.
\newblock arXiv:1802.09473.

\bibitem{Moresco:2018xdr}
M.~Moresco, R.~Jim{\'{e}}nez, L.~Verde, L.~Pozzetti, A.~Cimatti, and A.~Citro,
  ``{Setting the Stage for Cosmic Chronometers. I. Assessing the Impact of
  Young Stellar Populations on Hubble Parameter Measurements},'' {\em
  Astrophys. J.}, vol.~868, no.~2, p.~84, 2018.
\newblock arXiv:1804.05864.

\bibitem{Gomez-Valent:2018gvm}
A.~G{\'{o}}mez-Valent, ``{Quantifying the evidence for the current speed-up of
  the Universe with low and intermediate-redshift data. A more
  model-independent approach},'' {\em JCAP}, vol.~1905, no.~05, p.~026, 2019.
\newblock arXiv:1810.02278.

\bibitem{Qin:2019axr}
F.~Qin, C.~Howlett, and L.~Staveley-Smith, ``{The redshift-space momentum power
  spectrum – II. Measuring the growth rate from the combined 2MTF and 6dFGSv
  surveys},'' {\em Mon. Not. Roy. Astron. Soc.}, vol.~487, no.~4,
  pp.~5235--5247, 2019.
\newblock arXiv:1906.02874.

\bibitem{Shi:2017qpr}
F.~Shi {\em et~al.}, ``{Mapping the Real Space Distributions of Galaxies in
  SDSS DR7: II. Measuring the growth rate, clustering amplitude of matter and
  biases of galaxies at redshift $0.1$},'' {\em Astrophys. J.}, vol.~861,
  no.~2, p.~137, 2018.
\newblock arXiv:1712.04163.

\bibitem{Simpson:2015yfa}
F.~Simpson, C.~Blake, J.~A. Peacock, I.~Baldry, J.~Bland-Hawthorn, A.~Heavens,
  C.~Heymans, J.~Loveday, and P.~Norberg, ``{Galaxy and mass assembly: Redshift
  space distortions from the clipped galaxy field},'' {\em Phys. Rev.},
  vol.~D93, no.~2, p.~023525, 2016.
\newblock arXiv:1505.03865.

\bibitem{Blake:2013nif}
C.~Blake {\em et~al.}, ``{Galaxy And Mass Assembly (GAMA): improved cosmic
  growth measurements using multiple tracers of large-scale structure},'' {\em
  Mon. Not. Roy. Astron. Soc.}, vol.~436, p.~3089, 2013.
\newblock arXiv:1309.5556.

\bibitem{Blake:2011rj}
C.~Blake {\em et~al.}, ``{The WiggleZ Dark Energy Survey: the growth rate of
  cosmic structure since redshift z=0.9},'' {\em Mon. Not. Roy. Astron. Soc.},
  vol.~415, p.~2876, 2011.
\newblock arXiv:1104.2948.

\bibitem{Mohammad:2018mdy}
F.~G. Mohammad {\em et~al.}, ``{The VIMOS Public Extragalactic Redshift Survey
  (VIPERS): Unbiased clustering estimate with VIPERS slit assignment},'' {\em
  Astron. Astrophys.}, vol.~619, p.~A17, 2018.
\newblock arXiv:1807.05999.

\bibitem{Guzzo:2008ac}
L.~Guzzo {\em et~al.}, ``{A test of the nature of cosmic acceleration using
  galaxy redshift distortions},'' {\em Nature}, vol.~451, pp.~541--545, 2008.
\newblock arXiv:0802.1944.

\bibitem{Song:2008qt}
Y.-S. Song and W.~J. Percival, ``{Reconstructing the history of structure
  formation using Redshift Distortions},'' {\em JCAP}, vol.~0910, p.~004, 2009.
\newblock arXiv:0807.0810.

\bibitem{Okumura:2015lvp}
T.~Okumura {\em et~al.}, ``{The Subaru FMOS galaxy redshift survey (FastSound).
  IV. New constraint on gravity theory from redshift space distortions at
  $z\sim 1.4$},'' {\em Publ. Astron. Soc. Jap.}, vol.~68, no.~3, p.~38, 2016.
\newblock arXiv:1511.08083.

\bibitem{Camarena:2019rmj}
D.~Camarena and V.~Marra, ``{A new method to build the (inverse) distance
  ladder},'' {\em Mon. Not. Roy. Astron. Soc.}, vol.~495, no.~3,
  pp.~2630--2644, 2020.
\newblock arXiv:1910.14125.

\bibitem{Camarena:2019moy}
D.~Camarena and V.~Marra, ``{Local determination of the Hubble constant and the
  deceleration parameter},'' {\em Phys. Rev. Res.}, vol.~2, no.~1, p.~013028,
  2020.
\newblock arXiv:1906.11814.

\bibitem{Dhawan:2020xmp}
S.~Dhawan, D.~Brout, D.~Scolnic, A.~Goobar, A.~Riess, and V.~Miranda,
  ``{Cosmological model insensitivity of local $H_0$ from the Cepheid distance
  ladder},'' {\em Astrophys. J.}, vol.~894, no.~1, p.~54, 2020.
\newblock arXiv:2001.09260.

\bibitem{Benevento:2020fev}
G.~Benevento, W.~Hu, and M.~Raveri, ``{Can Late Dark Energy Transitions Raise
  the Hubble constant?},'' {\em Phys. Rev. D}, vol.~101, no.~10, p.~103517,
  2020.
\newblock arXiv:2002.11707.

\bibitem{Giani:2020fpz}
L.~Giani and E.~Frion, ``{Testing the Equivalence Principle with Strong Lensing
  Time Delay Variations},'' 2020.
\newblock arXiv:2005.07533.

\bibitem{Takahashi:2012em}
R.~Takahashi, M.~Sato, T.~Nishimichi, A.~Taruya, and M.~Oguri, ``{Revising the
  Halofit Model for the Nonlinear Matter Power Spectrum},'' {\em Astrophys.
  J.}, vol.~761, p.~152, 2012.
\newblock arXiv:1208.2701.

\bibitem{Chevallier:2000qy}
M.~Chevallier and D.~Polarski, ``{Accelerating universes with scaling dark
  matter},'' {\em Int. J. Mod. Phys.}, vol.~D10, pp.~213--224, 2001.
\newblock arXiv:gr-qc/0009008.

\bibitem{Linder:2002et}
E.~V. Linder, ``{Exploring the expansion history of the universe},'' {\em Phys.
  Rev. Lett.}, vol.~90, p.~091301, 2003.
\newblock arXiv:astro-ph/0208512.

\bibitem{Beutler:2011hx}
F.~Beutler, C.~Blake, M.~Colless, D.~H. Jones, L.~Staveley-Smith, L.~Campbell,
  Q.~Parker, W.~Saunders, and F.~Watson, ``{The 6dF Galaxy Survey: Baryon
  Acoustic Oscillations and the Local Hubble Constant},'' {\em Mon. Not. Roy.
  Astron. Soc.}, vol.~416, pp.~3017--3032, 2011.
\newblock arXiv:1106.3366.

\bibitem{Ross:2014qpa}
A.~J. Ross, L.~Samushia, C.~Howlett, W.~J. Percival, A.~Burden, and M.~Manera,
  ``{The clustering of the SDSS DR7 main Galaxy sample – I. A 4 per cent
  distance measure at $z = 0.15$},'' {\em Mon. Not. Roy. Astron. Soc.},
  vol.~449, no.~1, pp.~835--847, 2015.
\newblock arXiv:1409.3242.

\bibitem{TheLIGOScientific:2017qsa}
B.~Abbott {\em et~al.}, ``{GW170817: Observation of Gravitational Waves from a
  Binary Neutron Star Inspiral},'' {\em Phys. Rev. Lett.}, vol.~119, no.~16,
  p.~161101, 2017.
\newblock arXiv:1710.05832.

\bibitem{Creminelli:2017sry}
P.~Creminelli and F.~Vernizzi, ``{Dark Energy after GW170817 and GRB170817A},''
  {\em Phys. Rev. Lett.}, vol.~119, no.~25, p.~251302, 2017.
\newblock arXiv:1710.05877.

\bibitem{Ezquiaga:2017ekz}
J.~M. Ezquiaga and M.~Zumalacárregui, ``{Dark Energy After GW170817: Dead Ends
  and the Road Ahead},'' {\em Phys. Rev. Lett.}, vol.~119, no.~25, p.~251304,
  2017.
\newblock arXiv:1710.05901.

\bibitem{Uzan:2010pm}
J.-P. Uzan, ``{Varying Constants, Gravitation and Cosmology},'' {\em Living
  Rev. Rel.}, vol.~14, p.~2, 2011.
\newblock arXiv:1009.5514.

\bibitem{Audren:2012wb}
B.~Audren, J.~Lesgourgues, K.~Benabed, and S.~Prunet, ``{Conservative
  Constraints on Early Cosmology: an illustration of the Monte Python
  cosmological parameter inference code},'' {\em JCAP}, vol.~1302, p.~001,
  2013.
\newblock arXiv:1210.7183.

\bibitem{Lewis:2019xzd}
A.~Lewis, ``{GetDist: a Python package for analysing Monte Carlo samples},''
  2019.
\newblock arXiv:1910.13970.

\bibitem{Heavens:2017afc}
A.~Heavens, Y.~Fantaye, A.~Mootoovaloo, H.~Eggers, Z.~Hosenie, S.~Kroon, and
  E.~Sellentin, ``{Marginal Likelihoods from Monte Carlo Markov Chains},''
  2017.
\newblock arXiv:1704.03472.

\bibitem{DIC}
D.~J. Spiegelhalter, N.~G. Best, B.~P. Carlin, and A.~van~der Linde, ``Bayesian
  measures of model complexity and fit,'' {\em J. Roy. Stat. Soc.}, vol.~64,
  p.~583, 2002.

\bibitem{Schwarz1978}
G.~Schwarz, ``{Estimating the Dimension of a Model},'' {\em Ann. Stat.},
  vol.~6, pp.~461--464, 1978.

\bibitem{KassRaftery1995}
R.~E. Kass and A.~E. Raftery, ``{Bayes factors},'' {\em J. Amer. Statist.
  Assoc.}, vol.~90, no.~430, pp.~773--795, 1995.

\bibitem{Burnham2002}
K.~P. Burnham and D.~R. Anderson, {\em Model selection and multimodel
  inference}.
\newblock New York: Springer, 2002.

\bibitem{Nagata:2003qn}
R.~Nagata, T.~Chiba, and N.~Sugiyama, ``{WMAP constraints on scalar-tensor
  cosmology and the variation of the gravitational constant},'' {\em Phys.
  Rev.}, vol.~D69, p.~083512, 2004.
\newblock arXiv:astro-ph/0311274.

\bibitem{Acquaviva:2004ti}
V.~Acquaviva, C.~Baccigalupi, S.~M. Leach, A.~R. Liddle, and F.~Perrotta,
  ``{Structure formation constraints on the Jordan-Brans-Dicke theory},'' {\em
  Phys. Rev.}, vol.~D71, p.~104025, 2005.
\newblock arXiv:astro-ph/0412052.

\bibitem{Fritzsch:2012qc}
H.~Fritzsch and J.~Sol\`a, ``{Matter Non-conservation in the Universe and
  Dynamical Dark Energy},'' {\em Class. Quant. Grav.}, vol.~29, p.~215002,
  2012.
\newblock arXiv:1202.5097.

\bibitem{Muller:2007zzb}
J.~Muller and L.~Biskupek, ``{Variations of the gravitational constant from
  lunar laser ranging data},'' {\em Class. Quant. Grav.}, vol.~24,
  pp.~4533--4538, 2007.

\bibitem{Pogosian:2004wa}
L.~Pogosian, ``{Evolving dark energy equation of state and CMB/LSS
  cross-correlation},'' {\em JCAP}, vol.~04, p.~015, 2005.
\newblock arXiv:astro-ph/0409059.

\bibitem{Zucca:2019ohv}
A.~Zucca, L.~Pogosian, A.~Silvestri, Y.~Wang, and G.-B. Zhao, ``{Generalized
  Brans-Dicke theories in light of evolving dark energy},'' {\em Phys. Rev.},
  vol.~D101, no.~4, p.~043518, 2020.
\newblock arXiv:1907.07667.

\bibitem{Loureiro:2018pdz}
A.~Loureiro {\em et~al.}, ``{On The Upper Bound of Neutrino Masses from
  Combined Cosmological Observations and Particle Physics Experiments},'' {\em
  Phys. Rev. Lett.}, vol.~123, no.~8, p.~081301, 2019.
\newblock arXiv:1811.02578.

\bibitem{Shapiro:2000dz}
I.~L. Shapiro and J.~Solà, ``{Scaling behavior of the cosmological constant:
  Interface between quantum field theory and cosmology},'' {\em JHEP}, vol.~02,
  p.~006, 2002.
\newblock arXiv:hep-th/0012227.

\bibitem{Sola:2007sv}
J.~Solà, ``{Dark energy: A Quantum fossil from the inflationary Universe?},''
  {\em J. Phys.}, vol.~A41, p.~164066, 2008.
\newblock arXiv:0710.4151.

\bibitem{Shapiro:2009dh}
I.~L. Shapiro and J.~Solà, ``{On the possible running of the cosmological
  'constant'},'' {\em Phys. Lett.}, vol.~B682, pp.~105--113, 2009.
\newblock arXiv:0910.4925.

\bibitem{Babic:2004ev}
A.~Babic, B.~Guberina, R.~Horvat, and H.~Stefancic, ``{Renormalization-group
  running cosmologies. A Scale-setting procedure},'' {\em Phys. Rev.},
  vol.~D71, p.~124041, 2005.
\newblock arXiv:astro-ph/0407572.

\bibitem{Ward:2010qs}
B.~F.~L. Ward, ``{An estimate of $\Lambda$ in resummed quantum gravity in the
  context of asymptotic safety},'' {\em Phys. Dark Univ.}, vol.~2, pp.~97--109,
  2013.
\newblock arXiv:1008.1046.

\bibitem{Antipin:2017pbt}
O.~Antipin and B.~Melic, ``{Revisiting the decoupling effects in the running of
  the Cosmological Constant},'' {\em Eur. Phys. J.}, vol.~C77, no.~9, p.~583,
  2017.
\newblock arXiv:1703.10967.

\bibitem{Bahamonde:2017ize}
S.~Bahamonde, C.~G. Böhmer, S.~Carloni, E.~J. Copeland, W.~Fang, and
  N.~Tamanini, ``{Dynamical systems applied to cosmology: dark energy and
  modified gravity},'' {\em Phys. Rept.}, vol.~775-777, pp.~1--122, 2018.
\newblock arXiv:1712.03107.

\bibitem{Chen:1999qh}
X.-l. Chen and M.~Kamionkowski, ``{Cosmic microwave background temperature and
  polarization anisotropy in Brans-Dicke cosmology},'' {\em Phys. Rev.},
  vol.~D60, p.~104036, 1999.
\newblock arXiv:astro-ph/9905368.

\bibitem{Pinho:2018unz}
A.~M. Pinho, S.~Casas, and L.~Amendola, ``{Model-independent reconstruction of
  the linear anisotropic stress $\eta$},'' {\em JCAP}, vol.~11, p.~027, 2018.
\newblock arXiv:1805.00027.

\end{thebibliography}

\end{document}